\documentstyle[aps,epsf,rotate]{revtex}
\font\mb=msbm10

\newcommand{\pard}[2]{\frac{\partial #1}{\partial #2}}

\newcommand{\be}{\begin{equation}}
\newcommand{\ee}{\end{equation}}
\newcommand{\ba}{\begin{eqnarray*}}
\newcommand{\ea}{\end{eqnarray*}}
\newcommand{\bna}{\begin{eqnarray}}
\newcommand{\ena}{\end{eqnarray}}

\newcommand{\mpaa}{\begin{minipage}[t]{7.5cm}}
\newcommand{\mpea}{\end{minipage}}
\newcommand{\mpab}{\begin{minipage}[t]{8.5cm}}
\newcommand{\mpeb}{\end{minipage}}
\newcommand{\mpac}{\begin{minipage}[t]{6.6cm}}
\newcommand{\mpec}{\end{minipage}}
\newcommand{\mpad}{\begin{minipage}[t]{10cm}}
\newcommand{\mped}{\end{minipage}}
\begin{document}

\draft
\title{Microscopic chaos and transport\\
in thermostated dynamical systems}
\author{R.Klages}
\address{Max Planck Institute for the Physics of Complex Systems,\\
N\"othnitzer Str. 38,\\ 
D-01187 Dresden, Germany\\
E-mail: rklages@mpipks-dresden.mpg.de}
\date{\today}
\maketitle
\begin{abstract}
A fundamental challenge is to understand nonequilibrium statistical mechanics
starting from microscopic chaos in the equations of motion of a many-particle
system. In this review we summarize recent theoretical advances along these
lines. Particularly, we are concerned with nonequilibrium situations created
by external electric fields and by temperature or velocity gradients. These
constraints pump energy into a system, hence there must be some thermal
reservoir that prevents the system from heating up. About twenty years ago a
{\em deterministic and time-reversible modeling} of such thermal reservoirs
was proposed in form of Gaussian and Nos\'e-Hoover thermostats. This approach
yielded simple relations between fundamental quantities of nonequilibrium
statistical mechanics and of dynamical systems theory. The main theme of our
review is to critically assess the universality of these results. As a vehicle
of demonstration we employ the driven periodic Lorentz gas, which is a toy
model for the classical dynamics of an electron in a metal under application
of an electric field. Applying different types of thermal reservoirs to this
system we compare the resulting nonequilibrium steady states with each
other. Along the same lines we discuss an interacting many-particle system
under shear and heat. Finally, we outline an unexpected relationship between
deterministic thermostats and active Brownian particles modeling biophysical
cell motility.
\end{abstract}

\newpage

\tableofcontents

\newpage

\section{Introduction and outline}

Statistical mechanics endeavours to understand the origin of the macroscopic
properties of matter starting from the microscopic equations of motion of
single atoms or molecules. This program traces back to the founders of
statistical mechanics, Boltzmann, Maxwell and Gibbs, and for many-particle
systems in thermal equilibrium it was pursued with remarkable success, as is
demonstrated in many textbooks, see, e.g., Refs.\ \cite{Reif,Huang}. However,
in nonequilibrium situations, that is, for systems under constraints such as
external fields or by imposing temperature or velocity gradients, statistical
mechanical theories appear to be rather incomplete. We just mention that in
contrast to the equilibrium case there is no generally accepted definition of
a nonequilibrium entropy, and correspondingly yet there is no general
agreement on nonequilibrium ensembles that might replace the equilibrium ones
\cite{Penr79,EvMo90,stat99,Rue99,Gall99}.

Fresh input concerning these fundamental problems came from the side of
dynamical systems theory, in particular by work of mathematicians like Sinai,
Ruelle, Bowen and others over the past few decades. Indeed, {\em SRB
measures}\footnote{The acronym holds for the initials of Sinai, Ruelle, and
Bowen.} appear to provide good candidates for taking over the role of the
Gibbs ensemble in nonequilibrium
\cite{Gasp,Do99,Gall99,Rue99,Young02,Gall03b}.  Additionally, the advent of
powerful computers made it possible to solve numerically the nonlinear
equations of motion of many-particle systems and to investigate the interplay
between microscopic chaos in the collisions of the single particles and
transport properties on macroscopic scales in much more detail than it was
possible to the times of the founders of statistical mechanics.

The stage for this review is set by two basic approaches that evolved over the
past two decades trying to develop a concise picture of nonequilibrium
statistical mechanics by employing methods of dynamical systems theory. In the
following two sections we briefly summarize important features of these two
directions of research. We then sketch how the present review is embedded into
the existing literature, outline its contents and also say some words about
the style in which it is written.

\subsection{The Hamiltonian dynamical systems approach to nonequilibrium
steady states}

In recent work Gaspard, Nicolis and Dorfman studied nonequilibrium situations
by imposing specific boundary conditions onto spatially extended chaotic {\em
Hamiltonian} dynamical systems. A typical example are diffusion processes due
to concentration gradients at the boundaries. By this approach the macroscopic
transport properties of deterministic dynamical systems could be linked to the
underlying microscopic chaos in the equations of motion of the single
particles in two ways: The {\em escape rate approach} considers dynamical
systems with absorbing boundaries
\cite{GN,GB1,Gas93,GaBa95,DoGa95,GaDo95,Gasp,Do99}. Here the escape rate
determined by a statistical physical transport equation such as, for example,
the diffusion equation, is matched to the one resulting from solving the
Liouville equation of the dynamical system. This procedure yields simple
formulas linking transport coefficients to dynamical systems quantities, which
are here the positive Lyapunov exponents and the Kolmogorov-Sinai entropy of
the dynamical system, or respectively the fractal dimension of the associated
repeller of the open system.

A second, conceptually related approach applies to closed systems
with periodic boundary conditions and has been worked out for diffusion
\cite{GDG01,GCGD01} and for reaction-diffusion \cite{ClGa02} in
low-dimensional model systems. In this case the decay rate to thermal equilibrium as
obtained from the diffusion equation is related to the fractal dimension of
the corresponding hydrodynamic mode in the Liouville equation of the dynamical
system. The diffusion coefficient can thus be expressed as a function of the
system's largest Lyapunov exponent combined with the Hausdorff dimension of
this mode. Both approaches can consistently be derived by using Ruelle's
thermodynamic formalism \cite{GaDo95,GCGD01}.

A further interesting result obtained in this framework deals with the
parameter dependence of transport coefficients. By employing that statistical
physical transport coefficients can be computed from the eigenvalues of the
Liouville equation of the dynamical system it was found that, for a certain
class of chaotic dynamical systems, transport coefficients are fractal
functions of respective control parameters such as particle density,
temperature, or field strength
\cite{RKD,RKdiss,KlDo99,GaKl,HaGa01,GrKl02,KoKl02}. This phenomenon reflects
on a macroscopic scale the topological instability of the chaotic equations of
motion under parameter variation. Deterministic diffusion coefficients can
alternatively be computed from periodic orbit theory yielding elegant formulas
linking diffusion to the stability of the cycles of dynamical systems
trajectories \cite{CEG91,CvGS92,Gasp,CAMTV01}.

Much work has been devoted to understand the nonequilibrium entropy production
in simple chaotic model systems such as two-dimensional {\em multibaker maps}
and {\em periodic Lorentz gases}. T\'el, Vollmer, Breymann and Matyas
\cite{BrTV96,TeVB96,VTB97,BTV98,VTB98,TVS00,VTM00,MTV01,Voll02,VTB03} as well
as Gaspard, Tasaki, Dorfman and Gilbert
\cite{Gasp97a,GiDo99,TG99,TG00,GD00,GDG00,DGG02} have proposed new concepts
for defining coarse-grained Gibbs entropies leading to an entropy production
of nonequilbrium processes which is in agreement with irreversible
thermodynamics. The former group attributed the source of irreversible entropy
production to the chaotic mixing of the dynamical system and to the associated
loss of information due to the coarse graining. The latter authors argued that
the singularity of the SRB measures exhibited by these nonequilibrium systems
enforces a respective coarse graining for mathematical reasons. Hence, the
source of irreversible entropy production is identified with the fractal
character of these SRB measures. Cohen and Rondoni, on the other hand,
criticized both approaches due to the fact that the simple models analyzed so
far consist of moving point particles that do not interact with each other but
only with some fixed scatterers \cite{CoRo98,RC00,CR02,RC02}. In their view
these systems are non-thermodynamic models that do not allow to identify local
thermodynamic equilibrium or any proper source of thermodynamic entropy
production; however, see the replies of the criticized authors in Refs.\
\cite{Voll02,DGG02,MTV02,Gasp02,GND03}.

To conclude this brief outline of the Hamiltonian dynamical systems approach
to nonequilibrium steady states we refer to an interesting experiment that was
proposed and carried out by Gaspard et al.\ \cite{GBFS+98,BSF+01}. Its purpose
was to verify the existence of microscopic deterministic chaos in the dynamics
of a real interacting many-particle system. Long trajectories of a tracer
particle suspended in a fluid were recorded, and this data was used for
computing a coarse grained entropy that was argued to yield a lower bound for
the sum of positive Lyapunov exponents of the system. However, after this work
was published non-chaotic counterexamples were constructed yielding almost
indistinguishable results for the coarse-grained entropy
\cite{DC99,GS99,DC00}. This initiated ongoing discussions and motivated 
research on transport in non-chaotic models, which complements existing
results for chaotic systems \cite{LRB00,CFOKV00,CFVN02,ARV02,GNY02}.

\subsection{The thermostated dynamical systems approach to nonequilibrium
steady states}

A nontrivial limitation of the Hamiltonian approach to chaotic transport is
that it excludes nonequilibrium constraints generating a continuous flux of
energy into the system as, for example, the application of external
fields.\footnote{Note that the use of {\em Helfand moments} enables an
indirect treatment of such situations similar to the use of equilibrium time
correlation functions related to Green-Kubo formulas \cite{DoGa95,Gasp}.} Such
situations necessitate the modeling of an infinite dimensional {\em thermal
reservoir} that is able to continuously absorb energy in order to prevent a
subsystem from heating up \cite{Penr79,Rue99,Gall99,Do99,Rond02}. The need to
model these situations emerged particularly from {\em nonequilibrium molecular
dynamics computer simulations} that focus on simulating heat or shear flow of
many-particle systems or currents under application of external fields
\cite{EvMo90,Hoo91,Hess96,MoDe98,HoB99,DettS00,Mund00,TuMa00}.

A well-known example for modeling thermal reservoirs is provided by the {\em
Langevin equation} \cite{Lang08} yielding the interaction with a heat bath by
a combination of Stokes friction and stochastic forces
\cite{Wax54,Reif,Path88,KTH92,Zwan01}. Indeed, one way to derive generic types of
Langevin equations starts from a Hamiltonian formulation for a heat bath
consisting of infinitely many harmonic oscillators. This heat bath suitably
interacts with a subsystem that consists of a single particle
\cite{Zwan73,FK87,KTH92,SW99,Zwan01}. In the course of the derivation the detailed
bath dynamics is eliminated resulting in an equation of motion for the
subsystem that is {\em stochastic and non-Hamiltonian}. The Langevin equation
thus nicely illustrates Ruelle's statement ``if we want to study
non-equilibrium processes we have thus to consider an infinite system or
non-Hamiltonian forces'' \cite{Rue99b}.\footnote{For a related statement see,
e.g., Smale \cite{Sma80}: ``We would conclude that theoretical physics and
statistical mechanics should not be tied to Hamiltonian equations so
absolutely as in the past. On physical grounds, it is certainly reasonable to
expect physical systems to have (perhaps very small) non-Hamiltonian
perturbations due to friction and driving effects from outside energy
absorbtion. Today also mathematical grounds suggest that it is reasonable to
develop a more non-Hamiltonian approach to some aspects of physics.'' In his
interesting article Smale further suggests to ``revive the ergodic hypothesis
via introduction of a dissipative/forcing term'' into Hamiltonian equations of
motion, since in his view dissipative dynamical systems have a better chance
to be ergodic than Hamiltonian ones that usually exhibit profoundly
non-ergodic dynamics due to a mixed phase space.}

As we will argue in this review on various occasions, there is nothing
mysterious in modeling thermal reservoirs with non-Hamiltonian equations of
motion, see also, e.g., Refs.\
\cite{Penr79,Sma80,Rue99,ChLe95,Rue96,ChLe97} \cite{Gall99,Lieb99,Rond02}.
In case of thermostated systems the non-Hamiltonianity results
straightforwardly from projecting out spurious reservoir degrees of freedom,
as we will outline in this review on several occasions. Early nonequilibrium
molecular dynamics computer simulations actually employed stochastic models of
heat baths \cite{And79,CiTe80,SchSt78,TCG82,AT87,Nose91}, however, very soon
people started to look for alternatives. Infinite-dimensional Hamiltonian
thermal reservoirs can very well be modeled and analyzed analytically
\cite{EPRB99,EPRB99b,Zwan01}, but on a computer the number of degrees of
freedom must, for obvious reasons, remain finite. These constraints provided a
very practical motivation for constructing nonequilibrium steady states on the
basis of finite-dimensional, deterministic, non-Hamiltonian equations of
motion.

About twenty years ago Hoover et al.\ \cite{HLM82} and Evans \cite{Ev83}
independently and simultaneously came up with a strikingly simple
non-Hamiltonian modeling of a thermal reservoir, which they coined the {\em
Gaussian thermostat} \cite{EH83}. This scheme introduces a (Gaussian)
constraint in order to keep the temperature for a given subsystem strictly
constant in nonequilibrium at any time step. A few years later Nos\'e invented
a very related non-Hamiltonian thermal reservoir that was able to thermostat
the velocity distribution of a given sybsystem onto the canonical one in
equilibrium, and to keep the energy of a subsystem constant on average in
nonequilibrium \cite{Nose84a,Nose84b}. His formulation was simplified by
Hoover \cite{Hoov85} leading to the famous Nos\'e-Hoover thermostat
\cite{EvMo90,Hoo91,Hess96,MoDe98,HoB99,DettS00,Mund00,Rond02}. Suitable
adaptations of these schemes to nonequilibrium situations such as, e.g., shear
flows yielded results that were well in agreement with predictions of
irreversible thermodynamics and linear response theory
\cite{EvMo90,SEC98}. Hence, these thermostats became widely accepted 
tools for performing nonequilibrium molecular dynamics computer simulations.
Eventually, they were successfully applied even to more complex fluids such
as, for example, polymer melts, liquid crystals and ferrofluids
\cite{Hess96,HKL96,HABK97}, to proteins in water and to chemical
processes in the condensed matter phase \cite{TuMa00}.

Soon it was realized that this non-Hamiltonian modeling of thermal reservoirs
not only enabled to efficiently construct nonequilibrium steady states on the
computer but also that it made them amenable to an analysis by means of
dynamical systems theory
\cite{EvMo90,Hoo91,MaHo92,Mare97,TGN98,MoDe98,HoB99}. First of all,
in contrast to the stochastic Langevin equation Gaussian and Nos\'e-Hoover
thermostats preserve the deterministic nature of the underlying Newtonian
equations of motion. Even more, though the resulting dynamical systems are
dissipative, surprisingly the thermostated equations of motion are still
time-reversible hence yielding a class of systems characterized by the, at
first view, contradictory properties of being {\em time-reversible,
dissipative} and, under certain circumstances, even being {\em ergodic}
\cite{ChLe95,ChLe97,HKP96}. Computer simulations furthermore revealed that
subsystems thermostated that way contract onto {\em fractal attractors}
\cite{HHP87,MH87,Morr87,HoPo87,PoHo88,HMHE88,Morr89,PoHo89,Mo89a} with an {\em
average rate of phase space contraction that is identical to the thermodynamic
entropy production} \cite{HHP87,PoHo88,Ch1,Ch2}. This led researchers to
conclude that in thermostated dynamical systems the phase space contraction
onto fractal attractors is at the origin of the second law of thermodynamics
\cite{HHP87,Rue96,Rue97,Rue97b,HoB99,Rue99,Gall98,Gall99,Rue03}.

Interestingly, the average rate of phase space contraction plays the same role
in linking statistical physical transport properties to dynamical systems
quantities as the escape or decay rates in the Hamiltonian approach to
nonequilibrium \cite{TeVB96,Rue96,BTV98,GDG01}. The key observation is that,
on the one hand, the average phase space contraction rate is identical to the
sum of Lyapunov exponents of a dynamical system, whereas, on the other hand,
for Gaussian and Nos\'e-Hoover thermostats it equals the thermodynamic entropy
production. For thermostated dynamical systems this again furnishes a relation
between transport coefficients and dynamical systems quantities
\cite{MH87,PoHo88,ECM,Vanc,BarEC,Coh95,ECS+00,AK02}. A suitable
reformulation of these equations makes them formally analogous to the ones
obtained from the Hamiltonian approach to transport. These results were
considered as an indication for the existence of a specific backbone of
nonequilibrium transport in terms of dynamical systems theory
\cite{TeVB96,BTV98,Gasp,GDG01}.

Another interesting feature of Gaussian and Nos\'e-Hoover thermostated
dynamical systems is the existence of {\em generalized Hamiltonian and
Lagrangian formalisms} from which the thermostated equations of motion can be
deduced, which involve non-canonical transformations of the phase space
variables \cite{Nose84a,Nose84b,DeMo96,DeMo97a,MoDe98,Choq98}. Similarly to
Hamiltonian dynamics deterministically thermostated systems often share a
certain symmetry in the spectrum of their Lyapunov exponents known as the {\em
conjugate pairing rule}, which was widely studied in the recent literature
\cite{Dress88,PoHo88,ECM,SEM92,GKC94,Coh95,DMR95,DePH96,DeMo96b,LvBD97,DePo97,DePo97b}
\cite{DeMo97a,MoDe98,SEI98,BCP98,BLD98,WL98,Rue99,vBLD00,DK00,PvZ02,PvZ02b,Pan02,Mor02,TM02b}.
That is, all Lyapunov exponents of a given dynamical system can be grouped
into pairs such that each pair sums up to the same value, which in
nonequilibrium is non-zero. In most cases Lyapunov exponents of thermostated
systems can only be calculated numerically. However, for the periodic Lorentz
gas and related systems there is an elegant analytical approach by means of
kinetic theory developed by van Beijeren and Dorfman et al.\
\cite{BD95,vBDCP96,LvBD97,BDPD97,vZvBD98,BLD98,vBLD00}. 
Furthermore, in recent computer simulations of interacting many-particle
systems Posch et al.\ \cite{MPH98,PoHiS00,MP02,HPF+02,FHPH03} observed the
existence of {\em Lyapunov modes} in thermal equilibrium indicating that the
microscopic contributions to the Lyapunov instability of a many-particle fluid
form specific modes of instability, quite in analogy to the well-known
hydrodynamic modes governing macroscopic transport
\cite{EG00,NaMa01,TM02,TDM02,MaNa03}.

All these interesting properties inspired mathematicians to look at these
systems from a more rigorous point of view. A cornerstone is the proof by
Chernov et al.\ of the existence of Ohm's law for the periodic Lorentz gas
driven by an external electric field and connected to a Gaussian thermostat
\cite{ChLe95,ChLe97}. Another important development was the
{\em chaotic hypothesis} by Gallavotti and Cohen
\cite{GaCo95a,GaCo95b,Gall98,Gall99}, which was motivated by results from
computer simulations on thermostated dynamical systems
\cite{ECM93}.\footnote{The chaotic hypothesis, in its original formulation
\cite{GaCo95b}, states: {\em A reversible many-particle system in a stationary
state can be regarded as a transitive Anosov system for the purpose of
computing the macroscopic properties of the system.}}  This fundamental
assumption generalizes Boltzmann's ergodic hypothesis in summarizing
some general expectations on the chaotic nature of interacting many-particle
systems which, if fulfilled, considerably facilitate calculations of
nonequilibrium properties. 

Further promising achievements in the field of thermostated dynamical systems
are {\em fluctuation theorems} that establish simple symmetry relations
between positive and negative fluctuations of the nonequilibrium entropy
production. Again, such laws first came up in the framework of nonequilibrium
molecular dynamics computer simulations for thermostated systems, see Evans,
Cohen and Morriss \cite{ECM93}. Later on respective theorems were proven by
Gallavotti and Cohen starting from the chaotic hypothesis
\cite{GaCo95a,GaCo95b}.\footnote{For similarities and differences between
Evans-Cohen-Morriss and Gallavotti-Cohen fluctuation theorems see, e.g.,
Refs.\ \cite{CG99,Rond02,EvSe02}.} Related theorems for stochastic systems
were derived in Refs.\ \cite{Kur98,LS99}. In Ref.\ \cite{Mae99} it was argued
that fluctuation theorems can more generally be understood as an intrinsic
property of Gibbs measures\footnote{See, e.g., Refs.\ \cite{Beck,Do99} for
introductions to Gibbs measures.} defined in nonequilibrium
situations. Meanwhile fluctuation theorems have been verified for many
different systems in many different ways analytically
\cite{Do99,Rue99,RTV00,Jar00,EvSe02,JES03}, by computer simulations
\cite{BCL98,LRB00,EvSe02,SE03}, and in physical experiments
\cite{CiLa98,WSM+02}. It appears that fluctuation theorems belong 
to the rather few general results characterizing nonequilibrium steady states
very far from equilibrium thus generalizing Green-Kubo formulas and Onsager
reciprocity relations, which can be derived from them nearby equilibrium
\cite{Gall96,Gall98,Gall99,EvSe02,Gall03b}.

\subsection{Outline of this review}

This review focuses on the non-Hamiltonian approach to nonequilibrium steady
states employing deterministic and time-reversible thermostats. However, even
this part of chaotic transport theory became so large already that we had to
make a rather restrictive choice concerning subjects which we cover
here. Details of the Hamiltonian theory of transport will be discussed
whenever respective crosslinks can be established, which is primarily the case
concerning relations between transport coefficients, dynamical systems
quantities and nonequilibrium entropy production.

Our presentation attempts to be rather pedagogical. In Chapter II we motivate
thermostats in a very intuitive way that is particularly directed to 
non-experts. This motivation is supplemented by a more detailed analysis of the
Langevin equation from the point of view of modeling thermal reservoirs. We
then sketch briefly how to compute velocity distribution functions for a
subsystem interacting with a thermal reservoir that consists of arbitrarily
many degrees of freedom. This basic problem of equilibrium statistical
mechanics illustrates the importance of suitably projecting out reservoir
degrees of freedom, and the results will be used in Chapter VII for
systematically constructing thermal reservoirs modeling an arbitrary number of
degrees of freedom, as well as in Chapter VIII.

In the final section of Chapter II we define the periodic Lorentz gas, a
standard model in the field of chaos and transport. In the following chapters
this simple model will be driven by an external electric field and thermalized
by applying some generic types of thermostats that we introduce step by
step. The corresponding nonequilibrium steady states will be numerically
constructed and analyzed concerning their statistical and chaotic dynamical
properties.  Applying a variety of different thermal reservoirs to the same
model enables us to inquire about possible universal chaos and transport
properties of nonequilibrium steady states generated by different
thermostats. The question to which extent such universal properties exist
forms the main theme of our review.

This discussion is put forward in Chapter III by introducing the {\em Gaussian
thermostat}, which is the most simple and prominent example for modeling a
deterministic and time-reversible thermal reservoir. Note that this scheme
constrains the energy directly in the interior of a subsystem and not at some
boundaries, hence it is called a {\em bulk thermostat}. We summarize what we
consider to be the crucial properties of this class of thermostats as far as
connections between transport properties and dynamical systems quantities are
concerned.  

Chapter IV features a discussion of the {\em Nos\'e-Hoover thermostat}, from
which the Gaussian one is obtained as a special case, and some generalizations
of it. We motivate the Nos\'e-Hoover thermostat starting from the
(generalized) {\em Liouville equation} for dissipative dynamical systems which
we briefly derive. The Nos\'e-Hoover scheme is conceptually analogous to the
Gaussian one thus yielding analogous formulas relating chaos to
transport. However, despite this formal analogy the fractal structure of the
attractor of the Nos\'e-Hoover thermostated Lorentz gas changes differently
under parameter variation in comparison to the Gaussian thermostated
model. This is reflected in different bifurcation diagrams, and
correspondingly we obtain different field dependencies of the electrical
conductivity for both models.

In Chapter V we present a critical assessment of chaos and transport
properties of nonequilibrium steady states generated by these two well-known
and widely used thermostating schemes as presented up to this point. A key
feature is that the models of thermal reservoirs discussed so far exhibit a
built-in identity between phase space contraction and thermodynamic entropy
production. We take this opportunity to elaborate on the striking formal
analogy between three formulas relating transport coefficients to dynamical
systems quantities. Two of them are resulting from the Hamiltonian approach to
transport for closed and open systems, one is emerging from the thermostated
systems approach.

This enables us to pose more clearly the central question of this review,
namely, whether there are universal statistical and chaotic dynamical
properties of nonequilibrium steady states which are independent from the
specific type of thermostat used. Partly this issue has been discussed in the
literature under the label of {\em equivalence of nonequilibrium ensembles}
related to different thermostats
\cite{EvHo85,EvMo90,LBC92,SGB92,EvSa93,ChLe95,Gall96b,ChLe97,Gall97,CoRo98} \cite{SEC98,Rue99,vZ99,DettS00,ER02,BDL+02,HAHG03,Gall03b}.
Additionally, there is a line of work arguing for an equivalence of ensembles
between thermostated and non-thermostated time-discrete maps as far as
nonequilibrium entropy production is concerned
\cite{BrTV96,TeVB96,MoRo96,VTB97,BTV98,VTB98,GiDo99,GFD99,TVS00,VTM00,Voll02,VTB03}.

Our critical discussion of standard thermostating schemes motivates to look
for alternative models of thermal reservoirs, which will be introduced in the
remaining chapters. The main idea is to construct thermal reservoirs that do
not by default exhibit an identity between phase space contraction and entropy
production. In order to be comparable to Gaussian and Nos\'e-Hoover schemes
these alternative models must as well be deterministic and time-reversible,
and one has to show that they generate well-defined nonequilibrium steady
states. Sharing these properties they would provide counterexamples to the
claimed universality of the identity as concluded from the analysis of
standard Gaussian and Nos\'e-Hoover thermostats. Consequently, in this case
also the relations between transport coefficients and dynamical systems
quantities as derived for Gaussian and Nos\'e-Hoover thermostats would become
different.

A first class of such counterexamples is presented in Chapter VI by what we
call {\em non-ideal Gaussian and Nos\'e-Hoover thermostats}
\cite{RKH00}. It  follows a detailed discussion of their chaos and transport
properties, again for the example of the driven periodic Lorentz gas.  Chapter
VII starts by reviewing a well-known thermostat acting only at the boundaries of
a subsystem instead of in the bulk, which is known under the name of {\em
stochastic boundary conditions}
\cite{LeSp78,CiTe80,TCG82,GKI85,ChLe95,ChLe97,HP98,PH98}.  In order to
compare this thermal reservoir to Gaussian and Nos\'e-Hoover thermostats we
make stochastic boundaries deterministic and time-reversible leading to {\em
thermostating by deterministic scattering}, a scheme that contains stochastic
boundaries as a special case \cite{KRN00,RKN00,WKN99,RaKl02,Wag00}. As before,
we first apply this deterministic thermal reservoir to the driven periodic
Lorentz gas. However, in the final section of this chapter we also review
results for shear and heat flow in a many-particle hard-disk fluid
thermostated by deterministic scattering \cite{WKN99,Wag00}.

We will show that thermostating by deterministic scattering defines a second
class of systems exhibiting nonequilibrium steady states in which phase space
contraction is not necessarily equal to thermodynamic entropy
production. Consequently, as in case of non-ideal Gaussian and Nos\'e-Hoover
thermostats, there are no unique relations between transport coefficients and
dynamical systems quantities anymore. Furthermore, related to the fact that
this scheme defines boundary thermostats the spectra of Lyapunov exponents of
dynamical systems thermostated that way yield properties that are rather
different from the ones obtained for systems thermostated in the bulk by using
Gaussian or Nos\'e-Hoover thermostats. Further systems with nonequilibrium
steady states in which phase space contraction is not equal to entropy
production have been explored in Refs.\
\cite{Gasp97a,CoRo98,EPRB99b,DaNi99,Gasp,BR01,Rond02}.

We finish our discussion in Chapter VIII by pointing towards a surprising
connection between Nos\'e-Hoover thermostats and {\em active Brownian
particles} as introduced by Schweitzer and Ebeling et al.\
\cite{SET98,EST99,TSE99,EES+00}. The latter models are thought to mimick,
among others, the crawling of isolated biological cells on rough surfaces
which can be measured experimentally
\cite{FrGr90,SLW91,SchGr93,HLCC94,DDPKS03}. Active Brownian dynamics is
modeled by Langevin equations with velocity-dependent friction coefficients
that enable a Brownian particle to convert internal into kinetic energy and
{\em vice versa}. By using heuristic arguments we show that limiting cases of
such models reduce to conventional Nos\'e-Hoover dynamics. Establishing this
link to deterministic thermostats sheds some light onto the origin of
so-called crater-like velocity distribution functions as they were previously
observed in computer simulations for active Brownian particles. We argue that
they may emerge as superpositions of canonical and microcanonical velocity
distributions, a phenomenon that is nicely exemplified by the Nos\'e-Hoover
thermostat under parameter variations.

The main conclusion of our review, summarized in Chapter IX, is that {\em from
a dynamical systems point of view} conventional Gaussian and Nos\'e-Hoover
thermostats provide only a very specific access road to the modeling of
nonequilibrium steady states. A fundamental problem of these thermal
reservoirs is that they furnish a default identity between the rates of
average phase space contraction and thermodynamic entropy production, which is
at the heart of linking nonequilibrium thermodynamics to dynamical systems
theory in case of non-Hamiltonian equations of motion. Our analysis shows that
alternative, different types of thermostats lead to different such
relations. Correspondingly, the associated fractal attractors, the Lyapunov
spectra and even the field-dependent electrical conductivities may exhibit
very different properties. The last result concerning transport coefficients,
on the other hand, is contrasted by our example of an interacting
many-particle system under shear and heat flow thermostated at the boundaries.
Here all transport properties are very well in agreement with predictions from
irreversible thermodynamics and linear response theory. Still, almost all of
the chaotic dynamical properties of this system are profoundly different
compared to systems that were thermostated by Gaussian or Nos\'e-Hoover
schemes.

Consequently, one may suspect that the variety of different transport
properties obtained for the driven periodic Lorentz gas when thermalized with
different thermostats rather reflects the simplicity and the low
dimensionality of the model. In other words, for simple systems consisting of
non-interacting particles modifications of the equations of motion, e.g., by
applying different thermostats, may indeed profoundly change their macroscopic
transport properties. This aspect is particularly significant when the
dynamical systems are {\em topologically unstable} leading, for example, to
fractal transport coefficients, which are at variance with common expectations
from nonequilibrium thermodynamics. In contrast, for interacting many-particle
systems at least the {\em thermodynamic} properties appear to be rather
indepent from the type of thermostat used pointing towards an equivalence of
ensembles as discussed by other authors. However, we emphasize that we do {\em
not} find such an equivalence as far as the detailed {\em chaotic dynamical}
properties of many-particle systems are concerned.

We thus conclude that the quest for universal characteristics of
nonequilibrium steady states in dissipative chaotic dynamical systems is not
yet over.  At the moment the only candidate for a universal property of {\em
deterministically thermostated systems}, as far as chaos properties are
concerned, appears to be the {\em fractal structure of attractors}, whereas
any further chaos and, depending on the simplicity of the model, even
transport property might reflect the choice of the thermostat.  This poses a
challenge to find more general relations between chaos and transport
properties in thermostated dynamical system than discussed up to now.  For
such an endeavor one may want to start from some suitably coarse-grained
nonequilibrium entropy that does not measure details of phase space
contraction which appear to be spurious if compared to, say, the Clausius
entropy for which nothing else counts than the heat flux into the thermal
reservoir. Despite the obvious importance of entropy production we avoid
detailed discussions concerning the origin of the second law of
thermodynamics. We believe this goes considerably beyond the level of this
review, apart from the fact that there already exists a lot of profound
literature on this subject, see, e.g., Refs.\ \cite{stat99,Bric95,LiYn99} and
much further work mentioned in the course of this review.

As far as the style of discussion in this review is concerned we primarily
appeal to physical intuition. That is, we keep things as simple as possible
and and do not present more technical details than absolutely necessary. This
is reflected by the fact that there are only very few formulas but a lot of
text, and quite a number of figures. We presuppose some basic knowledge of
(nonequilibrium) statistical mechanics \cite{Reif,Huang} and of dynamical
systems theory \cite{Schu,ER,Ott,Beck,ASY97}. Mathematical concepts such as
SRB measures and Anosov systems are not employed explicitly, despite the fact
that all issues discussed in our review are intimately related to them.  These
objects do play a crucial role for building up a rigorous mathematical theory
of nonequilibrium steady states as outlined, e.g., in Refs.\
\cite{Gasp,Do99,Gall99,Rue99}. However, here we work on a level that is less
rigorous consisting of straightforward physical examples and demonstrations
plus some simple calculations and results from computer simulations. It is our
hope that this approach still suffices to make the reader familiar with what
we believe are some central problems in this field. Parts marked with a
$^*$ contain some more detailed information that may be skipped at a first
reading. Particularly Chapters I to IV should be understandable to a general
readership. Chapter V may be considered more difficult, however, it forms an
important core of this review. Chapter VI should again be more easy, whereas
Chapters VII and VIII are intermediate.

Concerning books and reviews that are closely related to the topic covered by
this work we may recommend Refs.\
\cite{EvMo90,Hess96,MoDe98,NiDa98,HoB99,DettS00,Rond02} for further reading. There
exists also a number of conference proceedings and related collections of
articles which the reader may wish to consult
\cite{MaHo92,Mare97,TGN98,Kark00,Sza00,GaOl02,chaotr03}. 
It is not our goal to come up with a complete list of references covering this
rapidly growing research area. Instead, we restricted ourselves to citing
articles that we feel are especially relevant to the problems highlighted in
this review. Still, this resulted in quite a large number of references. We
furthermore remark that this review does not intend to give a full {\em
historical} account of recent developments in chaos theory and nonequilibrium
statistical mechanics; for this purpose see, e.g., Refs.\
\cite{Bru76,Gle88,CAMTV01,Uff01,Voll02}. Only on certain occasions we
go a little bit into the historical depth. We finally remark that Chapters VI
and VII of this review summarize a recent series of papers by the author and
coworkers \cite{KRN00,RKN00,WKN99,RKH00,RaKl02,Wag00} originating from Ref.\
\cite{KRN00}. Sections VI.A and VIII.B, on the other hand, contain new results
that have not been published before.

\section{Motivation: coupling a system to a thermal reservoir}

In this chapter we build the bridge from the introductory section to the main
part of this review by introducing some basic concepts, models and notations.
We first outline, in a very heuristic way and from a very physical point of
view, what thermal reservoirs or so-called ``thermostats'' are and explain why
they are indispensable for studying steady states under typical nonequilibrium
conditions. 

We then briefly remind the reader of the well-known Langevin equation that may
also be thought of modeling the interaction of a subsystem (in this case a
single Brownian particle) with a thermal reservoir (the surrounding fluid). In
particular, we outline a short derivation of the Langevin equation from
Hamiltonian equations of motion in which the thermal reservoir is modeled as a
collection of non-interacting harmonic oscillators. This way we exemplify two
extreme cases of modeling thermal reservoirs, namely either by purely
Hamiltonian dynamics or, alternatively, in terms of some simple but
dissipative, irreversible and stochastic equations of motion. 

After this discussion of the detailed microscopic dynamics we elaborate on the
general functional forms of the velocity distribution functions of subsystem
plus thermal reservoir in thermal equilibrium. By means of very simple
statistical physical arguments we sketch how to calculate velocity
distributions for a subsystem of $d_s$ degrees of freedom which interacts with
a $d_r$-dimensional thermal reservoir in an ideal equilibrium situation. In
case of $d_r\to\infty$ these velocity distributions converge to their
canonical counterparts. This simple problem illustrates the importance of
projecting out reservoir degrees of freedom. The obtained information will be
used later on in order to construct thermal reservoirs, and for finding
necessary conditions concerning the existence of bimodal velocity
distributions. 

To the end of this chapter we define the periodic Lorentz gas, a standard
model in the field of chaos and transport, and give some indication of its
physical interpretations. In the remaining chapters we will apply different
types of thermal reservoirs to the periodic Lorentz gas under nonequilibrium
situations. The resulting nonequilibrium steady states we will compare with
each other concerning their statistical and chaotic dynamical properties.

\subsection{Why thermostats?} 

A common problem of thermostating is how to cool down a bottle of beer on
a hot summer day \cite{Reif}. One option is to simply put the bottle of beer
in a swimming pool. Microscopically, there emerges a transfer of energy
between the molecules in the bottle of beer composing a fluid at temperature
$T_B$ and the water molecules in the swimming pool at temperature
$T_P<T_B$. Here we define temperature operationally on the basis of
equipartitioning of energy onto all available degrees of freedom
\cite{Reif,Huang} connecting temperature with the average kinetic energy of a
system. Under the assumption that the water molecules are properly interacting
with each other, in the sense that such an equipartitioning can properly be
established, on average the excess energy related to the temperature
difference $T_B-T_P$ will flow across the surface of the beer bottle into the
surrounding fluid. Under the very same assumption, it will furthermore equally
distribute onto all the water molecules. However, since the number of water
molecules is extremely large the increase of the average temperature of the
pool, in terms of the average kinetic energy per particle, will be negligibly
small.

What we have sketched here is just the well-known process of classical thermal
equilibration between a sufficiently small subsystem and a surrounding thermal
reservoir, where both systems are initially at two different temperatures.
After equilibration both systems will be approximately at the same
temperature, $T_B\simeq T_P$. However, so far we have only discussed the
simple case of relaxation to thermal equilibrium when subsystem and reservoir
together form a closed system.

As a second, slightly more complicated example, think of a nonequilibrium
situation in an open system such as a light bulb connected to a battery. The
applied voltage will generate a current, and as an Ohmic resistance the thread
in the bulb will exhibit some Joule heating. In a naive microscopic picture
this heating may be understood as follows: The applied electric field will on
average accelerate the electrons in the resistance. Consequently, if there
were no mechanism for reducing their average kinetic energy the gas of
electrons simply heats up, and no stationary current exists. On the other
hand, the single electrons will collide with the atoms constituting the
resistance. During these collisions they will exchange energy with the atoms,
possibly in terms of exciting lattice modes and eventually causing the whole
resistance to heat up. However, across its surface the resistance allows an
average flow of energy into the surrounding medium by collisions between the
atoms of the thread and the atoms or molecules of the gas. In case of the
light bulb this flow of energy just causes the gas around the thread to
glow. 

The dissipation of energy into a thermal reservoir thus properly
counterbalances the pumping of energy into the system by the external electric
field and enables the system to evolve into a {\em nonequilibrium steady state
(NSS)}.  With that we mean that the statistical physical parameters describing
the system on macroscopic scales are constant in time, despite the fact that
the system is no longer in thermal equilibrium \cite{deGM84}. In case of our
example, the existence of a NSS implies that under the nonequilibrium
condition of applying an external electric field the electron gas eventually
exhibits a stationary current and that the temperature is constant. We note in
passing that the proper definition of a nonequilibrium temperature is a subtle
problem in itself \cite{Rugh97,MR99}, but operationally, again, the principle
of equipartitioning of energy might be used
\cite{PoHo88,EvMo90,MoDe98,HoB99}. If the existence of a NSS is due to the
action of a thermal reservoir we say the system is properly {\em
thermostated}.

In other words, thermostats are mechanisms by which the internal energy of a
many-particle system, and thus its temperature, can be tuned onto a specific
value. Any thermostat may be thought of being associated with some thermal
reservoir consisting of an infinite number of degrees of freedom thus being
large enough to absorb any amount of energy pumped into the system.
Thermostats can be applied in order to achieve equilibration to a global
thermal equilibrium or for sustaining a NSS in a nonequilibrium situation
where there is a flux of energy through the system, such as induced by
external fields or by imposing temperature or velocity gradients. In a similar
vein, other constraints instead of fixing the temperature may be imposed onto
a system leading, e.g., to constant pressure or constant stress ensembles
\cite{AT87,EvMo90,Nose91,Hess96,TuMa00}.

The first question we must focus on before we elaborate on NSS is therefore:
How can we suitably amend Newton's equations of motion in order to model an
energy dissipation into a thermal reservoir? The basic problems involved here
will be exemplified by means of the well-known Langevin equation. At this
level of discussion restrict ourselves to equilibrium situations where the
construction of thermal reservoirs may appear to be a rather technical
problem. However, our main objective is to eventually apply thermal reservoirs
as defined in equilibrium to nonequilibrium situations yielding NSS. In later
chapters we will inquire to which extent the detailed properties of NSS depend
on the detailed modeling of different thermal reservoirs, and to which extent
they might be universal.

\subsection{Stochastic modeling of thermal reservoirs: the Langevin equation}

The Langevin equation \cite{Lang08} provides a simple theoretical framework
for describing what is known as {\em Brownian motion}: A sufficiently small
particle immersed in a fluid performs irregular motion resulting from the
collisions between the atoms or molecules of the medium and the particle
\cite{Reif,Wax54,vK,Path88,Gall99}. For coal dust particles on the surface of
alcohol this phenomenon was observed by the Dutch physician J.\ Ingenhousz in
1785. However, it became known more widely only later on by the work of the
Scottish botanist R.\ Brown in 1827, who reported similar irregular movements
of pollen grains under a microscope. The name Brownian motion was coined by
A.\ Einstein in his famous work from 1905 describing this irregular motion in
terms of diffusion processes. A related approach was already proposed by the
French mathematician L.\ Bachelier in 1900 in order to understand the dynamics
of stock values. Einstein's work led J.B.\ Perrin to the experimental
measurement of the Avogadro number in 1908 and also motivated P.\ Langevin's
modeling of Brownian motion published in the same year.\footnote{For this
short historical note on Brownian motion we largely followed the nice
presentation in Ref.\ \cite{MeKl00}; for more details see, e.g., Refs.\
\cite{Bru76,Sta89}.} We first briefly summarize some fundamental features of
Langevin's stochastic approach to Brownian motion. In the following chapters
we will repeatedly come back to these characteristic properties by discussing
to which extent they are reproduced in deterministic chaotic models of
Brownian motion.

According to Langevin a Brownian particle is randomly driven by instantaneous
collisions with the surrounding particles. On the other hand, energy is
removed from the system by some bulk friction. In one dimension this dynamics
is modeled by amending Newton's equations of motion to
\bna
\dot{r}&=&v\nonumber\\
\dot{v}&=&-\alpha v+{\cal F}(t) \quad , \label{eq:lang1d}
\ena
where $r$ and $v$ denote the position and the velocity of the Brownian
particle. For convenience, here and in the following we set constants like the
mass $m$ of the particle and the Boltzmann constant $k$ equal to one. ${\cal
F}$ holds for white noise, that is, a $\delta$-correlated stochastic force
with zero mean modeling the random collisions with the fluid particles,
whereas $\alpha$ is a Stokes friction coefficient. In a steady state the
molecular stochastic forces and the friction balance each other leading to the
fluctuation-dissipation theorem \cite{Reif,Wax54,Path88,Zwan01}
\be
\alpha=\frac{1}{2T}\int_{-\infty}^{\infty}dt <{\cal F}(0) {\cal F}(t)>\quad
,\label{eq:fdt} 
\ee
where $<\ldots>$ denotes an ensemble average over the noise acting on the
moving particles. The expression $C_{\cal F}(t):=<{\cal F}(0){\cal F}(t)>$ is
an example of a correlation function. $\alpha$ is in turn related to the
diffusion coefficient $D$ of the Brownian particle according to the Einstein
relation
\be
D=T/\alpha \quad , \label{eq:einstvisc}
\ee
where $T$ is the temperature obtained from equipartitioning of energy,
$T=<v^2>$. $D$ is defined independently via the mean square displacement,
\be
D:=\lim_{t\to\infty}\frac{<\left[x(t)-x(0)\right]^2>}{2t}\quad .\label{eq:deinst}
\ee
Here the angular brackets denote an equilibrium ensemble average over moving
particles. Alternatively, $D$ can be obtained from the Green-Kubo formula for
diffusion
\be
D=\int_0^{\infty}dt\: <v(0)v(t)>\quad ,\label{eq:dgk}
\ee
which is an exact transformation of Eq.\ (\ref{eq:deinst}) \cite{Zwan01}. 

The Green-Kubo formula Eq.\ (\ref{eq:dgk}) and the fluctuation-dissipation
theorem Eq.\ (\ref{eq:fdt}) are formally analogous in relating a macroscopic
quantity characterizing the fluid to an integral over some correlation
function. $C_v(t):=<v(0)v(t)>$ is called the velocity autocorrelation function
of the moving particle. As can easily be shown by solving the Langevin
equation Eq.\ (\ref{eq:lang1d}) for $C_v$, the velocity autocorrelation
function decays exponentially in time thus ensuring that, in terms of the
Green-Kubo formula Eq.\ (\ref{eq:dgk}), the diffusion coefficient exists.

If one considers the Brownian particle as a subsystem and the surrounding
particles as an infinite dimensional thermal reservoir, the Langevin equation
precisely models the situation where a subsystem suitably interacts with a
thermal reservoir. However, so far we have motivated this equation purely
heuristically. It is therefore interesting to discuss a simple derivation of
the above Langevin equation that actually starts from a fully Hamiltonian
modeling of subsystem plus thermal reservoir in which the heat bath consists
of an infinite number of harmonic oscillators
\cite{Zwan73,FK87,KTH92,SW99,Zwan01}.\footnote{In Ref.\ \cite{SW99} this model
was attributed to Ford and Kac \cite{FK87}, however, it already appears at
least in the paper by Zwanzig \cite{Zwan73}; see also further references in
Ref.\ \cite{FK87}.} It nicely demonstrates what kind of simplifying
assumptions one has to make, starting from first principles, in order to
arrive at the simple modeling of thermal reservoirs in terms of Eq.\
(\ref{eq:lang1d}).

Let the Hamiltonian of the Brownian particle be
\be
H_s=\frac{v^2}{2} \label{eq:zwh1}
\ee
and let the heat bath of harmonic oscillators be
\be
H_B=\sum_j\left(\frac{v_j^2}{2} + \frac{\omega_j^2}{2}\left(x_j -
\frac{\gamma_j}{\omega_j^2}x\right)^2\right) \quad , \label{eq:zwh2}
\ee
where the oscillators exhibit a special coupling to the subsystem. Here
$\omega_j$ is the frequency of the $j$th oscillator and $\gamma_j$ the
strength of the coupling between the Brownian particle and the $j$th
oscillator. The Hamiltonian equations of motion for the combination of
subsystem plus thermal reservoir then read
\bna
\dot{x}&=&v \nonumber \\
\dot{v}&=&\sum_j\gamma_j\left(x_j-\frac{\gamma_j}{\omega_j^2}x\right)\label{eq:zwaneom1}\\
\dot{x_j}&=&v_j  \nonumber \\
\dot{v_j}&=&-\omega_j^2x_j+\gamma_jx \label{eq:zwaneom2} \quad .
\ena
The formal solution of the inhomogeneous differential equation of second order
Eq.\ (\ref{eq:zwaneom2}) reads
\be
x_j(t)=x_j(0)\cos(\omega_j t) + v_j(0)\frac{\sin(\omega_j t)}{\omega_j} +
\gamma_j\int_0^tds\: x(s)\frac{\sin(\omega_j(t-s))}{\omega_j} \label{eq:zwaneom2s}
\quad .
\ee
Performing integration by parts for the latter integral and thereafter putting
Eq.\ (\ref{eq:zwaneom2s}) into Eq.\ (\ref{eq:zwaneom1}) yields the formal
Langevin equation
\be
\dot{v}(t)=-\int_0^t ds\:K(s)v(t-s)+F(t) \quad . \label{eq:zwlang}
\ee
By comparing this equation to the former stochastic Langevin equation Eq.\
(\ref{eq:lang1d}) the first term on the right hand side is identified as some
nonlinear, non-Markovian friction containing the memory function
\be
K(t)=\sum_j\frac{\gamma_j^2}{\omega_j^2}\cos(\omega_j t)\quad . \label{eq:lmem}
\ee
The second term on the right hand side of Eq.\ (\ref{eq:zwlang}) must
consequently yield the ``noise'', which is given explicitly by
\be
F(t)=\sum_j\gamma_jv_j(0)\frac{\sin(\omega_j
t)}{\omega_j}+\sum_j\gamma_j\left(x_j(0)-\frac{\gamma_j}{\omega_j^2}x(0)\right)\cos(\omega_j
t) \quad . \label{eq:zwnoise}
\ee
Note that $F(t)$ is fully deterministic in depending on the initial conditions
of all heat bath variables.

In order to recover Eq.\ (\ref{eq:lang1d}) both expressions are now simplified
as follows: If the spectrum of frequencies ${\omega_j}$ is thought to be
continuous, the memory function Eq.\ (\ref{eq:lmem}) may be written as a
Fourier integral. By assuming convenient functional forms for the density of
states of $\omega_j$ and for the coupling coefficients $\gamma_j$ the memory
function can be replaced by a delta function, $K(t)\sim\delta(t)$.  In this
special case the Langevin equation Eq.\ (\ref{eq:zwlang}) is Markovian and the
ordinary Stokes friction of Eq.\ (\ref{eq:lang1d}) is recovered.

As far as the ``noise'' is concerned, according to Eq.\ (\ref{eq:zwnoise})
$F(t)$ is a linear function of the initial positions and velocities of all
bath oscillators that are furthermore independent degrees of freedom. Hence,
there are no correlations between these variables. If all initial conditions
are {\em sampled randomly} from a canonical distribution one can show that the
fluctuation-dissipation theorem takes the form \cite{Zwan01}
\be
<F(t)F(t')>=TK(t-t') \quad .\label{eq:zwanfdt}
\ee
Under the above assumption that the memory function is Markovian this equation
boils down to Eq.\ (\ref{eq:fdt}), or in other words, the fluctuations induced
by $F(t)$ turn out to be white noise. We see that, in this case, the noise is
put in ``by hand'' into Eq.\ (\ref{eq:zwlang}) by assuming a suitable
distribution of initial conditions for the heat bath variables.

In summary, we have discussed the connection between two extreme versions of
modeling the interaction of a subsystem consisting of a single particle with a
thermal reservoir: One employs the fully Hamiltonian, i.e., deterministic,
time-reversible and phase space preserving equations of motion Eqs.\
(\ref{eq:zwh1}) and (\ref{eq:zwh2}). To start this way may be considered
convenient from a microscopic point of view. However, note that the
statistical properties of the harmonic oscillator heat bath are pre-determined
by the choice of initial conditions supplemented by the specific form of the
memory kernel in Eq.\ (\ref{eq:zwanfdt}). The latter yields the decay of the
force correlations, hence the problem of introducing randomness is shifted
towards choosing a proper initial distribution for the oscillator variables. A
realistic heat bath, on the other hand, should generate a (canonical)
equilibrium distribution of position and velocity variables in a
self-contained way, that is, due to the action of the dynamical system
defining the bath and irrespective of any specific initial conditions. That
this is not the case for the harmonic oscillator heat bath one may consider as
quite a deficiency of the model. Another more practical disadvantage becomes
clear in nonequilibrium situations, where the sum over harmonic oscillators in
Eq.\ (\ref{eq:zwh2}) must be infinite in order to allow for the existence of a
NSS. Though taking this limit does not pose any crucial difficulty for
analytical investigations, the Hamiltonian scheme thus becomes very
inconvenient for computer simulations.

The alternative type of modeling consists of the very intuitive linear and
Markovian Langevin equation Eq.\ (\ref{eq:lang1d}). However, according to its
derivation from the harmonic oscillator heat bath strictly speaking these nice
properties emerge only because of profound approximations. Even more, the
dynamics eventually turns out to be non-Hamiltonian, that is, stochastic,
irreversible and, as a consequence of Stoke's friction, on average phase space
contracting; see also Ref.\ \cite{HoB99} for related arguments.

Hence, both types of heat bath models do not really capture the whole essence
of the problem. Correspondingly, there is plenty of room for generalizations
of modeling thermal reservoirs, in two directions: On the one hand, starting
from some Hamiltonian equations of motion or respective toy models of
dynamical systems one may wish to put the derivation of equations modeling the
action of a heat bath onto a small subsystem onto more general, rigorous
grounds. This approach was pursued, e.g., in a line of work by Beck et
al. \cite{BeRo87,Bec95,Bec96}. Here the analysis starts from a time-discrete
version of the Langevin equation Eq.\ (\ref{eq:lang1d}) in which randomness is
generated by a deterministic chaotic map thus trying to understand to which
extent deterministic chaos modeling reservoir degrees of freedom can mimick
stochastic noise. In a similar vein, Just and Kantz et al.\ considered chaotic
dynamical systems with time scale separation where the fast degrees of freedom
may be related to the action of a thermal reservoir
\cite{JKRH01,JGBRK03,KJBGR03}. Note that similar to Beck's approach they modeled
thermal reservoirs by dynamical systems that consist of only very few degrees
of freedom exhibiting a chaotic dynamics. By using projection operator
techniques for the phase space densities of the dynamical system variables
they studied to which extent the fast degrees of freedom can be replaced by
stochastic noise acting on the slow variables of respective subsystems, and
how the resulting equations of motion for the subsystems look like. A related
problem was investigated by Eckmann et al., who mathematically analyzed the
nonequilibrium properties of a chain of anharmonic oscillators coupled to two
heat baths at different temperatures modeled by Hamiltonian wave equations
\cite{EPRB99,EPRB99b}.

As far as we can tell it is extremely hard to derive rigorous results for the
dynamical properties of thermal reservoirs and the associated subsystems, even
in equilibrium situations, by starting from first principles. We are not aware
that yet there is a general theory for this problem. A major difficulty
appears to be that the precise characteristics of the thermal reservoirs
depend to quite some extent on the detailed nonlinear and non-Markovian
properties of the underlying equations of motion. Thus, typically one has to
employ profound approximations in order to come up with some general
statistical description, as was already indicated above. In nonequilibrium
situations the problem of characterizing steady states with respect to their
chaotic and dynamical properties becomes even more peculiar, which connects to
our respective discussion in the introduction.

As an alternative to a rigorous bottom-up approach one may therefore construct
a variety of simple and successively more sophisticated models of thermal
reservoirs, with the purpose to learn about the detailed statistical and
dynamical properties of the associated equilibrium and nonequilibrium steady
states generated by these models. One may then inquire about general
characteristics that are valid for all these models. We emphasize that this
top-to-bottom approach, of course, bears other risks. In particular, one may
too quickly conclude from the fascinating properties of a specific toy model
concerning universal features. But this can be cured by trying to construct
counterexamples.

Of course, this bottom-up approach towards the physics of steady states is of
a very heuristic nature. Still, in order to qualify as thermal reservoirs all
models have to fulfill certain basic criteria. In the following we will
require that, like ordinary equations of motion of moving particles, they are
deterministic, time-reversible, and that they are able to generate a NSS under
nonequilibrium conditions. 

However, some details are still missing. So far we have focused on the
equations of motion of subsystem plus thermal reservoir only. We did not
explicitly discuss the detailed properties of the associated probability
densities of subsystem and thermal reservoir and how they should look like, at
least in equilibrium situations. This will be performed in the following
section.

\subsection{Equilibrium velocity distribution functions for subsystem and thermal
reservoir}

In this part we shall look at a subsystem that is coupled to a thermal
reservoir from a purely statistical point of view. That is, we do not assume
any detailed knowledge about the underlying equations of motion. We are
interested only in the probability densities for the velocities of the
subsystem and of the thermal reservoir, which we call velocity distribution
functions, in the general case of thermal equilibrium. We do not discuss the
probability densities for the associated position coordinates of the degrees
of freedom, since the action of the thermal reservoir primarily concerns an
exchange of energy related to the velocities. Our goal is to assess how
different dimensionalities of subsystem and thermal reservoir, if the
reservoir is first thought to be finite dimensional, affect the functional
forms of the corresponding velocity distribution functions.

In order to illustrate the general case we start by discussing the most
trivial example, where subsystem and thermal reservoir each consist of only
one degree of freedom. Afterwards we switch to the general situation in which
there is a subsystem with $d_s$ degrees of freedom interacting with a thermal
reservoir of $d_r$ degrees of freedom, where typically $d_s<d_r$ and where the
total number of degrees of freedom is $d=d_s+d_r$.\footnote{This problem has
been posed as an exercise in the book of van Kampen on p.11 \cite{vK}.}

Let us consider a large collection of points in the full phase space of the
system, which in general may be restricted by some boundary conditions. The
equilibrium velocity distribution function $\rho\equiv\rho(v_1,v_2)$ for $d=2$
is then defined with respect to this ensemble of points, which move according
to the same equations of motion. It is the number of points one can find in
the phase space volume element $dv_1dv_2$ centered around the velocity vector
$(v_1,v_2)$ at time $t$ divided by the size of this volume element and by the
total number of points $N$ of the ensemble. By integrating over the whole
accessible phase space the velocity distribution is properly normalized to
one.

\begin{figure}[t]
\epsfxsize=8cm
\centerline{\hspace*{-1cm}\epsfbox{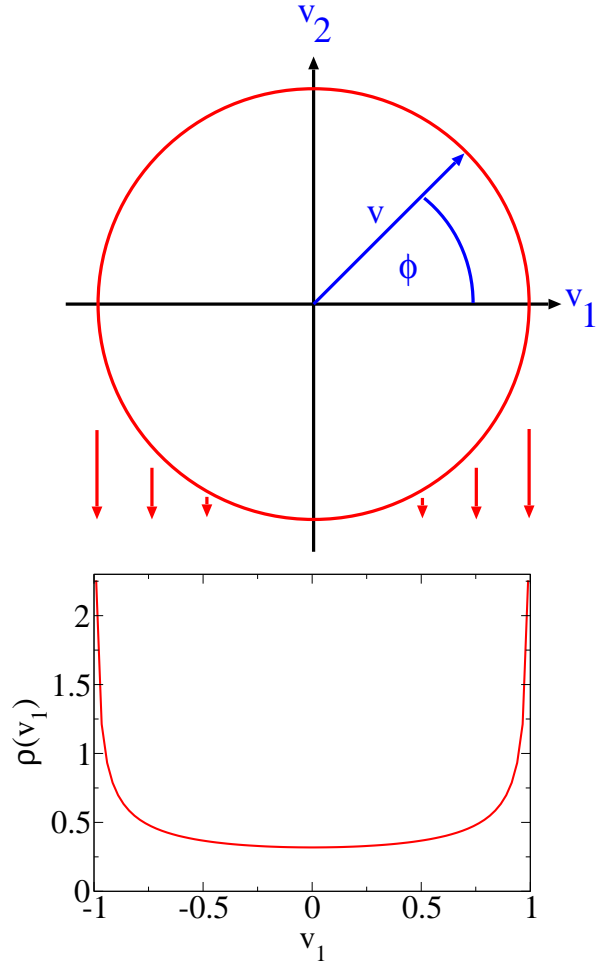}}
\caption{Projection of the two-dimensional microcanonical velocity distribution Eq.\
(\ref{eq:mcd2d}) onto one component. The density of points is uniform on a
circle of radius $v=1$. The velocity vector is composed of the two velocity
components $v_1$ and $v_2$, and its direction is determined by the polar angle
$\phi$. The lower figure represents the one-component velocity distribution
$\rho(v_1)$ Eq.\ (\ref{eq:circrho}) calculated from projecting out $v_2$ from
Eq.\ (\ref{eq:mcd2d}).}
\label{fig:circle}
\end{figure}

Let us now assume that we are in an equilibrium situation where
equipartitioning of energy holds between all degrees of freedom and that the
total energy of subsystem plus reservoir is conserved. For sake of simplicity
we may assume that there is only kinetic energy as, e.g., in a gas of hard
spheres. Hence, all points in velocity space should be uniformly distributed
on a circle of radius $v:=\sqrt{v_1^2+v_2^2}$, where for convenience we set
$v\equiv 1$, see Fig.\ \ref{fig:circle} for an illustration. Subsystem plus
reservoir thus obey the microcanonical distribution
\be
\rho(v_1,v_2)\sim\delta(1-v_1^2-v_2^2) \label{eq:mcd2d} \quad .
\ee
  
\vspace*{-0.5cm}
It is now straightforward to calculate the velocity distribution functions of
the subsystem, respectively of the thermal reservoir, which are obviously
identical, by projecting out one of the two degrees of freedom from this
microcanonical distribution. One way of how to do this is by using polar
coordinates, $v_1=\cos \phi$ and $v_2=\sin \phi$, where $\phi$ denotes the
polar angle in velocity space depicted in Fig.\ \ref{fig:circle}. The
probability density corresponding to $\phi$ is then simply
$\rho(\phi)=const$. On the other hand, $v_1$ is a function of
$\phi$. According to conservation of probability the probability densities for
$v_1$ and $\phi$ are related to each other by
\be
\rho(\phi)|d\phi|=\rho(v_1)|d v_1| \quad .
\ee
Hence, $\rho(v_1)$ is easily calculated in terms of $\rho(\phi)$ by
differentiating $v_1$ leading to
\be
\rho(v_1)=\frac{1}{\pi\sqrt{1-v_1^2}} \quad , \label{eq:circrho}
\ee
where the prefactor is obtained from normalization. As indicated in Fig.\
\ref{fig:circle}, this result matches to physical intuition: By projecting
the uniform distribution of points on the circle onto the $v_1$-axis it is
clear that there must be a higher density of points nearby the values of
$v_2=0$ on the circle, whereas around $v_1=0$ the density of points must be
minimal.

The main message of this trivial example is that non-uniform probability
densities, which may otherwise be considered as an indicator of some
non-Hamiltonian average phase space contraction, may very well coexist with a
uniform, microcanonical one as it characterizes ordinary Hamiltonian dynamical
systems. The obvious reason is that {\em one degree of freedom got projected
out} respectively, so we neglected some information by not looking at the full
combination of subsystem plus thermal reservoir.

\begin{figure}[t]
\epsfxsize=14cm
\centerline{\hspace*{-1cm}\epsfbox{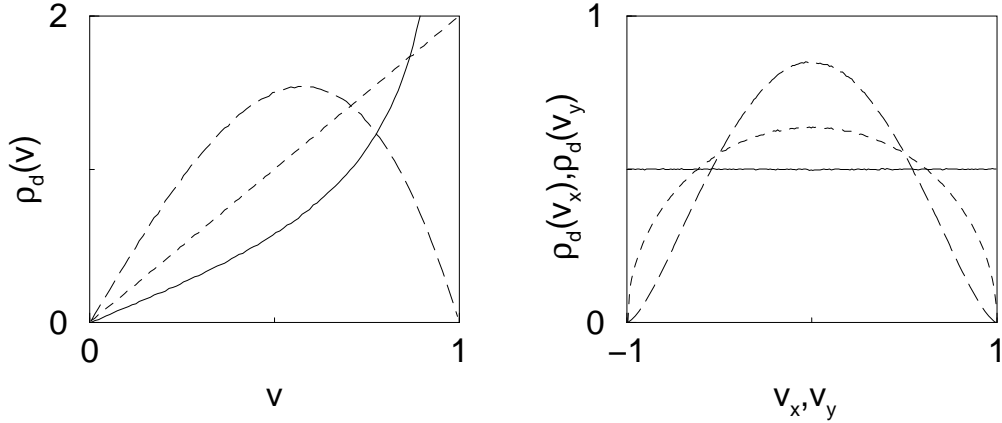}}
\caption{Left: equilibrium velocity distribution function $\rho_d(v)$ for the
absolute value $v$ of the velocity vector $(v_x,v_y)$, which may be thought to
represent a subsystem with two degrees of freedom coupled to a
$(d-2)$-dimensional thermal reservoir. Shown are $d=3$ (solid curve), $4$
(dashed curve) and $6$ (long dashed curve) at a kinetic energy of $E=0.5$ as
calculated from Eq.\ (\ref{eq:gausv}). Right: analogous results for the single
velocity components $v_x$, respectively $v_y$, calculated from Eq.\
(\ref{eq:gaus}).}
\label{fig:d346}
\end{figure}

Along the same lines and under the same conditions as described above, i.e.,
by assuming thermal equilibrium, equipartitioning of energy and a constant
total kinetic energy $E$, the general case of a $d_s$-dimensional subsystem
interacting with a $d_r$-dimensional thermal reservoir can now be studied.  In
this case the full velocity distribution function for subsystem plus thermal
reservoir obeys the microcanonical distribution
\be
\rho(v_1,\ldots,v_d)\sim\delta(2E-v_1^2-\ldots-v_d^2) \label{eq:mcd}
\ee
with $d=d_s+d_r$ and $v_k\:,\:k=1,\ldots,d$, as all the velocities. That is,
all points in velocity space are uniformly distributed on a hypersphere with
radius $\sqrt{2E}$. How do the corresponding velocity distributions of
subsystem and reservoir now look like for arbitrary numbers of degrees of
freedom $d$, $d_s$ and $d_r$?  Again, in order to calculate these distribution
functions one has to project out the respective number of degrees of freedom
from Eq.\ (\ref{eq:mcd}). Here we wish to focus on eliminating $d-1$ or $d-2$
degrees of freedom only. The solutions may be obtained along the same lines as
in the two-dimensional case by using generalized spherical coordinates.
However, we skip the calculations that are technically a bit more involved
\cite{RKN00}.  The results are as follows: By projecting out $d-1$ dimensions
the probability density for one velocity componet $v_1$ is calculated to
\begin{equation}
\rho_d(v_1) =
\frac{\Gamma(\frac{d}{2})}{\sqrt\pi\Gamma(\frac{d-1}{2})}
\frac{(2E-v_1^2)^\frac{d-3}{2}}{(2E)^\frac{d-2}{2}}
\quad , 
\label{eq:gaus}
\end{equation}
where $\Gamma(x)$ stands for the gamma function. This solution was already
known to Maxwell and Boltzmann \cite{Ma1879,Bo09}, however, in their
calculation they took a different starting point; for an alternative
derivation see also Ref.\ \cite{MPT98}. Projecting out $d-2$ dimensions yields
the probability density for the absolute value of the velocity vector of two
components $v_1$ and $v_2$ with $v=\sqrt{v_1^2+v_2^2}$ reading
\begin{equation}
\rho_d(v)=(d-2)v\frac{(2E-v^2)^\frac{d-4}{2}}{(2E)^{\frac{d-2}{2}}} \quad .
\label{eq:gausv}
\end{equation}
These solutions are depicted in Fig.\ \ref{fig:d346} for $d=3,4,6$. As can be
seen, for $d\gg1$ both types of velocity distributions start to converge
towards their canonical counterparts. This can be made precise by taking the
limit of $d\to\infty$ in Eqs.\ (\ref{eq:gaus}) and (\ref{eq:gausv}).  Using
equipartitioning of energy, $E=\sum_{k=1}^d v_k^2/2=Td/2$, where $T$ is the
temperature and $k_B\equiv 1$, we arrive at the well-known results
\be
\lim_{d\to\infty}\rho_d(v_1) = \frac{1}{\sqrt{2\pi T}}
e^{-v_1^2/2T}\quad , \label{eq:vxgauss}
\ee
respectively
\begin{equation}
\lim_{d\to\infty}\rho_d(v) =  \frac{1}{T}ve^{-\frac{v^2}{2T}} \quad .
\label{eq:vgauss}
\end{equation}
We have thus rederived the canonical distributions for $v_1$ and for $v$
starting from the microcanonical one of subsystem plus thermal reservoir by
projecting out an infinite number of reservoir degrees of freedom. Altogether
the above equations yield the full information how, according to elementary
equilibrium statistical mechanics, the velocity distribution functions of a
subsystem consisting of $d_s$ degrees of freedom shall look like if it
experiences ideal interactions with a $d_r$-dimensional thermal reservoir.

We remark that the velocity distribution Eq.\ (\ref{eq:vxgauss}) is reproduced
by an ensemble of particles obeying the stochastic Langevin equation Eq.\
(\ref{eq:lang1d}) in the limit of $t\to\infty$. The corresponding Markov
process modeling the stochastic behavior of the velocity of a Brownian
particle, which is characterized by a stationary Gaussian velocity
distribution, is well-known in more mathematical terms as an
Ornstein-Uhlenbeck process \cite{vK,Wax54}.
  
\vspace*{-6cm}
An interesting question is now to which extent the above ideal functional
forms of equilibrium velocity distributions are obtained for deterministic
dynamical systems of $d$ degrees of freedom when one looks at a respective
smaller number of degrees of freedom $d_s<d$. A very popular example of this
type of problem is the one- or two-dimensional harmonic oscillator coupled to
a suitable deterministic thermal reservoir. If the reservoir mimicks an
infinite number of degrees of freedom $d\to\infty$ the above derivations
suggest that the harmonic oscillator velocities should simply approach
canonical distributions. However, for such a behavior it is necessary that the
thermostated subsystem exhibits at least an ergodic dynamics, because only in
this case a single subsystem trajectory samples the whole phase space
appropriately in order to possibly generate a canonical distribution. If the
phase space is not uniquely accessible because of non-ergodicities the
resulting velocity distributions will strongly depend on initial conditions
and are typically not canonical. This discussion again indicates how
sensitively the interplay between subsystem and thermal reservoir may depend
on the detailed properties of the involved dynamical systems; for further
details see Section IV.D.

As we already mentioned, the emphasis of this review is on nonequilibrium
situations, and here thermal reservoirs are used as a tool to generate NSS
representing transport properties of the associated dynamical
system. Unfortunately, there is yet no analogue of microcanonical or canonical
ensembles for nonequilibrium processes. In other words, nothing is known about
how velocity distribution functions shall generally look like in
nonequilibrium situations. The standard approach is therefore to define a
thermal reservoir in an equilibrium situation by ``gauging'' it according to
the requirement that it generates a microcanonical or a canonical velocity
distribution for a respective subsystem connected to it. If this applies, a
suitable nonequilibrium situation may be created. 

For arbitrary deterministic dynamical subsystems it is by no means obvious or
guaranteed that the thermal reservoir still properly ``works'' under these
constraints, which intimately depends on the ergodic and chaotic dynamical
properties of subsystem and thermal reservoir.  If it does, the reservoir
should generate a NSS that is characterized by specific nonequilibrium
velocity distribution functions. This way, one can learn something about the
variety of different velocity distributions that may exist in different
nonequilibrium systems under application of specific models of thermal
reservoirs, rather than the other way around. Unless there is a general theory
of NSS starting from first principles, which at the moment is not the case, no
other approach appears to be feasible here. Of course, for modeling
nonequilibrium situations the associated thermal reservoirs must always mimick
an infinite number of degrees of freedom in order to prevent a subsystem from
heating up.

We finally remark that the above results for equilibrium velocity
distributions of subsystems plus finite-dimensional thermal reservoirs will be
used in Chapter VII in order to systematically construct a specific class of
deterministic, time-reversible thermal reservoirs that dissipate energy
according to inelastic collisions at the boundaries of a
subsystem. Furthermore, we will use these results in Chapter VIII by
discussing the origin of crater-like velocity distribution functions in models
of active Brownian particles.

\subsection{The periodic Lorentz gas}

As we just discussed, whether or not thermal reservoirs are able to generate a
specific velocity distribution for moving particles depends to quite some
extent on the detailed dynamical properties of the subsystem to which it is
applied. As an alternative to the problematic harmonic oscillator one may wish
to choose a subsystem that is by default chaotic thus providing a more
suitable starting point for the action of a thermal reservoir. Though chaotic
one may wish that, nevertheless, this system is well analyzed in the
mathematical literature. 

\begin{figure}[t]
\epsfxsize=8cm
\centerline{\epsfbox{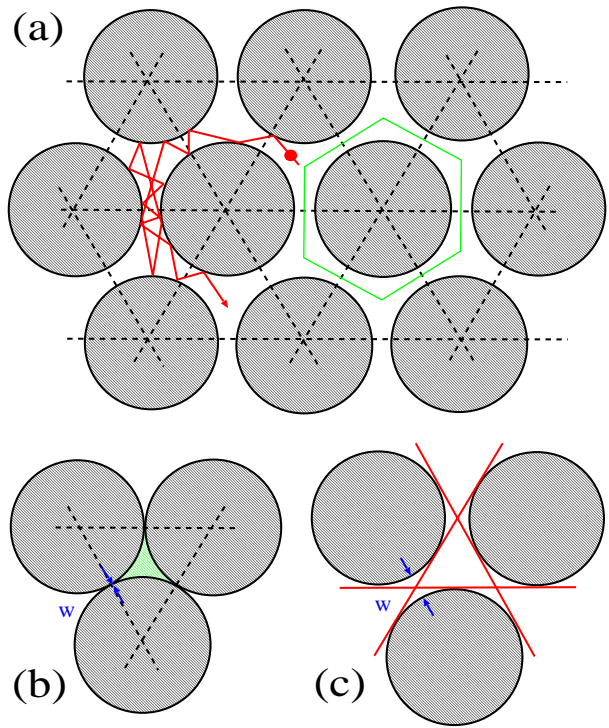}}
\vspace*{0.5cm}
\caption{Sketch of the two-dimensional periodic Lorentz gas where hard disks
of unit radius are situated on a triangular lattice. (a) contains the
hexagonal fundamental cell, or Wigner-Seitz cell, of this system. It also
shows the trajectory of a point particle moving with a constant velocity and
scattering elastically with the disks. (b) represents the limiting case of the
highest density of scatterers where the minimal distance $w$ between two
scatterers is zero. That is, the scatterers touch each other and there is no
diffusion. (c) depicts the other limiting case at a lower density where, for a
specific value of $w$, for the first time trajectories exist along which a
particle can move collision-free for an infinite time and thus sees an
``infinite horizon''. In this case the dynamics is ballistic and the diffusion
coefficient is infinite.}
\label{fig:plgasg}
\end{figure}

Such a paradigmatic toy model is the {\em Lorentz gas}. The two-dimensional
{\em periodic} version of it is sketched in Fig.\
\ref{fig:plgasg} and consists of hard disks of radius $R$ at density $n$ whose
centers are fixed on a triangular lattice in a plane. Between the scatterers a
point particle moves with velocity ${\bf v}$ and mass $m$ by performing a free
flight, whereas at collisions it exhibits specular reflections with the
disks. If no further external constraints are applied the kinetic energy of
the moving particle is constant, whereas the direction of the velocity changes
according to the collisions with the scatterers. The phase space thus consists
of the four variables $(x,y,v_x,v_y)$. However, because of energy conservation
there are only three independent variables, where one may replace $v_x$ and
$v_y$ by the angle of flight of the particle with the $x$-axis. For sake of
simplicity we set $R=m=v\equiv 1$.

In the original work by H.A.\ Lorentz \cite{Lo05}, to whom the model is
commonly attributed, the Lorentz gas consists of {\em randomly} distributed
{\em hard spheres}. It was invented by him in 1905 in order to model the
motion of classical electrons in metals, where the electrons do not interact
with each other but scatter elastically only with hard spheres mimicking the
atoms of the metal. However, as a laboratory device the two-dimensional
periodic version of this system was, in a way, already introduced about 30
years earlier by Sir Francis Galton in 1877 \cite{Galt77}. Hence, this model is
also sometimes referred to as the {\em Galton board}. Galton used his device,
which he called {\em the quincunx}, in lectures for demonstrating the
existence of a binomial distribution that results when particles falling under
the influence of a gravitational field are moving left or right through a
periodic board of scatterers. He actually claimed that the same probabilistic
laws existed in form of typical laws of heredity within populations of animate
beings \cite{Galt77}. His theme is therefore very different from Lorentz'
intention, who used his model to verify Drude's theory of classical electronic
transport in metals \cite{Drude00} by adding nonequilibrium constraints such
as an electric field and a gradient of temperatures.

Though Drude's and Lorentz' work provide some heuristic microscopic derivation
of Ohm's law, and of the law by Wiedemann and Frantz, it is well-known that
this classical approach is not suitable to yield explanations for other
fundamental properties of metals such as, e.g., their specific heat, which can
only be understood by using quantum mechanics \cite{IbLu96}. On the other hand
the two-dimensional Lorentz gas, the scatterers equipped with smooth
potentials, turned out to be very useful for understanding electronic
transport in {\em antidot lattices} under the action of electric and magnetic
fields. These low-dimensional semiconductor devices can be manufactured with
scatterers distributed both randomly and periodically in a plane. For small
enough field strength the Fermi wavelength is smaller than the lattice
constant. Hence, an electron exhibits an essentially classical transport
process, where in good approximation the single electrons do not interact with
each other; see Refs.\ \cite{Weis91,LKP91,WLR97} for some experimental as well
as Refs.\ \cite{Geis90,FGK92} for corresponding theoretical work. We thus
emphasize that much care has to be taken regarding the physical significance
of the Lorentz gas concerning a general understanding of transport in real
matter. The usefulness of this model for physically realistic situations
should best be assessed from case to case.

In this review we focus on the two-dimensional version of the Lorentz gas on a
triangular lattice, so if not said otherwise with ``Lorentz gas'' we denote
precisely this configuration of scatterers. The unit cell of this dynamical
system, which is called {\em Wigner-Seitz cell} in solid state physics, is a
hexagon, see Fig.\ \ref{fig:plgasg} (a). Note that alternatively the
scatterers may also be put on a square lattice, in which case the unit cell is
sometimes called a {\em Sinai billiard} \cite{Sin70}. If no further external
constraints are added there is only one control parameter, which is the
density of scatterers $n$ or alternatively the smallest distance between two
scatterers, the gap size $w$, as depicted in Fig.\ \ref{fig:plgasg} (b) and
(c). In the triangular case three different dynamical regimes are identified:
\begin{enumerate}
\item Particles are localized in space, i.e., they cannot leave a
unit cell, since all disks touch each other, see Fig.\ \ref{fig:plgasg}
(b). This is the high density regime with a minimal gap size $w$ between two
adjacent disks of $w=0$, and the diffusion coefficient is trivially zero.
\item There exist trajectories along which particles can move
ballistically and collision-free for an infinite time, i.e., under certain
initial conditions particles see ``infinite horizons'' in the Lorentz gas as
indicated by Fig.\ \ref{fig:plgasg} (c). This happens for
$w>4/\sqrt{3}-2\simeq 0.3094$ and marks the onset of a low-density regime in
which the diffusion coefficient is infinite.
\item In-between these two regimes the horizon for the moving particles is
finite. As has been proven in Refs.\ \cite{BuSi80a,BuSi80b,BuSiCh91,Chern99},
in this case a central limit theorem for the velocities exists and the
diffusion coefficient is finite. This is related to an exponential decay of
the velocity autocorrelation function of the moving particle
\cite{MaZw83,MaMa97} from which the existence of a diffusion coefficient
follows according to the Green-Kubo formula for diffusion Eq.\ (\ref{eq:dgk}).
\end{enumerate}
On this occasion we may refer to a further simple physical interpretation of
the periodic Lorentz gas reflecting these three different dynamical regimes:
The scattering of a point particle with a Lorentz gas disk may be looked at as
a two-body problem, where a particle of a reduced mass $\mu$ moves with a
relative velocity ${\bf v}$ and scatters, in this case, with a hard core
potential $V(R)$. By suitably rescaling the radii of the moving particle and
of the fixed scatterer, whose sum should be constant, and by periodically
continuing the unit cell including the moving particle, the dynamics may be
considered as the collective motion of some lattice modes in a periodic medium
such as a crystal. In this case the periodic Lorentz gas was coined the {\em
correlated cell model} \cite{AHW63,DePo96}.

According to the three different regimes discussed above there are also three
different regimes of lattice modes: In case 1 the lattice modes are again
localized, in case 2 dislocations are possible, whereas case 3 may be
interpreted as some kind of ``melting''.  This relation of the periodic
Lorentz gas to the correlated cell model was used to prove the existence of a
shear and a bulk viscosity in what was called a ``two-particle fluid''
\cite{BuSp}. Note also that, in kinetic theory, the random Lorentz gas is
well-known as a model for a gas mixture of heavy and light particles, where
the difference of the masses is large enough such that the heavy particles, in
good approximation, can be considered to be immobile with respect to the light
particles \cite{ChCo}.

The periodic Lorentz gas exhibits a number of important ``nice'' dynamical
properties.\footnote{For readers that are not too familiar with dynamical
systems theory this paragraph may be skipped; alternatively, for respective
definitions of dynamical systems quantities we refer to Refs.\
\cite{Schu,Ott,Gasp,Do99}.} First of all, it is Hamiltonian and thus deterministic,
area-preserving and time-reversible. Secondly, it has been proven that the
Lorentz gas is a K-system, which implies that it is mixing and ergodic
\cite{BuSi80a,BuSi80b,BuSiCh91,Chern99}. Furthermore, due to the defocusing
character of the hard disks\footnote{This type of scatterers is said to be
{\em convex}, or {\em dispersing}.} the Lorentz gas is a hyperbolic, chaotic
dynamical system. However, because of the hard walls it is not differentiable,
as is exemplified by the existence of tangent collisions, and therefore
strictly speaking it is not Anosov or Axiom A \cite{Gasp96,Gasp}. Finally,
corresponding to the three independent variables of the periodic Lorentz gas
there are three different Lyapunov exponents reflecting its hyperbolic
behavior: One is zero, which is the one parallel to the flow, the other two
are smaller respectively larger than zero, and because the system is phase
space preserving all Lyapunov exponents trivially sum up to zero.

We remark that for random configurations of scatterers in two and three
dimensions the Lyapunov exponents have been computed analytically by means of
kinetic theory, partly in comparison with results from computer simulations
\cite{BD95,LvBD97,BDPD97,DePo97,BLD98}. For the two-dimensional 
periodic Lorentz gas both on a quadratic and on a triangular lattice computer
simulation results for the Lyapunov exponents were presented in Refs.\
\cite{GaBa95,Gasp}. Random and periodic Lorentz gases are prototypical models
of so-called {\em particle billiards} that have been widely studied in the
field of chaos and transport both from the physical as well as from the
mathematical side of dynamical systems theory. Hence there exists a large
literature that the reader may wish to consult in order to learn about more
detailed properties, see, e.g., Refs.\ \cite{Gasp,Do99,Sza00} and further
references therein.

\subsection{Summary}

\begin{enumerate}
\item We started this chapter by elaborating on the physical meaning of a
thermal reservoir. This discussion was performed on a very elementary
heuristic level. The action of a thermal reservoir on a subsystem was first
outlined for thermal equilibrium. In nonequilibrium situations thermal
reservoirs are generally indispensable in order to generate NSS.

\item As a well-known example for modeling thermal reservoirs
we have heuristically introduced the simplest form of a stochastic Langevin
equation. We have also briefly summarized some of its most important
properties. It was then shown how this Langevin equation can be derived
starting from Hamiltonian equations of motion, where the heat bath is modeled
by a collection of non-interacting harmonic oscillators. On this basis we have
discussed deficiencies of both schemes as far as the modeling of thermal
reservoirs is concerned, and we have mentioned some more recent approaches
that try to go beyond this simple modeling.

\item On a more statistical level, we have outlined how to calculate the
equilibrium velocity distribution functions for a single velocity component as
well as for the absolute value of two velocity components for a subsystem that
is coupled to a $d_r$-dimensional thermal reservoir. In the limit of
$d_r\to\infty$ we recovered the corresponding well-known canonical
distributions, as was to be expected. We indicated that these ``ideal''
velocity distribution functions, according to equilibrium statistical
mechanics, may not by default be properly reproduced by an arbitrary
deterministic dynamical subsystem such as the harmonic oscillator, coupled to
some model of a thermal reservoir. Related problems for modeling NSS were also
briefly discussed.

\item Finally, we introduced a simple deterministic dynamical system, the
periodic Lorentz gas, which exhibits ``nice'' dynamical properties in an
equilibrium situation by being, among others, Hamiltonian, mixing and fully
chaotic. We started with a brief history of the (periodic) Lorentz gas,
discussed its physical significance and some possible physical interpretations
as far as physical reality is concerned. We classified its three different
dynamical regimes and very briefly summarized its most important mathematical
properties. In the following, the periodic Lorentz gas will serve as our
standard model in order to compare the action of different thermal reservoirs
with each other, particularly in nonequilibrium situations.

\end{enumerate}

\section{The Gaussian thermostat}

We now introduce a very popular deterministic and time-reversible modeling of
a thermal reservoir which is known as the {\em Gaussian thermostat}. We assign
this thermostating scheme to the periodic Lorentz gas driven by an external
electric field and summarize what we consider to be the most important
features of the resulting model from a dynamical systems point of view. The
properties we report are to a large extent typical for Gaussian thermostated
systems altogether. Specifically, we review a connection between thermodynamic
entropy production and the average phase space contraction rate as well as a
simple functional relationship between Lyapunov exponents and transport
coefficients. Furthermore, there exists a fractal attractor in the Gaussian
thermostated Lorentz gas which changes its topology under variation of the
electric field strength. Correspondingly, the electrical conductivity is an
irregular function of the field strength as a control parameter. We also
briefly elaborate on the existence of linear response in this model.

\subsection{Construction of the Gaussian thermostat}

In the following chapters the periodic Lorentz gas serves as a standard model
to which different thermal reservoirs are applied. A big advantage of the
Lorentz gas, say, in comparison to the harmonic oscillator is that it exhibits
deterministic chaos, which is a consequence of the defocusing geometry of the
scatterers. Hence, in contrast to the Langevin dynamics of Eq.\
(\ref{eq:lang1d}) we do not need to impose a stochastic force onto this model
in order to enforce Brownian motion-like spatial fluctuations for a moving
particle.\footnote{On the other hand, without applying an external field the
absolute value of the velocity for the moving particle is yet constant in the
Lorentz gas, which is at variance to ordinary Langevin dynamics. However, see
Chapter IV for respective further modifications of the Lorentz gas.} Instead,
some random-looking trajectories in position space are generated due to the
intrinsic Lyapunov instability of the system, see again Fig.\
\ref{fig:plgasg}. This is in turn intimately related to the mixing and
ergodic properties of the Lorentz gas. We thus replace stochasticity by
deterministic chaos in order to generate a Brownian motion-like dynamics, cp.\
also to our discussion in Section II.B.

As was outlined in the previous chapter, in a certain regime of densities of
scatterers the periodic Lorentz gas exhibits normal diffusion related to an
exponential decay of the velocity autocorrelation function, in analogy to the
diffusive properties of Langevin dynamics. In this review we do not study how
the Lorentz gas dynamics changes under variation of the density of scatterers,
see Refs.\ \cite{KlDe00,KlKo02}. Instead, we simply choose one specific
parameter value for the minimal distance between two scatterers, $w\simeq
0.2361$, as it is standard in the literature to ensure that a diffusion
coefficient exists \cite{MH87,LRM94,LNRM95,DeGP95,MDI96,DeMo96a}.

In order to drive this model out of equilibrium we apply an external electric
field that, if not said otherwise, is parallel to the $x$-axis. However, as
explained before, if there were no interaction with a thermal reservoir any
moving particle would on average be accelerated by the external field
consequently leading to an ongoing increase of energy in the system, and there
were no NSS. Therefore the model must be connected to some thermal reservoir.
In other words, it must be thermostated. The periodic Lorentz gas amended by
an external field and coupled to a suitable thermal reservoir is called the
{\em driven periodic Lorentz gas}.  This model has been widely studied in the
literature over the past fifteen years
\cite{MH87,HMHE88,HoMo89,HoMo92,Vanc,Ch1,Ch2,BarEC,LuBr93,LRM94,LNRM95,DMR95,DeGP95}
\cite{MDI96,DeMo96a,DeMo97b,DMR97,BGG97,CoRo98,MoDe98,HP98,HoB99,Do99,TVS00,BDL00,DettS00,MLL01,BDL+02,Voll02}
\cite{LLM02}.

Following our discussion of the Langevin equation one may remove energy from
the field-driven Lorentz gas according to
\bna
{\bf \dot{r}}&=&{\bf v}\nonumber\\ 
{\bf \dot{v}}&=&\mbox{{\boldmath $\varepsilon$}}-\alpha({\bf v}) {\bf v} 
\quad , \label{eq:eomdlg}
\ena
supplemented by the geometric boundary conditions imposed by the Lorentz gas
scatterers. With $\alpha\equiv\alpha({\bf v})=const.$ we have an ordinary
Stokes friction term, and the equations represent a deterministic variant of
the stochastic Langevin equation Eq.\ (\ref{eq:lang1d}) amended by an external
electric field $\mbox{{\boldmath $\varepsilon$}}$. Here and in the following
the electric charge $q$ of the particle is set equal to one. This viscous
Lorentz gas has been studied by computer simulations in Ref.\
\cite{HoMo92}.

For numerical solutions one has to discretize Eqs.\ (\ref{eq:eomdlg}) in
time. The second of the above two equations then reads
\be
{\bf v}(t+\Delta t)=\mbox{{\boldmath $\varepsilon$}}\Delta t + \tilde{\alpha}{\bf v}(t)
\quad , \quad 0<\Delta t\ll 1 \quad , \label{eq:vscal}
\ee
with $\tilde{\alpha}:=1-\alpha\Delta t$. For small enough $\Delta t$ there is
$0<\tilde{\alpha}<1$, and by keeping the time interval fixed the viscous force
amounts to a rescaling of the velocity periodically in time with a constant
scaling factor $\tilde{\alpha}$. This procedure can be made more efficient by
replacing $\tilde{\alpha}$ according to
\be
\tilde{\alpha}({\bf v}):=\sqrt{\frac{E}{E(t)}} \quad , \label{eq:velresc}
\ee
where $E$ is the target kinetic energy of the system and $E(t)$ the measured
kinetic energy at time $t$. The scaling factor now depends on the velocity of
the moving particle and may thus fluctuate in time. This elementary but rather
convenient numerical method to keep the energy fixed became well-known for
molecular dynamics computer simulations under the name of {\em velocity
rescaling} \cite{Ev83,AT87,HHP87,HVR95,Hess96,HKL96,HoKu97,DettS00}. Note that
both the Lorentz gas with a Stokes friction coefficient and with rescaled
velocities is deterministic but not time-reversible.

In order to further improve the efficiency of these thermostating mechanisms
one may start directly from a velocity-dependent friction coefficient in Eq.\
(\ref{eq:eomdlg}), $\alpha\equiv\alpha({\bf v})$.  Requiring energy
conservation at any time step implies $d{\bf v}^2/dt=0$ as a constraint on the
functional form of $\alpha({\bf v})$ and yields
\be
\alpha({\bf v})=\mbox{{\boldmath $\varepsilon$}}\cdot{\bf v}/v^2 \quad . \label{eq:alpdlg}
\ee
This method of thermostating was proposed simultaneously and independently by
Hoover and coworkers \cite{HLM82} and by Evans \cite{Ev83} in 1982.  Performed in
the limit of $\Delta t\to0$, the velocity rescaling Eqs.\ (\ref{eq:vscal}),
(\ref{eq:velresc}) is identical to the combination Eqs.\ (\ref{eq:eomdlg}),
(\ref{eq:alpdlg}) \cite{Ev83,HHP87,HVR95,Hess96,HoKu97,DettS00}. 

In contrast to ordinary Stokes friction that only reduces the kinetic energy,
the velocity-dependent version Eq.\ (\ref{eq:alpdlg}) pumps energy into the
system whenever the particle is moving opposite to the field.  This friction
is maximal when the velocity is parallel to the field, and it is zero when
velocity and electric field are perpendicular to each other. Trivially, it is
zero when the electric field is zero due to the fact that in the equilibrium
periodic Lorentz gas a particle is moving with a constant absolute value of
the velocity anyway. Still, in comparison to the friction of the
deterministic, generalized Langevin equation Eq.\ (\ref{eq:zwlang}) that was
obtained starting from the heat bath of harmonic oscillators, such a
generalization of ordinary Stokes friction may not be considered physically
unreasonable.

Alternatively, Eq.\ (\ref{eq:alpdlg}) can be derived from Gauss' principle of
least constraints \cite{EH83,EvMo90,MoDe98,Rond02}, which is a fundamental
principle of classical mechanics.\footnote{In contrast to d'Alembert's
principle, for the Gaussian version the virtual displacements are acceleration
terms instead of the positions of a moving particle.} For this reason a
velocity-dependent friction force that keeps the energy of a particle constant
at any time step was coined the {\em Gaussian thermostat}
\cite{EH83}. Correspondingly, the Gaussian thermostat applied to the
driven periodic Lorentz gas generates a microcanonical-like
distribution in velocity space \cite{MoDe98}.\footnote{Note that in
nonequilibrium the distribution of points on the energy shell is not uniform
anymore reflecting the existence of an average current.} 

In fact, here we are constraining only the kinetic energy of the moving
particle, therefore this version is sometimes further classified as the {\em
Gaussian isokinetic thermostat}.  For soft particles one may fix alternatively
the total internal energy consisting of the sum of kinetic plus potential
energy yielding the {\em Gaussian isoenergetic thermostat}, see Refs.\
\cite{EvMo90,MoDe98,DettS00,Rond02} for these denotations. Suitably amended
versions of the Gaussian thermostat can also be applied, for example, to
many-particle systems under shear, see Chapter VII for further discussions.

Similar to the case of Stoke's friction which models the loss of energy of a
moving particle into a surrounding viscous fluid one may associate with the
Gaussian friction variable Eq.\ (\ref{eq:alpdlg}) some imaginary thermal
reservoir that acts at any instant of time. Thus the Gaussian thermostat is
another example of a so-called {\em bulk thermostat}. We remark that this
physical picture contradicts the specific interpretation of the driven
periodic Lorentz gas as a model for noninteracting electrons moving in a
periodic crystal. Here one may expect that energy is only dissipated at the
collisions with a scatterer and not during a free flight. However, the action
of the Gaussian thermostat is conceptually fine if the Lorentz gas is
considered as a two-particle fluid that is immersed into another fluid serving
as a thermal reservoir. Let us furthermore remark that in Chapter VII we will
introduce boundary thermostats for the Lorentz gas, where a particle exchanges
energy with a thermal reservoir only at the collisions with a scatterer. On
this occasion we will discuss similarities and differences to what we find for
the Gaussian thermostated case summarized below.

Surprisingly, the equations of motion Eqs.\ (\ref{eq:eomdlg}) supplemented by
Eq.\ (\ref{eq:alpdlg}) are time-reversible consequently modeling the, at first
view, seemingly contradictory situation of a deterministic, time-reversible,
dissipative dynamical system \cite{HKP96}. As will be outlined in the
following, under certain conditions the Gaussian thermostated driven Lorentz
gas furthermore exhibits an ergodic and chaotic particle dynamics and
corresponding well-defined NSS. In contrast to standard Stoke's fricition and
to stochastic thermostats the Gaussian scheme thus enables to elaborate on the
fundamental problem how time-reversible, deterministic, microscopic equations
of motion may generate macroscopic irreversible transport in nonequilibrium
situations associated with energy dissipation.

\subsection{Fundamental relations between chaos and transport for Gaussian
thermostated dynamical systems}

We now analyse the Gaussian thermostated driven periodic Lorentz gas from a
dynamical systems point of view and summarize what we consider to be the
characteristic chaos and transport properties of this class of dynamical
systems.

\subsubsection{Phase space contraction and entropy production}

A first fundamental property of Gaussian thermostated systems is obtained from
computing their {\em average phase space contraction rate} defined by
\be
\kappa :=<\mbox{{\boldmath $\nabla$}}\cdot{\bf F}({\bf r},{\bf v})> \quad , \label{eq:psc}
\ee
where ${\bf F}({\bf r},{\bf v})$ stands for the equations of motion of the system
Eqs.\ (\ref{eq:eomdlg}) and the brackets denote an ensemble average. A
straightforward calculation, by taking into account that $\alpha$ is a
function of ${\bf v}$, yields
\be
\kappa =-<\alpha> \label{eq:peqa}
\ee
with $\alpha$ defined by Eq.\ (\ref{eq:alpdlg}), where $\kappa <0$. Let us now
rewrite this equation by introducing a nonequilibrium temperature $T$. If we
want to do so by using equipartitioning of energy \cite{EvMo90,MoDe98,HoB99}
we need to know about the number of degrees of freedom of the system.  The
velocity space is two-dimensional, but one may argue that the energy
constraint eliminates one degree of freedom of the system. However, this
argument is debatable, and actually there is quite an ambiguity in defining a
temperature for the driven Lorentz gas \cite{Ch1,Ch2}.\footnote{If we switch
back to the Lorentz gas without an external field we also, trivially, have
energy conservation. However, here nobody would claim that the number of
degrees of freedom is reduced respectively.}  By assuming that there is one
degree of freedom only we arrive at
\be
v^2=T \quad , \label{eq:eqoe}
\ee
with $k_B\equiv1$. We can now write
\be
-\kappa =\mbox{{\boldmath $\varepsilon$}}\cdot<{\bf v}>/T \quad , \label{eq:pepid}
\ee
where the right hand side is the familiar expression related to the Joule
heating derived from irreversible thermodynamics. Alternatively, one could
have started from Clausius' definition of entropy production in terms of the
heat transfer $dQ$ between subsystem and thermal reservoir, $\Delta S=\int
dQ/T$ \cite{deGM84,Huang,DettS00,Gall03b}, yielding the same
equation.\footnote{In the $d$-dimensional case and for $N$ particles there is
a prefactor of $Nd-1$ on the right hand side of Eq.\ (\ref{eq:peqa}), but this
factor cancels with a respective prefactor on the right hand side of Eq.\
(\ref{eq:eqoe}) such that the identity is preserved \cite{CoRo98}. Note also
that the Clausius form is only valid for quasistatic processes in, at least,
local thermodynamic equilibrium.} We thus arrive at the fundamental result
that, for Gaussian thermostated systems, there is an {\em identity between the
absolute value of the average rate of phase space contraction and
thermodynamic entropy production}
\cite{HHP87,PoHo87,PoHo88,EvMo90,ChLe95,ChLe97,Ch1,Ch2,VTB97,BTV98,GiDo99,HoB99,DettS00,VTB03}.
A third way to obtain this identity starts from Gibbs' definition of entropy
production, as will be discussed in Chapter V.

\subsubsection{Lyapunov exponents and transport coefficients}
Note that Eq.\ (\ref{eq:pepid}) provides a crucial link between dynamical
systems properties, in terms of phase space contraction, and thermodynamics,
in terms of entropy production. By employing this link a second fundamental
property for Gaussian thermostated systems is easily derived as follows: First
of all, there is an identity between the average phase space contraction of a
dynamical system and the sum of Lyapunov exponents \cite{ER,Gasp}. For the
driven Lorentz gas under consideration two Lyapunov exponents are zero, due to
the energy constraint and assessing the neutral direction parallel to the
trajectory of a moving particle. From the remaining two exponents one is
positive and one is negative, at least for small enough field strength, see,
for example the numerical results of Refs.\
\cite{LNRM95,DGP95,DMR95,DeMo96a,MDI96,MoDe98}. Consequently we have
\be
\kappa =\lambda_++\lambda_- \quad . \label{eq:pscl}
\ee
Let us now replace the average current by introducing the field-dependent
electrical conductivity $\sigma(\varepsilon)$ according to\footnote{Strictly
speaking this quantity may rather be called the {\em mobility} of a particle,
in distinction from the conductivity that involves the charge and the density
per unit volume of the mobile electrons of a conductor
\cite{Wan66,Reif}. However, in our case there is no difference.}
\be
<{\bf v}>=:\sigma(\varepsilon)\mbox{{\boldmath $\varepsilon$}}\quad , \label{eq:ohm}
\ee
where $\varepsilon$ stands for the absolute value of the field strength. For
sake of simplicity here we consider only the case of the field being parallel
to the $x$-axis. Because of the symmetry of the two-dimensional model
$\sigma(\varepsilon)$ is then a scalar quantity, whereas for general direction
of the field strength one has to employ a tensorial formulation
\cite{LNRM95}. Replacing the average current $<{\bf v}>$ on the right hand
side of Eq.\ (\ref{eq:pepid}) by Eq.\ (\ref{eq:ohm}) and the average phase
space contraction rate $\kappa$ on the other side by Eq.\ (\ref{eq:pscl}) we
arrive at a simple relation between the conductivity $\sigma(\varepsilon)$ and
the sum of Lyapunov exponents of the dynamical system,
\be
\sigma(\varepsilon)=-\frac{T}{\varepsilon^2}(\lambda_+(\varepsilon)+\lambda_-(\varepsilon))
\quad . \label{eq:lsr} 
\ee
This important connection between transport coefficients and dynamical systems
quantities was first reported by Posch and Hoover for a many-particle system
with an external field \cite{PoHo88} and was called the {\em Lyapunov sum
rule} later on \cite{ECM}. For the driven periodic Lorentz gas its existence
was rigorously proven in Refs.\ \cite{Ch1,Ch2}. Eq.\ (\ref{eq:lsr}) can also
be related to periodic orbits \cite{Vanc} and was studied in much detail
numerically \cite{BarEC,LNRM95,DGP95}.  Functional forms that are completely
analogous to Eq.\ (\ref{eq:lsr}) also hold for many-particle systems and for
other transport coefficients such as viscosity \cite{ECM,EvMo90,Coh95,SEC98}
and thermal conductivity \cite{AK02} under the necessary condition that heat
is dissipated by Gaussian and related thermostats \cite{MoDe98}.

The Lyapunov sum rule moved particularly into the center of interest starting
from the work by Evans, Cohen and Morriss \cite{ECM} who introduced a further
simplification of it for systems with many degrees of freedom. The observation
was that for dissipative many-particle systems the spectrum of Lyapunov
exponents may exhibit a remarkable symmetry property
\cite{PoHo88,Morr89,ECM}. This was first reported for the case of a uniform
viscous damping by Dressler \cite{Dress88} and was later on called the {\em
conjugate pairing rule}: If all non-zero Lyapunov exponents are ordered
according to their values they can be grouped into pairs such that each pair
sums up to precisely the same value. This pairing rule is well-known for
Hamiltonian dynamical systems, however, here the Lyapunov exponents appear in
pairs that always sum up to zero. Indeed, in phase space preserving dynamical
systems it can be shown that conjugate pairing is a direct consequence of the
symplecticity of the Hamiltonian \cite{ER,Meis92,AbMa94,Gasp}.

If conjugate pairing holds for the very different class of dissipative,
thermostated systems it enables to replace the sum over the full spectrum of
Lyapunov exponents which appears on the right hand side of Eq.\ (\ref{eq:lsr})
by the sum over one pair of exponents only. For higher-dimensional dynamical
systems this drastically simplifies the computation of transport coefficients
according to Eq.\ (\ref{eq:lsr}), since the total number of different
exponents is here extremely large and the full spectrum is very hard to
extract numerically. Instead, for a conjugate pair one may just choose the
largest and the smallest Lyapunov exponent, where the largest exponent is
typically not too difficult to get \cite{BGGS80b,WSSV85,Ott}. The smallest
one, in turn, is best computed by reversing the direction of time
\cite{HTP88,ECM,Coh95,vBDCP96,LvBD97,SEC98,SEI98,TM02b}. 

However, note that the reversed trajectories still have to be constrained to
the invariant set associated with the forward dynamics. As we will further
explain in the following section, in dissipative systems this set is usually a
{\em fractal attractor} that becomes a {\em fractal repeller} under time
reversal, and it requires some additional numerical efforts to keep the
trajectories in this now unstable region. From a computational point of view
the Lyapunov sum rule Eq.\ (\ref{eq:lsr}) supplemented by conjugate pairing
thus does not appear to provide a more efficient mean for computing transport
coefficients than other existing standard procedures. For example, along the
lines of Eq.\ (\ref{eq:lsr}) it were easier to compute the average value of
the thermostating variable instead of any Lyapunov exponent, see Eqs.\
(\ref{eq:peqa}) and (\ref{eq:pscl}).

Conjugate pairing was analytically proven to hold under certain conditions
that particularly concern the pecific form of the interparticle potential and
the special type of deterministic thermostat used
\cite{DeMo96b,DeMo97a,MoDe98,WL98,Rue99,Pan02}; see also the kinetic theory
calculations for the three-dimensional Gaussian isokinetically
thermostated driven random Lorentz gas in Refs.\
\cite{LvBD97,BLD98,vBLD00} and the respective comparison to computer
simulations \cite{DePo97}. Conjugate pairing in the three-dimensional driven
periodic Lorentz gas, again connected to a Gaussian isokinetic thermostat, was
numerically corroborated in Refs.\ \cite{DMR95,MDI96}. 

For interacting many-particle systems under shear the situation concerning the
validity of conjugate pairing appears to be considerably less clear, see
Refs.\ \cite{ECM,Coh95,GKC94,MoDe98} and the very recent discussion in Refs.\
\cite{Mor02,TM02b,PvZ02b,PvZ02}; see also Refs.\
\cite{DePH96,SEM92,SEI98} for numerical studies of various thermostated
systems in which conjugate pairing appears to hold, respectively does not
appear to hold.  Generally, it seems to be easier to give sufficient
conditions under which conjugate pairing does {\em not} hold: one example is a
three-dimensional periodic Lorentz gas under electric and magnetic fields for
which there is, or is not, conjugate pairing depending on the specific
direction of the two fields \cite{DK00}. Another generic class of
counterexamples to conjugate pairing is provided by systems that are
thermostated at the boundaries \cite{PoHo88,DePo97b,Wag00}. The latter models
indicate that conjugate pairing is intimately related to the thermostat acting
homogeneously and `democratically' on all particles in the bulk of a system
\cite{MoDe98,ECS+00}, whereas inhomogeneous thermostating at the boundaries prevents
the Lyapunov exponents to come in conjugate pairs. In Chapter VII we will more
explicitly discuss counterexamples of the latter type.

\subsubsection{Fractal attractors characterizing nonequilibrium steady states}

How does it fit together that {\em time-reversible} microscopic equations of
motion yield {\em irreversible} macroscopic transport in a NSS? As was first
discussed by Hoover et al.\ \cite{HHP87,MH87} and by Morriss \cite{Morr87},
for this type of driven thermostated systems reversibility and irreversibility
are intimately linked to each other by the existence of a {\em fractal
attractor} in phase space; see Refs.\
\cite{Morr87,Mo89a,EvMo90,PIM94} for a sheared two-particle system, Refs.\
\cite{PoHo87,PoHo88,Morr89,PoHo89,HoPo94,DePo97b,Wag00} for many-particle systems under
various nonequilibrium conditions, Refs.\
\cite{MH87,HMHE88,HoMo89,Vanc,HoMo92,Ch1,Ch2,LuBr93,LRM94,LNRM95}
\cite{DeGP95,DeMo96a,DMR97,MoDe98,HP98,KRN00,RKN00,RKH00,RaKl02} for the
driven periodic Lorentz gas and Refs.\
\cite{HHP87,KBB90,Hoo91,TeVB96,PH97,PH98,GiDo99,GFD99,HoB99,TVS00,Voll02} for
related models.\footnote{Here we are using the term {\em attractor} according
to the operational definition of Eckmann and Ruelle
\cite{ER}: Let there be points in phase space generated by the
equations of motion of a dynamical system. The attractor is the set on which
these points accumulate for large times. For more rigorous mathematical
definitions see, e.g., Refs.\
\cite{ER,Gasp,GH,Dev89}. Sometimes the additional term {\em strange} is
employed to further characterize attractors
\cite{RT71}. However, its use appears to be quite ambiguous in the 
literature, partly indicating the existence of a chaotic dynamics on the
attracting set \cite{Beck,Schu,ER,Lev}, partly being synonymous for the
attractor exhibiting a fractal structure \cite{Ott,GOPY84}. Hence, here we
avoid this denotation.}

\begin{figure}[t]
\epsfxsize=6cm
\centerline{\epsfbox{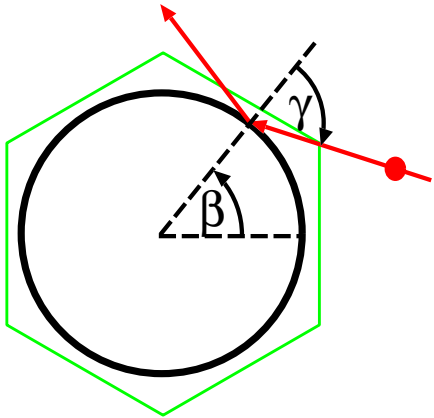}}
\vspace*{0.5cm} 
\caption{Phase space coordinates describing the collision of a point particle with a
disk in case of the two-dimensional periodic Lorentz gas. $\beta$ denotes the
position of the colliding particle on the circumference of the disk, whereas
the impact angle $\gamma$ is related to the direction of the velocity of the
moving particle before the collision.}
\label{plko}
\end{figure}

We now outline characteristic features of the attractor of the driven periodic
Lorentz gas. For convenience the temperature of the thermostat is fixed at
$T=1$. In order to simplify the analysis a Poincar\'e surface of section is
considered. That is, we introduce a further constraint in position space by
representing the dynamics in terms of the phase space coordinates at a
collision with a disk only. These collisions are defined by the position on
the circumference of the disk according to the angle $\beta$ and by the impact
angle $\gamma$ related to the direction of the velocity of the moving
particle, see Fig.\ \ref{plko}. The set $(\beta,\sin\gamma)$ forms the {\em
Birkhoff coordinates} of a particle billiard \cite{Gasp}. A convenient
property of the Birkhoff coordinates is that in the Hamiltonian case at field
strength $\varepsilon=0$ the invariant probability density
$\rho(\beta,\sin\gamma)$ associated with these coordinates is simply uniform
in the whole accessible phase space.

\begin{figure}[t]
\epsfxsize=12cm
\centerline{\epsfbox{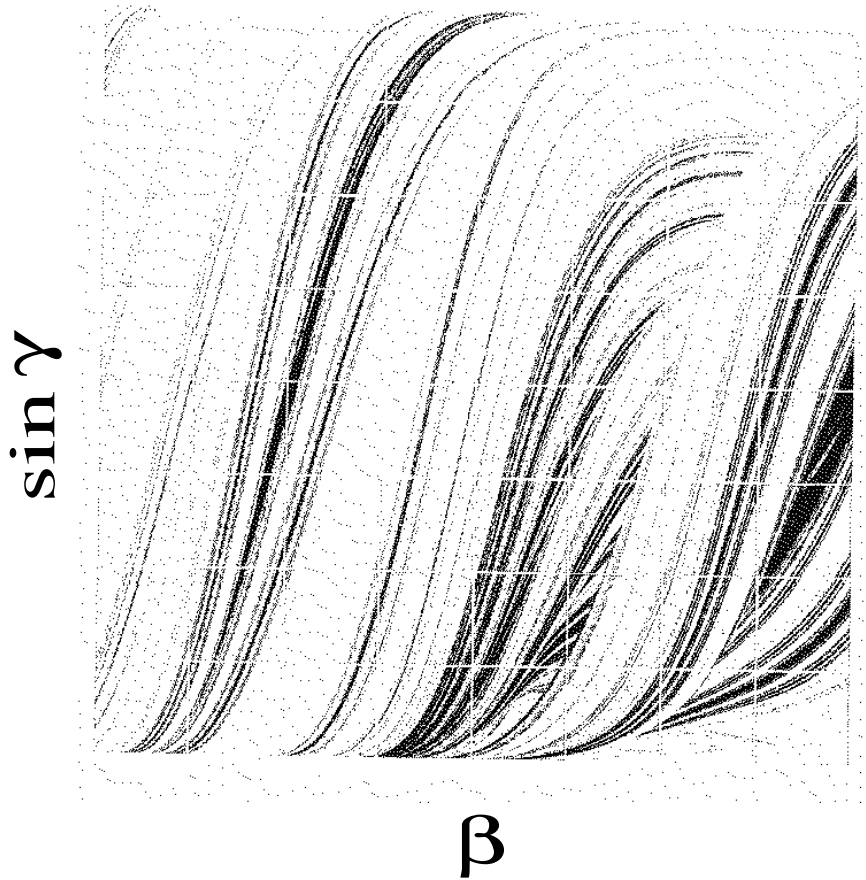}}
\vspace*{0.3cm} 
\caption{Attractor for the Gaussian thermostated periodic Lorentz gas driven
by an electric field of strength $\varepsilon=1.5$ which, here, is parallel to
the $y$-axis. The temperature is $T=1$, $\beta$ and $\sin\gamma$ are defined
in Fig.\ \ref{plko}. The figure, originally being a 1m-square, is from Ref.\
\protect\cite{HoMo89}. For related results see Refs.\ \protect\cite{MH87,HMHE88,DeGP95,HoB99}.}
\label{gatt}
\end{figure}

In contrast to that, Fig.\ \ref{gatt} shows that in case of dissipation
generated by an electric field and counterbalanced by a Gaussian thermostat
the phase space density contracts onto a complicated fractal-looking
object. Numerical analysis provides evidence that the attractor shown in Fig.\
\ref{gatt} is {\em multifractal}
\cite{Ott,Beck,Lev}: Although the two-dimensional phase space spanned by the
Birkhoff coordinates is fully covered for small enough field strength yielding
a box-counting dimension of two in computer simulations, the higher-order
Renyi dimensions such as the information dimension and the correlation
dimension are non-integer and slightly smaller than two indicating an
inhomogeneous fractal folding of the phase space
\cite{HoMo89,DeMo96a,DettS00}. 

However, note that here we are discussing the attractor in the reduced phase
space defined by the Poincar\'e surface of section only. As far as the whole
accessible phase space corresponding to the three independent phase space
variables of the model is concerned there is a proof that the Hausdorff
dimension of this fractal set is less than three and identical to the
information dimension for non-zero but small enough field strength
\cite{Ch1,Ch2}, which appears to be consistent with computer simulation
results \cite{DeGP95}.

The basic structure of the attractor shown in Fig.\ \ref{gatt} can roughly be
understood as follows: points in phase space with an impact angle of
$\gamma=\pm\pi/2$ are reminiscent of {\em tangent collisions} where a moving
particle just passes a disk. These points are due to the hard disk geometry
and yield discontinuities in the equations of motion which separate regions in
phase space corresponding to different types of trajectories.\footnote{With
different types of trajectories we mean that trajectories of different regions
hit different scatterers in the spatially extended periodic Lorentz gas, as
can be quantitatively assessed in terms of a symbolic dynamics by assigning
different symbols to different scatterers, see, e.g., Refs.\
\cite{LRM94,LNRM95,Gasp96,Gasp,MoDe98} for further details.}  Calculating
higher iterates of these lines of discontinuity from the equations of motion
yields complicated boundaries in phase space which match to the structures
depicted in Fig.\ \ref{gatt}, see Refs.\ \cite{Gasp96,Gasp} for the field-free
Lorentz gas and Refs.\ \cite{MH87,HMHE88,BDL00} for the driven case.  We
remark that this procedure is intimately related to the construction of {\em
Markov partitions} for the periodic Lorentz gas
\cite{BuSi80a,BuSi80b,BuSiCh91,Chern99}. Note that the topology of these
structures changes under parameter variation, which has direct consequences
for physical quantities such as the electrical conductivity of the system, as
we will discuss in the following section.

\begin{figure}[t]
\epsfxsize=10cm
\centerline{\epsfbox{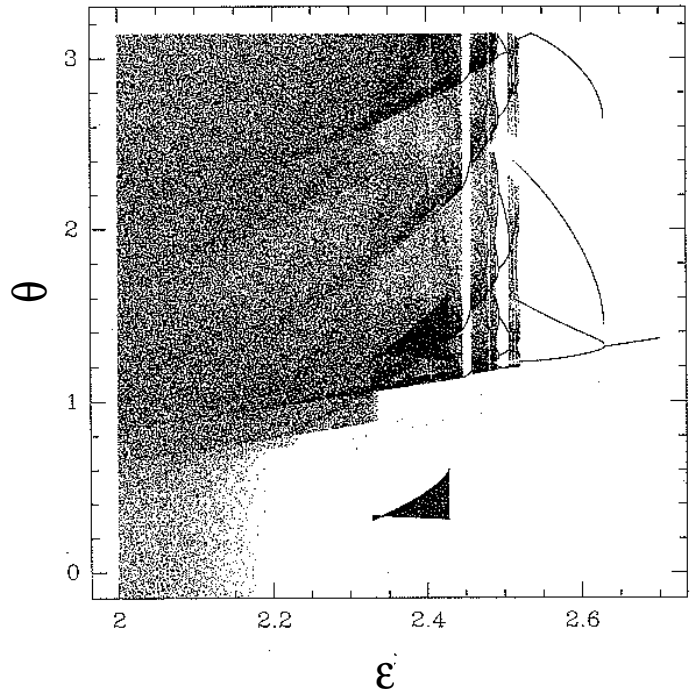}}
\vspace*{0.3cm} 
\caption{Bifurcation diagram for the Gaussian thermostated periodic Lorentz
gas driven by an electric field of strength $\varepsilon$ which, here, is
again parallel to the $x$-axis. $\theta$ is the angle between the $x$-axis and
the velocity of the moving particle after the collision and is a simple
function of $\gamma$ as defined in Fig.\ \ref{plko}. The temperature is set to
$T=1$. The figure is from Refs.\ \protect\cite{LRM94,LNRM95,MoDe98}.}
\label{gbifu}
\end{figure}

The bifurcation diagram Fig.\ \ref{gbifu} indicates how the attractor changes
by increasing the field strength $\varepsilon$
\cite{MH87,HoMo89,LRM94,LNRM95,DeMo96a,MoDe98}. Here we show results for the
angle $\theta$ which is the angle between the horizontal $x$-axis and the
velocity of the moving particle after the collision. Consequently, $\theta$ is
a simple function of the coordinate $\gamma$. Results for $\beta$ and $\gamma$
are analogous. This figure reveals a complicated bifurcation scenario: By
increasing the field strength the attractor first experiences a {\em crisis}
\cite{Ott} at which it suddenly collapses onto a subset in phase space
\cite{LRM94,LNRM95,DeMo96a,MoDe98}. For higher values of the field strength
there are Feigenbaum-like bifurcations showing a complicated interplay between
periodic windows and chaotic parameter regions. Around $\varepsilon=2.4$ there
is even an example of an elliptic fixed point in phase space. 

Note that for small enough field strength, that is, before the first crisis
occurs, the system is ergodic, see Refs.\
\cite{MH87,HoMo89,LRM94,LNRM95,DeMo96a,MoDe98} for numerical explorations and
Refs.\ \cite{Ch1,Ch2} for a proof. In the course of the bifurcations the
system exhibits a complex transition scenario from ergodic and chaotic to
non-ergodic and non-chaotic behavior. For high enough field strength the
attractor eventually collapses onto a single periodic orbit. This orbit is
such that a particle moves parallel to the field during a free flight, whereas
at a collision it ``creeps'' along the circumference of a disk
\cite{MH87,LRM94,LNRM95}.

For Gaussian-type thermostated dynamical systems the existence of multifractal
attractors in respective Poincar\'e sections appears to be typical. This
applies not only to a single-particle dynamics but also to interacting
many-particle systems under general nonequilibrium conditions, see the long
list of references cited at the beginning of this section for some
examples. The fractality also survives in chaotic dynamical systems where the
single particles interact with each other via smooth potentials
\cite{PoHo87,PoHo88,PoHo89,Morr89,PH97,HoKu97} and in systems that are
thermostated by a Stokes friction coefficient
\cite{HoMo92} or by some constant restitution coefficient at the collisions
with a scatterer \cite{LuBr93}. Important questions are whether the existence
of fractal attractors is a universal property of NSS in chaotic dynamical
systems connected to thermal reservoirs, irrespective of the specific
thermostat used, and to which extent the structure of the attractor depends on
the specific type of thermostat.

\subsubsection{Electrical conductivity and linear response for the Gaussian
thermostated driven periodic Lorentz gas}

\begin{figure}[t]
\epsfxsize=10cm
\centerline{\rotate[r]{\epsfbox{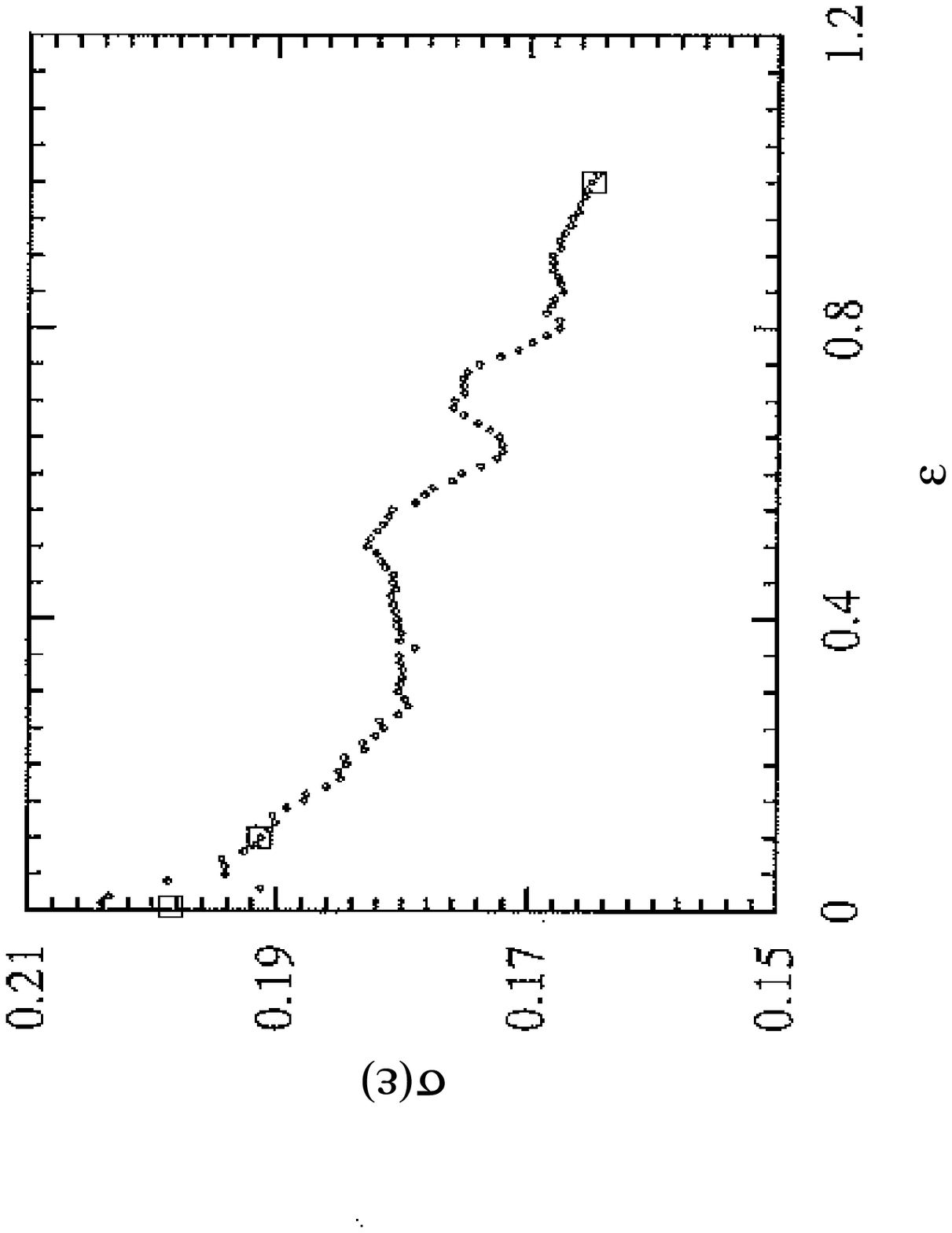}}}
\caption{Electrical conductivity $\sigma(\varepsilon)$ for the Gaussian thermostated
periodic Lorentz gas driven by an electric field of strength $\varepsilon$
which is parallel to the $x$-axis. The temperature is again $T=1$. The small
diamonds are data points computed from higher precision simulations, see Ref.\
\protect\cite{LNRM95} for further details which also contains the original
figure. Note the nonlinear response of this quantity and the irregular
structure on fine scales.}
\label{gcond}
\end{figure}

In the final part of this chapter we focus on the transport properties of the
Gaussian thermostated driven periodic Lorentz gas by studying the electrical
conductivity $\sigma(\varepsilon)$ introduced in Eq.\ (\ref{eq:ohm}). Here we
only discuss the ergodic regime of the model at field strengths
$\varepsilon\ll2$ for which $\sigma(\varepsilon)$ is uniquely defined. In this
case it can be obtained from computer simulations in form of the time average
over the velocity of a single particle sampled from a long trajectory. If the
system is non-ergodic the situation is more complicated, since the values for
the conductivity then depend on the choice of initial conditions in phase
space. Results for $\sigma(\varepsilon)$ in the ergodic regime calculated from
computer simulations \cite{LNRM95} are shown in Fig.\
\ref{gcond}. Related results have been obtained, also from simulations, in
Refs.\ \cite{MH87,DeGP95,BDL00,DettS00}; for calculations based on periodic
orbits see Refs.\ \cite{Vanc,DeMo97b}, for the conductivity of a driven
Lorentz gas thermostated by a Stokes friction coefficient see Ref.\
\cite{HoMo92}, and for conduction in a modified many-particle Lorentz
gas see Refs.\ \cite{BarEC,BGG97,BDL+02}.
  
Two aspects are remarkable in this figure: first of all, the validity of Ohm's
law implies that the field-dependent conductivity is simply constant,
$\sigma(\varepsilon)\equiv const$. However, here this quantity clearly
decreases on average thus indicating that, for the range of field strengths
shown in the figure, the response is already in the nonlinear regime. Indeed,
even for the smallest values of the field strength which are accessible to
simulations no linear response behavior is visible in this figure. These
results \cite{LNRM95} are in agreement with simulations performed by other
authors \cite{MH87,DeGP95,BDL00,DettS00}. On the other hand, in the limit of
$\varepsilon\to0$ the conductivity appears to converge to the correct value of
the diffusion coefficient obtained from the Einstein formula\footnote{This
equation is easily obtained from the Einstein relation Eq.\
(\ref{eq:einstvisc}) if an external electric field $\varepsilon$ is added to
the (one-dimensional) Langevin equation Eq.\ (\ref{eq:lang1d}). In a NSS the
average over the time derivative of the velocity on the left hand side of Eq.\
(\ref{eq:lang1d}) must be zero and the average over the noise on the right
hand side is by definition also zero, hence $E-\alpha <v>=0$. By using Eq.\
(\ref{eq:ohm}) one arrives at $\sigma=1/\alpha$ \cite{Reif,Path88}. Note that
for the Gaussian thermostated driven Lorentz gas Eqs.\ (\ref{eq:eomdlg}),
(\ref{eq:alpdlg}) the validity of these relations is not obvious.}
\be
D/T=\sigma(\varepsilon)\quad (\varepsilon\to0) \quad , \label{eq:dts}
\ee
which is $D\simeq 0.196$ for $T=1$
\cite{MaZw83,BarEC,GaBa95,MoRo94,DeGP95,BDL00,KlDe00}. This numerically
corroborates the analytical proof of the Einstein relation presented in Refs.\
\cite{Ch1,Ch2}. For sake of completeness let us remark that the conductivity
also depends on the density of scatterers. As expected, for higher densities
the response appears to be getting more linear in the regime of field
strengths shown in Fig.\ \ref{gcond} \cite{DeGP95}. However, to decide about
linearity or nonlinearity is still very ambiguous, since this depends on the
scale according to which the results are plotted; see also Fig.\ 4 in Ref.\
\cite{DettS00}.

A second fact makes the situation even more complicated, namely the existence
of irregularities in the conductivity as a function of the field strength in
form of non-monotonocities on fine scales, as was first reported in Ref.\
\cite{MH87}. The existence of such irregularities has also been verified by
cycle expansion calculations for higher values of the field strength
\cite{DeMo97b} and by other computer simulations \cite{LNRM95,DeGP95,BDL00},
hence they are not artifacts representing numerical imprecisions. Indeed, this
phenomenon extends the discussion of irregular and fractal transport
coefficients outlined in the introduction
\cite{RKD,RKdiss,KlDo99,GaKl,HaGa01,GrKl02,KoKl02} to the case of
thermostated systems: by using Birkhoff coordinates the periodic Lorentz gas
can be linked to maps having the same properties as the ones exhibiting
fractal transport coefficients
\cite{Gasp96,Gasp,GFD99,GaKl,TVS00,Voll02}. It was thus early conjectured that
the field-dependent conductivity of the driven Lorentz gas too exhibits
fractal properties \cite{RKD,RKdiss,KlDo99}. For one-dimensional maps with
broken symmetry yielding drift-dependent currents it was shown that the
conductivity is typically\footnote{i.e., for all control parameters except of
Lebesgue measure zero} a nonlinear fractal function of control parameters
\cite{GrKl02,GilDo03}. Similar results were obtained numerically for
the parameter-dependent diffusion coefficient in the field-free periodic
Lorentz gas \cite{KlDe00,KlKo02} and in related billiards of Lorentz-gas type
\cite{HaKlGa02}. However, to rigorously prove the existence of irregular
structures in transport coefficients is generally a hard task, even for the
most simple systems \cite{GrKl02}, as is exemplified by a recent analysis for
the driven Lorentz gas that was not conclusive in this respect \cite{BDL00}.

Apart from these numerical assessments of the Lorentz gas conductivity there
are seminal mathematical results proving Ohm's law for this model
\cite{Ch1,Ch2}. However, this proof does not yield any precise upper bound for
the validity of the range of linear response and may thus rather be considered
as an existence proof. Indeed, simple heuristic arguments quantifying such
bounds as discussed in Refs.\ \cite{Do99,Ch2} lead to wrong values for the
driven periodic Lorentz gas in comparison to simulation results
\cite{RKN00}. According to these mathematical and numerical analyses one may
conclude that a regime of linear response exists in the driven periodic
Lorentz gas which, however, is so small in the field strength that, up to now,
it could not properly be detected in computer simulations. Even then, an open
problem is the possible existence of fractal-like irregularities on fine
scales, which for the driven Lorentz gas one may expect to persist as well in
the limit of $\varepsilon\to0$ \cite{GrKl02}. If this is the case Ohm's law
may strictly speaking only hold with respect to performing a suitable coarse
graining for the conductivity over small but finite subintervals of the field
strength.

These problems strongly remind of the famous van Kampen objections against
linear response theory \cite{vK71}. It is not our goal here to give a full
account of the debate around van Kampen's arguments. For that we may refer,
e.g., to Refs.\ \cite{MECB89,EvMo90,KTH92,Ch2}, to the nice outline in Ref.\
\cite{Do99} and to further references therein. Instead, we will 
highlight certain points that we believe are important for an understanding of
the interplay between chaos and transport in the driven periodic Lorentz gas.

Van Kampen criticizes derivations of the laws of linear response that assume
microscopic linearity in the equations of motion. By taking this assumption
serious he arrives at a simple estimate for the range of validity of linear
response which is $10^{-18}V/cm$ for electrons in a conductor.\footnote{We
remark that in Ref.\ \cite{BDL00} a number for the range of validity has been
given for the driven periodic Lorentz gas, ``possibly as small as of
$10^{-20}$'', whose order of magnitude is close to the original estimate by
van Kampen \cite{vK71}, however, without indicating how this value was
calculated for this specific system.}  Van Kampen uses this unrealistically
small value to illustrate his point that {\em linearity of the microscopic
motion is entirely different from macroscopic linearity}. In other words, the
question is how to reconcile the ubiquitous microscopic nonlinearity in the
equations of motion of a dynamical system with the macroscopic linearity in
the response of the same system. According to van Kampen {\em it is the
randomization of the microscopic variables that enables one to eliminate them
from the macroscopic picture}, whereas {\em in linear response theory the
previous history is not forgotten through randomization, but ignored through
linearization}. Curiously enough, van Kampen discusses just the Galton board
under external forces as an example. By formally solving the Boltzmann
equation that is based on Boltzmann's {\em molecular chaos} assumption he
concludes that {\em the nonlinear deviations of the microscopic motions
somehow combine to produce a linear macroscopic response}; see also Refs.\
\cite{Ch2,CoRo98,KlDe00} on discussions of that point.

To us, the existence proof of linear response for the driven periodic Lorentz
gas given in Refs.\ \cite{Ch1,Ch2} exemplifies how to reconcile macroscopic
linearity with microscopic nonlinearity. On the other hand, results from
computer simulations and related work
\cite{MH87,LNRM95,DeGP95,DeMo97b,BDL00} still point back to van 
Kampen's first objection how to quantitatively identify the regime of validity
of linear response.  Thus, the driven periodic Lorentz gas appears to provide
a good illustration for the difficulty formulated by van Kampen that both the
existence and the size of the regime of linear response may not be trivially
guaranteed in a nonlinear dynamical system; see also Ref.\
\cite{GrKl02} for another example which is even more drastic but also
more abstract than the Lorentz gas. A proper theoretical approach thus needs
to take into account that, for microscopic chaotic dynamics, the macroscopic
response of the system may very sensitively depend on the special character of
the system. However, let us emphasize that the periodic Lorentz gas shares
specific properties such as being low-dimensional and spatially periodic, which
may not be considered to be typical for gases or fluids that are generally
composed of many interacting particles. In random Lorentz gases, or in systems
with more degrees of freedom, there is no evidence that the response exhibits
problems of the type as described by van Kampen.

\subsection{Summary}

\begin{enumerate}

\item Using the mixing and chaotic periodic Lorentz gas as a simple model system
saves us from adding noise in order to generate Brownian motion-like
trajectories of a moving particle. This model thus provides an alternative to
analyzing the interplay between subsystems and thermal reservoirs by means of
stochastic Langevin equations. By furthermore making the friction coefficient
velocity-dependent, under the constraint that the total kinetic energy of the
system is constant at any time step, we arrive at the formulation of the {\em
Gaussian thermostat} for the periodic Lorentz gas driven by an external
electric field. Surprisingly, this combination of subsystem plus modeling of a
thermal reservoir is still deterministic and time-reversible.  It is also
non-Hamiltonian reflecting that the resulting dynamical system is
dissipative. The Gaussian thermostat is an example of a thermal reservoir that
instantaneously acts in the bulk of a dynamical system.

\item For Gaussian thermostated dynamical systems in nonequilibrium situations
there is an identity between the absolute value of the rate of phase space
contraction and thermodynamic entropy production in terms of the Clausius
entropy, respectively in terms of Joule's heat, by applying an external
electric field. This identity represents a crucial link between dynamical
systems quantities related to phase space contraction and thermodynamic
quantities related to irreversible entropy production.

\item As a consequence of this identity the electrical conductivity is a
simple function of the sum of Lyapunov exponents of the dynamical system,
which is known as the {\em Lyapunov sum rule}. Similar formulas hold for other
transport coefficients. Gaussian thermostated dynamical systems with more
degrees of freedom, such as interacting many-particle systems, may furthermore
exhibit a specific symmetry in the spectrum of Lyapunov exponents, which
enables to split the spectrum into {\em conjugate pairs} of exponents all
summing up to the same value. If conjugate pairs exist the Lyapunov sum rule
can be drastically simplified. However, a necessary condition for this
symmetry appears to be that the thermostat acts in the bulk of the system,
whereas for boundary thermostats there are no such conjugate pairs. And even
for bulk thermostats the precise conditions under which conjugate pairing
holds are not perfectly clear and under active discussion.

\item Gaussian thermostated dynamical systems are characterized by the
existence of {\em multifractal attractors} in the Poincar\'e sections of the
phase space. For the Gaussian thermostated driven periodic Lorentz gas the
fractal structure of the attractor is particularly determined by sequences of
tangent collisions reflecting strong dynamical correlations. Under variation
of the external electric field strength the topology of the attractor changes
exhibiting a complex bifurcation scenario.

\item Computer simulations yield that the electrical conductivity of this model
is a nonlinear function of the electric field strength, which furthermore
displays irregularities on fine scales that might be of a fractal origin. Ohms
law is proven to hold for sufficiently small field strengths which supposedly
were not yet accessible to computer simulations. The problem of detecting the
regime of linear response reminds to some of van Kampen's objections against
the validity of linear response in chaotic dynamical systems.

\end{enumerate}

\section{The Nos\'e-Hoover thermostat}

We construct and analyse a second fundamental form of deterministic and
time-reversible thermal reservoirs known as the {\em Nos\'e-Hoover
thermostat}. To motivate this scheme we need a Liouville equation that also
holds for dissipative dynamical systems, which we introduce at the beginning
of this chapter. We then focus on the chaos and transport properties of the
Nos\'e-Hoover thermostated driven periodic Lorentz gas by characterizing the
associated NSS. For this purpose we first check whether there holds an
identity between the average phase space contraction rate and thermodynamic
entropy production as it was obtained for the Gaussian thermostat. We then
discuss computer simulation results for the attractor and for the electrical
conductivity of the Nos\'e-Hoover driven periodic Lorentz gas by comparing
them to their Gaussian counterparts. Finally, we briefly elaborate on
subtleties resulting from the construction of the Nos\'e-Hoover thermostat,
and we outline how this scheme can be generalized leading to a variety of
additional, similar thermal reservoirs.

\subsection{The Liouville equation for dissipative dynamical systems}

Here we particularly follow the presentations of Refs.\ \cite{EvMo90,Do99}.
For sake of simplicity we shall formally restrict ourselves to the case of a
one-particle system such as the Lorentz gas. Let the position ${\bf r}$ and
the velocity ${\bf v}$ of the point particle be defined by some equations of
motion. We wish to consider an ensemble of particles moving in the same
dynamical system. That is, we take a large collection of points in the
accessible phase space of the system, which in general may be restricted by
(geometric) boundary conditions, cp.\ also to Section II.C for related
definitions. Let $\rho\equiv\rho(t,{\bf r},{\bf v})$ be the distribution
function, or probability density, associated with the ensemble of moving
particles. It is the number of points of the ensemble which one can find in
the phase space volume element $d{\bf r}d{\bf v}$ centered around the position
$({\bf r},{\bf v})$ at time $t$ divided by the size of this volume element and
by the total number of points $N$ of the ensemble. By integrating over the
whole accessible phase space this function is properly normalized to one. Let
us consider the fraction $N(t)/N$, where $N\equiv N(0)$ is the number of
ensemble members at time zero while $N(t)$ is the number of points at time
$t$. The total change of this fraction of ensemble points in a volume element
$V$ with surface $A$ in the phase space obeys the balance equation
\be
\frac{d}{dt}\frac{N(t)}{N}=\int_Vd{\bf r}d{\bf v}\pard{\rho}{t}+\int_Ad{\bf
A}\cdot({\bf F}\rho)
\quad ,
\label{eq:partbal}
\ee
where $({\bf \dot{r}},{\bf \dot{v}})^*={\bf F}\equiv{\bf F}({\bf r},{\bf v})$
stands for the flux given by the equations of motion of a dynamical system
such as, e.g., Eqs.\ (\ref{eq:eomdlg}), and $^*$ denotes the transpose.  The
first integral stands for the source and the second one for the flux term. If
we assume that once an ensemble of particles is chosen the number of points is
conserved by the equations of motion, i.e., that there is no vanishing of
points in phase space due to chemical reactions or related mechanisms, the
left hand side must be zero at any instant of time. Applying Gauss' divergence
theorem to the flux term, $\int_Ad{\bf A}\cdot({\bf F}\rho)=\int_Vd{\bf
r}d{\bf v}\:\mbox{{\boldmath $\nabla$}}\cdot({\bf F}\rho)$ with
$\mbox{{\boldmath $\nabla$}}:=(\partial{{\bf r}},\partial{{\bf v}})^*$,
enables us to combine both integrals of Eq.\ (\ref{eq:partbal}) leading to the
differential formulation
\cite{Path88,Do99,EvMo90}
\be
\pard{\rho}{t}+\mbox{{\boldmath $\nabla$}}\cdot({\bf F}\rho)=0 \label{eq:gliouv2}
\ee
or alternatively
\be
\frac{d\rho}{dt}+\rho(\mbox{{\boldmath $\nabla$}}\cdot{\bf F})=0 \quad . \label{eq:gliouv}
\ee
Note that the continuity equation Eq.\ (\ref{eq:partbal}) formulates only the
conservation of the {\em number of points} in phase space. Hence, Eqs.\
(\ref{eq:gliouv2}), (\ref{eq:gliouv}) are valid even if the dynamical system
is not phase space {\em volume} preserving. In turn, for Hamiltonian dynamical
systems it is $\mbox{{\boldmath $\nabla$}}\cdot{\bf F}=0$ and the well-known
Liouville equation for volume-preserving dynamics is recovered
\cite{Reif,Huang,Path88,Do99},
\be
\frac{d\rho}{dt}=0\quad . \label{eq:liouv}
\ee
This equation expresses the fact that according to Hamiltonian dynamics an
ensemble of points moves like an incompressible fluid in phase space, which is
often referred to as the {\em Liouville theorem} in the literature
\cite{Reif,Huang}.

On this occasion some historical remarks are necessary. In his original work
\cite{Liou38} Liouville did not care at all about Hamiltonian dynamics. He
just started from the equations of motion for an arbitrary dynamical system
and derived the evolution equation for its Jacobian determinant, from which
Eq.\ (\ref{eq:gliouv}) follows in a straightforward way. Hence, as pointed out
by Andrey \cite{And86},\footnote{see also the footnote on p.69 of Ref.\
\cite{Gasp} for a related brief historical note} Eq.\ (\ref{eq:gliouv}) should
rather simply be called ``the Liouville equation''. However, in textbooks
usually only Eq.\ (\ref{eq:liouv}) is derived and called ``the Liouville
equation'' \cite{Reif,Huang,Do99}, probably because Liouville's work was
mainly applied to Hamiltonian dynamics. Respectively, in the more specialized
literature Eq.\ (\ref{eq:gliouv}) is often denoted as ``the generalized
Liouville equation'' \cite{Gerl73,Stee79,Stee80,And85,Rams86,HoB99,Rams02}. In
this review we will stick to the historically correct denotation and simply
call Eq.\ (\ref{eq:gliouv}) ``the Liouville equation''.

Alternative derivations of the Liouville equation in its most general form can
be found in Refs.\ \cite{Gerl73,Stee79,Stee80,And86,Rams02,Serg03}, where we
should highlight the seminal work of Gerlich \cite{Gerl73}. Recently Tuckerman
et al.\ \cite{TMK97,TMM99} derived a seemingly different (generalized)
Liouville equation by claiming that Eq.\ (\ref{eq:gliouv}) is not correct, see
Refs.\ \cite{Mund00,TuMa00} for summaries of this approach. The resulting
controversy about the correct form of the Liouville equation
\cite{HEPHM98,Reim98,TMK98,ESHH98,TMBK98,TMM99} was eventually resolved by
Ramshaw \cite{Rams02} starting from the observation that Eq.\
(\ref{eq:gliouv}) is not covariant \cite{Rams86}. However, covariance can be
achieved by incorporating a respective coordinate transformation into the
Liouville equation, which is generally a function of the metric. Along these
lines the connection between Eq.\ (\ref{eq:gliouv}) and the form of the
Liouville equation considered in Refs.\ \cite{TMK97,TMK98,TMM99} can be
established. Applying this transformation both formulations turn out to be
completely equivalent, however, for many purposes Eq.\ (\ref{eq:gliouv})
appears to be the more convenient choice \cite{Rams02}.

The Gaussian thermostated driven Lorentz gas provides a simple example where
the divergence in Eq.\ (\ref{eq:gliouv}) does not disappear. It may thus be
replaced by Eq.\ (\ref{eq:peqa}) leading to
\be
\pard{\rho}{t}=\alpha\rho \quad . \label{eq:liouvdplg}
\ee
By employing this equation the {\em Gibbs entropy production} for the Gaussian
thermostated driven Lorentz gas can be computed right away, which furnishes an
alternative derivation of the identity Eq.\ (\ref{eq:pepid}) between phase
space contraction and thermodynamic entropy production
\cite{Do99,DettS00}. This and further important implications of the Liouville
equation concerning irreversibility and entropy production will be discussed
in Section V.B.

We are now set up to introduce a second fundamental type of deterministic
thermal reservoirs generalizing the Gaussian scheme. For this construction the
Liouville equation Eq.\ (\ref{eq:gliouv}) will be used as a convenient
starting point.

\subsection{Construction of the Nos\'e-Hoover thermostat}

\subsubsection{Heuristic derivation}

As a crucial property of the Gaussian thermostated driven Lorentz gas the
velocity distribution of the moving particle is intimately related to the
microcanonical one, that is, the system is constrained onto an energy shell in
phase space. Here we introduce a generalized thermostating scheme that, under
certain conditions, enables to transform the velocities of a dynamical system
onto a {\em canonical distribution} in thermal equilibrium. Our heuristic
derivation follows the presentations in Refs.\ \cite{Hoov85,Hoo91}; see also a
respective short note in Ref.\ \cite{HoHo86} and a formal generalization of
this argument in Refs.\ \cite{KBB90,BuKu90}.  We remark that quantum
mechanical formulations of deterministic and time-reversible thermal
reservoirs can be obtained along the same lines \cite{Kus93,MS01,KLA02,MS03}.

Let us start again from the driven Lorentz gas equations of motion Eqs.\
(\ref{eq:eomdlg}) that already exhibit a coupling with a thermal reservoir in
form of the friction variable $\alpha$. Let us now assume that $\alpha$ is not
necessarily a simple analytical function of ${\bf v}$ like the Gaussian
thermostat Eq.\ (\ref{eq:alpdlg}), but that it is determined by some unknown
differential equation instead. Formally, the equations of motion then read
\be
({\bf \dot{r}},{\bf \dot{v}},\dot{\alpha})^*={\bf F}({\bf r},{\bf v},\alpha)
\quad ,
\ee
and the generalized Liouville equation Eq.\ (\ref{eq:gliouv}) can be written
as
\be
\pard{\rho}{t}+\dot{{\bf r}}\pard{\rho}{{\bf r}}+\dot{{\bf v}}\pard{\rho}{{\bf v}}+
\dot{\alpha}\pard{\rho}{\alpha}+\rho\left[\pard{\dot{{\bf r}}}{{\bf r}}+\pard{\dot{{\bf v}}}{{\bf v}}+
\pard{\dot{\alpha}}{\alpha}\right]=0 \quad .
\ee
In contrast to the Gaussian thermostat, which was designed to keep the kinetic
energy constant at any time step specifically in nonequilibrium, here our
strategy is to define the thermostating mechanism first in thermal
equilibrium. The goal is to come up with a dissipation term that transforms
the velocity distribution of the subsystem onto the canonical one. In a second
step we then create a nonequilibrium situation. Here we will check whether our
thermostat is able to generate a NSS, and if so we will study what its precise
properties are. In other words, our goal is to find a differential equation
for $\alpha$ that is consistent with the existence of a canonical velocity
distribution for an ensemble of subsystem particles. For this purpose we make
the ansatz \cite{Nose84a,Nose84b,PoHo88,PoHo89,EvMo90,MoDe98,HoB99}
\be
\rho(t,{\bf r},{\bf v},\alpha)\equiv const.\
\exp\left[-\frac{v^2}{2T}-(\tau\alpha)^2\right]\quad , \label{eq:fans}
\ee
where both the distribution for the velocities of the subsystem and for the
thermal reservoir variable $\alpha$ are assumed to be canonical, cp.\ to Eq.\
(\ref{eq:vxgauss}).\footnote{For an $N$-particle system with $d$ degrees of
freedom the exponent compiling the total energy of subsystem and thermal
reservoir reads respectively
$E_{total}/T=(\sum_iv_i^2/2+NdT\tau^2\alpha^2/2)/T$
\cite{Nose84a,Nose84b,PoHo88,PoHo89,EvMo90,MoDe98,HoB99}.} According to
equipartitioning of energy $T$ is identified as the temperature of the desired
canonical distribution, and in thermal equilibrium this temperature should be
identical to the temperature of the associated thermal reservoir. Furthermore,
since $\alpha$ has a dimension of $1/second$ we needed to introduce the new
quantity $\tau>0$ with $[\tau]=s$ in order to make the exponent of the second
term dimensionless.

We now combine Eq.\ (\ref{eq:fans}) with the equations of motion for a single
free particle, see Eqs.\ (\ref{eq:eomdlg}) with $\varepsilon=0$. Feeding these
equations as well as the ansatz for $\rho$ into Eq.\ (\ref{eq:gliouv}) yields
for $\alpha$ the differential equation
\be
\alpha
\frac{v^2}{T}-\alpha\dot{\alpha}2\tau^2-2\alpha+\pard{\dot{\alpha}}{\alpha}=0\quad
.
\ee
This equation can be further simplified by {\em ad hoc} restricting to the
case $\partial\dot{\alpha}/\partial\alpha=0$. We then arrive at the simple
solution
\be
\dot{\alpha}=\frac{v^2-2T}{\tau^22T} \quad . \label{eq:anh}
\ee
This differential equation for the thermostating variable $\alpha$, together
with the equations of motion Eqs.\ (\ref{eq:eomdlg}), defines the so-called
{\em Nos\'e-Hoover thermostat} for the driven periodic Lorentz gas. The first
version of such a thermostat transforming onto canonical distributions was
developed by Nos\'e in 1984 \cite{Nose84a,Nose84b}. His original derivation
proceeded along different lines by starting from a generalized Hamiltonian
formalism, see Section IV.C.2. Furthermore, it featured an at least for
practical purposes spurious differential equation related to another reservoir
variable, which was eliminated by Hoover
\cite{Hoov85}. This simplified version of the Nos\'e thermostat was coined the
Nos\'e-Hoover thermostat \cite{EvHo85}. In first explorations it was
particularly applied to the harmonic oscillator, and the combined system was
respectively called the {\em Nos\'e-Hoover oscillator}
\cite{Hoov85,PHV86,HoHo86,MKT92,Nose93,HoHo96,PH97,HoKu97,LaLe03}.

As for the Gaussian thermostat, the Nos\'e-Hoover scheme can be used {\em
isokinetically} by only constraining the kinetic energy of particles, see
above, or {\em isoenergetically} by constraining the total internal energy of
soft particles consisting of kinetic plus potential energy
\cite{MoDe98,DettS00,Rond02}.  It can also be adapted to drive
interacting many-particle systems under nonequilibrium conditions into NSS
\cite{Nose84a,Nose84b,PoHo88,PoHo89,EvMo90,Hoo91,Hess96,MoDe98,HoB99,DettS00,Mund00,Rond02}.
The results obtained from such simulations are typically well in agreement
with predictions from irreversible thermodynamics and linear response theory
\cite{EvMo90,SEC98}. Hence, like the Gaussian modeling of a thermal reservoir
the Nos\'e-Hoover scheme became a widely accepted useful tool for performing
nonequilibrium molecular dynamics computer simulations not only of simple but
also of more complex fluids such as polymer melts, liquid crystals and
ferrofluids \cite{Hess96,HKL96,HABK97}, of proteins in water and of chemical
processes in the condensed matter phase \cite{TuMa00}.

\subsubsection{Physics of this thermostat}

First of all, like the equations of motion Eqs.\ (\ref{eq:eomdlg}),
(\ref{eq:alpdlg}) of the Gaussian thermostated Lorentz gas the Nos\'e-Hoover
counterparts Eqs.\ (\ref{eq:eomdlg}), (\ref{eq:anh}) are as well deterministic
and time-reversible but non-Hamiltonian. However, in contrast to the
microcanonical Gaussian thermostat Nos\'e-Hoover was constructed for
generating a canonical velocity distribution. More than Gaussian constraint
dynamics, Nos\'e-Hoover thermostated systems may thus be considered as
deterministic and time-reversible variants of Langevin dynamics, cp.\ Eqs.\
(\ref{eq:eomdlg}), (\ref{eq:anh}) to Eq.\ (\ref{eq:lang1d}).  

For interacting many-particle systems the Nos\'e-Hoover thermostat
conveniently works in thermal equilibrium
\cite{Nose84a,EvHo85,PoHo89}. However, similar to the Gaussian thermostat this
is not the case for the Lorentz gas without external fields, which is due to a
malfunctioning of the friction coefficient related to the hard disk collisions
\cite{RKH00}. Nevertheless we are safe in studying the driven Lorentz gas
because, in analogy to the Gaussian thermostat, as soon as an electric field
is added this problem disappears.

Following this comparison with Langevin dynamics one may inquire about the
existence of a fluctuation-dissipation theorem such as Eq.\ (\ref{eq:fdt}) for
equilibrium Nos\'e-Hoover dynamics. But as we already discussed in Section
III.A, for dissipative deterministic dynamical systems such as the Gaussian or
the Nos\'e-Hoover thermostated Lorentz gas there is no separation between
stochastic forces and dissipation matching to the respective terms in the
Langevin equation Eq.\ (\ref{eq:lang1d}). Actually, for both deterministic
thermostats the velocity-dependent friction coefficient is itself a
fluctuating variable, in addition to the fact that the position space Lorentz
gas dynamics is also nonlinear and fluctuating. Hence, a balance equation like
Eq.\ (\ref{eq:fdt}) between a constant friction coefficient and stochastic
forces cannot exist; however, for more general fluctuation-dissipation
relations in case of deterministic dynamics see Ref.\ \cite{Rue99}.

In order to better understand the action of the Nos\'e-Hoover fricitional
force let us look at the ensemble averaged version of Eq.\ (\ref{eq:anh}),
\be
<\dot{\alpha}>=\frac{<v^2>-2T}{\tau^22T} \quad .
\ee
If the thermostat acts properly in equilibrium the average of the fluctuations
of the thermostating variable $\alpha$ must be zero and consequently
$<v^2>=2T$ on the right hand side, which corresponds to equipartitioning of
energy for a system with two degrees of freedom.\footnote{Since here we do not
have Gaussian constraints we may safely assume the existence of two degrees of
freedom for the subsystem.} $T$ thus serves as a control parameter
representing the temperature of the thermal reservoir. Choosing some $T$ the
thermostat variable $\alpha$ acts accordingly in order to thermalize the
subsystem onto the same temperature. Consequently, in contrast to the Gaussian
thermostat the proper temperature is not attained instantaneously but with
respect to an ensemble average over a canonical velocity distribution, as we
stipulated with our ansatz Eq.\ (\ref{eq:fans}). That is, for a Lorentz gas
particle the kinetic energy is allowed to fluctuate in time around a mean
value of $<v^2>/2$.

As in case of the Gaussian thermostat, $\alpha$ supplemented by the respective
dynamics for this variable represents both the coupling of the subsystem to
the thermal reservoir as well as the action of the thermal reservoir
itself. In a way, one may think of $\alpha$ as a dynamical variable on which
all hypothetical degrees of freedom of the thermal reservoir are projected
upon just for the purpose of dissipating energy from the subsystem. Along this
line of reasoning $\tau$ may then be interpreted as the {\em reservoir
response time} with respect to exchanging energy between the subsystem and the
thermal reservoir. $\tau$ thus serves as another control parameter determining
the efficiency of the interaction. Note that this control parameter does not
exist for the Gaussian thermostat.

Let us look at the two limiting cases of Eq.\ (\ref{eq:fans}) with respect to
a variation of $\tau$: For $\tau\to\infty$ obviously $\alpha\to const.$, and
one recovers the familiar case of Newton's {\em irreversible} equations of
motion with constant Stokes' friction. On the other hand, making $\tau$
smaller implies that there is a more immediate response of the reservoir to
fluctuations of the kinetic energy in the subsystem. Consequently, in the
limit of $\tau\to0$ there must eventually be instantaneous control leading
back to the constraint of keeping the kinetic energy constant at any time step
as formulated by the Gaussian thermostat. This can be seen by moving $\tau^2$
from the denominator on the right hand side of Eq.\ (\ref{eq:anh}) to the left
hand side and performing the limit $\tau\to0$ enforcing
$v^2=2T$.\footnote{However, note that for a vanishing external field
$\varepsilon=0$ the friction coefficient of the Gaussian thermostat Eq.\
(\ref{eq:alpdlg}) is by default zero, at least for the driven periodic Lorentz
gas. Without external forces the respective solution for Nos\'e-Hoover should
therefore also go to zero for $\tau\to0$. Thus, in detail the relation between
the Gaussian and the Nos\'e-Hoover thermostat is more intricate, see also the
discussion in Chapter VIII.} For more detailed studies of how Nos\'e-Hoover
dynamics depends on the value of the coupling constant $\tau$ see Refs.\
\cite{Nose84b,Hoov85,EvHo85,HoHo86,Nose91,Nose93,HVR95,KBB90}.

The Gaussian thermostat may furthermore be classified as a feedback mechanism
that is {\em differential} in time, reflecting the existence of an explicit
functional form for the velocity-dependent friction coefficient. The
Nos\'e-Hoover equations can be seen as an {\em integral} version according to
the formal solution of the differential equation for the friction variable
\be
\alpha(t)=\alpha(0)+\int_0^tds\: \dot{\alpha}(s)
\ee
with $\dot{\alpha}$ being determined by Eq.\ (\ref{eq:anh})
\cite{HoHo86,HHP87,HVR95}. Interestingly, the existence of these two 
limiting cases shows up in the specific functional form of the velocity
distribution under variation of $\tau$: For $\tau\to0$ and small enough field
strength $\varepsilon$ this function must evidently approach a microcanonical
velocity distribution $\rho(v)\sim\delta(2E-v^2)$, whereas for $\tau\to\infty$
it yields a canonical one. Consequently, between these two limiting cases
there must exist a superposition of the two different shapes in form of a {\em
crater-like} velocity distribution with a dip at the place of the former
maximum of the canonical distribution.

\begin{figure}[t]
\epsfxsize=10cm
\centerline{\epsfbox{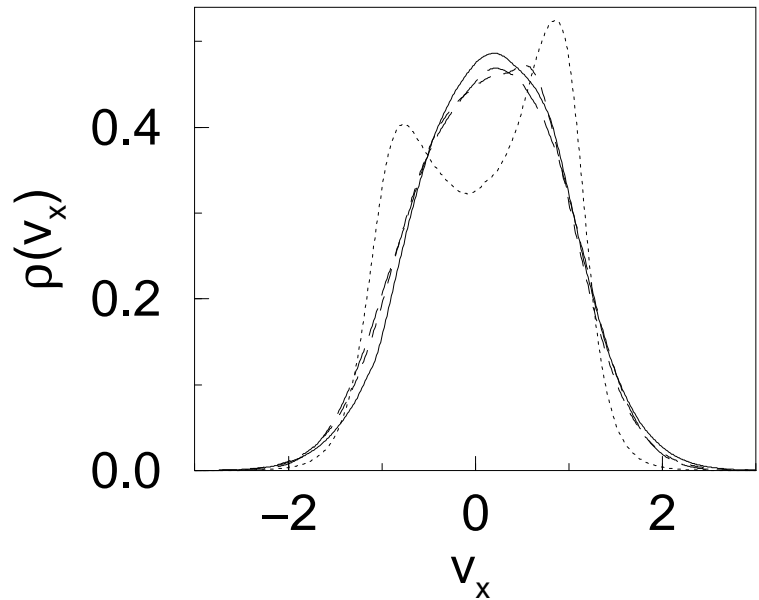}}
\caption{Velocity distributions $\varrho(v_x)$ for an ensemble of
moving point particles in the thermostated driven periodic Lorentz gas for an
electric field of strength $\varepsilon=0.5$, which is parallel to the
$x$-axis, at temperature $T=0.60029$: Nos\'e-Hoover thermostat with a reponse
time of the thermal reservoir of $\tau^2=0.01$ (dotted curve), $\tau^2=1$
(dashed curve), $\tau^2=1000$ (long dashed curve) in comparison to the
application of the deterministic boundary thermostat (``thermostating by
deterministic scattering'') discussed in Chapter VII (solid curve). Note the
transition from a microcanonical-like to a canonical distribution under
variation of $\tau$. The figure is from Ref.\
\protect\cite{RKH00}.}
\label{fig:rhonhtds}
\end{figure}

Indeed, right this transition is depicted in Fig.\ \ref{fig:rhonhtds} where
$\rho(v_x)$ for a Nos\'e-Hoover thermostated driven periodic Lorentz gas is
represented for three different values of $\tau$ \cite{RKH00}; for related
results concerning another, different dynamical system thermostated by
Nos\'e-Hoover see Ref.\ \cite{Nose91}. Note the symmetry breaking in Fig.\
\ref{fig:rhonhtds} because of the external electric field, however, one
clearly recognizes the approach towards a microcanonical distribution for
$\tau\to0$.\footnote{Curiously, in case of {\em non-ergodic} Nos\'e-Hoover
dynamics similar microcanonical-like and canonical-like distributions can be
obtained with respect to different choices of initial conditions for the
ensemble of particles, see the preprint of Ref.\ \cite{RKH00} for an example.}
We will come back to a more explicit discussion of this interesting point in
Chapter VIII, where it will be shown that crater-like velocity distribution
functions that are very similar to the one discussed here play an important
role for so-called active Brownian particles which are thought to model, among
others, biological cell motility.

\subsection{Properties of the Nos\'e-Hoover thermostat}

\subsubsection{Fundamental relations between chaos and transport}

For the Gaussian thermostated driven Lorentz gas the crucial relation linking
thermodynamics to dynamical systems theory was the identity Eq.\
(\ref{eq:pepid}) between the average phase space contraction rate and the
thermodynamic entropy production in form of Joule's heat. Let us now check
whether the Nos\'e-Hoover thermostat fulfills this identity as well. For this
purpose we calculate the energy balance between subsystem and thermal
reservoir \cite{PoHo88,RKH00} by starting from the Hamiltonian for the
combined system, see the exponent in Eq.\ (\ref{eq:fans}),
\be
H=v^2/2+T\tau^2\alpha^2\quad . \label{eq:enbal}
\ee
If the system is properly thermostatd yielding a NSS, the average of the time
derivative of the total energy must be zero, $<dH/dt>=0$. Differentiating Eq.\
(\ref{eq:enbal}) and replacing the derivatives by the equations of motion
Eqs.\ (\ref{eq:eomdlg}), (\ref{eq:anh}) yields
\be
\frac{<\mbox{{\boldmath $\varepsilon$}}\cdot{\bf v}>}{T}=2<\alpha>\quad \label{eq:nhprep}
\ee
The right hand side equals minus the average phase space contraction rate
$\kappa$ defined by Eq.\ (\ref{eq:psc}), whereas the left hand side is
obviously Joule's heat. We have thus corroborated the identity Eq.\
(\ref{eq:pepid}) for Nos\'e-Hoover dynamics.

Alternatively, in order to check for the identity one may start directly from
the Nos\'e-Hoover equations of motion Eqs.\ (\ref{eq:eomdlg}),
(\ref{eq:anh}). Multiplying the differential equation for the velocity with
${\bf v}$ and replacing the velocity squared on the right hand side by Eq.\
(\ref{eq:anh}) one arrives at
\be
{\bf \dot{v}}\cdot{\bf v}={\bf \varepsilon}\cdot{\bf
v}-\alpha(\dot{\alpha}\tau^22T+2T) \quad .
\ee
One may now collect all terms related to fluctuations on the left hand side by
considering their statistical averages,
\be
<\frac{d}{dt}\left(\frac{1}{2}v^2+\tau^2T\alpha^2\right)>=<{\bf
\varepsilon}\cdot{\bf v}-\alpha 2T> \quad .
\ee
The left hand side is just identical to the average of the Hamiltonian Eq.\
(\ref{eq:enbal}) and must be zero in a NSS, see above, hence the right hand
side yields again the identity. This derivation is useful if the Hamiltonian
related to the thermostat is not known in advance, see Section VI.C for some
examples.

Based on the identity Eq.\ (\ref{eq:nhprep}) the Lyapunov sum rule can be
derived in the same way as explained in Section III.B.2. For four nontrivial
phase space variables it reads
\be
\sigma(\varepsilon)=-\frac{T}{\varepsilon^2}\sum_{i=1}^4\lambda_i(\varepsilon)
\quad . \label{eq:lsrnh} 
\ee
Hence, with respect to relations between chaos and transport the Nos\'e-Hoover
thermostat belongs to precisely the same class of thermal reservoirs as the
Gaussian one.

\subsubsection{$^*$Generalized Hamiltonian formalism for the Nos\'e-Hoover
thermostat}

A crucial feature of Hamiltonian systems is that the dynamics is phase space
volume preserving, which follows from their symplectic structure
\cite{Meis92,AbMa94}. On the other hand, as was shown for the Gaussian
thermostat in Section III.B.1 and as we just verified for the Nos\'e-Hoover
thermostat, the application of such deterministic thermostating mechanisms
leads to an average phase space contraction in nonequilibrium. From that point
of view it is surprising that for Gaussian and Nos\'e-Hoover dynamics a {\em
generalized Lagrangian and Hamiltonian formalism} exists in which this
dynamics is phase space volume preserving. Here we outline some main ideas of
the Hamiltonian approach for the example of the isokinetic Nos\'e-Hoover
thermostat.

Let us consider the very general case of a subsystem where a particle of unit
mass moves under the influence of a potential $u({\bf x})$.  Let the
microscopic dynamics of this particle be described by the phase space
coordinates $({\bf r},{\bf v})$.  The dynamics of the thermal reservoir is
modeled by only one degree of freedom in terms of the coordinates
$(r_0,v_0)$. All these representations may be called {\em physical}
coordinates, since $({\bf r},{\bf v})$ represents the actual position and the
velocity of some particle. For the {\em decoupled} system the formal
Hamiltonian is
\be
H_0({\bf r},{\bf v},r_0,v_0)=E({\bf v},v_0)+U({\bf r},r_0) \quad , \label{eq:hdec}
\ee
where $E({\bf v},v_0)=m{\bf v}^2/2+Qv_0^2/2$ stands for the kinetic energy of
particle plus reservoir and $U({\bf r},r_0)=u({\bf r})+2Tr_0$ for the
potential energy of both systems. In formal analogy to the particle's kinetic
energy, the new quantity $Q$ is sometimes called the {\em mass} of the
reservoir, and $T$ is identified with the reservoir
temperature.\footnote{Strictly speaking the choice of the potential term for
the reservoir, and fixing the number of degrees of freedom associated with it,
is a bit more subtle \cite{Choq98,MoDe98}.}  The Hamiltonian equations of
motion for the decoupled system then read
\bna
{\bf \dot{r}}&=&{\bf v} \nonumber\\ 
{\bf \dot{v}}&=&-\pard{u}{{\bf r}} \nonumber\\
\dot{r_0}&=&Qv_0\nonumber \\
\dot{v_0}&=&-2T \quad . \label{eq:hdeceom}
\ena
The problem is now two-fold: firstly, to find a Hamiltonian that 
generalizes Eq.\ (\ref{eq:hdec}) in suitably representing the {\em coupled}
system, and secondly, to derive the Nos\'e-Hoover equations of motion, such as
the special case Eqs.\ (\ref{eq:eomdlg}), (\ref{eq:anh}) for the driven
Lorentz gas, from this new Hamiltonian. In a slightly modified setting, this
problem first appeared in Nos\'e's groundbreaking work \cite{Nose84a,Nose84b}
leading to the formulation of the so-called Nos\'e Hamiltonian. Here we follow
a simplified approach as it was proposed later on in Refs.\
\cite{DeMo97a,Choq98}.

Since the Nos\'e-Hoover equations of motion are not Hamiltonian it is clear
that a Hamiltonian formulation of this dynamics can only be achieved by
choosing a new set of {\em generalized} coordinates, for which we may write
$({\bf R},{\bf V})$ in case of the subsystem and $(R_0,V_0)$ in case of the
reservoir. Now we implement the so-called {\em exponential coupling ansatz}
for the Hamiltonian in these generalized coordinates,
\be
H({\bf R},{\bf V},R_0,V_0):=e^{-R_0}E({\bf V},V_0)+e^{R_0}U({\bf R},R_0)\quad .
\ee
These exponential prefactors define a {\em nonlinear, multiplicative coupling}
between subsystem and thermal reservoir in generalized coordinates. Physically
speaking, they suitably rescale the total energy of the system thus serving to
keep the energy of the full system constant on average, whereas in
nonequilibrium it would increase to infinity otherwise. The different signs of
the exponents are adjusted to the different types of kinetic and potential
energy. We can now derive Hamilton's equations of motion from this new
Hamiltonian to
\bna
{\bf \dot{R}}&=&e^{-R_0}m{\bf V} \nonumber\\
{\bf \dot{V}}&=&-e^{R_0}\pard{u}{{\bf R}} \nonumber\\
\dot{R_0}&=&e^{-R_0}QV_0\nonumber\\
\dot{V_0}&=&2(e^{-R_0}E({\bf V},V_0)-e^{R_0}T) \quad , \label{eq:hcoupeom}
\ena
where we have imposed that $H({\bf R},{\bf V},R_0,V_0)\equiv 0$. Matching the
first two equations to the corresponding ones in physical coordinates Eqs.\
(\ref{eq:hdeceom}) suggests the following transformation between physical and
generalized coordinates of the subsystem,
\bna
{\bf R}&=&{\bf r} \nonumber\\
{\bf V}&=&e^{R_0}{\bf v} \quad . \label{eq:gctra1}
\ena
Making the same choice for the reservoir coordinates,
\bna
R_0&=&r_0 \nonumber\\
V_0&=&e^{R_0}v_0 \quad , \label{eq:gctra2}
\ena
we obtain a complete set of equations for transforming between physical and
generalized coordinates. Identifying $Q\equiv 2T\tau^2$ and $\alpha\equiv
v_0Q$ the Nos\'e-Hoover equations Eqs.\ (\ref{eq:eomdlg}), (\ref{eq:anh}) for
the driven Lorentz gas are precisely recovered from Eqs.\
(\ref{eq:hcoupeom}). Let us emphasize at this point that the transformation
between the different coordinates given by Eqs.\ (\ref{eq:gctra1}),
(\ref{eq:gctra2}) is strictly {\em noncanonical} not preserving the phase
space volume. Therefore, one should carefully distinguish between traditional
Hamiltonian formulations of classical mechanics and the {\em generalized}
Hamiltonian formalism outlined above.

The idea to use a generalized Hamiltonian formalism in order to define the
action of a thermal reservoir was pioneered by Nos\'e \cite{Nose84a,Nose84b}
who constructed his thermostating mechanism along these lines.  In Refs.\
\cite{Nose91,Nose93,Hoov85,EvHo85,PHV86,EvMo90} this scheme was applied to
different types of subsystems. Later on Dettmann and Morriss showed that there
also exist generalized Hamiltonian formulations for the Gaussian isokinetic
\cite{DeMo96} and isoenergetic \cite{Dett99} thermostat, see Ref.\
\cite{HMHE88} for a pre-version of this approach. They also came up with a
modified Nos\'e Hamiltonian leading more straightforwardly to the
Nos\'e-Hoover equations than starting from Nos\'e's original formulation
\cite{DeMo97a,HoB99}, as is sketched in the derivation above. The generalized
Lagrangian approach was outlined by Choquard in Ref.\ \cite{Choq98}, see also
this reference as well as Refs.\ \cite{MoDe98,DettS00} for summaries
concerning the subject of this section. We remark that the symplectic
properties of these generalized Hamiltonians appear to be intimately related
to the validity of the conjugate pairing rule discussed in Section III.B.2.

\subsubsection{Attractors, bifurcation diagrams and electrical conductivities
for the Nos\'e-Hoover thermostated driven periodic Lorentz gas}

As we have shown in Sections III.B.1 and IV.C.1, both the Gaussian and the
Nos\'e-Hoover thermostated driven Lorentz gas exhibit an identity between the
average phase space contraction rate and thermodynamic entropy production. For
the Nos\'e-Hoover case we now discuss further chaos and transport properties,
which are the fractality of the corresponding attractor, the structure of the
associated bifurcation diagram and the response to an electric field in terms
of the field-dependent conductivity. Particularly, we elaborate on
similarities and differences of these properties in comparison to the
respective figures shown for the Gaussian thermostat. The computer simulation
results for the Nos\'e-Hoover thermostated driven periodic Lorentz gas
presented here are from Ref.\ \cite{RKH00}.

\begin{figure}[t]
\epsfxsize=9cm
\epsfysize=8cm
\centerline{\epsfbox{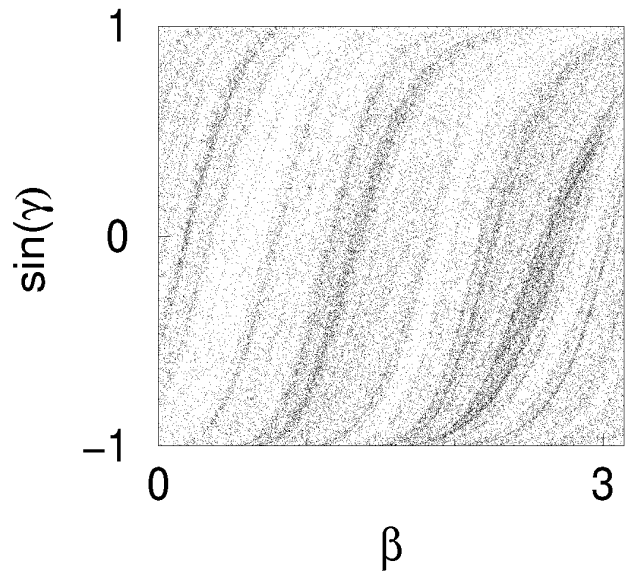}}
\vspace*{0.3cm} 
\caption{Attractor for the Nos\'e-Hoover thermostated periodic Lorentz gas
driven by an electric field of strength $\varepsilon=1$ which is parallel to
the $x$-axis. $\beta$ and $\sin\gamma$ are the same as in Fig.\ \ref{gatt} and
are also defined in Fig.\ \ref{plko}. The temperature is $T=0.7740645$. The
reservoir coupling parameter $\tau$, see Eq.\ (\ref{eq:anh}), has the value
$0.1$. The results are from Ref.\ \protect\cite{RKH00}.}
\label{attrnh}
\end{figure}

Fig.\ \ref{attrnh} depicts a projection of the attractor of the Nos\'e-Hoover
thermostated driven periodic Lorentz gas onto the phase space in Birkhoff
coordinates.\footnote{Note that, in contrast to the Gaussian thermostated
driven periodic Lorentz gas, Fig.\ \ref{attrnh} does not represent a
Poincar\'e surface of section: For Nos\'e-Hoover the kinetic energy is not
kept constant but fluctuates, hence this plot is composed of points with
different absolute values of the velocity of the colliding particle.} Some
repeated folding in phase space is clearly visible indicating that, as in case
of the Gaussian thermostat, most probably this attractor is again of
a multifractal nature. However, for Nos\'e-Hoover yet no values for a
(fractal) dimension quantitatively assessing this structure were
computed.\footnote{but see Section VII.D.2 for a very similar attractor for
which the Kaplan-Yorke dimension was found to be fractal} By comparing this
attractor to the one obtained from Gaussian thermostating Fig.\
\ref{gatt} one realizes that the structure of both sets is essentially the
same. This is not too surprising since both systems share the same geometry,
and as we discussed in Section III.B.3 the main features of this structure are
induced by the geometry of the scatterers. That the Nos\'e-Hoover thermostat
attempts to transform the system onto a canonical distribution explains why
the attractor related to the Nos\'e-Hoover thermostat appears to be a
smoothed-out version of the Gaussian counterpart. These results suggest that
the appearance of fractal attractors is also typical for Nos\'e-Hoover
thermostatd systems. Numerical computations of fractal dimensionality losses
in many-particle systems thermostated by Nos\'e-Hoover confirm this statement
\cite{HHP87,PoHo88,PoHo89,AK02,KLA02,HPA+02,PH03}.

\begin{figure}[t]
\epsfxsize=17cm
\centerline{\epsfbox{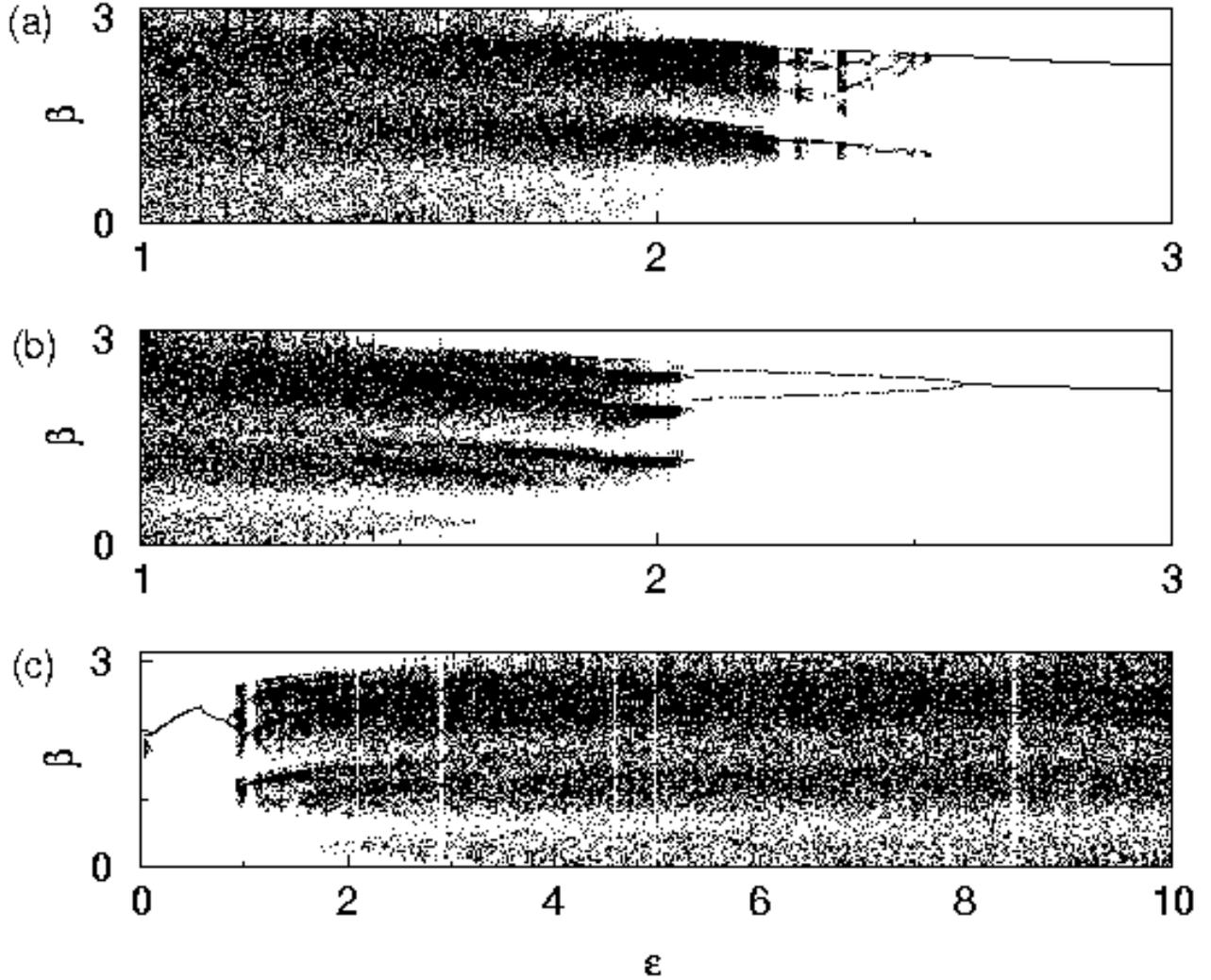}}
\vspace*{0.3cm} 
\caption{Bifurcation diagrams for the Nos\'e-Hoover thermostated periodic
Lorentz gas driven by an electric field of strength $\varepsilon$ which is
parallel to the $x$-axis. $\beta$ is defined in Fig.\ \ref{plko}. The
temperature is fixed to $T=0.5$ and the values of the reservoir coupling
parameter $\tau$, see Eq.\ (\ref{eq:anh}), are: (a) $\tau=0.1$, (b)
$\tau\simeq 31.6$ $(\tau^2=1000)$ and (c) the dissipative limit of
$\tau\to\infty$ corresponding to a constant friction coefficient $\alpha$
whose value was chosen to be $1$. The figure is from Ref.\ \protect\cite{RKH00}.}
\label{bifunh}
\end{figure}
Fig.\ \ref{bifunh} contains bifurcation diagrams for the Nos\'e-Hoover
thermostated driven periodic Lorentz gas at three different values of the
reservoir coupling constant $\tau$, see Eq.\ (\ref{eq:anh}). For quick
response of the reservoir, $\tau\ll1$, the attractor is phase space filling up
to $\varepsilon<2$, which is reminiscent of the results for the Gaussian
thermostated Lorentz gas presented in Fig.\ \ref{gbifu}. Note that Fig.\
\ref{gbifu} depicts the angle $\theta$ as a function of the field strength
$\varepsilon$, whereas Fig.\ \ref{bifunh} displays the angle $\beta$,
respectively. However, both bifurcation diagrams exhibit qualitatively the
same behavior. This is in agreement with the discussion of Section III.B.2
where we argued that the Nos\'e-Hoover thermostat yields Gaussian constraint
dynamics in the limit of $\tau\to0$.

In agreement to a naive physical reasoning, by making the response time of the
reservoir larger the attractor collapses onto periodic orbits at smaller field
strengths than for smaller response times, see Fig.\ \ref{bifunh}
(b). However, in detail the situation is much more intricate \cite{RKH00}, as
is already indicated by Fig.\ \ref{bifunh} (c) that displays the dissipative
limit of $\tau\to\infty$. Here the bifurcation diagram depends on the value
for the constant friction coefficient $\alpha$. In any case, for all values
studied numerically one observes a kind of inverted bifurcation scenario
compared to the Gaussian thermostated model and to Figs.\
\ref{bifunh} (a) and (b) in that the attractor starts with periodic orbits and
covers the whole accessible phase space only for higher field strengths
\cite{RKH00}. Fig.\ \ref{bifunh} thus shows that for the Nos\'e-Hoover
thermostated driven periodic Lorentz gas the specific structure of the fractal
attractor sensitively depends on the choice of the reservoir coupling
parameter $\tau$; see also Refs.\
\cite{Nose84b,Hoov85,EvHo85,HoHo86,Nose91,Nose93,HVR95,KBB90}
for studies concerning the variation of $\tau$ in other systems and for the
impact on dynamical systems properties.

\begin{figure}[t]
\epsfxsize=15cm
\centerline{\epsfbox{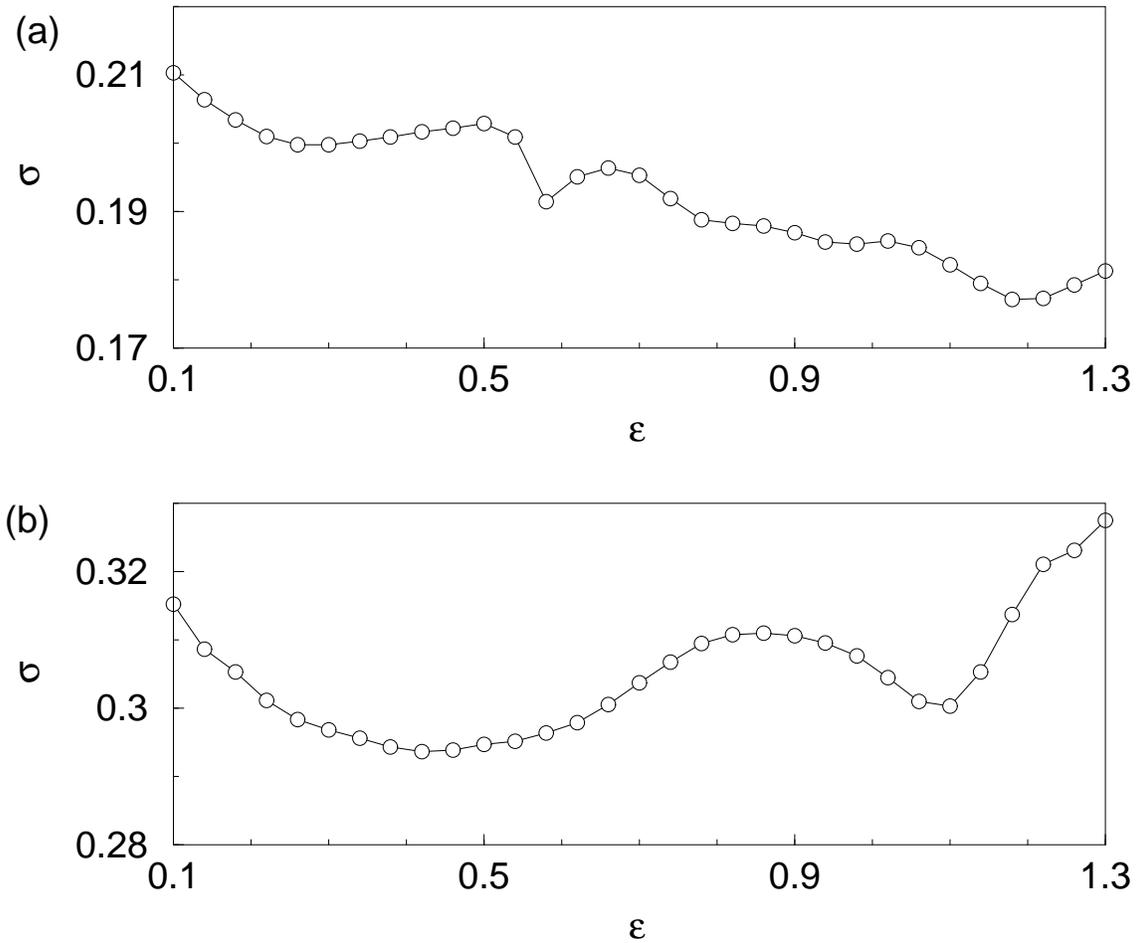}}
\vspace*{0.3cm} 
\caption{Electrical conductivity $\sigma(\varepsilon)$ for the Nos\'e-Hoover
thermostated periodic Lorentz gas driven by an electric field of strength
$\varepsilon$ which is parallel to the $x$-axis. The temperature is $T=0.5$
and the values of the reservoir coupling parameter $\tau$, see Eq.\
(\ref{eq:anh}), are: (a) $\tau=0.1$ and (b) $\tau\simeq 31.6$
$(\tau^2=1000)$. The numerical uncertainty of each point is less than the
symbol size. The figure is from Ref.\ \protect\cite{RKH00}.}
\label{condnh}
\end{figure}

As one might suspect already with respect to these bifurcation diagrams, there
is no unique result for the field-dependent electrical conductivity
$\sigma(\varepsilon)$ at different values of the response time $\tau$ either,
not even for small field strengths $\varepsilon\ll 2$. Fig.\ \ref{condnh}
presents computer simulation results for the two parameter values of $\tau$
studied in Figs.\ \ref{bifunh} (a) and (b). The conductivity for $\tau=0.1$
looks like a smoothed-out version of the Gaussian thermostat conductivity
depicted in Fig.\ \ref{gcond}, whereas the curve at $\tau\simeq31.6$ appears
to be more stretched out along the $\varepsilon$-axis. In the latter case it
is not even clear whether the conductivity is globally decreasing or
increasing. There is little to say concerning the validity of the Einstein
relation Eq.\ (\ref{eq:dts}) in the limits of $\tau>0$ and $\varepsilon\to0$
as discussed for the Gaussian version in Section III.B.4, since the
Nos\'e-Hoover thermostat does not work properly for the Lorentz gas in thermal
equilibrium, see Section IV.B.2. Hence, the field-free diffusion coefficient
cannot be computed independently.

That both field-dependent Nos\'e-Hoover conductivities are significantly less
irregular than the Gaussian thermostat result of Fig.\
\ref{gcond} may be understood with respect to the more ``stochastic''
nature of the Nos\'e-Hoover thermostat. According to its definition it
modifies the billiard dynamics more profoundly than in case of the Gaussian
constraint by transforming onto canonical velocity distributions, see again
Fig.\ \ref{fig:rhonhtds} for the probability densities at these
parameters. Nevertheless, both conductivities still exhibit pronounced
irregularities in form of non-monotonicities, and there are no indications for
a regime of linear response, at least not in the numerically accessible region
of $\varepsilon>0.1$. In other words, even the Nos\'e-Hoover thermostat
featuring a canonical velocity distribution does not lead to a more profound
appearance of a regime representing Ohm's law.

In summary, from a statistical mechanical point of view we have the rather
unpleasant result that not only dynamical systems properties such as
bifuraction diagrams but also transport properties such as the field-dependent
electrical conductivity of the deterministically thermostated, driven periodic
Lorentz gas strongly depend on details of the specific way of
thermostating. However, as in case of the discussion of van Kampen's criticism
in Section III.B.4 one may speculate that this is rather a consequence of the
small number of degrees of freedom and of the spatial periodicity of this
model.

\subsection{$^*$Subtleties and generalizations of the Nos\'e-Hoover thermostat}

\subsubsection{Necessary conditions for Nos\'e-Hoover thermostating and
generalizations of this scheme}

As we briefly discussed in Sections III.A and IV.B.2, both the Gaussian and
the Nos\'e-Hoover thermostat do not work properly for the periodic Lorentz gas
without external fields: In this case the Gaussian thermostat is trivially
non-existent, since the kinetic energy of the moving particle is conserved by
definition anyway, whereas the Nos\'e-Hoover thermostat does not act
appropriately because of the hard disk collisions. Controlling the two
velocity variables {\em separately} by Nos\'e-Hoover eliminates this problem
\cite{RKH00}. In any case, as outlined in the previous sections for both
thermostats a small electric field resolves these intricacies leading to
well-defined NSS. Apart from that, for interacting many-particle systems it
was confirmed numerically that both the Gaussian
\cite{HLM82,EH83,PoHo88} and the Nos\'e-Hoover thermostat
\cite{Nose84a,EvHo85,PoHo89} work correctly even in thermal
equilibrium. These examples suggest that the thermostated subsystem must
fulfill certain necessary conditions in order attain canonical velocity
distributions.

In the literature this problem became well-known by studying both the {\em
Nos\'e-Hoover oscillator}, which is a harmonic oscillator subject to a
Nos\'e-Hoover thermostat, and a respectively thermalized particle moving in a
double-well potential. Already in early computer simulations of the
Nos\'e-Hoover oscillator it was observed that this system did not attain a
canonical velocity distribution in thermal equilibrium
\cite{Hoov85,PHV86,HoHo86}. This is due to the  fact that the unperturbed
harmonic oscillator is non-chaotic and perfectly integrable. Hence, under
application of some thermal reservoir its regular phase space structure may
just be getting deformed, and there is no reason why the resulting subsystem
should become fully ergodic.

However, if such a previously regular subsystem is not driven to ergodic
behavior by the action of some thermostat the whole phase space will not be
sampled appropriately. As a consequence, the resulting velocity distribution
may strongly depend on initial conditions and is typically not the canonical
one
\cite{Hoov85,PHV86,HoHo86,KBB90,BuKu90}
\cite{Ham90,Hoo91,Wink92,MKT92,LHHa93,Nose93,WKR95,HoHo96,HoKu97,Mund00,LiTu00,HHI01,SeFe01,LaLe03}.
The same applies to the Nos\'e-Hoover thermostated double-well potential
\cite{KBB90,LHHa93,WKR95,BLL03}. Jellinek and Berry \cite{Jell88,JeBe89}
and Kusnezov et al.\ \cite{KBB90,BuKu90} thus emphasized that chaos and
ergodic behavior are necessary conditions in order to thermalize a subsystem
onto a canonical velocity distribution. This indicates that a thermostated
subsystem should strictly speaking be mixing \cite{ArAv68,Schu,Gasp,Do99}
implying ergodicity and, under rather general conditions, also chaotic
behavior \cite{Dev89,BBCDS92}.

These difficulties motivated the construction of a large variety of
generalizations of the original Nos\'e-Hoover thermostat
\cite{Ho89,KBB90,BuKu90,BuKu90b,Ham90,Hoo91,Wink92,MKT92,LHHa93,WKR95,HoHo96,PH97,HoKu97,BLL99,Mund00,LiTu00}
\cite{LaLe03,Bran00,BrWo00,HHI01,SeFe01,BLL03}.
The formal basis for these efforts was laid out again by Jellinek and Berry
\cite{Jell88,JeBe88} and by Kusnezov et al.\ \cite{KBB90,BuKu90}: Jellinek and
Berry showed that there is the additional freedom of changing the {\em
multiplicative} coupling between subsystem and thermal reservoir in
generalized coordinates while still being consistent with Nos\'e's basic idea
of transforming onto canonical distributions.\footnote{In more detail,
starting from the Nos\'e Hamiltonian they showed that changing the scaling
rules between the physical and the generalized variables, cp.\ to Eqs.\
(\ref{eq:gctra1}), (\ref{eq:gctra2}), plus possibly also changing a time
scaling between physical and generalized time (that we did not discuss in
Section IV.C.2) is not at variance with canonical velocity distributions being
the solutions of the respective Liouville equation Eq.\ (\ref{eq:gliouv}).}
Kusnezov et al., in turn, pointed out that further {\em additive}
contributions to Hamilton's equations of motion in generalized variables, see
Eqs.\ (\ref{eq:hcoupeom}), are possible, again without contradicting the
existence of canonical velocity distributions. Both approaches will typically
lead to thermostated equations of motion that are still time-reversible but
more nonlinear than the ones corresponding to the original Nos\'e-Hoover
scheme. Thus Nos\'e-Hoover represents just one choice of infinitely many for
modeling a deterministic thermal reservoir transforming onto canonical
distributions.

Starting from this important conclusion alternatives to Gaussian and
Nos\'e-Hoover dynamics were explored by many authors. Their strategy was to
improve the Nos\'e-Hoover scheme by constructing a thermostat that is able to
conveniently thermalize the harmonic oscillator, or the double-well problem,
onto a canonical velocity distribution in equilibrium. A straightforward
generalization along the lines of Nos\'e-Hoover is the thermalization of
higher even moments of the subsystem's velocity distribution onto the
canonical ones by means of additional thermostating variables
\cite{Ho89,JeBe89,HoHo96,PH97,HoKu97,LiTu00}. Another formal option is to
thermalize the position coordinates of a moving particle leading to the cubic
coupling scheme \cite{KBB90,BuKu90}. Enforcing the virial theorem, Hamilton
used a linear scaling of the position coordinates \cite{Ham90}, which later on
was combined with some multiplicative coupling \cite{LHHa93}. A thermalization
of odd and even particle velocities, partly supplemented by using higher
powers of the corresponding friction coefficients, was also explored and was
argued to represent a deterministic, time-reversible analogue of the
stochastic Langevin equation \cite{BuKu90b}. More recently, a directional
thermostat called {\em twirler} stirring the angular momentum of a moving
particle was introduced and applied in suitable combination with Gaussian and
Nos\'e-Hoover thermostats \cite{HeMo02,He03}.

Another scheme that became rather popular is the Nos\'e-Hoover chain
thermostat \cite{MKT92,TuMa00} that couples the differential equation for the
Nos\'e-Hoover friction coefficient additively to another Nos\'e-Hoover
friction coefficient, and so on. However, this solution was assessed to be
unstable out of equilibrium \cite{ESHH98,Bran00} leading to another
modification of it \cite{Bran00}.  Some multiplicative extensions of
Nos\'e-Hoover along the lines of Jellinek's formalism have been worked out by
Winkler et al.\ \cite{Wink92,WKR95} and became further amended in Ref.\
\cite{BrWo00}. Another, in a way, multiplicative variant was discussed in
Ref.\ \cite{SeFe01}. 

We finally mention the so-called Nos\'e-Poincar\'e method representing a
scheme that is in-between the original approaches by Nos\'e and
Nos\'e-Hoover. The resulting dynamics is identical to Nos\'e-Hoover, however,
it is generated by symplectic equations of motion that enable the application
of symplectic integration algorithms, which considerably increases the
numerical precision \cite{BLL99,BLL03}. A further extension of this approach
allowing to model generalized types of Nos\'e-Hoover dynamics along these
lines, which appears to share quite some similarities with Jellinek's ideas
\cite{Jell88,JeBe88}, was recently proposed in Ref.\ \cite{LaLe03}.

It is not the purpose of this review to give a full account of the specific
physical and numerical advantages and disadvantages of all these different
thermostating schemes. Here we may refer, e.g., to Refs.\
\cite{Jell88,HoHo96,PH97,HoKu97,Bran00} that partly review and criticize these
more recent methods. Following Jellinek, Branka and Wojciechowski
\cite{BrWo00} particularly endeavored to systematically explore the
numerical practicability and efficiency of these different classes of
thermostats. They arrived at essentially ruling out many of the up to now
existing solutions. Further, more physical constraints contradicting some of
the methods mentioned above will be briefly summarized in the following
section.

However, in our view one may even go so far to question the philosophy
underlying most of these novel constructions in that a `good' thermostat
should, by all means, enforce a canonical velocity distribution even for
regular dynamical systems such as the harmonic oscillator. The other way
around, one may argue that the harmonic oscillator is rather unsuited to be
properly thermalized at all just because of its inherent
regularity. Correspondingly, a successful transformation of this dynamics onto
a canonical velocity distribution indicates that the original dynamics of this
system is profoundly destroyed according to the action of the thermal
reservoir. This raises the question to which extent the resulting dynamics
represents merely the action of the thermostat and whether anything is left at
all of the original characteristics of the previously regular
subsystem. Nevertheless, we suspect that for practical computational purposes
it is still desirable to have a thermostat at hand which is capable to always
create a canonical velocity distribution irrespective of the detailed
properties of the original dynamical system to which it is applied.

This discussion clearly demonstrates that the world of deterministic and
time-reversible thermostats does not only consist of Gaussian and
Nos\'e-Hoover thermostats. Obviously, there exists a microcosmos of variants
of them. This should be taken into account if one attempts to come to general
conclusions concerning the second law of thermodynamics and the universality
of chaotic and transport properties of NSS based on the analysis of Gaussian
and Nos\'e-Hoover thermostats only.

\subsubsection{Applying thermal reservoirs to nonequilibrium situations}

In Section II.C we have argued that thermal reservoirs should generally be
constructed in equilibrium situations. In this case the statistical ensembles
are well-defined and the corresponding velocity distribution functions for
subsystem and thermal reservoir are known exactly. After their definition in
thermal equilibrium one may apply these reservoirs for thermalizing a
subsystem under nonequlibrium constraints, by expecting that the respective
thermostat still works sufficiently well such that a proper NSS is
created. One may now learn something new about the NSS resulting for a
particular subsystem by analyzing the corresponding nonequilibrium
distribution functions as well as the associated chaos and transport
properties, which are generally not known in advance.

However, first of all, it is not guaranteed that a thermal reservoir which
properly acts in thermal equilibrium also functions in nonequilibrium by
generating a NSS. Actually, it appears that most of the generalizations of the
Nos\'e-Hoover thermostat listed above have only been tested in thermal
equilibrium so far. Some of them have already been criticized for not working
properly under nonequilibrium constraints \cite{ESHH98,Bran00,BrWo00}. Apart
from that, only a few studies of nonequilibrium situations are available for
these generalized reservoirs \cite{PH97,HoKu97,Bran00,BrWo00}. The same
applies to the chaotic dynamical properties of subsystems connected to these
different thermal reservoirs even in thermal equilibrium, see only Refs.\
\cite{KBB90,MKT92,PH97,HHI01} for results of some specific cases.

Secondly, a thermal reservoir should always remain in thermal equilibrium,
even if the subsystem is under nonequilibrium constraints, and it should only
control the temperature of the subsystem, or possibly respective higher
moments of it. Consequently, a thermal reservoir should only act on the even
moments of the velocity distribution of a subsystem. If a thermal reservoir
were defined to also control odd moments of the velocity its action goes
beyond simple temperature control and may change the dynamics of the subsystem
profoundly. For example, in the driven periodic Lorentz gas the first moment
of the velocity yields the current of the subsystem in nonequilibrium. Thus, a
thermostat that constrains this quantity would simply pre-determine the
current.\footnote{Note, however, that a thermal reservoir may be defined such
that it is comoving with the current, in order to thermalize the subsystem
onto the proper temperature in the comoving frame; see Refs.\
\cite{PoHo88,SEI98} and also the discussion of a thermostated shear flow in
Section VII.E, respectively Refs.\ \cite{WKN99,Wag00}.}  In the same vein, a
thermal reservoir acting onto position coordinates may pre-determine, or at
least profoundly influence, the respective moments of the position coordinates
hence affecting transport coefficients like diffusion and higher-order Burnett
coefficients. Consequently, on the basis of physical grounds one may wish to
constrain the action of thermal reservoirs to even velocity moments only.  If
one follows this argumentation, this already rules out schemes such as the
ones presented in Refs.\ \cite{KBB90,BuKu90b,BuKu90,Ham90,LHHa93}.

In conclusion, all what one can demand is that a thermal reservoir properly
acts in a thermal equilibrium situation by thermalizing a subsystem onto a
pre-determined velocity distribution. In nonequilibrium the general
requirement should be the existence of a NSS at a certain temperature, whose
properties as resulting from the action of this thermal reservoir may then be
studied respectively.

\subsection{Summary}

\begin{enumerate}

\item Starting from the conservation of the number of points in phase
space we have derived the Liouville equation for dissipative dynamical
systems. The Hamiltonian version of this Liouville equation, which typically
appears in textbooks, is obtained from it under the additional assumption of
conservation of phase space volume.

\item By employing the general form of the Liouville equation and by
requiring that canonical distributions for the velocities of a subsystem and
for a reservoir variable exist, we arrived at a simple differential equation
determining the velocity-dependent friction coefficient. This equation defines
the action of the {\em Nos\'e-Hoover thermostat}, supplemented by a respective
friction term in the original equations of motion of the subsystem.

In complete analogy to the Gaussian thermostat, Nos\'e-Hoover thermostated
equations of motion are deterministic and time-reversible but
non-Hamiltonian. However, in contrast to the Gaussian reservoir the
Nos\'e-Hoover scheme, by construction, attempts to transform the subsystem
velocities onto a canonical distribution in thermal equilibrium. Hence, it
yields a dynamics that is more similar to Langevin's theory than the one
resulting from the Gaussian constraint of constant energy. Another difference
of the Nos\'e-Hoover thermostat compared to the Gaussian one is the appearance
of an additional control parameter that may be interpreted as the response
time of the thermal reservoir regarding its interaction with a subsystem. In
case of infinitely slow response the Stokes friction coefficient is recovered,
whereas for infinitesimally quick response the Gaussian constraint is
approached. A variation of this control parameter changes the shapes of the
corresponding velocity distributions that may accordingly be composed of
superpositions of microcanonical and canonical densities.

\item The Nos\'e-Hoover thermostat belongs to the same class of deterministic
and time-reversible thermal reservoirs as the Gaussian one as far as an
identity between the average rate of phase space contraction and thermodynamic
entropy production is concerned. Respectively, there is also an analogous
relation between transport coefficients and Lyapunov exponents for
Nos\'e-Hoover dynamics.

For both thermal reservoirs there exist generalized Hamiltonian and Lagrangian
formulations. For the isokinetic Nos\'e-Hoover thermostat we have outlined how
to derive the respective equations of motion from a Hamiltonian in generalized
variables via non-canonical transformations.

Like the Gaussian thermostated driven periodic Lorentz gas, the Nos\'e-Hoover
version is as well characterized by a fractal attractor of a similar type. The
specific structure of this attractor depends on the value of the reservoir
response time. For large enough response times the bifurcations displayed by
this fractal attractor are typically very different from the ones observed for
the Gaussian reservoir. The same applies to the field-dependent electrical
conductivities in case of Nos\'e-Hoover, which still show irregularities on
fine scales and no indication of linear response in the numerically accessible
regime of the field strength.

\item Finally, we have summarized problems with Gaussian and particularly
with Nos\'e-Hoover thermostats in equilibrium and nonequlibrium
situations. For regular dynamical systems such as the harmonic oscillator or a
particle moving in a double-well potential it is well-known that the standard
Nos\'e-Hoover thermostat is not able to thermalize these systems onto
canonical velocity distributions in thermal equilibrium. A necessary condition
is that the thermostated system must be mixing, respectively chaotic and
ergodic. In order to achieve this goal the Nos\'e-Hoover scheme can be
generalized yielding more nonlinear equations of motion than the original
method. However, such ``stronger'' thermostating forces affecting more
profoundly the original dynamics of the subsystem to be thermostated pose the
question to which extent the original dynamics of the subsystem still plays a
role at all. In any case, our discussion clearly shows that a great variety of
deterministic and time-reversible thermal reservoirs exists. This fact one may
particularly want to take into account with respect to suspected
universalities of NSS as concluded from applications from Gaussian and
Nos\'e-Hoover thermostats only.

\end{enumerate}

\section{Summary and criticism of Gaussian and Nos\'e-Hoover thermostats}

The approach to nonequilibrium transport reviewed in the previous chapters
yields NSS on the basis of non-Hamiltonian equations of motion. This dynamics
results from employing deterministic and time-reversible thermal reservoirs
that control the temperature of a subsystem under nonequilibrium
constraints. The Gaussian and the Nos\'e-Hoover thermostat constructed before
serve as two famous examples of such mechanisms. Both reservoirs have been
applied to the periodic Lorentz gas driven by an external electric field, and
the resulting NSS have been analyzed with respect to their chaos and transport
properties. In the following we focus onto what we consider to be generic
properties of NSS associated with Gaussian and Nos\'e-Hoover dynamics. Our
summary is accompanied by a critical discussion of these fundamental features
as far as a possible universal description of NSS, irrespective of the
specific type of thermostat, is concerned. This assessment will set the scene
for the subsequent two chapters.

\subsection{Non-Hamiltonian dynamics for nonequilibrium steady states}

Constructing NSS requires either to start from a Hamiltonian that models a
thermal reservoir of infinitely many degrees of freedom or to use
non-Hamiltonian equations of motion, see Section I.B. As an example, in
Section II.B we have derived the stochastic Langevin equation from a
Hamiltonian defining a thermal reservoir that consists of infinitely many
harmonic oscillators. The resulting equation turned out to be dissipative,
non-deterministic and non time-reversible thus providing a well-known example
of non-Hamiltonian dynamics. This dynamics particularly results from the fact
that the equations of motion of the infinitely many reservoir degrees of
freedom were eliminated in the course of the derivation, supplemented by some
further approximations.

Actually, as we have demonstrated in Section II.C, projecting out spurious
reservoir degrees of freedom typically\footnote{A counterexample was provided
by subsystem and reservoir living altogether in three dimensions and
projecting out two of them, see Fig.\ \ref{fig:d346}.} yields a non-uniform
probability density for the thermostated subsystem while the combined
distribution of subsystem plus thermal reservoir is still uniform,
respectively microcanonical. The non-uniformity of the projected equilibrium
density may be taken as an indication of some phase space contraction in the
associated variables which is not present in the original phase space volume
preserving Hamiltonian dynamics of the full system. Indeed, in case of the
Langevin equation and for Gaussian and Nos\'e-Hoover dynamics the equations of
motion of the respectively thermostated subsystem are always non-Hamiltonian.

We thus argue that the non-Hamiltonian character results from conveniently
simplifying the equations of motion of subsystem plus reservoir by considering
the relevant degrees of freedom only while neglecting others. Hence, such a
non-Hamiltonian formulation is in principle very well compatible with a
Hamiltonian description of the complete combination of subsystem plus
reservoir.

\begin{figure}[t]
\epsfysize=13cm
\centerline{\rotate[r]{\epsfbox{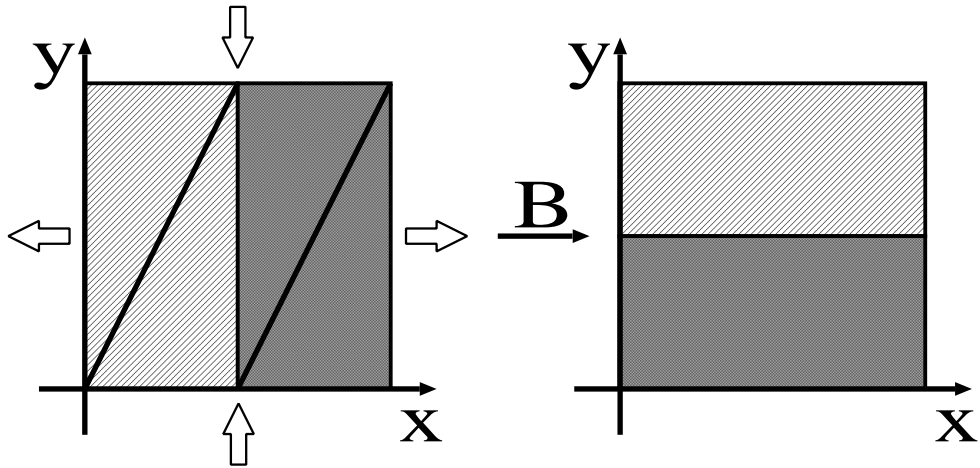}}}
\vspace*{0.2cm} 
\caption{Sketch of the baker map defined by Eq.\ (\ref{eq:baker}): a unit square is
strechted horizontally and squeezed vertically by preserving its unit
area. The resulting rectangle is cut in the middle, and both parts are put on
top of each other forming again a unit square. The two diagonal lines in the
left part of the figure represent the one-dimensional Bernoulli shift acting
onto the $x$ variable only, which determines the dynamics of the baker map
along the horizontal axis.}
\label{fig:baker}
\end{figure}

Since the use of non-Hamiltonian dynamics for modeling NSS was
\cite{EyLe92,Coh92} and is \cite{LLM02} often criticized, we illustrate our
arguments by means of a very simple example. Let us consider the
two-dimensional baker map acting onto the unit square,
\begin{equation}
B(x,y)  = \left\{ 
\begin{array}{r@{\quad,\quad}l}
(2x,(y+1)/2) & 0\le x<\frac{1}{2} \\ 
(2x-1,y/2) & \frac{1}{2} \le x\le 1
\end{array}
\right. \quad . \label{eq:baker}
\end{equation}
This map squeezes the unit square along the vertical direction while
stretching it horizontally such that the area is preserved. The resulting
rectangle is cut in the middle, and both parts are put on top of each other
yielding the dynamics depicted in Fig.\
\ref{fig:baker}. The corresponding equations of motion read
\be
(x_{n+1},y_{n+1})=B(x_n,y_n) \quad ,
\ee
where $n \in\hbox{\mb N}$ holds for the discrete time. 
As already remarked in Refs.\ \cite{Gasp,GND03}, it appears that the baker map
was first defined and studied by Seidel in 1933 \cite{Sei33} who, in turn,
gives informal credit to Birkhoff for its invention. Later on this map was
considered by Hopf \cite{Ho37}. Eventually, it became very popular starting
from the discussion by Arnold and Avez \cite{ArAv68}, who also coined the
name.\footnote{Both Seidel and Hopf used the baker map as an example for a
system that is metrically transitive, or mixing, respectively. Arnold and Avez
proved that this map is Bernoulli exhibiting the property of a K-system with
positive metric entropy. Seidel originally discussed a baker with
stretching/squeezing factors of $10$, respectively $1/10$.} The baker map is
one of the most simple chaotic dynamical systems serving as a standard model
in many textbooks on nonlinear dynamics, see, e.g., Refs.\
\cite{Schu,Ott,Gasp,Do99}. Starting from the work by Gaspard \cite{PG1} the
baker dynamics became extremely popular in form of an analytically tractable
model for nonequilibrium transport, the spatially extended so-called
multibaker map
\cite{Gas93,TG2,TeVB96,VTB97,Gasp97a,BTV98,VTB98,GaKl,GiDo99,GFD99,TG99,TG00}
\cite{VTM00,RTV00,GDG00,GD00,TVS00,GDG01,Voll02,WD02,GaOl02,GilDo03,GND03,WD03,VTB03}.\footnote{For
multibaker maps the unit cell of the baker map is periodically continued along
the $x$-axis. The single cells are then coupled with each other by mapping,
say, the left vertical strip of the left part in Fig.\ \ref{fig:baker} into
the next cell to the right while mapping the right vertical strip into the
next cell to the left, and so on.} Even so-called thermostated multbaker maps
have been constructed and analyzed from the point of view of mimicking the
action of Gaussian and Nos\'e-Hoover thermostats
\cite{TeVB96,VTB97,BTV98,VTB98,GiDo99,GFD99,VTM00,RTV00,GD00,VTB03}. However, in this
review we restrict ourselves to the original formulation of Gaussian and
Nos\'e-Hoover dynamics in terms of differential equations only. Corresponding
time-discrete dissipative models are reviewed particularly in Refs.\
\cite{TVS00,Voll02,Do99}.

That the baker map Eq.\ (\ref{eq:baker}) is area-preserving can be confirmed
by computing its Jacobian determinant. The map is furthermore invertible and,
even more, shares a specific symmetry property that is usually referred to as
{\em reversibility} in time-discrete maps.\footnote{\label{fn:trev}A map $B$
is called {\em reversible} if there exists an involution $G, GG=1$, in phase
space reversing the direction of time, $BGB=G$ \cite{Robe}. Thus reversibility
in maps is more than the existence of an inverse. However, note that such a
reversibility does not necessarily imply time-reversibility \cite{DK00} and
that there is a controversy of how to define time-reversibility in maps
\cite{Robe,TG2,HKP96,HP98b,BTV98,GFD99,Gasp,Voll02}.} These properties
are analogous to respective features in time-continuous Hamiltonian dynamical
systems.\footnote{However, note that the baker map does not share the property
of being symplectic, in contrast, for example, to the standard map
\cite{Ott,Meis92}.} It is not our purpose to give a detailed account concerning
the dynamical systems properties of the baker map.\footnote{Let us mention
only some very basic characteristics: Apart from being mixing and a K-system
implying chaotic behavior the baker map is also hyperbolic
\cite{Do99}. Strictly speaking it is not Anosov because it is not
differentiable at the point of discontinuity \cite{Gasp}.} We just wish to
convey that it is a widely studied standard model mimicking, at least to some
extent, a Hamiltonian chaotic dynamical system on the level of time-discrete
equations of motion.

In order to illustrate our above discussion about the origin of
non-Hamiltonian dynamics we now project out one variable of the baker map.
This can be performed starting from the Liouville equation for the baker
map\footnote{The time-discrete Liouville equation is usually called {\em
Frobenius-Perron equation} in dynamical systems theory
\cite{Do99,Gasp,Ott}.} by simply integrating over the $y$-variable, see Ref.\
\cite{Do99} for the detailed calculation. However, $x$ and $y$ are already nicely
decoupled in Eq.\ (\ref{eq:baker}) with respect to the direction of the
projection. One may therefore immediately conclude that the time-discrete
Liouville equation determining the evolution of the probability density along
the $x$-axis only must be governed by the action of the one-dimensional map
\be
x_{n+1}\equiv B_x(x_n)=2x_n \quad\mbox{mod}\quad 1 \quad ,
\ee
This {\em Bernoulli shift} is included in Fig.\ \ref{fig:baker}. Its Jacobian
determinant is equal to two, hence this map is uniformly expanding and does
not preserve the phase space volume. The baker map thus provides a simple
example of a system whose complete two-dimensional dynamics is
area-preserving, whereas by projecting out one variable the associated
one-dimensional equations of motion are dissipative.

Curiously, the equilibrium density of the Bernoulli shift is still uniform, as
for the original baker map, whereas in Section II.C the probability densities
obtained from projecting out reservoir degrees of freedom were typically
non-uniform. In order to more closely connect to these results one may simply
rotate the unit square of the baker map by $45$ degrees
\cite{HP98b}. Now the density projected upon the horizontal line by
integrating over the vertical axis will be non-uniform, whereas the full baker
map is still area-preserving. Unfortunately, in contrast to the non-rotated
case here the dynamics along the horizontal, respectively the vertical axes
are not decoupled anymore,\footnote{This is related to the fact that we are
not projecting onto the unstable manifold only, which is usually the adequate
choice for performing a projection.} and it is not straightforward to extract
equations of motion governing the dynamics along the $x$-axis only.

This simple example demonstrates again that non-Hamiltonian dynamics may fit
very well together with Hamiltonian equations of motion for the complete
system consisting of subsystem plus thermal reservoir. In such a case a
non-Hamiltonian dynamics just results from conveniently projecting out
superfluous (reservoir) degrees of freedom such as the passive $y$-variable in
the baker map, see also Rondoni et al.\ \cite{Rond02,ER02} for an analogous
argumentation. In our view a non-Hamiltonian description thus emerges very
naturally in equations designed to model NSS, and in this respect Gaussian and
Nos\'e-Hoover thermostats are not particularly artificial or unusual.

On the other hand, one should keep in mind that Gaussian and Nos\'e-Hoover
reservoirs have not been derived starting from a purely Hamiltonian dynamics,
see, e.g., again the derivation of the Langevin equation in Section II.B for
comparison. Moreover, in contrast to the projected baker dynamics and to the
Langevin equation the former two types of dynamics are even
time-reversible. Of course, for Gauss and Nos\'e-Hoover there exists the
generalized Hamiltonian approach discussed in Section IV.C.2, however, it only
represents a formal analogy and not a rigorous Hamiltonian derivation. We
therefore emphasize again that Gaussian and Nos\'e-Hoover dynamics yield only
a {\em heuristic modeling} of NSS rather than providing a theory starting from
first principles. Hence, one cannot rule out in advance that the properties of
dynamical systems thermostated that way may depend on specificalities of these
thermal reservoirs. This should be taken into account when searching for
universal properties of NSS along the lines of this approach.

\subsection{Phase space contraction and entropy production}

The periodic Lorentz gas driven by an external electric field yielded NSS
after connecting this model to either the Gaussian or to the Nos\'e-Hoover
thermostat. Both types of thermostated systems were characterized by an
identity between the absolute value of the average rate of phase space
contraction and the thermodynamic entropy production in terms of Joule's heat,
see Sections III.B.1 and IV.C.1. This identity represents a basic property of
ordinary Gaussian and Nos\'e-Hoover dynamics and furnishes a crucial link
between dynamical systems quantities and transport properties. Interestingly,
even the stochastic Langevin equation Eq.\ (\ref{eq:lang1d}) supplemented by
an external electric field exhibits this identity, as one can easily
verify. However, in contrast to Gaussian and to Nos\'e-Hoover dynamics this
equation is less amenable to an analysis in terms of dynamical systems theory
because of the stochastic forces involved. 

We now show that this identity is in fact linked to a more general
relationship between phase space contraction and entropy production, which
holds irrespective of the specific type of dynamical system considered. In our
presentation we follow Ref.\ \cite{And85}, for other derivations see
particularly Refs.\ \cite{Ev85,EvMo90,Gasp97a,Do99,Gall99,Rue99,DettS00,ER02}.

Let us start from the general dynamical system
\be
{\bf \dot{x}}={\bf F}({\bf x}) \quad , \label{eq:gds}
\ee
where ${\bf x}$ and ${\bf F}$ denote vectors in a $k$-dimensional phase space
${\bf \Gamma}\subset\hbox{\mb R}^k$, $k\in \hbox{\mb N}$. For convenience here
we shall not distinguish between position and velocity components. One may now
inquire about the average rate of entropy production in this system by using
{\em Gibbs definition of entropy} \cite{Penr79,EvMo90,Rue99,Gasp,DettS00}
\be
S_G:=-\int_{{\bf \Gamma}}d{\bf \Gamma}\: \rho\ln\rho \quad . \label{eq:gent}
\ee
Here $\rho\equiv\rho(x_1,x_2,\ldots,x_k)$ denotes the distribution function of
the {\em complete} dynamical system Eq.\ (\ref{eq:gds}) determined by the
Liouville equation Eq.\ (\ref{eq:gliouv2}) and $d{\bf \Gamma}:=dx_1dx_2\ldots
dx_k$ represents a volume element of the phase space ${\bf \Gamma}$. In order
to calculate the Gibbs entropy production we differentiate Eq.\
(\ref{eq:gent}) by employing the Leibnitz rule, which yields
\be
\frac{dS_G}{dt}=-\int_{{\bf \Gamma}}d{\bf \Gamma}\:(1+\ln\rho)\pard{\rho}{t} -
\sum_i\int_{{\bf \Gamma}'}d{\bf \Gamma}' \rho\ln\rho\:F_i|^{ }_{\partial\Gamma_i}
\ee
with $d{\bf \Gamma}':=dx_1\ldots dx_{i-1}dx_{i+1}\ldots dx_k$ and $F_i$
defined at the boundary $\partial\Gamma_i$. The derivative
$\partial{\rho}/\partial{t}$ can be substituted by using the Liouville
equation Eq.\ (\ref{eq:gliouv2}) leading to
\be
\frac{dS_G}{dt}=\int_{{\bf \Gamma}}d{\bf \Gamma}\:\rho\:\mbox{{\boldmath $\nabla$}}\cdot{\bf F} +
\int_{{\bf \Gamma}}d{\bf \Gamma}\:\rho\ln\rho \:\mbox{{\boldmath $\nabla$}}\cdot{\bf F}
 + \int_{{\bf \Gamma}}d{\bf \Gamma}\:(1+\ln\rho)\:{\bf F}\cdot\mbox{{\boldmath
 $\nabla$}}\rho - \sum_i\int_{{\bf \Gamma}'}d{\bf \Gamma}' \rho\ln\rho\:F_i|^{
 }_{\partial\Gamma_i} \quad .
\ee
Performing integration by parts the second term on the right hand side of the
above equation precisely cancels with the last two terms in the same
equation. The final result thus reads
\be
\frac{dS_G}{dt}=\int_{{\bf \Gamma}}d{\bf \Gamma}\: \rho\mbox{{\boldmath $\nabla$}}\cdot{\bf F}
\label{eq:prgep} \quad ,
\ee
which means that the rate of Gibbs entropy production is always identical to
the average rate of phase space contraction defined by Eq.\
(\ref{eq:psc}). This important identity was noted by Gerlich
\cite{Gerl73} and was subsequently discussed in various settings by other
authors \cite{Dobb76a,Dobb76b,Stee79,And82}. A very lucid and clear statement
of this identity is due to Andrey \cite{And85}. In the context of Gaussian and
Nos\'e-Hoover dynamics the importance of the relationship Eq.\
(\ref{eq:prgep}) was particularly emphasized in Refs.\
\cite{Ev85,BrTV96,HHP87,HoPo87,PoHo88,Rue96,Rue97b,Gall99} \cite{Rue99,HoB99,Do99,Rue03}.

As we have mentioned at the beginning of Chapter I, it appears that there is
no generally accepted definition of a thermodynamic entropy in nonequilibrium
situations. Some authors considered the Gibbs entropy Eq.\ (\ref{eq:gent}) to
be a suitable candidate for such a nonequilibrium entropy
\cite{EvMo90,Gall99,Rue99,HoB99}. However, in our view using this concept
causes a number of problems.

First of all, Eq.\ (\ref{eq:prgep}) tells us that for phase space volume
preserving dynamics the Gibbs entropy production is strictly zero. This fact
is problematic \cite{NiDa98,DettS00}, for example, in case of diffusion in
Hamiltonian dynamical systems under concentration gradients imposed by flux
boundary conditions \cite{Gasp97a,Gasp}, see also Ref.\
\cite{EPRB99b} for a Hamiltonian modeling of a heat flow under temperature
gradients. In both nonequilibrium situations there is a well-defined
irreversible entropy production while the Gibbs entropy production is strictly
zero, since there is no non-Hamiltonian thermostating.

Secondly, dissipative dynamical systems usually exhibit a {\em negative}
average rate of phase space contraction. Specific examples are provided by the
Gaussian and the Nos\'e-Hoover thermostated driven Lorentz gas, see Eqs.\
(\ref{eq:peqa}) and (\ref{eq:nhprep}). According to Eq.\ (\ref{eq:prgep}) the
Gibbs entropy production is also negative reflecting a contraction of the
system onto a subset in phase space such as a fractal attractor. The decrease
of the Gibbs entropy thus measures the fact that the dynamical system is
getting ``more ordered'' in phase space. However, a negative thermodynamic
entropy production in NSS appears to be in contradiction with the second law
of thermodynamics.

In our earlier derivations of Sections III.B.1 and IV.C.1, which specifically
concerned the Gausssian and the Nos\'e-Hoover thermostated Lorentz gas, we
circumvented this sign problem, since we did not start from the Gibbs entropy
production. Instead, we directly computed the average rates of phase space
contraction by realizing afterwards that the {\em absolute} values yielded the
thermodynamic entropy production in form of Joule's heat. This enabled us to
``invert the sign'' and to establish the identity without entering into the
discussion concerning a negative Gibbs entropy production.

A more physical interpretation of the sign reversal may be obtained as follows
\cite{Ch1,Rue96,Do99,HoB99}: Let us consider a balance equation for the total
entropy production of a subsystem under nonequilibrium conditions, in analogy
to Eq.\ (\ref{eq:partbal}). The source term stands for the entropy production
within the subsystem, whereas the flow term holds for the average heat
transfer from the subsystem to the thermal reservoir. The reservoir fully
absorbs the entropy produced by the subsystem such that in a NSS the average
entropy production of subsystem plus thermal reservoir is zero. Hence, from
the side of the thermal reservoir the entropy flux must be {\em positive}
causing a positive entropy production in the reservoir. This entropy
production may be identified as the relevant one in the sense of irreversible
thermodynamics thus yielding the sign reversal. However, in detail this
explanation is debatable \cite{NiDa96,Gasp97a,Gasp,NiDa98,GiDo99,DaNi99},
because we did not identify a positive entropy production {\em within} the
subsystem itself.

More refined approaches resolve the sign problem by using methods of coarse
graining for the phase space densities as they enter into the definition of the
Gibbs entropy Eq.\ (\ref{eq:gent})
\cite{BrTV96,NiDa96,VTB97,Gasp97a,VTB98,BTV98,NiDa98}
\cite{GiDo99,TG99,DaNi99,Gasp,GDG00,GD00,TG00,TVS00,DGG02,Voll02,MaNe03,VTB03}. 
Such a coarse grained formulation may be motivated by a second basic
deficiency of the Gibbs entropy \cite{NiDa96,Gasp97a,DaNi99,GiDo99,Gasp}: As
we have shown in Sections III.B.3 and IV.C.3, in deterministically
thermostated systems the phase space density typically contracts onto fractal
attractors. However, the fractal structure implies that the probability
densities characterizing the associated NSS must be singular and
non-differentiable. Consequently, they are no well-defined mathematical
objects anymore.\footnote{In contrast, the corresponding probability {\em
measures} may still be well-defined. Typically, these are SRB measures being
smooth along unstable but fractal along stable manifolds
\cite{Gasp,Do99,Gall99,Rue99,Young02,Gall03b}.}  Correspondingly, the
integration in Eq.\ (\ref{eq:gent}) is also not well-defined
anymore.\footnote{In this case the integral should be replaced by a
Lebesgue-Stieltjes integral over the invariant measure \cite{Do99,Gasp}.}  As
a consequence of the formation of these fractal structures it was thus
proposed by many authors to use {\em coarse-grained} Gibbs entropies in order
to arrive at a suitable concept of a nonequilibrium entropy, see the long list
of references cited above. Up to now these methods have mostly been worked out
for time-discrete dynamical systems such as multibaker-maps for which
different types of coarse graining have been explored. Concerning further
details we may refer, e.g., to Refs.\
\cite{GiDo99,Gasp,TVS00,Voll02} providing recent summaries of this issue.

In the light of this criticism it may appear rather surprising that for the
Gaussian and for the Nos\'e-Hoover thermostated driven Lorentz gas the
ordinary Gibbs entropy production can nevertheless be fully identified with
the thermodynamic entropy production in terms of Joule's heat subject to a
sign reversal. A related fact that further contributes to this surprise is
that Gibbs' formulation assesses the entropy of a system with respect to {\em
all} phase space variables and with respect to their {\em complete}
deterministic dynamics in time. This is at variance to concepts like
coarse-grained Gibbs entropies or Boltzmann entropies. Here there is already
a loss of information, because these quantities are not employing the complete
phase space densities or respective probability measures as determined by the
Liouville equation. In a similar vein, a heat flux from a subsystem to a
thermal reservoir will usually involve only a specific fraction of all phase
space variables. From the point of view of Clausius definition of entropy
production, see Section III.B.1, the Gibbs entropy may thus measure spurious
contributions in comparison to a thermodynamic entropy production by assessing
phase space coordinates that are not involved in any heat transfer to a
thermal reservoir.

All these physical arguments support again the use of a respectively coarse
grained entropy that only assesses the physically relevant contributions to a
thermodynamic entropy production. We believe that such a formulation of a
nonequilibrium entropy should be compatible with the one of Clausius in terms
of a heat transfer. From that point of view there is no reason why phase space
contraction, or the respective Gibbs entropy production, should always be
identical to the thermodynamic one. One may suspect that such an identity may
rather be characteristic of the specific family of thermostats investigated so
far.

This hypothesis will be verified by presentings generic families of
counterexamples where the identity between phase space contraction and
thermodynamic entropy production, respectively between Gibbs and thermodynamic
entropy production, is not fulfilled. In order to qualify as counterexamples
alternative models of thermal reservoirs must share some basic physical
properties with Gaussian and Nos\'e-Hoover thermostats, such as being
dissipative, deterministic and time-reversible. Furthermore, there should be
an adequate physical interpretation of the thermostating mechanism. Such
counterexamples will be introduced and analyzed in the following two chapters.

\subsection{Transport coefficients and dynamical systems quantities}

If there exists an identity between the average phase space contraction rate
and thermodynamic entropy production, it is straightforward to link transport
coefficients with dynamical systems quantities. Interestingly, for the
field-dependent conductivity $\sigma(\varepsilon)$ of both the Gaussian and
the Nos\'e-Hoover thermostated driven periodic Lorentz gas one obtains the
same functional relationship reading
\be
\sigma(\varepsilon)=-\frac{T}{\varepsilon^2}\sum_{i=1}^N\lambda_i(\varepsilon)
\quad , 
\label{eq:tklt} 
\ee
cp.\ to Eqs.\ (\ref{eq:lsr}) and (\ref{eq:lsrnh}), where $N$ denotes the total
number of field-dependent Lyapunov exponents $\lambda_i(\varepsilon)$ and $T$
stands for the temperature. The existence of such a simple functional
relationship between chaos quantities and transport coefficients is typical
for Gaussian and Nos\'e-Hoover thermostated dynamical systems, see Sections
III.B.2 and IV.C.1 for further details.

Surprisingly, by using the escape rate approach to chaotic transport outlined
in Section I.A, independently very similar formulas were derived for the very
different class of {\em open Hamiltonian} dynamical systems, where there is an
escape of phase space points due to absorbing boundaries. This second
fundamental approach to NSS does not require any modeling of thermal
reservoirs, hence by default it does not involve any phase space
contraction. We will now provide a brief summary concerning formulas linking
transport coefficients to dynamical systems quantities as they emerge from
both the escape rate and the thermostated systems approach. The resulting
different but related formulas will be compared in detail and
possible crosslinks between them will be critically assessed.

For sake of simplicity we mainly restrict ourselves again to the periodic
Lorentz gas. We start with the driven Lorentz gas under application of a
Gaussian thermostat for which $N=2$ in Eq.\ (\ref{eq:tklt}). In order to
connect to the escape rate approach we reformulate this equation as follows
\cite{Gasp,GDG01,Rue96}: For the Gaussian thermostated Lorentz gas the sum
of Lyapunov exponents on the right hand side consists only of a positive one,
$\lambda_+(\varepsilon)$, and a negative one, $\lambda_-(\varepsilon)$, see
Eq.\ (\ref{eq:lsr}). Both Lyapunov exponents are defined with respect to the
invariant measure of the dynamical system which is concentrated on the
field-dependent fractal attractor $\cal A_{\varepsilon}$. We now indicate this
dependence explicitly by writing
$\lambda_{\pm}(\varepsilon)\equiv\lambda_{\pm}({\cal A_{\varepsilon}})$.

As was proven in Refs.\ \cite{Ch1,Ch2}, under certain conditions there holds
the {\em Pesin identity} \cite{ER,Ott,Beck,Gasp,Do99} between the
Kolmogorov-Sinai entropy $h_{KS}({\cal A_{\varepsilon}})$ and the positive
Lyapunov exponent on the attractor, $h_{KS}({\cal
A_{\varepsilon}})=\lambda_+({\cal A_{\varepsilon}})$, for the Gaussian
thermostated driven Lorentz gas. Having this in mind we rewrite the right hand side
of Eq.\ (\ref{eq:tklt}) as
\bna
-\sum_{i=1}^2\lambda_i({\cal A_{\varepsilon}})&=&-\lambda_+({\cal
A_{\varepsilon}})-\lambda_-({\cal A_{\varepsilon}}) 
\nonumber\\ 
&=&-h_{KS}({\cal A_{\varepsilon}})-\lambda_-({\cal A_{\varepsilon}}) \nonumber \\
&=&|\lambda_-({\cal A_{\varepsilon}})|-h_{KS}({\cal A_{\varepsilon}}) \label{eq:tdspes}
\ena
The domain of the escape rate approach is particularly the description of
diffusion processes, see Section I.A. Accordingly, since the Einstein relation Eq.\
(\ref{eq:dts}) holds for the Gaussian thermostated driven Lorentz gas
\cite{Ch1,Ch2} the electrical conductivity $\sigma(\varepsilon)$ on the left
hand side of Eq.\ (\ref{eq:tklt}) may be replaced by the diffusion coefficient
$D$, $D=T\sigma(\varepsilon)\:(\varepsilon\to0)$. This yields as a final
result \cite{Gasp,GDG01,Rue96,BrTV96}
\be
D=\lim_{\varepsilon\to0}\left(\frac{T}{\varepsilon}\right)^2\left(|\lambda_-({\cal
A_{\varepsilon}})|-h_{KS}({\cal A_{\varepsilon}})\right) \quad . \label{eq:tkltf}
\ee
A corresponding formula from the escape rate approach to chaotic diffusion is
obtained as follows
\cite{GN,GB1,Gas93,GaBa95,DoGa95,GaDo95,RKdiss,KlDo99,Gasp,Do99}:
We consider a slab of the periodic Lorentz gas with reflecting or periodic
boundaries, say, in the vertical direction while in the horizontal direction
we use absorbing boundaries a distance $L$ apart from each other. We do not
apply any external field but just look at the diffusion of an initial ensemble
of point particles along the $x$-axis. Choosing the density of scatterers such
that the periodic Lorentz gas is normal diffusive, see Section II.D, we expect
an exponential decrease of the number $N$ of particles in time $t$ according
to
\be
N(t)=N(0)\exp(-\gamma_{esc}t) \quad , 
\ee
where $N(0)$ stands for the initial number of particles at time zero. Solving
the one-dimensional diffusion equation with absorbing boundary conditions
yields
\be
D=\lim_{L\to\infty}\left(\frac{L}{\pi}\right)^2\gamma_{esc} \quad
, \label{eq:dkesc}
\ee
that is, the diffusion coefficient is obtained in terms of the {\em escape
rate} which strictly speaking depends on $L$,
$\gamma_{esc}\equiv\gamma_{esc}(L)$. Let us now assume that the {\em Pesin
identity for open systems} holds for the open periodic Lorentz
gas,\footnote{So far this identity is only proven for Anosov diffeomorphisms,
see, e.g., Refs.\
\cite{ER,Beck,Do99,Gasp} for outlines concerning this generalization of
Pesin's identity and further references therein.} 
\be
\gamma_{esc}=\lambda_+({\cal R}_L)-h_{KS}({\cal R}_L) \quad . 
\ee
Here $\gamma_{esc}\equiv\gamma_{esc}({\cal R}_L)$ denotes the {\em escape
rate} of particles with respect to the fractal repeller ${\cal R}_L$.
Combining this equation with Eq.\ (\ref{eq:dkesc}) leads to the fundamental
result
\be
D=\lim_{L\to\infty}\left(\frac{L}{\pi}\right)^2\left(\lambda_+({\cal
R}_L)-h_{KS}({\cal R}_L)\right) \quad . \label{eq:tkler}
\ee
Obviously, the functional forms of Eqs.\ (\ref{eq:tkltf}) and (\ref{eq:tkler})
are just the same. However, we emphasize again that both the types of
dynamical systems considered and the approaches by which these equations have
been derived are very different.

On the other hand, for both derivations the key was to identify a quantity
providing a link between chaos and transport properties. For thermostated
systems this role is played by the average phase space contraction rate
$\kappa$ defined in Eq.\ (\ref{eq:psc}). In case of open Hamiltonian systems
the escape rate $\gamma_{esc}$ serves for the same purpose. This analogy was
first seen by Breymann, T\'el and Vollmer \cite{BrTV96} and by Ruelle
\cite{Rue96}. These authors suggested to define a Gibbs entropy for open dynamical
systems, in parallel to the Gibbs entropy of closed but thermostated, phase
space contracting dynamical systems, related to the concept of conditionally
invariant measures. Here the probability measures of particles remaining in
the system, respectively the corresponding probability densities, are
renormalized in time in order to make up for the loss of absorbed
particles. Along these lines the Gibbs entropy production for open systems can
be identified with the escape rate, $dS_G/dt=\gamma_{esc}$. On the other hand,
for closed systems we have shown in the previous section that the Gibbs
entropy production is equal to the average rate of phase space contraction,
see Eq.\ (\ref{eq:prgep}). Hence, if the rates of Gibbs entropy production for
closed and open systems were the same both Eqs.\ (\ref{eq:tkltf}) and
(\ref{eq:tkler}), that apply to very different dynamical systems, yielded the
same transport coefficients in terms of dynamical systems quantities. This was
indeed shown to be the case for some simple model systems, that is,
(multi)baker maps, under periodic, absorbing and flux boundary conditions thus
suggesting an equivalence of these very different nonequilibrium ensembles
\cite{MoRo96,VTB97,VTB98,GiDo99,GFD99,GD00,TVS00,Voll02,VTB03}.

T\'el et al.\ have furthermore considered the situation of a hybrid system
that is both open and dissipative \cite{TeVB96,TVS00,Voll02}. By solving the
corresponding Fokker-Planck equation with absorbing boundaries and employing
again the Pesin identity for open systems they arrived at the generalized
formula (see also Ref.\ \cite{Gasp})
\be
D\left(\frac{\pi}{L}\right)^2+\frac{\sigma^2\varepsilon^2}{4D}=\sum_{\lambda_i>0}\lambda_i({\cal
R}_L)-h_{KS}({\cal R}_L)\quad (L\to\infty , \varepsilon\to0) \quad . \label{eq:tklgen}
\ee
Carrying out the limit of $\varepsilon\to0$ eliminates the second term on the
left hand side yielding precisely Eq.\ (\ref{eq:tkler}) of the escape rate
approach if applied to the Lorentz gas. On the other hand, performing
$L\to\infty$ eliminates the first term on the left hand side. The conductivity
may then be replaced again by Einstein's formula Eq.\ (\ref{eq:dts}). Taking
into account that for the Gaussian thermostat the temperature $T$ should be
replaced by $T\equiv T/2$, see Section III.B.1, and assuming that in the limit
of $\varepsilon\to0$ the spectrum of Lyapunov exponents exhibits conjugate
pairing, $\sum_{\lambda_i>0}\lambda_i=\sum_{\lambda_i<0}|\lambda_i|$, one
recovers the result for the thermostated Lorentz gas Eq.\ (\ref{eq:tkltf}).

We finally outline a third basic approach linking transport coefficients to
dynamical systems quantities as recently formulated by Gilbert et al.\
\cite{GDG01,GCGD01,ClGa02}. Here particularly diffusion in closed
volume-preserving dynamical systems has been considered. Starting from the
diffusion equation with periodic boundary conditions the solution for the
diffusion coefficient reads
\be
D=\lim_{L\to\infty}\left(\frac{L}{2\pi}\right)^2\gamma_{dec} \quad
. \label{eq:dkdec}
\ee
This result is in formal analogy to Eq.\ (\ref{eq:dkesc}), however, here
$\gamma_{dec}$ denotes the {\em decay rate} by which an equilibrium state is
approached. By solving the Liouville equation of the respective dynamical
system $\gamma_{dec}$ can be related to the second largest eigenvalue of the
Liouville operator associated with the corresponding hydrodynamic mode of
diffusion of the dynamical system, see, e.g., Refs.\ \cite{RKdiss,KlDo99}. The
crucial observation is now that this eigenvalue is linked to the Hausdorff
dimension of the respective mode. Thus, the diffusion coefficient in Eq.\
(\ref{eq:dkdec}) can be written as a function of the fractal dimension of the
second largest eigenmode of the Liouville equation \cite{GDG01,GCGD01}. For
systems with two degrees of freedom which are periodically continued in one
direction over a length $L$ one gets
\be
D=\lim_{L\to\infty}\left(\frac{L}{2\pi}\right)^2\lambda_+\left(d_H(2\pi/L)-1\right)
\label{eq:tkhdm}
\quad ,
\ee
where the Hausdorff dimension $d_H$ is a function of the length of the system,
or respectively of the wavenumber $k=2\pi/L$ of the corresponding hydrodynamic
mode. $\lambda_+$ denotes again the positive Lyapunov exponent of the
system. This approach has been worked out in detail for a multibaker map
\cite{GDG01} as well as for diffusive and reactive-diffusive billiards of
Lorentz gas type \cite{GCGD01,ClGa02}.

In order to compare Eq.\ (\ref{eq:tkhdm}) with the previous two formulas Eqs.\
(\ref{eq:tkltf}) and (\ref{eq:tkler}) one may employ the Kaplan-Yorke formula,
respectively Young's formula, which in case of the two-dimensional periodic
Lorentz gas with a constrained kinetic energy reads
\cite{ER,Ott,GDG01,Ch1}\footnote{We remark that so far this formula has only been
proven for two-dimensional Anosov diffeomorphisms with an ergodic measure on
compact manifolds, see Refs.\ \cite{ER,Ott} for further literature.}
\be
d_I=2+h_{KS}/|\lambda_-| \quad . \label{eq:kyc}
\ee
This formula links the information dimension $d_I$ of a fractal set to the
Kolmogorov-Sinai entropy $h_{KS}$ and to the negative Lyapunov exponent
$\lambda_-$ of the corresponding dynamical system. By using this equation Eq.\
(\ref{eq:tkltf}) can be rewritten as
\be
D=\lim_{\varepsilon\to0}\left(\frac{T}{\varepsilon}\right)^2\lambda_+({\cal
A_{\varepsilon}})\left(3-d_I({\cal A_{\varepsilon}})\right) \quad
. \label{eq:tkltfd}
\ee
It appears that this formula was first derived in Ref.\
\cite{ECS+00}. Analogously, Eq.\ (\ref{eq:tkler}) reads
\be
D=\lim_{L\to\infty}\left(\frac{L}{\pi}\right)^2\lambda_+({\cal
R}_L)\left(3-d_I({\cal R}_L)\right) \quad . \label{eq:tklerd}
\ee
In both cases the information dimension $d_I$ is furthermore identical to the
Hausdorff dimension $d_H$ of the respective fractal set and may be replaced
respectively \cite{Ch1,Ch2,GaBa95}.

In conclusion, all three equations Eqs.\ (\ref{eq:tkhdm}), (\ref{eq:tkltfd})
and (\ref{eq:tklerd}) relate the diffusion coefficient to the largest positive
Lyapunov exponent of the corresponding dynamical system times a term
containing the information dimension of the associated fractal structure,
which in Eq.\ (\ref{eq:tkhdm}) is a fractal hydrodynamic mode, in Eq.\
(\ref{eq:tkltfd}) a fractal attractor and in Eq.\ (\ref{eq:tklerd}) a fractal
repeller.\footnote{In order to make this analogy even closer one may further
introduce partial codimensions for the fractal structures in Eqs.\
(\ref{eq:tkltfd}) and (\ref{eq:tklerd}), however, this does not appear to be
possible for the hydrodynamic mode in Eq.\ (\ref{eq:tkhdm}) \cite{GDG01}.}
Hence, on this level there is quite a formal analogy even between all three
approaches. One may thus indeed wonder whether these formulas form a kind of
general backbone of nonequilibrium transport in terms of dynamical systems
theory \cite{BrTV96,TeVB96,Rue96}. We remark that, in addition, there exist
very interesting, simple formulas by which transport coefficients can be
calculated in terms of periodic orbits, see, e.g., Ref.\
\cite{Vanc} for the electrical conductivity and Refs.\
\cite{CEG91,CvGS92,Gasp,CAMTV01} for the diffusion coefficient. However, these
formulas are conceptually rather different from the ones discussed above.
Therefore we do not discuss them here by instead referring to the respective
literature, see, e.g., Ref.\ \cite{CAMTV01} for an overview.

Despite their striking formal analogy one may not overlook that all three
equations concern very different physical settings
\cite{Coh92,DettS00,GDG01}: Eq.\ (\ref{eq:tklerd}) applies to diffusion in open
Hamiltonian dynamical systems without external fields, where the link between
chaos and transport is formed by the rate assessing the escape from the
fractal repeller. Eq.\ (\ref{eq:tkhdm}) concerns diffusion in closed
Hamiltonian dynamical systems, again without using external fields, however,
here there is no repeller. Instead, the fractality of the hydrodynamic mode of
diffusion is assessed. In contrast to these two relations, Eq.\
(\ref{eq:tkltfd}) originally started from the current generated in a
non-Hamiltonian dynamical system under application of external fields in
combination with a specific thermostat. Here the link between chaos and
transport is provided by the average phase space contraction rate onto the
fractal attractor. Particularly in case of Eqs.\ (\ref{eq:tkltfd}) and
(\ref{eq:tklerd}) the physical situations involved remain inherently
different. 

That both equations can nevertheless be derived from a `master formula' is
nicely demonstrated by the hybrid equation Eq.\ (\ref{eq:tklgen}). However,
the left hand side of this equation represents obviously just the sum of the
left hand sides of Eqs.\ (\ref{eq:tkltf}) and (\ref{eq:tkler}) supplemented by
the prefactors from the respective right hand sides. If we add the
prefactor-free right hand sides of these two equations as well we obtain, by
using the original formulation in the first line of Eq.\ (\ref{eq:tdspes})
instead of the right hand side of Eq.\ (\ref{eq:tkltf}), $-\lambda_+({\cal
A_{\varepsilon}})-\lambda_-({\cal A_{\varepsilon}})+\lambda_+({\cal
R}_L)-h_{KS}({\cal R}_L) \:(L\to\infty,
\varepsilon\to\infty)$. By assuming that, in the above limits,
$\lambda_{\pm}({\cal A_{\varepsilon}})=\lambda_{\pm}({\cal R}_L)$ and that, as
before, conjugate pairing holds, $\lambda_+=-\lambda_-$, we recover the right
hand side of Eq.\ (\ref{eq:tklgen}) for the periodic Lorentz gas. Hence, one
may argue that Eq.\ (\ref{eq:tklgen}) represents an additive combination of
both formulas. That a more intricate, common root of these two formulas going
beyond this equation exists, possibly even combining these two relations with
the third one in terms of hydrodynamic modes, appears to be very unlikely to
us. 

We may furthermore emphasize again that the validity of Eq.\ (\ref{eq:tkltf})
stands and falls with the existence of the identity between phase space
contraction and thermodynamic entropy production. In the previous section we
have already casted doubt on the general validity of this identity for
dissipative systems. Indeed, the two key quantities linking thermodynamics to
chaotic dynamics discussed above measure quite different physical processes:
The escape rate merely assesses the absorption of the number of phase space
points, or particles, at some boundary, whereas the phase space contraction
rate is defined with respect to the full details of the dynamical system in
all variables. As we have outlined to the end of the previous section it is
thus conceivable that the phase space contraction rate contains spurious
information as far as thermodynamic entropy production and transport processes
are concerned. We are not aware that yet there are any counterexamples
questioning the validity of the formulas emerging from the escape rate and the
closed Hamiltonian systems approach. However, as we will argue in the
remaining chapters, for the thermostated systems approach the situation
appears to be much more delicate.

Concerning practical applications for the calculation of transport
coefficients it seems that these three formulas do not provide more efficient
computational schemes than, say, Einstein formulas or Green-Kubo relations. As
far as we can tell, in most cases dynamical systems quantities such as
Lyapunov exponents and Kolmogorov-Sinai entropies are more difficult to
compute than statistical averages defined within the framework of common
nonequilibrium statistical mechanics only. With respect to a simplification of
the Lyapunov sum rule Eq.\ (\ref{eq:tklt}) for higher-dimensional dynamical
systems we may recall that the conjugate pairing rule of Lyapunov exponents is
not universal for NSS in dissipative dynamical systems. Boundary thermostats
and systems under electric and magnetic fields provide counterexamples, see
Section III.B.2 for more details.

Irrespective of these rather practical concerns, we may emphasize that the
existence of these three formulas linking transport coefficients to dynamical
systems quantities provides highly interesting results from a fundamental
physical and dynamical systems point of view, and the formal similarity of
these different equations remains very remarkable.

\subsection{Fractal attractors characterizing nonequilibrium steady states} 

Another crucial property of Gaussian and Nos\'e-Hoover thermostated systems is
the existence of fractal attractors underlying NSS. These objects form a
fundamental link between the microscopic non-Hamiltonian equations of motion
that are, for this type of systems, still deterministic and time-reversible
while irreversible transport is exhibited on macroscopic
scales. Quantitatively, the existence of these attractors manifests itself in
the average phase space contraction rate serving as a link between chaos and
transport in dissipative systems, see our discussion in the previous sections.

In Sections III.B and IV.C we have studied the Gaussian and the Nos\'e-Hoover
thermostated driven periodic Lorentz gas as paradigmatic examples for this
class of thermostated systems. For the Gaussian type numerical and analytical
results provided evidence that in the full accessible phase space the
Hausdorff dimension of the attractor is non-integer. In Poincar\'e surfaces of
section, or in respective projections of the phase space, the attractor turned
out to be even multifractal; see Figs.\ \ref{gatt} and
\ref{attrnh} for plots depicting the associated fractal folding. Supported by
results for many other Gaussian and Nos\'e-Hoover thermostated systems it was
thus conjectured that the existence of fractal attractors, and of the
associated singular probability measures, is generic for non-Hamiltonian
dynamical systems in NSS subject to deterministic and time-reversible thermal
reservoirs.

How the topology of these attractors changes under variation of the electric
field strength was assessed by means of bifurcation diagrams, as discussed in
the same sections. For this purpose the positions of the colliding particles
at the Lorentz gas disk were plotted as functions of the electric field
strength, see Figs.\ \ref{gbifu} and \ref{bifunh}. For both the Gaussian and
the Nos\'e-Hoover thermostated driven periodic Lorentz gas these diagrams
revealed intricate bifurcation scenarios. On the other hand, the specific form
of these scenarios depends intimately on the type of thermal reservoir used
and is already very different for Gaussian and Nos\'e-Hoover dynamics. Even
more, in case of Nos\'e-Hoover thermostating different bifurcation diagrams
were obtained for different values of the coupling strength between subsystem
and thermal reservoir. Hence, there is no universality concerning the specific
change of the topology of these attractors under variation of the field
strength. Whether the mere existence of bifurcations, at least, is typical for
the thermostated driven periodic Lorentz gas will be clarified in the
following two chapters.

However, instead of applying Gaussian and Nos\'e-Hoover schemes, thermal
reservoirs can be mimicked by introducing {\em stochastic interactions}
between subsystem and reservoir particles. Here we briefly sketch various
existing numerical and analytical methods of how to model this second
fundamental class of thermostats. We then connect to our discussion concerning
the possible universality of fractal attractors in NSS.

One way to introduce a stochastic reservoir is by using a Langevin equation,
as was discussed in detail in Section II.B. For molecular dynamics computer
simulations this approach was first implemented by Schneider and Stoll
\cite{SchSt78}, see also later work in Ref.\ \cite{Nose91,PHH00}. Andersen
\cite{And79} proposed a variant of this stochastic bulk thermostat by considering
an interacting many-particle system, where at certain time intervals the
velocity of a randomly selected particle is chosen randomly from a canonical
velocity distribution \cite{AT87,Jell88}.

Alternatively, a stochastic sampling of velocities from canonical velocity
distributions can be performed at the boundaries of a subsystem. Such {\em
stochastic boundary conditions} were proposed by Lebowitz and Spohn in order
to mathematically analyze a three-dimensional random Lorentz gas under a
temperature gradient \cite{LeSp78}. For molecular dynamics computer
simulations this thermostating scheme was implemented by Ciccotti and
Tenenbaum \cite{CiTe80}, again in order to model thermal gradients, see also
Refs.\ \cite{TCG82,ChLe95,ChLe97,PH98,WKN99} for later use of this method in
the context of computer simulations. Stochastic boundary conditions will be
introduced in full detail later on in Section VII.A. However, for the
following discussion it is not necessary to know about such technical details.

A particularly important result for the class of stochastic boundary
thermostated models was reported by Goldstein, Kipnis and Ianiro
\cite{GKI85}. They studied a system of Newtonian particles maintaining a
heat flux due to a temperature field that is modeled by stochastic boundaries
and varies with the position at the boundary. A mathematical analysis of this
system yielded that there exists a unique invariant probability measure that
is absolutely continuous \cite{ER,Do99} with respect to the Lebesgue
measure. In other words, in this case the measure is not singular and hence
not fractal. This result was claimed to be generic for stochastically
thermostated systems \cite{EyLe92,Coh92} thus contradicting the universality
of fractal attractors as conjectured from studying Gaussian and Nos\'e-Hoover
thermostated systems \cite{HHP87,HoB99}.

The claim that the physically relevant probability measures are generally
smooth in stochastically thermostated systems has been doubted by Hoover et
al.\ on the basis of numerical explorations: In Ref.\ \cite{HP98} these
authors studied a driven periodic Lorentz gas thermalized by a `hybrid'
thermostat consisting of deterministic and stochastic boundaries. Computations
of the information dimension of the fractal attractor by evaluating phase
space projections such as Fig.\ \ref{attrnh} indicated a slight deviation from
an integer value. Furthermore, they considered a one-dimensional Hamiltonian
model for heat conduction (the {\em ding-a-ling} model) with
reservoirs defined by stochastic boundaries \cite{PH98}. They also
investigated an interacting many-particle system under an external field
\cite{PHH00} (the {\em color conductivity model} \cite{EvMo90}). In
the latter two cases they computed the information dimension according to the
Kaplan-Yorke conjecture, respectively Young's formula, cp.\ to Eq.\
(\ref{eq:kyc}) of the previous section. The numerical results yielded again
deviations from integer values. However, this dimension formula relies on the
computation of Lyapunov exponents that in turn necessitate to define a
Jacobian quantifying the interaction of the subsystem with the stochastic
boundaries. To us it appears that the definition which these authors used for
the Jacobian is generally ill-defined, hence we cannot consider the numerical
results of Refs.\ \cite{PH98,PHH00} to be conclusive.\footnote{In Ref.\
\cite{HP98} the phase space contraction at a stochastic boundary is defined by
the second equation below Fig.\ 2. For this purpose an equality between phase
space contraction and entropy production has been {\em stipulated} for this
type of system. On the other hand, a natural assumption is just the
conservation of phase space probability (or points in phase space) at the
boundary, $dr dv \rho(r,v)=dr' dv' \rho(r',v')$, where $(r,v)$ represent
position and velocity of a particle before a collision, $(r',v')$ the same
variables after a collision, and $\rho$ is the respective probability
distribution. A Jacobian at the collision is then straightforwardly defined by
$|J|=|dr'dv'/drdv|=\rho(r,v)/\rho(r',v')$. If one follows this argument one
arrives at the result of Ref.\ \cite{HP98} {\em only} (i) if both
distributions before and after the collisions are identified with canonical
distributions which is natural after a collision, but which is not so clear
before a collision, cp., e.g., to Ref.\ \cite{WKN99}; and (ii) if both
distributions before and after a collision have precisely the same variance in
terms of a temperature, $T=T'$. However, both distribution are only expected
to be the same, with the same temperature, in the hydrodynamic limit, whereas
systems of finite length as handled on the computer one may expect to exhibit
temperature jumps \cite{KLA02}. For this reason we consider the validity of
the results reported in Refs.\ \cite{PH98,PHH00} concerning the information
dimension to be debatable.} Thus, whether attractors in stochastically
thermostated systems are generally smooth or fractal, and to which extent this
property depends on the specific type of subsystem considered, still remains a
very open question to us.

Apart from these considerations we may emphasize again that both types of
thermal reservoirs define very different classes of dynamical systems:
Gaussian and Nos\'e-Hoover thermostats keep the dynamics deterministic and
time-reversible, whereas stochastically thermostated systems render the
equations of motion non-deterministic and non-time reversible. The controversy
\cite{GKI85,EyLe92,Coh92,HP98,PH98,PHH00,Gall03b} concerning the type of
invariant measure that emerges under application of these two very different
types of thermal reservoirs thus boils down to the question which type of
thermostat one considers to be more `physical'. We furtmermore remark that,
connected to our discussion in Section V.A, Nicolis and Daems
\cite{NiDa96,NiDa98,DaNi99} and Ruelle \cite{Rue97} argued for the positivity of the
thermodynamic entropy production in stochastically thermostated systems.

\subsection{Nonlinear response in the thermostated driven periodic Lorentz gas}

According to the numerical results for the thermostated driven periodic
Lorentz gas the topology of the attractors does not only depend on the
specific type of thermal reservoir that has been applied, but it is also very
sensitive to variations of the external field. From that point of view it is
not too surprising that also the field-dependent electrical conductivities of
the Gaussian and the Nos\'e-Hoover thermostated Lorentz gas are not the same,
as we have discussed in Sections III.B.4 and IV.C.3, see Figs.\ \ref{gcond}
and \ref{condnh}. Furthermore, even for Nos\'e-Hoover only the Lorentz gas
conductivities still depend on the value for the coupling parameter between
subsystem and reservoir, in the same way as the bifurcation diagrams of the
different attractors.

On the other hand, all conductivities jointly exhibit a profoundly nonlinear
response for the numerically accessible values of the field strength
$\varepsilon>0$. For the Gaussian thermostated Lorentz gas there must exist a
regime of linear response in the limit of very small fields according to the
mathematical proof by Chernov et al. However, the respective range of field
strengths close to zero appears to be so small that up to now computer
simulations could not really corroborate its existence. This reminds to some
extent to van Kampen's objections concerning the validity of linear response,
who argued that a linear response is not trivially guaranteed for nonlinear
chaotic dynamical systems. He emphasized that there might be a nontrivial
interplay between microscopic nonlinearity and macroscopic linearity yielding
a quantitatively negligibly small regime of linear response.

In addition to these problems concerning coarse functional forms, all
field-dependent conductivities exhibit irregularities on fine scales that are
not due to numerical errors. Such irregular transport coefficients are in fact
well-known from very related classes of dynamical systems, which share the
properties of the periodic Lorentz gas of being deterministically chaotic,
low-dimensional and spatially periodic. For very simple types of such systems
it could be shown that these transport coefficients are even of a fractal
nature. For thermostats of Nos\'e-Hoover type the irregularities appear to be
smoothed-out, which may be understood with respect to the fact that here the
kinetic energy fluctuates according to a canonical velocity
distribution. Imposing instead a constant kinetic energy onto the system by
means of a Gaussian thermostat these irregularities look considerably more
``fractal-like''.

In summary, to us it still remains an open question to {\em quantitatively}
identify a regime of linear response in the Gaussian thermostated driven
periodic Lorentz gas, either analytically or numerically. One may furthermore
ask whether such a regime of linear response may also be expected for
applications of the Nos\'e-Hoover and possibly of other types of
thermostats. On the other hand, one may suspect that the existence of
irregularities on finer scales of the field-dependent conductivity is rather
typical for deterministically thermostated driven periodic Lorentz gases. To
which extend this holds true, and particularly how this goes together with the
expected linear response for very small fields, are important open questions.

This points back again to the question concerning the equivalence of
nonequilibrium ensembles. Unfortunately, in low-dimensional deterministic and
periodic dynamical systems like the Lorentz gas the conductivity obviously
reflects the specific type of thermal reservoir that has been applied. This
appears to be at variance with such an equivalence that is expected to yield
the same conductivities. Related findings have been reported in Ref.\
\cite{BDL+02}. In order to get more clear about this point it will be
important to apply further types of thermal reservoirs to the Lorentz gas
dynamics in order to check for similarities and differences, which we will do
in the following two chapters.

Finally, as for the hotly debated fractality of attractors in NSS one may
expect that the irregular or even fractal structure of parameter-dependent
transport coefficients may be getting more regular by imposing stochasticity
onto the system. This can be performed by either distributing the scatterers
randomly in space, by using a stochastic thermal reservoir as discussed above,
or by imposing additional noise onto a deterministic system. For thermostated
driven {\em random} Lorentz gases field- and density-dependent Lyapunov
exponents and diffusion coefficients have indeed already been calculated
analytically and numerically
\cite{BD95,vBDCP96,LvBD97,DePo97,BDPD97,BLD98,vBLD00} 
verifying such a smooth dependence on parameters. Additionally, for simple
low-dimensional maps it has been studied how the fractal structure of a
parameter-dependent diffusion coefficient is affected while imposing different
types of perturbations on them in time or in space \cite{RKla01a,RKla01b}. As
expected the fractality on arbitrarily fine scales disappears, however,
irregularities still survive in form of smoothed-out oscillations. This
irregular structure turns out to be very persistent against random
perturbations, that is, typically rather strong random perturbations are
needed to make the transport completely random walk-like.\footnote{An
interesting exception is provided by a specific type of quenched disorder
leading to a Golosov random walk. In this case an originally normal diffusive
process may immediately become anomalous, that is, a diffusion coefficient
does not exist anymore \cite{Rado96}.}

Hence, there may be complicated scenarios between the two limiting cases of
completely deterministic and completely stochastic diffusion, and the
application of external randomness significantly affects fractal properties of
the type as discussed above. Generally, one may expect that problems with
linear response and the irregularity of parameter-dependent transport
coefficients are rather specific to low-dimensional dynamical systems. For
interacting many-particle systems, for example, it is natural to assume that
all transport coefficients are smooth in their physical parameters and that
there are broad regimes of linear response, as predicted by standard
nonequilibrium thermodynamics. Examples of such systems will be discussed in
Chapter VII.

\subsection{Summary}

\begin{enumerate}

\item Gaussian and Nos\'e-Hoover dynamics provide prominent examples for
a non-Hamiltonian modeling of NSS. By employing the well-known baker map we
demonstrated how volume-preserving ``Hamiltonian'' dynamics may go together
with dissipative equations of motion and corresponding non-uniform probability
densities. The key is the elimination of reservoir degrees of freedom by
projecting out respective variables from the full equations of motion. Thus,
there is nothing mysterious about using a non-Hamiltonian description of NSS.

\item Gaussian and Nos\'e-Hoover thermostats typically feature an identity between
thermodynamic entropy production and the average rate of phase space
contraction. However, there exists an even more general identity relating
phase space contraction to entropy production in terms of the Gibbs entropy
which is completely independent of the type of modeling. On the other hand, in
case of a Hamiltonian approach to NSS the Gibbs entropy production is simply
zero. For this and other reasons one may doubt whether the Gibbs entropy
should be considered as a correct definition of a nonequilibrium entropy being
compatible with nonequilibrium irreversible thermodynamics.

Dissipative dynamical systems in NSS exhibit a negative average rate of phase
space contraction. Due to the abovementioned identity the Gibbs entropy
production is negative as well. This appears to be at variance with the second
law of thermodynamics. However, here the Gibbs entropy can be saved by
applying methods of coarse graining leading to an inversion of the sign.

To consider a coarse grained entropy production is furthermore suggested by
the fact that the invariant measures of deterministically thermostated systems
are typically singular. Still, the decisive question is whether the Gibbs
entropy always yields the correct thermodynamic entropy production in
nonequilibrium as, for example, compared to the Clausius entropy. For Gaussian
and Nos\'e-Hoover thermostats this holds true, but there is no reason why this
should generally apply to other types of thermal reservoirs.

\item Having an identity between phase space contraction and entropy
production at hand enables one to link transport coefficients to dynamical
systems quantities. Surprisingly, the escape rate approach to Hamiltonian
dynamical systems yields a formula that is very analogous to the one obtained
for dissipative dynamics: In both cases a transport coefficient is related to
a difference between sums of Lyapunov exponents and the Kolmogorov-Sinai
entropy on some fractal structure. In case of thermostated systems this
structure consists of a fractal attractor, whereas for open Hamiltonian
systems it is a fractal repeller.

There is even a third formula of a similar type derived for closed Hamiltonian
dynamical systems, which connects the Hausdorff dimension of the hydrodynamic
mode governing a transport process and a respective Lyapunov exponent to the
corresponding transport coefficient. By performing suitable transformations
the previous two formulas can be compared to the latter one. All resulting
equations involve fractal dimensions of respective fractal sets and Lyapunov
exponents thus revealing a striking formal analogy.

On the other hand, these equations are obtained from three very different
approaches to NSS, and correspondingly the physical meaning of the involved
quantities is rather different. We described a `master formula' linking two of
the three equations, however, its additivity suggests that there is no further
reduction onto a common `root'. In addition, the formula derived for
thermostated systems stands and falls with the validity of the identity
between phase space contraction and entropy production, which in turn appears
to be doubtful. In this respect the formulas emerging from the two Hamiltonian
approaches seem to be on safer grounds.

\item We commented upon the suspected typicality of fractal
attractors in dissipative NSS. For Gaussian and Nos\'e-Hoover thermostated
systems there is a wealth of work corroborating the fractality of attractors.
However, at least for the driven periodic Lorentz gas the topology of these
fractal structures depends not only on the type of thermal reservoir used, but
also on the variation of control parameters such as the coupling strength
between subsystem and reservoir and the electric field strength. This is
exemplified by bifurcation diagrams exhibiting complicated bifurcation
scenarios for the attractor under variation of these parameters.

Alternatives to deterministic and time-reversible thermal reservoirs are
provided by the stochastic Langevin equation and by stochastic boundary
conditions. For a specific stochastically thermostated system it has been
proven that the associated invariant measure is smooth hence contradicting the
ubiquituous existence of fractal attractors in NSS. This was contrasted by
numerical computations for other stochastically thermostated systems yielding
again fractal objects, however, partly the numerical concepts employed appear
to be debatable. Whether attractors in stochastically thermostated systems are
typically smooth or fractal, or whether this actually depends on the specific
type of system, remains an important open question.

\item A similar reasoning applies to the field-dependent conductivities of
the thermostated driven periodic Lorentz gas. As in case of attractors, their
functional forms reflects the specific type of thermal reservoir that has been
applied. Numerically, no regimes of linear response could be detected for the
driven periodic Lorentz gas, despite a proof for the existence of such a
regime in the Gaussian thermostated case. In addition, all conductivities
showed pronounced irregularities on finer scales as functions of the field
strength which partly may be of a fractal origin. How the latter property goes
together with a suspected linear response and to which extent all these
properties hold for arbitrary thermal reservoirs remains to be clarified.

From a thermodynamic point of view these intricate dependencies of transport
properties on parameter variations and on the type of reservoir are rather
undesirable questioning the equivalence of nonequilibrium ensembles. One may
thus attribute such nontrivial characteristics to the simplicity and low
dimensionality of the model. This does not imply that the model itself is
unphysical, since systems with these properties very well exist in
nature. Imposing randomness onto such dynamics indeed smoothes out these
irregularities. Considering a higher-dimensional system composed of
interacting many particles is expected to yield the same consequences.

\end{enumerate}

In summary, a major question concerning the theory of transport in dissipative
systems might be formulated as follows: How general are the results obtained
from Gaussian and Nos\'e-Hoover thermostats for NSS of dissipative dynamical
systems? That is, to which extent are these properties universal in that they
hold for other thermostating schemes as well, or do they depend on the
specific way of thermostating?

The remaining chapters of this review will be devoted to shed further light
onto this fundamental question. Our approach will be carried out in two steps:
A first step is to invent alternative, possibly more general thermostating
schemes which are not fully identical to the previous class of methods but
share certain characteristics with them, such as being deterministic and
time-reversible and leading to NSS. In the following we present two generic
examples of such methods. In a second step we study whether, by using these
schemes, the same results for quantities characterizing chaos and transport
are obtained as exemplified by Gaussian and Nos\'e-Hoover thermostats.

\section{Gaussian and Nos\'e-Hoover thermostats revisited}

In our previous discussions we have identified important crosslinks between
chaos and transport in Gaussian and Nos\'e-Hoover thermostated models. Some
authors even conjectured these relations to be universal for dissipative
dynamical systems altogether. It is thus illuminating to look at simple
generalizations of ordinary Gaussian and Nos\'e-Hoover dynamics, which we
define again for the driven periodic Lorentz gas. We then analyze these
thermostating schemes along the same lines as discussed previously, that is,
by analytically and numerically studying the transport and dynamical systems
properties of the respectively thermostated driven periodic Lorentz gas.
Finally, we compare the results obtained from our analysis to the ones
characterizing conventional Gaussian and Nos\'e-Hoover schemes as discussed in
Chapters III to V.

\subsection{Non-ideal Gaussian thermostat}

Consider again the Gaussian thermostated driven periodic Lorentz gas defined
by Eqs.\ (\ref{eq:eomdlg}), (\ref{eq:alpdlg}) with an external field that is
parallel to the $x$ direction, $\mbox{{\boldmath
$\varepsilon$}}\equiv(\varepsilon_x,0)^*$. We now distinguish between the
interactions of subsystem and thermal reservoir parallel and perpendicular to
the field direction by making the parallel coupling field-dependent,
\bna
\dot{r}_x&=&v_x \nonumber \\
\dot{v}_x&=&\varepsilon_x-\alpha v_x-\alpha \varepsilon_x v_x \nonumber\\
\dot{r}_y&=&v_y \nonumber \\
\dot{v}_y&=&-\alpha v_y \quad , \label{eq:eg}
\ena
where $\alpha\equiv\alpha(v_x,v_y)$. The difference to the Gaussian case is
reminiscent by the third term on the right hand side of the second
equation. This additional term conveniently adjusts the action of the
thermostat to the anisotropy of the external field thus providing a physical
justification for its introduction. Correspondingly, one may expect that this
version thermalizes the driven Lorentz gas more efficiently than the
conventional, as we may now call it, Gaussian thermostat Eqs.\
(\ref{eq:eomdlg}), (\ref{eq:alpdlg}). Formally, this generalization just
amounts in making the conventional coupling constant field-dependent and then
expanding it in the field strength $\varepsilon_x$ in form of a power series,
$\alpha\equiv(1+\varepsilon_x)\alpha$. The conventional Gaussian case is
obviously recovered in zeroth order while Eqs.\ (\ref{eq:eg}) include the
first order.

The explicit functional form for $\alpha$ can be calculated in complete
analogy to Section III.A from the requirement of energy conservation yielding
\be
\alpha(v_x,v_y)=\frac{\varepsilon_x v_x}{v^2+\varepsilon_xv_x^2}\quad
. \label{eq:aeg} 
\ee
Comparing this result to the corresponding one for the conventional Gaussian
thermostat Eq.\ (\ref{eq:alpdlg}) identifies the second term in the
denominator as a new contribution. By using the definition of Eq.\
(\ref{eq:psc}) the average phase space contraction rate is calculated to
\be
\kappa=-<\alpha>-<\frac{2\varepsilon_x^2v_xv_y^2}{(v^2+\varepsilon_xv_x^2)^2}>
\label{eq:preg} \quad .
\ee
Again, a comparison of $\kappa$ with the result for the conventional Gaussian
thermostat Eq.\ (\ref{eq:peqa}) is instructive: apart from the correction
already contained in $\alpha$, Eq.\ (\ref{eq:preg}) features a second term for
the phase space contraction rate. All these additional terms become negligible
in the limit of $\varepsilon_x\to0$ thus approximately recovering the
conventional Gaussian thermostat. But the point is that according to Eq.\
(\ref{eq:preg}) there is clearly no identity between phase space contraction
and thermodynamic entropy production anymore. Indeed, some entropy production
related to Joule's heat, $dS=\varepsilon_x<v_x>/T$, see Eq.\ (\ref{eq:pepid}),
is still reminiscent in Eqs.\ (\ref{eq:aeg}), (\ref{eq:preg}), however, it
does not show up in form of a simple functional relationship to the average
phase space contraction rate.

We may thus refer to the coupling in terms of the conventional Gaussian
thermostat defined by Eqs.\ (\ref{eq:eomdlg}), (\ref{eq:alpdlg}) as an {\em
ideal coupling}, in the sense that it yields this simple identity, whereas we
call the coupling Eqs.\ (\ref{eq:eg}), (\ref{eq:aeg}) {\em non-ideal} in that
it does not preserve the identity. Consequently, we call the thermostating
scheme defined by Eqs.\ (\ref{eq:eg}), (\ref{eq:aeg}) the {\em non-ideal
Gaussian thermostat}.\footnote{This scheme was orally presented on occasion of
the conference on {\em Microscopic chaos and transport in many-particle
systems} in Dresden, August 2002 \cite{Kla03}.}

\begin{figure}[t]
\epsfysize=15cm
\centerline{\epsfbox{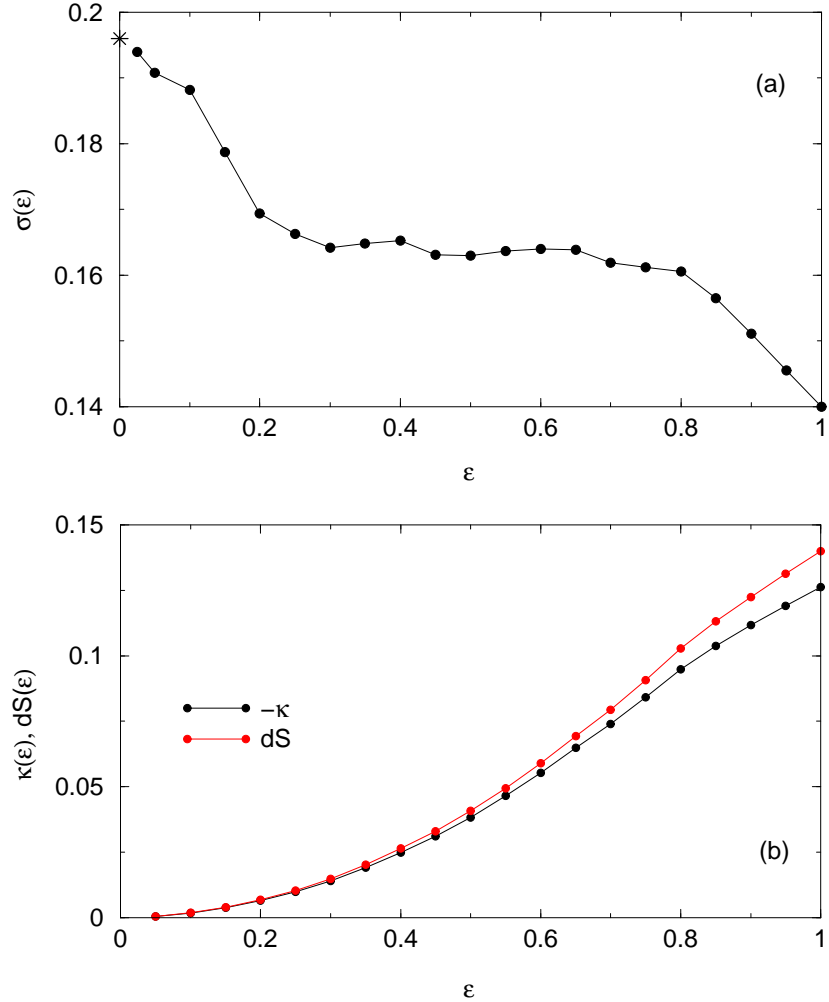}}
\vspace*{-1.5cm} 
\caption{Computer simulation results for the driven periodic Lorentz gas
equipped with the non-ideal Gaussian thermostat Eqs.\ (\ref{eq:eg}) at the
temperature $T=1$: (a) Field-dependent conductivity $\sigma(\varepsilon)$ as a
function of the field strength $\varepsilon\equiv \varepsilon_y$. The star
denotes the result in the limit of $\varepsilon\to0$ according to the Einstein
formula Eq.\ (\ref{eq:dts}). (b) Comparison between the average phase space
contraction rate $\kappa(\varepsilon)$ and the thermodynamic entropy
production $dS(\varepsilon)$.}
\label{prepega}
\end{figure}

We now discuss computer simulation results for the driven periodic Lorentz gas
thermostated that way \cite{KR99}. Note that for technical reasons here the
field of strength $\varepsilon$ was chosen to be parallel to the $y$-axis,
$\varepsilon\equiv\varepsilon_y$,\footnote{Numerical results indicated that
for $\mbox{{\boldmath $\varepsilon$}}||x$ the dynamics suffered from more
creeping orbits than for $\mbox{{\boldmath $\varepsilon$}}||y$.} by adjusting
the thermostating scheme in Eqs.\ (\ref{eq:eg}) respectively. Fig.\
\ref{prepega} (a) shows results for the conductivity $\sigma(\varepsilon)$ of
the non-ideal Gaussian thermostated driven Lorentz gas. The fluctuations on
small scales are less irregular than for the conductivity of the conventional
Gaussian thermostat Fig.\ \ref{gcond}, which probably reflects the more
efficient coupling with respect to the direction of the field. However, the
oscillations seem to be less smooth than the ones for the Nos\'e-Hoover case
shown in Fig.\ \ref{condnh}. Again there is no visible regime of linear
response for $\varepsilon\to0$. Curiously, the nonlinearities conspire to form
an approximately linear response regime at intermediate values of the field
strength; see, e.g., Ref.\ \cite{CoRo98} for a brief appreciation of such
phenomena. Reassuringly, for small enough field strength the conductivity
approaches the value indicated by a star, which was calculated from the
respective diffusion coefficient in the field-free case according to the
Einstein formula Eq.\ (\ref{eq:dts}), see Section III.B.4.

Fig.\ \ref{prepega} (b) depicts numerical results for the average phase space
contraction rate Eq.\ (\ref{eq:preg}) and for the thermodynamic entropy
production according to $dS(\varepsilon)=<\varepsilon
v_y/v^2>=\sigma(\varepsilon)\varepsilon^2/T$. These results quantitatively
confirm the non-identity between both quantities as already suggested by Eq.\
(\ref{eq:preg}). Note that in analogy to the irregularity of
$\sigma(\varepsilon)$ at least the entropy production $dS(\varepsilon)$ should
yield irregularities on fine scales as well. However, this structure might be
largely suppressed by the quadratic factor in $\varepsilon$. It would be
interesting to investigate whether $\kappa$ and $dS$ exhibit different types
of such irregularities corresponding to more intricate higher-order deviations
between them.

Since there is no identity between phase space contraction and entropy
production there is trivially no simple relation between the conductivity and
the Lyapunov exponents such as the Lyapunov sum rule Eq.\ (\ref{eq:lsr}), cp.\
to its derivation for the conventional Gaussian thermostat in Section
III.B.2. Indeed, for the average phase space contraction rate Eq.\
(\ref{eq:preg}) it is not even obvious how to single out the conductivity
without performing suitable approximations. For the two-dimensional Lorentz
gas a conjugate pairing of Lyapunov exponents is guaranteed by
default. However, one may conjecture that this property also holds in case of
higher-dimensional non-ideal Gaussian thermostated systems. It is furthermore
natural to assume that a fractal attractor exists for the respectively
thermostated driven periodic Lorentz gas which is qualitatively of the same
type as the one for the conventionally Gaussian thermostated model Fig.\
\ref{gatt}. These questions as well as how the respective non-ideal
bifurcation scenario looks like remain to be investigated.

\subsection{Non-ideal Nos\'e-Hoover thermostat}

Precisely the same reasoning as for the non-ideal Gaussian thermostat can be
applied for constructing a non-ideal Nos\'e-Hoover scheme. We illustrate this
by summarizing results from Ref.\ \cite{RKH00}. Without loss of generality the
coordinate system is chosen such that the direction of the field with strength
$\varepsilon$ is parallel to the $x$-axis. Indeed, with $\alpha\equiv
(1+\varepsilon_x)\alpha$ it is easy to see that the heuristic derivation for
the conventional case provided in Section IV.B.1 must not be repeated, since
all other functional forms remain precisely the same. This implies that even
with a field-dependent $\alpha$ the respectively thermostated system is still
transformed onto the same canonical distribution Eq.\ (\ref{eq:fans}) as in
case of the conventional Nos\'e-Hoover thermostat.

For our definition of the non-ideal Nos\'e-Hoover thermostat we thus simply
combine Eqs.\ (\ref{eq:eg}) with $\alpha$ defined by Eq.\ (\ref{eq:anh})
yielding 
\bna
\dot{r}_x&=&v_x \nonumber \\
\dot{v}_x&=&\varepsilon_x-\alpha v_x-\alpha \varepsilon_x v_x \nonumber \\
\dot{r}_y&=&v_y \nonumber \\
\dot{v}_y&=&-\alpha v_y \nonumber \\ 
\dot{\alpha}&=&\frac{v^2-2T}{\tau^22T} \quad.  \label{eq:enh} 
\ena
The conventional Nos\'e-Hoover thermostat Eqs.\ (\ref{eq:eomdlg}),
(\ref{eq:anh}) is recovered from these equations as a special case in thermal
equilibrium, $\varepsilon_x\to0$.

That the strength of the coupling between particle and reservoir is indeed
properly adjusted to the anisotropy induced by the field is made more explicit
as follows: Eqs.\ (\ref{eq:enh}) can be rewritten by defining two
field-dependent friction coefficients, $\alpha_x=(1+\varepsilon_x)\alpha$ and
$\alpha_y\equiv\alpha$, governed by
\be
\dot{\alpha}_x=(v^2/2T-1)(1+\varepsilon_x)/\tau^2
\ee
and 
\be
\dot{\alpha}_y=(v^2/2T-1)/\tau^2\quad ,
\ee
respectively. For each velocity component there is consequently a separate
reservoir response time according to $\tau_x:=\tau/\sqrt{1+\varepsilon_x}$ and
$\tau_y\equiv\tau$. Correspondingly, with increasing field strength the
response time parallel to the field decreases thus making the action of the
thermostat parallel to the field more efficient compared to the coupling
perpendicular to the field.

In analogy to the previous section and to Sections III.B, IV.C, we shall now
discuss the chaos and transport properties of the non-ideal Nos\'e-Hoover
thermostated driven periodic Lorentz gas. Starting from Eq.\ (\ref{eq:psc})
the phase space contraction rate of this thermostated system is obtained to
\be
\kappa=-(2+\varepsilon_x)<\alpha>\quad . \label{eq:enhpr}
\ee
Performing an analogous calculation as in Section IV.C.1, i.e., starting from
the energy balance Eq.\ (\ref{eq:enbal}) and requiring that the average time
derivative of the total energy is zero, by using Eqs.\ (\ref{eq:enh}) we
arrive at the equation
\begin{equation}
\frac{\varepsilon_x<v_x>}{T}=2<\alpha>+\frac{\varepsilon_x<v_x^2\alpha>}{T}
\quad , \label{eq:enhep} 
\end{equation} 
which should again be compared to the result of the conventional case Eq.\
(\ref{eq:nhprep}). Obviously, as in case of the non-ideal Gaussian thermostat
Eq.\ (\ref{eq:preg}) there is a new second term that does not allow for an identity
between the average phase space contraction rate and thermodynamic entropy
production. In fact, if $v_x^2$ and $\alpha$ were independent quantities and
if equipartition of energy were fulfilled, $<v_x^2>=T$, then the identity
would be recovered from Eq.\ (\ref{eq:enhep}). However, first of all,
according to computer simulations $v_x^2$ and $\alpha$ are no independent
quantities. Secondly, $<v_x^2>=T$ is only strictly fulfilled in thermal
equlibrium, thus the identity cannot hold.

Some numerical results for the phase space contraction rate $\kappa$ and for
the thermodynamic entropy production $dS$ are presented in Table \ref{preptab}
at different values of $\tau$ and $\varepsilon\equiv\varepsilon_x$ yielding
quantitative evidence for these deviations. An interesting aspect is that both
functions apparently cross each other, which is at variance with the numerical
results for the non-ideal Gaussian thermostat shown in Fig.\
\ref{prepega}, where $-\kappa$ appears to provide a lower bound for $dS$. The
reason for this phenomenon is not yet understood and may deserve further
investigations.

In analogy to the non-ideal Gaussian thermostat, Eq.\ (\ref{eq:enhep}) does
not allow to recover the Lyapunov sum rule Eq.\ (\ref{eq:lsrnh}) as it holds
for the conventional Nos\'e-Hoover thermostat: The electrical conductivity may
be suitably introduced on the left hand side of Eq.\ (\ref{eq:enhep}), and by
using Eq.\ (\ref{eq:enhpr}) one may also replace the average friction
coefficient in the first term on the right hand side by the average phase
space contraction rate, respectively by the sum of Lyapunov exponents. Still,
there remains the friction coefficient in the second term on the right hand
side, which cannot be eliminated without making further assumptions. In any
case, the resulting expression is significantly different from any ordinary
Lyapunov sum rule as it holds for conventional Gaussian and Nos\'e-Hoover
thermostats. Thus, the non-ideal Nos\'e-Hoover thermostat provides another
counterexample against the universality of simple relations between chaos and
transport in dissipative dynamical systems as discussed specifically in
Section V.C.

No Lyapunov exponents have yet been computed for the non-ideal Nos\'e-Hoover
thermostated driven periodic Lorentz gas. However, since this scheme defines
also a bulk thermostat we would expect that conjugate pairing holds for the
four Lyapunov exponents of this system.

In Section IV.C.2 we have outlined a generalized Hamiltonian formalism for the
conventional Nos\'e-Hoover thermostat. Along similar lines a derivation for
the non-ideal Nos\'e-Hoover thermostat can be performed
\cite{Kla03}. The same holds for the non-ideal Gaussian thermostat in
comparison to the conventional one. But these appear to be rather technical
aspects such that we do not go into further detail here.

\begin{table}[tb]
\begin{center}
\begin{tabular}{c|cccc}
$\varepsilon$ & $\tau=1$ && $\tau\simeq31.6$ & \\ &
$-\kappa$ & $dS$ & $-\kappa$ & $dS$ \\
\hline
0.5 & 0.152 & 0.145 & 0.145 & 0.147 \\ 1.0 & 0.547 & 0.561 & 0.567 &
0.592 \\ 1.5 & 1.240 & 1.366 & 1.256 & 1.391
\end{tabular}
\vspace*{0.3cm}
\caption{Average phase space contraction rate $\kappa$, see Eq.\
(\ref{eq:enhpr}), and thermodynamic entropy production $dS=\varepsilon
<v_x>/T$, $\varepsilon\equiv\varepsilon_x$, for the non-ideal Nos\'e-Hoover
thermostated driven periodic Lorentz gas Eqs.\ (\ref{eq:enh}). As a value for
the temperature we have $T=0.5$. The numerical error is about $\pm 0.001$. The
data are from Ref.\
\protect\cite{RKH00}.}
\label{preptab}
\end{center}
\end{table}

The structure of the attractor of the non-ideal Nos\'e-Hoover thermostated
Lorentz gas, if projected onto Birkhoff coordinates, see Section III.B.3, is
qualitatively more ``smoothed out'' than the one of the conventionally
Gaussian thermostated version Fig.\ \ref{gatt}. On the other hand, it is still
a bit more detailed than the one corresponding to the attractor of the
conventional Nos\'e-Hoover thermostat Fig.\ \ref{attrnh}. Here we do not show
the attractor generated by Eqs.\ (\ref{eq:enh}) but refer to Ref.\
\cite{RKH00} for a respective figure and for further details.

\begin{figure}[t]
\epsfxsize=17cm
\centerline{\epsfbox{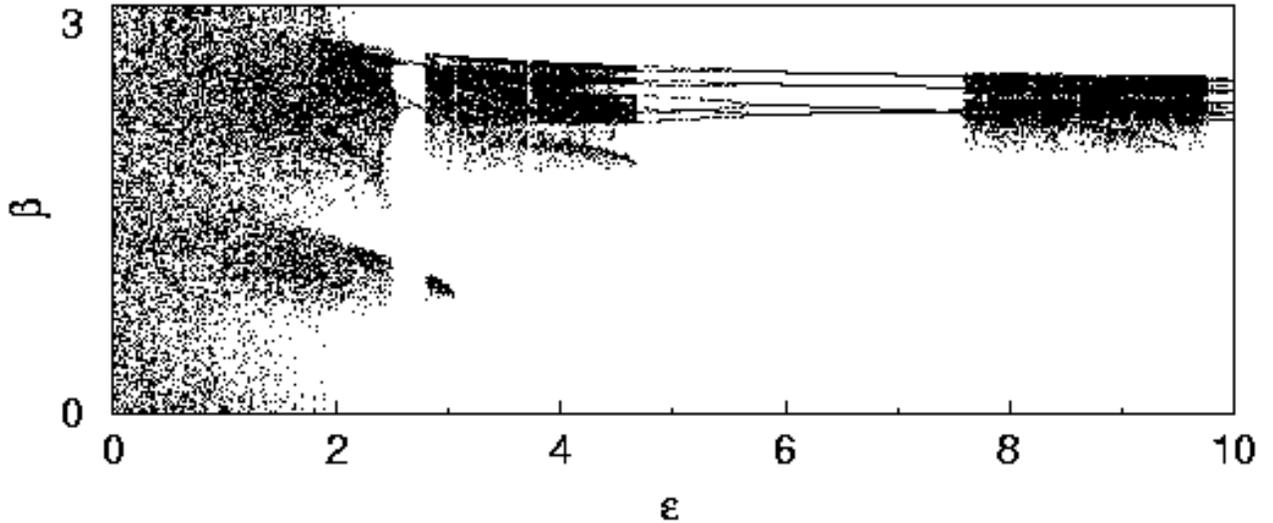}}
\caption{Bifurcation diagram for the non-ideal Nos\'e-Hoover thermostated
periodic Lorentz gas driven by an electric field of strength $\varepsilon$
which is parallel to the $x$-axis. $\beta$ is defined in Fig.\ \ref{plko}.
The temperature is $T=0.5$, the reservoir coupling parameter $\tau=1$. The
figure is from Ref.\ \protect\cite{RKH00}.}
\label{bifunhe}
\end{figure}

Instead, we present a bifurcation diagram for the non-ideal Nos\'e-Hoover
thermostated driven periodic Lorentz gas. Fig.\ \ref{bifunhe} may be compared
to the one of the conventionally Nos\'e-Hoover thermostated case Fig.\
\ref{bifunh} (a). Although the response time $\tau$ is one order of magnitude smaller
for the conventional case than for the non-ideal version thus indicating a
more efficient coupling to the thermal reservoir, the action of the non-ideal
version appears to be similarly efficient concerning a full covering of the
phase space, which in both cases breaks down approximately at
$\varepsilon\simeq 2$. This is in full agreement with the physical motivation
for introducing the field-dependent coupling as discussed in the previous
section in that it should enhance the efficiency of the thermostat's action
compared to conventional schemes. A bifurcation diagram for the conventional
Nos\'e-Hoover case at $\tau=1$ can be found in the material accompanying Ref.\
\cite{RKH00} and confirms our above assessment.

The parameter-dependent electrical conductivity for the non-ideal version has
not yet been computed, however, based on our discussion of the previous cases
it is not unreasonable to expect that the results will just yield another
highly nonlinear, irregular curve.

\subsection{$^*$Further alternatives to conventional Gaussian and Nos\'e-Hoover dynamics}

In order to conclude this chapter we briefly summarize whether various
mechanisms generating NSS fulfill the crucial identity between the average
rate of phase space contraction and thermodynamic entropy production discussed
before. To a large extent, these mechanisms have already been encountered on
previous occasions, see in particular Sections IV.D and V.B.

Cohen and Rondoni showed that even the Gaussian {\em isoenergetic} thermostat
applied to a many-particle system provides an identity between phase space
contraction and entropy production only in the thermodynamic limit of a large
number of particles \cite{CoRo98,Rond02}.

Benettin and Rondoni introduced a specific time-reversible rescaling of the
single velocity components of a moving particle at the boundary of a system
while keeping the total kinetic energy fixed \cite{BR01,Rond02}. This
mechanism may be considered as an intermediate case between a straightforward
velocity rescaling, see Section III.A, and an ordinary Gaussian isokinetic
thermostat. The complete dynamical system is defined such that it exhibits a
symmetry breaking thus generating a current in a NSS. Similarly to the
non-ideal thermostats discussed in the previous two sections, for small enough
bias the average phase space contraction rate is getting proportional to the
thermodynamic entropy production while for larger bias this relationship
becomes inherently nonlinear. Related dissipation mechanisms have been
introduced earlier by Zhang and Zhang \cite{ZhZh92} and in particular by
Chernov and Lebowitz \cite{ChLe95,ChLe97}, see also Section VII.E.1 where we
will discuss the example of a shear flow in a NSS. However, Chernov and
Lebowitz argued for an identity according to their method.

As far as the generalized Nos\'e-Hoover thermostats sketched in Section IV.D
are concerned, whether or not they exhibit an identity can easily be checked
by following the second approach outlined in Section IV.C.1 for the example of
the conventional Nos\'e-Hoover thermostat. It turns out that controlling
higher even moments along the lines of Nos\'e-Hoover
\cite{Ho89,JeBe89,HoHo96,PH97,HoKu97,LiTu00}, the Nos\'e-Hoover chain
thermostat \cite{MKT92,HoKu97,TuMa00} and the deterministic formulation of a
Langevin-like equation according to Bulgac and Kusnezov \cite{BuKu90b} all
preserve the identity. These schemes are furthermore all time-reversible thus
sharing fundamental properties with the class of conventional Gaussian and
Nos\'e-Hoover thermostats.

The cubic coupling scheme proposed by Kusnezov et al.\
\cite{KBB90,BuKu90}, on the other hand, is time-reversible but does not show
the identity. In this case the non-identity is due to a thermalization of the
position coordinates creating an additional phase space contraction that does
not contribute to an entropy production in terms of Joule's heat. In a way,
this method exemplifies our assessment formulated in Section V.B that the
phase space contraction of a dynamical system may involve contributions that,
from the point of view of a thermodynamic entropy production, simply do not
matter. Unfortunately, as we discussed already in Section IV.D.2 the physical
interpretation of thermalizing position coordinates in the cubic coupling
scheme remains rather unclear. Hence, this modeling may not be considered as
an unambiguous counterexample concerning the identity.

The generalized Nos\'e-Hoover thermostats introduced by Hamilton et al.\
\cite{Ham90,LHHa93} and by Winkler et al.\ \cite{Wink92,WKR95} are all
time-reversible but do not fulfill the identity either. However, here it is
not yet clear whether these schemes work conveniently in nonequilibrium
situations. The methods proposed by Branka et al.\ \cite{Bran00,BrWo00} and by
Sergi et al.\ \cite{SeFe01} are already not time-reversible and hence do not
qualify as counterexamples for Gaussian and Nos\'e-Hoover dynamics.

Apart from considering the action of thermal reservoirs leading to dissipative
dynamical systems we may recall that there are other ways in order to create
NSS which do not involve any phase space contraction thus also providing
counterexamples concerning an identity. For example, Gaspard studied diffusion
under concentration gradients, respectively dynamical systems under flux
boundary conditions\cite{Gasp97a,Gasp}. Eckmann et al.\ considered a model for
a heat flow by using a fully Hamiltonian thermal reservoir consisting of
infinitely many planar waves \cite{EPRB99b}. All these studies are reminiscent
of the Hamiltonian approach to NSS. In the latter case the system actually
reminds of the infinite Hamiltonian modeling of the Langevin equation, see
Section II.B, rather than the thermostating approach. In any case, in both
situations there is by default no phase space contraction and consequently no
identity. Similarly, van Beijeren and Dorfman argued for a non-existence of
this identity by using a Lorentz gas model for a heat flow \cite{vBD00},
however, their argument appears to be debatable.

Finally, Nicolis and Daems analyzed the general setting of dynamical systems
perturbed by noise by deriving a balance equation for the entropy production
\cite{NiDa96,NiDa98,DaNi99}. They found correction terms to an identity
between phase space contraction and entropy production which they identified
as functions of the noise strength. But in this case the dynamical systems
considered are again not time-reversible anymore because of the noise and thus
belong to a different class than conventional Gaussian and Nos\'e-Hoover
thermostats.

\subsection{Summary}

\begin{enumerate}

\item We introduced a straightforward generalization of the conventional Gaussian
thermostat, again for the example of the driven periodic Lorentz gas. This
thermostating scheme is by default deterministic and time-reversible but is
constructed such that it does not yield an identity between the average rate
of phase space contraction and thermodynamic entropy production. We coined
this method the {\em non-ideal Gaussian thermostat}. Since there is no
identity there is also no Lyapunov sum rule relating transport coefficients to
dynamical systems quantities. The field-dependent conductivity of this model
is again highly nonlinear and different from the one found in other
thermostated driven periodic Lorentz gases.

\item In the same vein, we introduced a {\em non-ideal Nos\'e-Hoover
thermostat} that also does not exhibit the identity and hence no Lyapunov sum
rule. In both cases, the non-identity resulted from modifying the coupling
between subsystem and thermal reservoir by suitably adjusting it to the
symmetry breaking caused by the external electric field. This even increased
the efficiency of the non-ideal thermostats compared to their conventional
colleagues as could be seen, for example, by means of matching bifurcation
diagrams.

\item We finally summarized various other existing approaches towards the
construction of NSS by inquiring whether or not they feature an identity
between phase space contraction and entropy production. Apart from Hamiltonian
modelings of NSS, it appears that there are many other examples of
thermostating schemes not providing such an identity. However, in most cases
they have not yet been studied respectively in nonequilibrium situations.

\end{enumerate}

\section{Stochastic and deterministic boundary thermostats}

Gaussian and Nos\'e-Hoover schemes belong to the class of so-called {\em bulk
thermostats}, that is, they remove energy from a subsystem during the
originally free flight of a particle. {\em Boundary thermostats}, on the other
hand, change the energy of a particle only at the collisions with the
boundaries of the subsystem.\footnote{Of course, the action of Gaussian and
Nos\'e-Hoover thermostats can be restricted to some boundary layers of a
subsystem thus bridging the gap between bulk and boundary thermostats as has
been done in Ref. \cite{PoHo89}.}

In this chapter we construct and analyze fundamental types of the latter class
of thermostats. We start with well-known stochastic versions, which are
so-called {\em stochastic boundary conditions} as they have already been
mentioned in Section V.D. We then show how to make stochastic boundaries
deterministic and time-reversible by using a simple chaotic map. This recent,
new modeling of a thermal reservoir was called {\em thermostating by
deterministic scattering} in the literature. The equations of motion resulting
from this method are still deterministic and time-reversible hence sharing
fundamental properties with Gaussian and Nos\'e-Hoover schemes. However, both
the construction of this deterministic boundary thermostat and the dynamics
associated with it are generally very different in comparison to conventional
bulk thermostats.

An analysis of models thermostated by deterministic scattering thus enables us
to learn more about possible universal properties of nonequilibrium steady
states in dissipative dynamical systems. We demonstrate this by studying again
the transport and dynamical systems properties of a respectively thermostated
driven periodic Lorentz gas. Our findings are compared to the ones reported
previously for Gaussian and Nos\'e-Hoover thermostats. We finally briefly
discuss results for a hard-disk fluid under nonequilibrium conditions, which is
also thermostated at the boundaries.

\subsection{Stochastic boundary thermostats}

\begin{figure}[t]
\epsfxsize=6cm
\centerline{\rotate[r]{\epsfbox{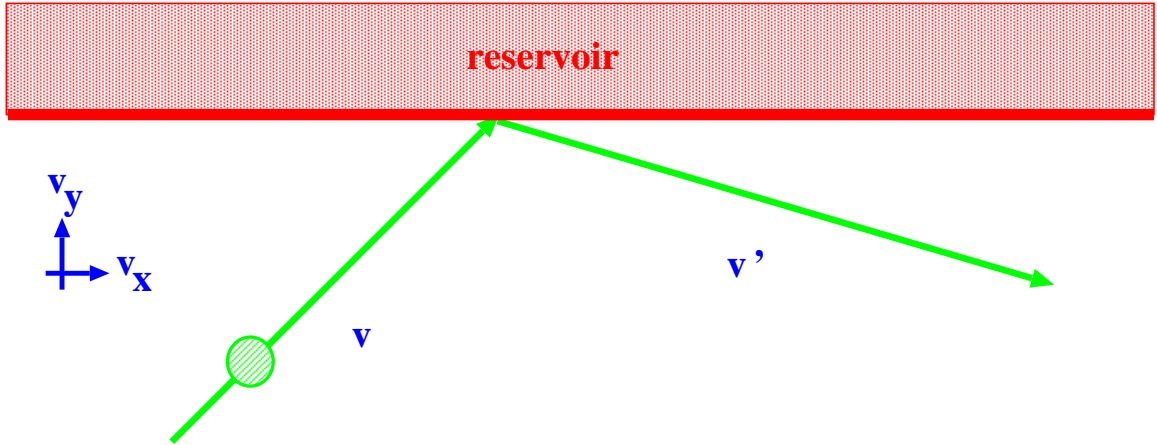}}}
\vspace*{0.2cm} 
\caption{Setup for defining stochastic boundary conditions: a point particle of mass
$m=1$ moving with velocity ${\bf v}$ collides inelastically with a flat
wall. The velocity after the collision is denoted by ${\bf v'}$. The
difference in the kinetic energy $(v'^2-v^2)/2$ is transfered to, or from, a
thermal reservoir associated with the wall. ${\bf v'}$ is obtained according
to Eq.\ (\ref{eq:sbctr}), see the text.}
\label{fig:sbc}
\end{figure}

Let us consider a point particle moving in a plane, which interacts with a
thermal reservoir through collisions at a flat wall, see Fig.\
\ref{fig:sbc}. If the reservoir is in thermal equilibrium at a certain
temperature, the reservoir degrees of freedom will be distributed according to
a canonical distribution. Consequently, the subsystem particle will be
thermalized at the collision with respect to a Gaussian velocity
distribution. If we assume that the memory of the particle is completely lost
at the collision, the velocity after the collision can be randomly sampled
from the respective velocity distribution of the thermal reservoir. This is
the physical essence of what is called {\em stochastic boundary conditions},
see also our brief account of them in Section V.D.

Casting this physical picture into an equation it must be taken into account
that an impenetrable wall breaks the symmetry of the subsystem. That is, there
are different velocity distributions parallel and perpendicular to the wall:
Parallel to the wall the particle's outgoing velocity component $v'_x$, see
Fig.\ \ref{fig:sbc}, should be sampled simply from the Gaussian distribution
for reservoir velocities parallel to the wall. However, in the perpendicular
direction an observer who is sitting at the wall will measure a non-zero
average flux of reservoir, respectively subsystem particles to the wall. This
flux is due to the fact that all these particles can approach the impenetrable
wall from one side only. Consequently, the symmetry of the corresponding
distributions of velocity components perpendicular to the wall is broken as
well. This is properly modeled by multiplying the bulk velocity distribution
of reservoir, respectively subsystem, with the absolute value of the
perpendicular velocity component, $|v_y|$.

In summary, if the velocity distribution of the thermal reservoir is canonical
at temperature $T$ an observer at the boundary measures for the two velocity
components $v_x\in[-\infty,\infty),v_y\in[0,\infty)$ of a reservoir particle
colliding with the boundary \cite{LeSp78,CiTe80,TCG82,ChLe95,ChLe97,WKN99}
\be
\rho(v_x,v_y)=(2\pi
T^3)^{-1/2}|v_y|\exp\left(-\frac{v_x^2+v_y^2}{2T}\right)\quad . \label{eq:sbc}
\ee
The subsystem particle must then exhibit precisely the same distribution of
outgoing velocities, under the conditions that its velocities before and after
the collision are not correlated and that the reservoir distribution is not
modified.

In practice, these scattering rules are implemented as follows \cite{WKN99}:
draw two independently and identically distributed random numbers $\zeta
,\xi\in[0,1]$ from a uniform density $\rho(\zeta,\xi)=1$ by using a suitable
random number generator. The transformation rules of how to get from these
uniform distributions to the Gaussian ones of Eq.\ (\ref{eq:sbc}) can be
calculated from the requirement of phase space conservation.\footnote{For a
two-dimensional Gaussian without a flux term the same reasoning in polar
coordinates leads to the well-known Box-M\"uller algorithm
\cite{PFTV92}.}  For the outgoing velocities $v'_x,v'_y\in[0,\infty)$ this condition reads
\bna
\rho(\zeta)\rho(\xi)\left|\frac{d\zeta 
d\xi}{dv'_xdv'_y}\right|&=&\sqrt{\frac{2}{\pi
T^3}}|v'_y|\exp\left(-\frac{{v'}_x^2+{v'}_y^2}{2T}\right) \nonumber\\
&=&\left|\frac{\partial \mbox{\boldmath${\cal T}$ }(v'_x,v'_y)}{\partial
v'_x\partial v'_y}\right| \quad . \label{eq:sbcce}
\ena
Integration yields for the transformation ${\cal T}$ the functional form
\bna 
(v'_x,v'_y) &=& \mbox{\boldmath${\cal T}$ }^{-1}
(\zeta,\xi)\nonumber \\  
&=&\sqrt{2T}\left(\mbox{erf}^{-1}(\zeta),
\sqrt{-\ln(\xi)}\right) \quad , \label{eq:sbctr}
\ena 
which completes the algorithm.

\subsection{Deterministic boundary thermostats, or thermostating by deterministic scattering}

From the point of view of dynamical systems theory the problem with stochastic
boundaries is that they are not deterministic and
time-reversible. Consequently, they do not enable a detailed analysis of
transport and dynamical systems properties of dissipative systems as outlined
in the previous chapters. However, this problem can rather straightforwardly
be resolved as follows \cite{KRN00,RKN00,WKN99}: Let ${\cal
M}:[0.1]\to[0,1]\:,\: (\zeta,\xi)\to{\cal M}(\zeta,\xi)$ be a two-dimensional
deterministic map with an inverse ${\cal M}^{-1}$. Loosely speaking, the
random number generators in Eq.\ (\ref{eq:sbctr}) are then replaced by the
action of this map, however, in form of the following scattering rules
\be
(|v_x'|,|v_y'|)=
\left\{\begin{array}{l@{\:;\:}l}
\mbox{\boldmath${\cal T}$ }^{-1}\circ{\cal
M}\circ\mbox{\boldmath${\cal T}$ } (|v_x|,|v_y|)& v_x\ge 0\\ 
\mbox{\boldmath${\cal T}$ }^{-1}\circ{\cal
M}^{-1}\circ\mbox{\boldmath${\cal T}$ } (|v_x|,|v_y|)& v_x<0
\end{array}\right. \label{eq:tds}
\ee
with $\mbox{\boldmath${\cal T}$ }^{-1}:[0,1]\times[0,1]\to
[0,\infty)\times[0,\infty)$. This prescription is not yet complete. First of
all, we have to further specify the sign of the outgoing velocities.  We do
this for the geometry presented in Fig.\ \ref{fig:sbc} by requiring that
particles going in with positive (negative) tangential velocity go out with
positive (negative) tangential velocity. For the perpendicular component we
require that $v_y$ always changes its sign. This distinction between positive
and negative velocities, supplemented by the respective use of ${\cal M}$ and
${\cal M}^{-1}$ in the equations, renders the collision process
time-reversible. This is easily verified by using the collision rules Eq.\
(\ref{eq:tds}) and reversing the direction of time after a collision. Finally,
to avoid any artificial symmetry breaking in numerical experiments, the
application of ${\cal M}$ and ${\cal M}^{-1}$ in Eq.\ (\ref{eq:tds}) should be
alternated with respect to the position of the scattering process at the wall.

It remains to characterize the map ${\cal M}$ of this equation in more
detail. In order to compare with Gaussian and Nos\'e-Hoover thermostats we
require that this map is deterministic and reversible; for a definition of the
latter property see footnote number \ref{fn:trev}. Furthermore, ${\cal M}$
should be mixing \cite{ArAv68,Schu,Gasp,Do99} implying ergodicity and, under
rather general conditions, also chaotic behavior \cite{Dev89,BBCDS92} in the
sense of exhibiting a positive Lyapunov exponent \cite{Ott}. Preferably, the
map should also be uniformly hyperbolic \cite{ER,Ott,Gasp,Do99}, and for
convenience it should generate a uniform probability density in phase
space. All these properties, except the last one, are necessary conditions for
enabling a proper thermostating. That is, the whole phase space of the map
should be sampled such that, in combination with the transformation ${\cal T}$
according to Eq.\ (\ref{eq:tds}), the canonical distribution can be reproduced
completely on the basis of these scattering events; see also Section IV.D.1
for a related discussion of necessary conditions regarding a proper
functioning of deterministic thermostats. 

Apart from these constraints the precise functional form of the map appears to
be arbitrary. By choosing different maps one may model different particle-wall
interactions. In this respect the map ${\cal M}$ together with the
tranformation ${\cal T}$ plays the role of a scattering function containing
detailed information about the microscopic scattering process of a
particle. Fixing the functional form of the map completes our set of
scattering rules, which has been called {\em thermostating by deterministic
scattering} \cite{KRN00,RKN00,WKN99,RaKl02}.

If not indicated otherwise, in the following we will choose for ${\cal M}$ the
probably most simple example of a map fulfilling all of the above
requirements. This is the two-dimensional {\em baker map} as it has been
introduced and explicitly discussed in Section V.A, see Eq.\ (\ref{eq:baker})
and the corresponding Fig.\ \ref{fig:baker}. However, we remark that other
simple maps such as the cat map, respectively for general values of the
control parameter the sawtooth map, as well as the standard map
\cite{Schu,Ott,Meis92,ASY97}, both for large enough values of their
control parameters, have also been used within this scheme already
\cite{WKN99,KR99,Wag00}.

\subsection{$^*$Stochastic and deterministic boundary thermostats from first principles}

There are two crucial ingredients in the approach outlined above leading to
stochastic and deterministic boundary thermostats: Firstly, the knowledge of
the equilibrium velocity distribution functions of a thermal reservoir with
infinitely many degrees of freedom is required. Secondly, the energy transfer
at the boundaries must be defined in form of suitable microscopic collision
rules.

In both respects specifically our derivation of thermostating by deterministic
scattering is quite heuristic. That is, we started from essentially
pre-defined stochastic boundaries and conveniently modified them by choosing
new collision rules. However, a systematic construction from first principles
should rather proceed the other way around by first leading to the general
solution, which is the deterministic case, and then recovering the stochastic
case by eliminating any correlations at the collisions with the wall. Such an
approach should provide more insight into the detailed mechanism and into the
conditions for a proper functioning of this thermostat. Here we outline in
three steps how such a systematic construction is performed \cite{KRN00,RKN00}:\\

{\bf 1. equilibrium velocity distribution functions:}

Our first goal is to construct a model for a thermal reservoir consisting of
an arbitrary but finite number of degrees of freedom. For this purpose we need
to know the velocity distribution functions of a subsystem consisting of $d_s$
degrees of freedom, which interacts with a thermal reservoir of $d_r<\infty$
degrees of freedom in thermal equilibrium. With $d=d_s+d_r$ we denote the
total number of degrees of freedom, where $d_s<d_r$, and we may assume that
there is only kinetic energy. In Section II.C we have summarized how to
calculate the finite-dimensional velocity distributions of the subsystem by
giving some exact results; see Eq.\ (\ref{eq:gaus}) for the equilibrium
distribution of one velocity component $v_1$ and Eq.\ (\ref{eq:gausv}) for the
distribution of the absolute value of the velocity vector of two components
$v_1$ and $v_2$ with $v=\sqrt{v_1^2+v_2^2}$.

Note that the corresponding distribution functions are bulk distributions
only, whereas, for defining a boundary thermostat, we need to know the
functional forms of these distributions at the boundaries of the thermal
reservoir, respectively of the subsystem. This bulk-boundary transformation
may be performed heuristically such as outlined in the previous section. Ref.\
\cite{RKN00} provides a more microscopic derivation of this relation
by means of calculating the average time of flight between two collisions and
the average collision length. In any case, all these arguments eventually
amount in a multiplication of the finite-dimensional bulk densities Eqs.\
(\ref{eq:gaus}), (\ref{eq:gausv}) with a ``flux'' factor of $v_1$,
respectively of $v$, supplemented by a suitable renormalization of the
associated velocity distributions.\\

{\bf 2. defining a proper coupling:}

Knowing these fundamental functional forms of the subsystem's boundary
equilibrium velocity distribution functions, the next step is to define a
suitable microscopic coupling between subsystem and thermal reservoir. The
coupling should be such that it establishes an energy transfer between
subsystem and reservoir degrees of freedom, which generates the desired
velocity distribution. In order to preserve the Hamiltonian character of the
microscopic equations of motion of the subsystem to be thermostated we demand
that this coupling is deterministic and time-reversible. In some sense, there
should furthermore be conservation of phase space volume, as will be explained
later on.

A first idea of how such a coupling may look like can be obtained from the
{\em rotating disk model}: Let us consider a unit cell of the field-free
periodic Lorentz gas, see the geometry of Fig.\ \ref{fig:plgasg}, where the
fixed disk rotates with an angular velocity $\omega$. Altogether we have thus
three degrees of freedom, two for the moving point particle and one for the
rotating disk.  Assuming that the particle is elastically reflected
perpendicular to the disk reduces the problem of an energy transfer between
particle and disk to the problem of two elastically colliding masses on a
line. The corresponding equations of motion can easily be derived and are of
the form of a nonlinear two-dimensional map governing the exchange between
translational and rotational energy.

This rotating disk model was originally proposed in Ref.\ \cite{RKN00} but
became more popular only later on by applying it to a periodic Lorentz gas under a
temperature gradient, see Ref.\ \cite{MLL01} and more recent studies in Refs.\
\cite{LLM02,BuKh03}. Here all disks of the spatially extended Lorentz gas
rotate with different angular velocities thus mimicking a thermal reservoir
of, in principle, arbitrary dimensionality. Interactions between these
different rotational degrees of freedom are provided by the moving particle
carrying energy from one disk to another. To some extent, this Hamiltonian
modeling of a thermal reservoir reminds of the one employing an arbitrary
number of harmonic oscillators which was used for deriving the stochastic
Langevin equation, see Section II.B. It thus suffers from the same deficiency
that, at least numerically, it cannot conveniently be applied to
nonequilibrium situations involving external fields, because in this case an
infinite number of reservoir degrees of freedom is needed in order to
continuously remove energy from the subsystem.

Hence we follow a different path leading to a more abstract and more general
model of a thermal reservoir. This approach is further motivated by a second
problem with the rotating disk model: Computer simulations for a single
rotating disk under periodic boundary conditions show that only for specific
values of the associated control parameters this coupling reproduces the
equilibrium velocity distributions Eqs.\ (\ref{eq:gaus}), (\ref{eq:gausv}) at
$d=3$ \cite{RKN00,LLM02}. This reflects the fact that generally the dynamics
of this system is very complicated indicating non-ergodic behavior in a mixed
phase space consisting of islands of stability and chaotic layers. As a
consequence, not for all values of the control parameters there is
equipartitioning of energy, which is needed for a microcanonical distribution
for subsystem plus thermal reservoir \cite{Reif}. In turn, only the presence
of equipartitioning appears to guarantee a proper energy transfer between
subsystem and reservoir degrees of freedom providing a proper thermalization.

In other words, an arbitrary map ${\cal M}$ defining the collision rules of a
moving particle with a disk, such as provided by the rotating disk model, must
generally be {\em enforced} to yield equipartitioning of energy, respectively
the desired equilibrium velocity distributions for reservoir and
subsystem.\footnote{In Ref.\ \cite{LLM02} it was argued that relating
different rotational degrees of freedom to different disks also resolves this
problem.} This can be achieved by, first of all, choosing a map ${\cal M}$
that is sufficiently `nice', in the sense of fulfilling the list of necessary
conditions already provided in the previous section. The map must further be
combined with a second map ${\cal T}$ according to Eq.\ (\ref{eq:tds}) in
order to ensure that the desired functional forms of the equilibrium velocity
distributions are attained. This map ${\cal T}$ is defined in complete formal
analogy to Eqs.\ (\ref{eq:sbcce}), (\ref{eq:sbctr}) in replacing the Gaussian
of Eq.\ (\ref{eq:sbcce}) by the finite-dimensional ``target'' velocity
distribution function determined by Eq.\ (\ref{eq:gaus}).

As an example, let us consider a single Lorentz gas scatterer under periodic
boundary conditions. We modify the collision rules according to the above
reasoning such that this system mimicks a rotating disk-like situation with
$d_s=2$ for the moving particle and $d_r=1$ associated with the disk, thus
$d=3$ altogether. For the map ${\cal M}$ we choose again the baker map Eq.\
(\ref{eq:baker}). The target equilibrium velocity distributions are determined
by Eqs.\ (\ref{eq:gaus}), (\ref{eq:gausv}) at $d=3$. The transformation ${\cal
T}$ is then calculated from the prescription used in Eq.\ (\ref{eq:sbcce}) if
the Gaussian of this equation is replaced by the respective functions of Eqs.\
(\ref{eq:gaus}), (\ref{eq:gausv}) at $d=3$. Here one needs to make a choice
concerning the coordinate system in which the velocity variables shall be
defined. For the Lorentz gas disk one may choose, e.g., a local coordinate
system with respect to the point at the collision. Some further details are
explained in the following section. The resulting expression for ${\cal T}$ is
finally combined with the map ${\cal M}$ according to
Eq. (\ref{eq:tds}). Computer simulations of this inelastic Lorentz gas
scatterer provide evidence that, indeed, the desired velocity distributions
are generated by this model \cite{RKN00}.

Obviously, the general idea underlying this method is not restricted to
$d=3$. According to the transformation ${\cal T}$, all what we need to know is
the proper equilibrium distribution function of a thermal reservoir associated
with $d_r$ degrees of freedom while interacting with a subsystem of $d_s$
degrees of freedom. In other words, we do not need to care about any explicit
equations of motion defining $d_r$ degrees of freedom in order to model a
thermal reservoir. From now on we therefore carefully distinguish between the
{\em number of degrees of freedom} of the thermal reservoir that we mimick,
and the {\em number of dynamical variables} which are actually involved in
modeling this reservoir. For the collision rules that we are currently
discussing the action of the thermal reservoir may be represented by one
dynamical variable only: Let $k$ represent the absolute value of the velocity
vector of all reservoir degrees of freedom. The value of $k$ is then defined
via energy conservation for subsystem plus reservoir, $E=v^2/2+k^2/2$, where
$v$ is the absolute value of the velocity of a subsystem particle before a
collision. Accordingly, the value $v'$ after a collision is generated by the
collision rules Eq.\ (\ref{eq:tds}), with a suitable choice for ${\cal T}$,
and the corresponding value $k'$ of the reservoir variable is again obtained
from energy conservation. Thus, $\Delta k:=k'-k$ yields the energy transfer
between subsystem and reservoir at a collision.

We add that this general formalism might be used to obtain deterministic and
time-reversible collision rules for two colliding granular particles instead
of using normal and tangential restitution coefficients according to which the
microscopic dynamics is dissipative, non-Hamiltonian, and not reversible.  A
finite-dimensional thermal reservoir then mimicks $d_r$ internal degrees of
freedom of some granular material storing kinetic energy from which it can be
restored at a collision, compare, e.g., to microscopic models of inelastic
collisions as discussed in Refs.\ \cite{AGZ98,BSHP96}. However, note that we
did not yet arrive at the transformation represented by Eq.\ (\ref{eq:sbctr}),
since our collision rules still concern a finite dimensional thermal
reservoir only.\\

{\bf 3. modeling an infinite dimensional reservoir:}

In a general nonequilibrium situation only a thermal reservoir consisting of
an infinite number of degrees of freedom is able to continuously absorb energy
and to possibly generate a NSS. For the equilibrium velocity distributions
Eqs.\ (\ref{eq:gaus}), (\ref{eq:gausv}) this implies that we need to take the
limit of $d\to\infty$. As was shown in Section II.C, employing
equipartitioning of energy we then arrive at the well-known canonical
distributions Eqs.\ (\ref{eq:vxgauss}), (\ref{eq:vgauss}). In this limit the
total energy of the thermal reservoir is not well-defined anymore,
consequently it cannot be represented by a dynamical variable $k$ as
introduced for the finite-dimensional case discussed above. However, this
variable can easily be dropped, since it does not appear explicitly in the
collision rules.

By using now the canonical velocity distribution Eq.\ (\ref{eq:vxgauss}) we
arrive at Eq.\ (\ref{eq:sbcce}) in order to define the transformation ${\cal
T}$. Neglecting any correlations in the map ${\cal M}$ before and after a
collision, the stochastic boundary conditions Eqs.\ (\ref{eq:sbctr}) are
recovered from their deterministic counterpart Eq.\ (\ref{eq:tds}) as a
special case. Hence, we may say that this approach provides a derivation of
stochastic boundaries from first principles. In fact, thermostating by
deterministic scattering was first constructed by following this bottom-up
approach \cite{KRN00,RKN00}. Only subsequently it was realized that the final
outcome may be considered as a deterministic generalization of stochastic
boundaries along the top-to-bottom approach of Section VII.A
\cite{WKN99}. This completes our derivation of a modeling of stochastic and
deterministic boundaries ``from first principles''.\\

We conclude this section by taking up the thread from Sections II.C and V.A,
in which we provided a simple explanation for the origin of an average phase
space contraction in thermostated nonequilibrium systems. In fact, the
previous bottom-up approach towards deterministic boundary thermostats nicely
illustrates the respective reasoning as follows: Let us start in an
equilibrium situation with constant total energy and let the number
$d=d_s+d_r$ of degrees of freedom of subsystem plus thermal reservoir for a
moment again be finite. As we explained in Section II.C, in this case an
ensemble of points representing the combination of subsystem plus reservoir is
uniformly distributed on a hypersphere in velocity space indicating that there
is no average phase space contraction. However, the projected-out velocity
distributions of the subsystem are typically not uniform anymore, cp.\ to
Eqs.\ (\ref{eq:gaus}), (\ref{eq:gausv}).

Correspondingly, by calculating the Jacobian determinant for the collision
rules Eqs.\ (\ref{eq:tds}) {\em only} on the basis of respective functional
forms of ${\cal T}$ for given $d$, one finds that the subsystem dynamics alone
is not phase space conserving at a single collision \cite{Kla03}. This should
not come as a surprise, since at any collision there is an exchange of kinetic
energy with the reservoir. Nevertheless, in case of an equilibrium situation
the {\em average} phase space contraction rate for the subsystem is still zero
reflecting the fact that on average there is no energy transfer between
subsystem and reservoir. This was also confirmed numerically for the
respectively modified Lorentz gas.

The apparent phase space contraction at a collision simply results from the
fact that so far we looked at the dynamics of the subsystem only. In a
complete analysis the dynamics of the thermal reservoir must be taken into
account as well. This dynamics is adequately represented by the variable $k$,
which, here, is well-defined again because of $d<\infty$ and can be obtained
via energy conservation. The complete dynamics thus actually consists of the
set of variables $({\bf v},k)$. By looking at the Jacobian determinant of the
combined system $({\bf v},k)$ one can indeed verify that the complete dynamics
is always locally phase space conserving. Consequently, there is no
contradiction between a locally dissipative dynamics of the subsystem alone
and a phase space preserving Hamiltonian character of the full equations of
motion for subsystem plus thermal reservoir.

Let us now apply these arguments to a nonequilibrium situation, e.g., by
imposing an external field onto the subsystem. The field will continuously
increase the kinetic energy of a particle moving in the subsystem, and the
thermal reservoir will continuously absorb this energy through collisions by
storing it onto infinitely many degrees of freedom according to
equipartitioning of energy. During the free flight of the particle there is no
action of the thermal reservoir, and the subsystem is trivially
Hamiltonian. At a collision we are using precisely the same collision rules as
defined above for which we have argued, in case of $d<\infty$, that there is
no loss of phase space volume either. Consequently, by properly looking at
subsystem {\em plus} thermal reservoir we conclude that the whole dynamics is
phase space conserving, and in this respect Hamiltonian-like, in
nonequilibrium as well. This is not in contradiction to the fact that there is
an average flux of energy from the subsystem to the thermal reservoir, which
counterbalances the generation of an average current parallel to the external
field. These arguments should also be valid by making the number of reservoir
degrees of freedom arbitrarily large, that is, in the limit of $d\to\infty$.

We finally remark that, although for thermostating by deterministic scattering
the equations of motion appear to be compatible with a Hamiltonian dynamics,
as argued above, a Hamiltonian formulation of this scheme is currently not
known.

\subsection{Deterministic boundary thermostats for the driven periodic Lorentz gas}

Let us now apply the scheme of thermostating by deterministic scattering to
the periodic Lorentz gas by modeling a thermal reservoir mimicking an {\em
infinite number} of degrees of freedom as outlined in the previous
section. That is, the originally elastic collisions of the moving particle
with the hard disk are made {\em inelastic} allowing for an energy transfer at
a collision without changing the geometry of the system. Note that the unit
cell with a single scatterer is periodically continued by applying periodic
boundary conditions. In physical terms one may thus think of the arbitrarily
many degrees of freedom that we are now going to associate with the disk as
mimicking, e.g., different lattice modes in a crystal, which remove energy
from a colliding particle.

This modification of the standard Lorentz gas is performed by adapting the
collision rules Eqs.\ (\ref{eq:baker}), (\ref{eq:sbctr}) and (\ref{eq:tds}) to
the circular geometry of a Lorentz gas disk. For this purpose we replace the
usual Cartesian coordinates by the tangential, respectively the normal
component of the velocity of the colliding particle in a local coordinate
system at the scattering point. That is, we write $v_{\parallel}\equiv v_x$
and $v_{\perp}\equiv v_y$ in Eqs.\ (\ref{eq:sbctr}), (\ref{eq:tds}).
Alternatively, one may wish to choose local polar coordinates
\cite{KRN00,RKN00}.

We are now prepared to study the chaos and transport properties of the driven
periodic Lorentz gas thermostated by deterministic scattering and to compare
the results with the previous ones obtained for ideal and non-ideal Gaussian
and Nos\'e-Hoover thermostats. Note that thermostating by deterministic
scattering was constructed in order to generate a canonical velocity
distribution in thermal equilibrium. In this respect this boundary thermostat
forms a counterpart to the Nos\'e-Hoover thermostat acting in the bulk. In
fact, our previous Fig.\ \ref{fig:rhonhtds} depicts some representative
nonequilibrium velocity distribution functions for the driven periodic Lorentz
gas thermostated both by Nos\'e-Hoover and by deterministic scattering. For
moderate to large response times the Nos\'e-Hoover distributions are indeed
very similar to the one obtained from thermostating by deterministic
scattering.

Concerning other chaos and transport properties we will show in the following
that a comparison is more non-trivial. First we check analytically for an
identity between phase space contraction and entropy contraction. We then show
numerical results for the attractor both resulting from deterministic and from
stochastic boundary thermostats. In the deterministic case we further discuss
the associated bifurcation diagram as well as the electrical
conductivity. Finally, we elaborate on the spectrum of Lyapunov exponents.  In
this section we particularly summarize results published in Refs.\
\cite{KRN00,RKN00,RaKl02}.

\subsubsection{Phase space contraction and entropy production}

In contrast to bulk thermostats such as Gauss or Nos\'e-Hoover, for a driven
periodic Lorentz gas thermostated at the boundaries the phase space volume is
conserved during free flights between collisions. As was briefly mentioned in
the previous section, any phase space contraction can thus only be generated
by the collision rules defined in terms of the composition of maps Eqs.\
(\ref{eq:baker}), (\ref{eq:sbctr}) and (\ref{eq:tds}). For maps any change of
the phase space volume is assessed by the Jacobi determinant, see Section V.A.
In case of thermostating by deterministic scattering adapted to the Lorentz
gas geometry this quantity may be denoted by $|dv'_{\parallel}
dv'_{\perp}|/|dv_{\parallel} dv_{\perp}|$. Here the numerator is composed of
the two velocity components after a collision and in the denominator there are
the corresponding velocity components before a collision with the disk. If a
particle collides $n\in\mbox{\mb N}$ times the product of Jacobian
determinants evolves in time according to
\bna
\prod_{i=1}^n \frac{|dv'_{i,\parallel} dv'_{i,\perp}|}{|dv_{i,\parallel} dv_{i,\perp}|} 
&\equiv& \exp\left(\sum_{i=1}^n\ln \frac{|dv'_{i,\parallel}
dv'_{i,\perp}|}{|dv_{i,\parallel} dv_{i,\perp}|}\right) \nonumber \\  
&=& \exp\left(n<\ln \frac{|dv'_{\parallel} dv'_{\perp}|}{|dv_{\parallel}
dv_{\perp}|}>\right) \quad .
\ena
For the last step it was assumed that the system is ergodic, which enables to
replace the time average by an ensemble average over colliding particles. In
this expression
\be
\kappa_b:=<\ln \frac{|dv'_{\parallel} dv'_{\perp}|}{|dv_{\parallel}
dv_{\perp}|}> \label{eq:pscmap} 
\ee
is denoted as the average exponential rate of phase space contraction per unit
time, where the index $b$ indicates the definition at a boundary. This
quantity is obviously zero if the map is volume preserving and deviates from
zero otherwise. Hence, for time-discrete dynamics such as the collision map
Eq.\ (\ref{eq:tds}) under consideration $\kappa_b$ replaces the average phase
space contraction rate $\kappa$ of time-continuous dynamics defined by Eq.\
(\ref{eq:psc}).\footnote{More explicitly, $\kappa_b$ naturally shows up in the
formal solution for the probability distribution function of the time-discrete
Liouville (Frobenius-Perron) equation in dissipative maps. This is in full
analogy to solving the time-continuous Liouville equation Eq.\
(\ref{eq:gliouv}), where the average of the divergence plays the same role.}

For the periodic Lorentz gas thermostated by deterministic scattering
$\kappa_b$ can be calculated from the collision rules Eqs.\ (\ref{eq:baker}),
(\ref{eq:sbctr}) and (\ref{eq:tds}) to \cite{RaKl02}
\be
\kappa_b=\frac{<v'^2>-<v^2>}{2T} \quad ,
\label{eq:pcontds}
\ee
where $v'$ and $v$ are the absolute values of the velocities after and before
a collision, respectively. $\kappa_b$ is thus identical to minus the average
outward flux of kinetic energy to the reservoir $dQ:=(<v^2>-<v'^2>)/2$ divided
by the temperature $T$, where $T$ derives from Eq.\ (\ref{eq:sbctr}). This
result holds in equilibrium as well as in nonequilibrium. As in case of
Gaussian and Nos\'e-Hoover thermostats, Eq.\ (\ref{eq:pcontds}) may now be
compared to the entropy production $dS=dQ/T_r$ in terms of the heat transfer
between subsystem and thermal reservoir, cp.\ to Section III.B.1, where $T_r$
denotes the temperature of the thermal reservoir, leading to
\be
\kappa_b=-\frac{T_r}{T} dS \quad . \label{eq:idtds}
\ee
In this case, the question concerning an identity between $dS$ and $\kappa_b$
boils down to the problem whether the temperature $T$ coming from Eq.\
(\ref{eq:sbctr}) can always be identified with the actual temperature of the
thermal reservoir $T_r$. In equilibrium, that is, without applying an external
field ${\bf \varepsilon}$, it is $T=T_r$ for the periodic Lorentz gas, which
follows from the definition of this thermostating scheme outlined in Section
VII.A to C. However, in this situation Eq.\ (\ref{eq:idtds}) is trivially zero
on both sides anyway.

On the other hand, to define a proper temperature in a nonequilibrium
situation is very problematic, see also our brief discussion in Section
II.A. For our model one can argue at least along two different lines in order
to define a proper reservoir temperature: One reasoning is that the
temperature of a thermal reservoir consisting of an infinite number of degrees
of freedom should never change under whatsoever nonequilibrium conditions and,
hence, should always be the same as in thermal equilibrium \cite{Hoo02}. This
implies $T_r\equiv T$ and according to Eq.\ (\ref{eq:idtds}) the identity
between phase space contraction and entropy production, as discussed for ideal
Gaussian and Nos\'e-Hoover thermostats, holds as well.

However, for all thermostats analyzed in this review it is well-known that
generally the nonequilibrium temperature of the subsystem, if defined via the
kinetic energy of the moving particle on the basis of equipartitioning of
energy, is not necessarily equal to the equilibrium reservoir temperature $T$
that is contained in the equations of motion
\cite{PoHo88,PoHo89,EvMo90,SEI98,RKH00}. In case of boundary thermostats this
implies that identifying $T_r$ with $T$ typically yields temperature jumps at
the boundaries \cite{KLA02}. Such discontinuities in the temperature profile
are at variance with a local equipartitioning of energy between the reservoir
degrees of freedom and the degrees of freedom of the colliding particle.

In contrast to the first definition outlined above, this motivates to simply
{\em assume} such an existence of a local equipartitioning of energy in a
NSS. Under this assumption the reservoir temperature can be defined by
computing the average kinetic energy of a particle at the moment of the
collision \cite{WKN99,KRN00,RKN00,RaKl02}. This definition of a nonequilibrium
temperature thus amounts in smoothly extrapolating the temperature profile
from the bulk of the subsystem to the boundary of the reservoir. An
application of this method to the boundary thermostated driven periodic
Lorentz gas yields $T_r>T$ for an electric field with strength $\varepsilon>0$
\cite{RaKl02}. Consequently, by following a temperature definition that is
based on a no temperature-slip assumption at the boundary the identity is not
recovered. In our view, both definitions of a reservoir temperature can be
defended. Therefore, the issue about the existence or non-existence of the
identity may be considered as undecided in this case.

As far as the Lyapunov sum rule introduced in Section III.B.2 is concerned,
the situation is simpler. Replacing $\kappa_b$ on the left hand side of Eq.\
(\ref{eq:idtds}) by the sum of Lyapunov exponents we may employ
$dS=\varepsilon <v_x>/T_r$ on the right hand side, which is the Joule heat
that the particle gains during a free flight. Using furthermore
$\sigma(\varepsilon)=<v>/\varepsilon$ we obtain
\be
\sigma(\varepsilon)=-\frac{T}{\varepsilon^2}\sum_{i=1}^4\lambda(\varepsilon)
\quad , \label{eq:lsrtds} 
\ee
which, formally, is precisely the Lyapunov sum rule derived for the ideal
Gaussian and Nos\'e-Hoover thermostats, cp.\ to Eqs.\ (\ref{eq:lsr}),
(\ref{eq:lsrnh}). The only ambiguity is, again, whether $T$ shall be
identified with the temperature of the thermal reservoir or whether it shall
be considered as a parameter in nonequilibrium by which the reservoir
temperature $T_r$ can be tuned accordingly, as discussed above. In any case,
whether or not there holds an identity between phase space contraction and
entropy production, here there exists a simple relation between the
conductivity and the Lyapunov exponents of the system.

\begin{figure}[t]
\epsfxsize=16cm
\epsfysize=7.5cm
\centerline{\epsfbox{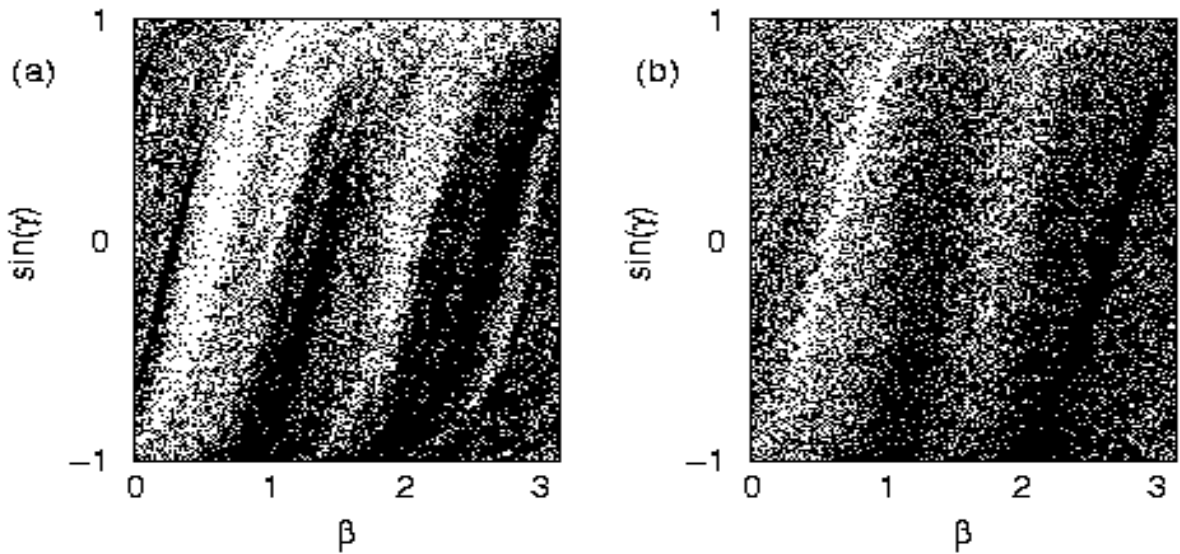}}
\caption{Attractor for the periodic Lorentz gas driven by an electric field of strength
$\varepsilon=1$ which is parallel to the $x$-axis. The reservoir equilibrium
temperature is $T=0.5$, $\beta$ and $\sin\gamma$ are defined in Fig.\
\ref{plko}. In (a) the system is thermostated by deterministic scattering, see
Eqs.\ (\ref{eq:baker}), (\ref{eq:sbctr}) and (\ref{eq:tds}), in (b) it is
thermostated by stochastic boundaries, see Eq.\ (\ref{eq:sbctr}) and
explanations. The results are from Refs.\ \protect\cite{KRN00,RKN00}.}
\label{fig:attrtds}
\end{figure}

\subsubsection{Attractors, bifurcation diagram and electrical conductivity}

Fig.\ \ref{fig:attrtds} shows two attractors for the driven periodic Lorentz
gas thermostated at the boundaries, where all results presented for this
thermostat are at $T=0.5$. Part (a) displays simulation results for the
deterministic boundaries Eqs.\ (\ref{eq:baker}), (\ref{eq:sbctr}) and
(\ref{eq:tds}) adapted to the Lorentz gas geometry. Part (b) shows analogous
results for stochastic boundaries, where the baker map Eq.\ (\ref{eq:baker})
has been replaced by a random number generator. Both figures may be compared
to the previous attractors Fig.\
\ref{gatt} for the Gaussian thermostated Lorentz gas and Fig.\ \ref{attrnh}
for Nos\'e-Hoover.

Clearly, the deterministic boundary solution exhibits a fractal-like folding,
in analogy to the ones displayed for the Gaussian and for the Nos\'e-Hoover
thermostat. In Fig.\ \ref{fig:attrtds} (a) this structure just appears to be a
bit more smoothed-out than for the Nos\'e-Hoover case. On the other hand,
apart from a few non-uniformities on a coarse scale no fine structure is
visible anymore for the attractor obtained from stochastic boundaries
suggesting that Fig.\ \ref{fig:attrtds} (b) does not represent a fractal set.

For deterministic boundaries the dimensionality of the attractor in the
four-dimensional phase space was computed numerically by means of the
field-dependent Kaplan-Yorke dimension $D_{KY}(\varepsilon)$ that was briefly
mentioned in Section V.C \cite{RaKl02}. The results show that the
dimensionality monotonously decreases from $D_{KY}(0)=4$ to $D_{KY}(1)\simeq
3.7$ quantitatively confirming the fractality of the attractor in Fig.\
\ref{fig:attrtds} (a).  For Fig.\ \ref{fig:attrtds} (b) respective
computations are more subtle\footnote{The Kaplan-Yorke dimension requires the
computation of the full spectrum of Lyapunov exponents. Therefore, in case of
stochastic boundaries one has to compute Lyapunov exponents for a
stochastically perturbed dynamics; see also our discussion in Section V.D for
respective difficulties.} and remain to be done. We emphasize that such
computations would be very interesting, since according to our discussion in
Section V.D it is not yet clear whether or not stochastically thermostated
systems typically exhibit fractal attractors. 

\begin{figure}[t]
\epsfxsize=15.5cm
\epsfysize=7cm
\centerline{\epsfbox{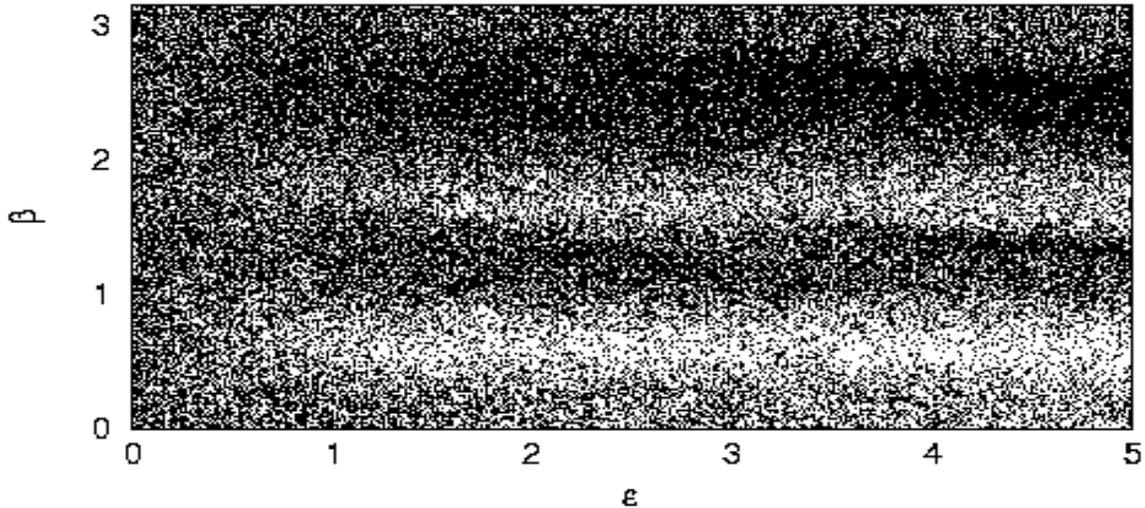}}
\caption{Bifurcation diagram for the periodic Lorentz gas 
thermostated by deterministic scattering and driven by an electric field of
strength $\varepsilon$ which is parallel to the $x$-axis. The reservoir
equilibrium temperature is $T=0.5$, $\beta$ is defined in Fig.\
\ref{plko}. The figure is from Refs.\ \protect\cite{KRN00,RKN00}.}
\label{fig:bifutds}
\end{figure}

The bifurcation diagram for the driven periodic Lorentz gas thermostated by
deterministic boundaries is depicted in Fig.\ \ref{fig:bifutds}. Again, it
should be compared to the previous counterparts Fig.\ \ref{gbifu} for the
ideal Gaussian thermostat, Fig.\ \ref{bifunh} for the ideal Nos\'e-Hoover and
Fig.\ \ref{bifunhe} for the non-ideal Nos\'e-Hoover thermostat. Remarkably, in
case of deterministic scattering there is no bifurcation scenario at all. That
is, Fig.\ \ref{fig:bifutds} neither shows a contraction of the attractor onto
periodic orbits nor a breakdown of ergodicity for higher field strengths as it
occurs for Gauss and Nos\'e-Hoover. Similar results have been obtained for
other choices of projections in phase space as discussed in Ref.\
\cite{RKN00}. These results indicate that deterministic boundaries more
strongly regularize the dynamics of the subsystem than Gauss and
Nos\'e-Hoover.  Furthermore, the different bifurcation diagrams clearly show
that for simple systems such as the periodic Lorentz gas the deformations of
the attractor under variation of the field strength intimately depend on the
specific type of thermostating.

\begin{figure}[t]
\epsfxsize=10cm
\centerline{\epsfbox{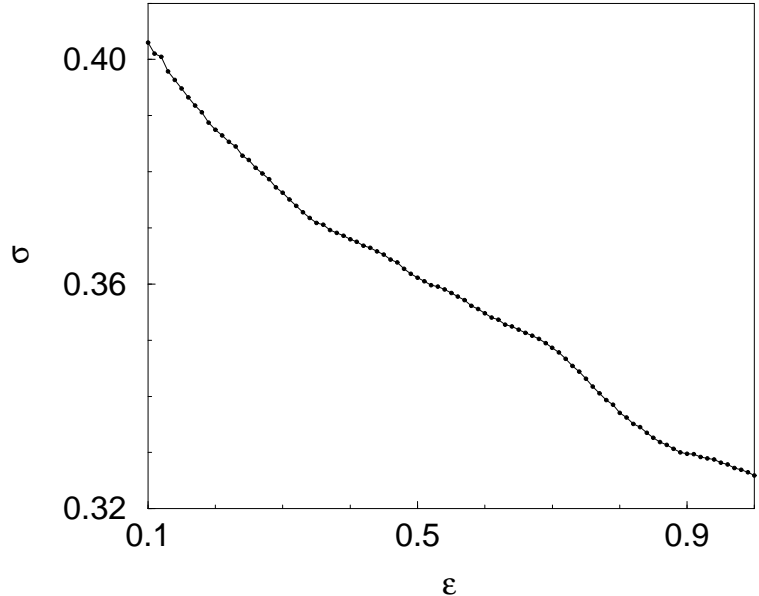}}
\caption{Electrical conductivity $\sigma(\varepsilon)$ for the periodic Lorentz gas 
thermostated by deterministic scattering and driven by an electric field of
strength $\varepsilon$ which is parallel to the $x$-axis. The reservoir
equilibrium temperature is $T=0.5$. Each point has a numerical uncertainty that
is less than the symbol size. The results are from Refs.\ \protect\cite{KRN00,RKN00}.}
\label{fig:condtds}
\end{figure}

Fig.\ \ref{fig:condtds} finally contains the field-dependent electrical
conductivity for the driven periodic Lorentz gas deterministically
thermostated at the boundaries. In comparison to the conductivities
corresponding to the ideal Gaussian, Nos\'e-Hoover and to the non-ideal
Nos\'e-Hoover thermostats shown in Figs.\ \ref{gcond}, \ref{condnh} and
\ref{prepega} (a), respectively, this curve looks rather smooth. As in the
other cases, there is no indication of a regime of linear response in the
numerically accessible regime of field strengths above $\varepsilon\ge
0.1$. On the other hand, again some wiggles on fine scales are visible
suggesting that even for a bifurcation diagram such as Fig.\
\ref{fig:bifutds} the corresponding conductivity might be, to some extent, an irregular
function on fine scales. Note that the diffusion coefficient of the field-free
periodic Lorentz gas thermostated by deterministic scattering may be different
from the one for fully elastic scatterers and has not yet been
computed. Hence, we do not know the prospective limiting value of
$\sigma(\varepsilon)$ for $\varepsilon\to0$ according to the Einstein formula
Eq.\ (\ref{eq:dts}).

\subsubsection{Lyapunov exponents}

By applying our deterministic boundary thermostat the kinetic energy is not
kept constant at any instant of time, in analogy to Nos\'e-Hoover, but in
contrast to the Gaussian thermostat.  Consequently, this system is
characterized by four Lyapunov exponents in the two-dimensional periodic
Lorentz gas thermostated by deterministic scattering. Numerical computations
of this spectrum of Lyapunov exponents turn out to be rather delicate due to
the inelasticity at the collisions and require some more intricate adjustments
compared to standard methods \cite{RaKl02}.

\begin{figure}[t]
\epsfxsize=12cm
\centerline{\rotate[r]{\epsfbox{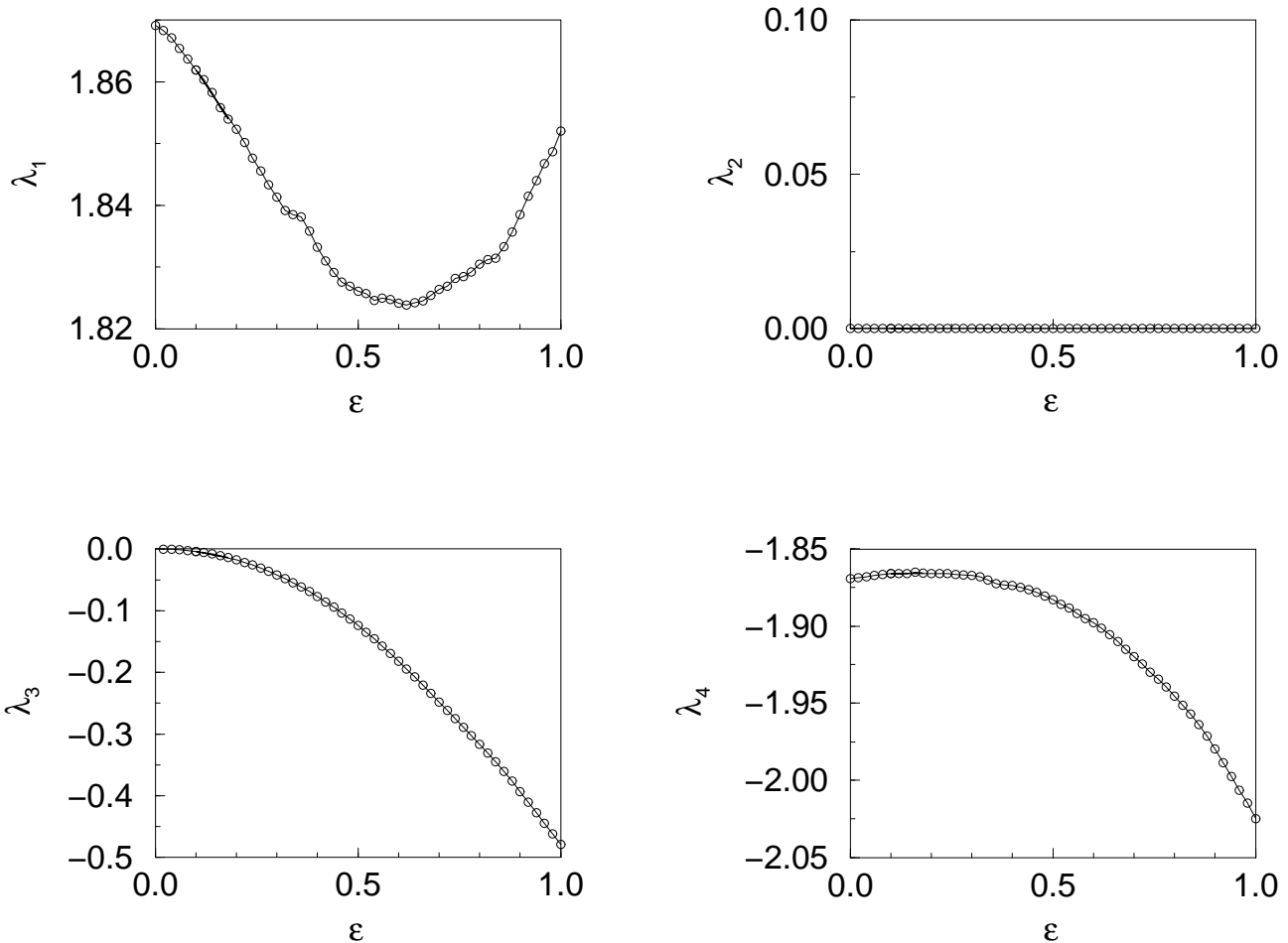}}}
  
\vspace*{0.8cm}
\caption{The four Lyapunov exponents $\lambda_i(\varepsilon)$ for the periodic
Lorentz gas thermostated by deterministic scattering and driven by an electric
field of strength $\varepsilon$ which is parallel to the $x$-axis. The
reservoir equilibrium temperature is $T=0.5$. The numerical uncertainty of
each point is less than $5\cdot10^{-4}$. The results are from Ref.\
\protect\cite{RaKl02}.}
\label{fig:lyaptds}
\end{figure}

For zero field numerical results corroborate that conjugate pairing holds with
two zero, one negative and one positive Lyapunov exponent. However, in
nonequilibrium there is no pairing of Lyapunov exponents anymore as shown in
Fig.\ \ref{fig:lyaptds}. This is in sharp contrast to the Lyapunov spectrum
associated with bulk thermostats such as Gauss and Nos\'e-Hoover but appears
to be typical for boundary thermostats, see our discussion in Section
III.B.2. More precisely, one Lyapunov exponent is zero reflecting phase space
conservation parallel to the flow. Two become negative and decrease
monotonously in the field strength similar to power laws. Their functional
forms remind of the non-zero exponents associated with the Gaussian
thermostated driven Lorentz gas \cite{MDI96}. Indeed, as we have shown, for
both thermostats there holds the Lyapunov sum rule, see Eqs.\ (\ref{eq:lsr}),
(\ref{eq:lsrtds}). This equation, in turn, requires that the Lyapunov
exponents are quadratic at least for small enough field strength.

The single positive Lyapunov exponent, on the other hand, displays a more
intricate non-monotonous behavior under variation of the field
strength. According to Pesin's theorem, see Section V.C, it must be equal to
the Kolmogorov-Sinai entropy suggesting the following heuristic understanding
of its functional form \cite{RaKl02}: For small field strength the particle
attempts to move parallel to the field, hence the Kolmogorov-Sinai entropy
decreases. However, by increasing the field strength the system starts to heat
up in the bulk, consequently the Kolmogorov-Sinai entropy increases
again. Such a behavior was not observed for the Gaussian thermostated Lorentz
gas, where the Kolmogorov-Sinai entropy always decreases monotonically
\cite{DeGP95,DMR95}, which is probably due to constraining the energy in the bulk.

\subsection{Hard disk fluid under shear and heat flow}

Up to now we have applied different schemes of thermostating to the driven
periodic Lorentz gas only, which is a one-particle system that derives its
statistical properties from the chaotic scattering of a moving point particle
with a fixed hard disk. We deliberately chose a very simple model in order to
focus on the impact that different thermostats have on this system. From this
point of view, an important question is to which extent our findings
concerning relations between chaos and transport still hold for many-particle
systems that include interactions between different moving particles. Suitably
amended periodic Lorentz gases of this type were already investigated in
Refs.\ \cite{BarEC,BGG97,BDL+02}. Here we skip this intermediate step and
immediately consider interacting many-particle systems for which hard disk
fluids under nonequilibrium conditions are typical examples.

In a way, by looking at these systems we are back to the roots of
thermostating: As explained in Chapters I to III, deterministic thermostats
were originally designed as efficient tools within the framework of molecular
dynamics computer simulations for driving large, interacting many-particle
systems into well-defined NSS that are consistent with the laws of
thermodynamics. That deterministically thermostated systems furthermore
exhibit interesting chaos properties was only realized later on and
subsequently motivated more specific studies of simple models such as the
Nos\'e-Hoover oscillator and the thermostated driven Lorentz gas.

Here we move the other way around from simple to more complex
systems. Particularly, we want to check whether for interacting many particles
there is an identity between phase space contraction and entropy production,
whether there are simple equations relating transport coefficients to Lyapunov
exponents, and whether NSS are still characterized by the existence of fractal
attractors.

As an example, we discuss a hard disk fluid under shear and heat flow. We
start by briefly sketching various known methods for modeling a shear flow. A
well-known example are the {\em SLLOD} equations of motion representing the
application of the conventional Gaussian thermostat for obtaining a sheared
NSS. We then outline the construction of {\em Maxwell daemon boundaries} for
modeling a hard disk fluid under shear. Finally, we discuss in more detail
results for heat and shear flows generated according to thermostating by
deterministic scattering, where we draw upon Refs.\
\cite{WKN99,Wag00}. Interesting questions are whether these systems yield an
identity between phase space contraction and thermodynamic entropy production
and, correspondingly, whether there are Lyapunov sum rules for the respective
transport coefficients. We furthermore summarize numerical results for the
Lyapunov spectrum and for the Kolmogorov-Sinai entropy of this system.

\subsubsection{Homogeneous and inhomogeneous modelings of shear and heat flows}

Modeling and understanding shear and heat flows features as a prominent
problem in nonequilibrium statistical mechanics. This is reminiscent by the
existence of a huge literature on this subject. It is certainly not our goal
to give a full account of all these works, in particular since many important
aspects are nicely summarized in a number of reviews
\cite{AT87,EvMo90,Hess96,HKL96,HoB99,Mund00}. However, by mainly focusing on
shearing we briefly describe at least some important approaches that have been
pursued and to give a short account of investigations along the lines of
connecting chaos and transport.

Probably the most naive way to shear a many-particle system is by using walls
that move with certain velocities \cite{Hess96,Mund00}. Early attempts
along these lines trace back to work by Hoover and Ashurst, see, e.g., Ref.\
\cite{HoAs75} and also Refs.\ \cite{LBC92,SGB92} for later investigations. Posch and
Hoover used such {\em sliding boundaries} in combination with a Nos\'e-Hoover
thermostat restricted to these boundaries \cite{PoHo89}.

In contrast to this inhomogeneous driving from the boundaries, {\em
Lees-Edwards} or {\em sliding brick} boundary conditions act homogeneously in
the bulk of the system \cite{AT87,EvMo90,Hess96,HKL96,HoB99,Mund00,Coh95}. The
idea is to suitably displace copies of a unit cell of particles in space such
that a linear shear profile is imposed onto the system. Lees-Edwards
boundaries alone already yield a planar Couette flow for not too high shear
rates. However, it turned out that this method could be improved by combining
it with a Gaussian thermostat. In fact, the need to conveniently simulate
sheared fluids and solids originally motivated the invention of the Gaussian
scheme, cp.\ also to Section III.A \cite{HLM82,Ev83}. Further refinements of
the approach consisting of homogeneous shearing and Gaussian bulk
thermostating eventually led to the formulation of the {\em
SLLOD}\footnote{Concerning the origin of this name, which is not an acronym,
see, e.g., Ref.\ \cite{Mund00} and further references therein.} equations of
motion
\cite{EvMo90,ECM,SEM92,Coh95,Hess96,MECB89,SEC98,SEI98,Mund00}.

Curiously, by applying these shear flow algorithms to fluids consisting of
hard spheres shear-induced microstructures were emerging in the simulations
for high enough shear rates. That is, the particles were aligning themselves
parallel to the direction of the flow in form of strings, or blocks
(``plugs''). These orderings became known as the {\em string phase},
\cite{EvMo86,EvMo90,Mund00}, respectively the {\em plug phase}
\cite{LoHe90,HeLo90,HKL96,HABK97}. Partly it was argued that these phases
represent artifacts of the homogeneous thermostating algorithms described
above, since by construction these methods are adapted to the shear flow
profile in a self-contained way \cite{EvMo86,EvMo90}. This led to the
invention of more refined {\em profile unbiased} thermostating schemes ({\em
PUT}) for which, partly, such structures were not observed anymore. Further
refinements of PUT thermostats, on the other hand, reproduced these
flow-induced positional orderings again \cite{LoHe90,HeLo90}. To some extent,
this problem reminds of our previous discussion starting from Chapter III,
where we have shown that the application of different thermostats to the
driven periodic Lorentz gas may yield different NSS. Here we appear to have a
similar situation in the nonlinear regime of a sheared and thermostated
many-particle system.

More recent work focused onto the connection between momentum transport and
chaos properties of shear flows. It may not come as a surprise that, again,
the SLLOD equations furnish an identity between phase space contraction and
entropy production \cite{Coh97}. Consequently, there is also a Lyapunov sum
rule as already discussed in Section III.B.2 \cite{ECM}. The most simple
sheared systems allowing more detailed analyses are two-particle systems
\cite{LaHo83} for which the existence of a shear viscosity could be proven
\cite{BuSp}. Morriss et al.\ \cite{Morr87,Morr89,Mo89a,PIM94} extensively
computed the complete spectra of Lyapunov exponents for such systems by
showing that the phase space in these models contracts onto fractal
attractors. Further numerical computations of Lyapunov spectra for sheared
many-particle systems under application of conventional thermostating
algorithms have been performed in Refs.\
\cite{PoHo89,ECM,SEI98,SEM92,ECS+00}. Note that the existence of a conjugate
pairing rule even in homogeneously thermostated shear flows appears to be very
delicate topic, as we have already outlined in Section III.B.2. We finally
mention that even a multibaker map modeling a shear flow has been constructed
and analyzed in Refs.\ \cite{TVM01,MTV01}.

Alternatively, Chernov and Lebowitz \cite{ChLe95,ChLe97} proposed a
many-particle shear flow model exhibiting a well-defined NSS without using a
conventional thermostat. They considered a two-dimensional system of hard
disks confined in a square box of length $L$ with periodic boundary conditions
along the $x$-axis, i.e., the left and right sides at $x=\pm L/2$ are
identified. At the top and bottom sides of the box, $y=\pm L/2$, they
introduced rigid walls where the disks are reflected according to certain
scattering rules. The disks interact among themselves by impulsive hard
collisions so that the bulk dynamics is purely conservative.

Shear is now introduced by imposing collision rules at the boundaries which
turn the angle $\theta$ that the velocity vector of a particle moving towards
the wall forms with the wall into an outgoing angle $\theta':=c\theta$,
$0<c<1$. These ``Maxwell daemon-like'' boundaries mimic the impact of a
shear force at the particle-wall interaction and enforce that particles move
favorably parallel to the wall. Since energy is conserved by this operation
the total energy of the particles is strictly kept constant thus reminding of
the action of a Gaussian thermostat. Similar constraints have also been
studied by other authors in different settings \cite{ZhZh92,BR01}.

Chernov and Lebowitz used both time-reversible and irreversible formulations
of such scattering rules. In both cases the resulting NSS of the hard
disk-fluid were in full agreement with hydrodynamics. As a consequence of the
specific scattering rules at the boundaries again the total phase space volume
was not conserved. However, in this case the authors argued for an identity
between phase space contraction and entropy production. Subsequent numerical
studies of the dynamical instability of this system performed by other authors
\cite{DePo97} clearly showed that, as usual for boundary thermostats, the
model exhibited no conjugate pairing rule. On the other hand, by using the
full spectrum of Lyapunov exponents again evidence was provided for the
existence of a fractal attractor in nonequilibrium. In further work the
fluctuations of the entropy production of this model were investigated
\cite{BCL98,BL01}.

As far as heat flows are concerned, general approaches to model the
application of temperature gradients are summarized in Refs.\
\cite{HoAs75,HoB99,EvMo90,AT87,SEC98,KLA02,HAHG03}. A mathematical analysis of
Fourier's law in some simple stochastic and Hamiltonian systems was provided
in Refs.\ \cite{LeSp78,EPRB99,EPRB99b,MNV03}. Numerical implementations
considered both stochastic \cite{TCG82,CiTe80} and deterministic boundaries,
see, e.g., Refs.\ \cite{PH98,HPA+02,AK02,KLA02,HAHG03,PH03}. Bulk thermostats
were tested in Refs.\ \cite{HMH84,HoKr86}.  In more recent work even a heat
flow in a periodic Lorentz gas with rotating disks was studied
\cite{MLL01,LLM02}. We remark that, apart from these works, there exists a
vivid discussion about the validity of Fourier's law in more simple models,
partly also elaborating on the role of microscopic chaos, see Ref.\
\cite{LLP03} and further references therein. However, here we will touch the
heat flow case only very briefly by restricting ourselves to a specific
many-particle system.

In the following we present results for a system consisting of a large number
of hard disks that collide elastically with each other. This hard disk fluid
is sheared and thermostated at the boundaries. The shear is modeled by moving
walls, or sliding boundaries, imposing a local shear force onto the
particles. In order to pump energy out of this system we use thermostating by
deterministic scattering at the walls.

\subsubsection{Shear and heat flows thermostated by deterministic scattering}

We consider $N$ hard disks that are confined between two flat walls as
described above for the Maxwell daemon model. However, here we change the
scattering rules at both walls according to Eqs.\ (\ref{eq:baker}),
(\ref{eq:sbctr}) and (\ref{eq:tds}), cp.\ to Fig.\ \ref{fig:sbc} and the
respective discussion. Associating different temperatures $T^u$ and $T^d$ to
the upper and the lower wall, respectively, enables us to study the case of a
heat flow from a hot to a cold reservoir \cite{WKN99}.

For the simulations the length of the box was chosen to be $L=28$ implying
that the volume fraction occupied by $N=100$ hard disks of radius $r=1/2$ is
$\rho=0.1$. The temperature gradient was set according to $T^d=1$ with
$T^u=1.5$ or $T^u=2$. In both cases the temperature profile was found to be
approximately linear, apart from boundary effects, and the kinetic energy of
the particles was equipartitioned between the two degrees of freedom. As
discussed in Section VII.D.1, assigning $T^u$ and $T^d$ as they appear in Eq.\
(\ref{eq:sbctr}) to be the reservoir temperatures yields temperature jumps at
the walls. This can be avoided by redefining the two reservoir temperatures
according to smoothly extrapolating from the bulk to the boundaries under the
assumption of equipartitioning of energy at the walls. In the following we
take the latter point of view. Computing the thermal conductivity from
simulations, the numerical results matched well to the predictions from
Enskog's kinetic theory \cite{RC96,G78}. This agreement improved by driving
the system into the hydrodynamic limit. Hence, thermostating by deterministic
scattering is able to generate a heat flow in a linear response regime that is
in full agreement with hydrodynamics.

In a next step, the identity between phase space contraction and thermodynamic
entropy production was checked for this situation. We do not go into too much
detail at this point, see Ref.\ \cite{WKN99} for an explicit
discussion. However, there was good agreement between both quantities as
obtained from computer simulations, which again improved in the hydrodynamic
limit, so in this case the identity was confirmed. We furthermore note that
the ingoing and outgoing fluxes at the walls were approaching local
thermodynamic equilibrium in the hydrodynamic limit.

Starting from the same setup we now want to model the case of a shear
flow. For this purpose we choose $T^u=T^d\equiv 1$. In addition, we wish that
the upper and lower wall move with constant velocities $d$, respectively $-d$,
opposite to each other. At a collision a hard disk must then experience the
shear force
\be
{\cal S}_d(v_x,v_y)=(v_x + d,v_y) \label{eq:shfo}
\ee
with $d$ being positive (negative) at the upper (lower) wall. The main problem
is to suitably combine $S_d$ with the action of the thermal reservoir at the
wall. In Ref.\ \cite{WKN99} three different options for linking $S_d$ with the
thermostating rules Eqs.\ (\ref{eq:baker}), (\ref{eq:sbctr}) and
(\ref{eq:tds}) were explored. We first discuss
\be
(v_x',v_y')=\\
\qquad\left\{\begin{array}{l@{\:;\:}l}{\cal
S}_d\circ\mbox{\boldmath${\cal T}$}^{-1}\circ{\cal  
M}\circ\mbox{\boldmath${\cal T}$}\circ {\cal S}_d (v_x,v_y)&
v_x\ge\pm d\\ 
{\cal S}_d\circ\mbox{\boldmath${\cal T}$}^{-1}\circ{\cal
M}^{-1}\circ\mbox{\boldmath${\cal T}$}\circ{\cal S}_d (v_x,v_y)&
v_x< \pm d\end{array}\right. \label{eq:tdsshear} \quad .
\ee
Requiring time-reversibility enforces us to apply $S_d$ symmetrically before
and after the thermalization. Performing simulations for $d=0.05$ and $d=0.1$
the analysis proceeded along the same lines as for the heat flow: It was found
that the system exhibited a linear shear profile for the average velocity
parallel to the $x$-direction and a quadratic temperature profile between the
walls, both as expected from hydrodynamics. Again, there were temperature
jumps at the walls if the reservoir temperature $T_r$ in nonequilibrium was
identified with the equilibrium reservoir temperature $T$. A comparison of the
viscosity computed from simulations with Enskog's theory
\cite{G78,ChLe95,ChLe97} yielded good agreement in the hydrodynamic limit,
which is demonstrated in Table \ref{tab:visc}. Thus, the combined mechanism of
boundary thermostating and shear Eq.\ (\ref{eq:tdsshear}) generated a shear
flow exhibiting a linear response regime in agreement with hydrodynamics.

We are now prepared to discuss the relation between phase space contraction
and entropy production for this model. The average phase space contraction
rate $\kappa_b$ at the walls is defined in complete analogy to Eq.\
(\ref{eq:pscmap}),
\be
\kappa_b:=<\ln \frac{|dv_x'dv_y'dx'dy'|}{|dv_xdv_ydxdy|}> \quad .
\ee
Calculating the Jacobian determinant from the collision rules Eq.\
(\ref{eq:tdsshear}) yields
\begin{eqnarray} 
\kappa_b^{u/d}&=&\frac{1}{2T^{u/d}}\left<v_x'^2+v_y'^2-v_x^2-v_y^2-<v_x'>^2+<v_x>^2\right>\nonumber
\\ 
 & & +\frac{1}{2T^{u/d}}\left<<v_x'>^2-<v_x>^2- 2d(v_x'+v_x)\right> \quad , \label{eq:pscshear}
\end{eqnarray}
where the angular brackets denote ensemble averages over moving particles. The
temperatures $T^{u/d}$ are the reservoir equilibrium temperatures at the upper
(lower) walls defined in Eq.\ (\ref{eq:tdsshear}).

Eq.\ (\ref{eq:pscshear}) was already decomposed such that it can be compared
to the thermodynamic entropy production of this system. For this purpose we
consider again the Clausius form of entropy production in terms of the average
heat transfer at the walls reading
\be
dS = -\frac{1}{2T_r^{u/d}}\left<v_x'^2+v_y'^2-v_x^2-v_y^2
-<v_x'>^2+<v_x>^2\right> \quad . \label{eq:epshear}
\ee
\begin{table}
\begin{center}
\begin{tabular}{l|llll}
&N=100&N=200&N=400&N=800 \\ \hline
$d$=0.05 &0.9616&0.9904&1.0081&1.0382\\
$d$=0.1 &0.9702&1.001&1.0226&1.0232\\ 
\end{tabular}
\vspace*{0.3cm}
\caption{Results for the fraction $\eta_{exp}/\eta_{th}$ in a sheared fluid consisting
of $N$ hard disks. The shear viscosity $\eta_{exp}$ was obtained from computer
simulations, whereas $\eta_{th}$ was calculated from Enskog's kinetic
theory. The system is driven by boundaries moving with velocities $\pm d$ and
thermostated by deterministic scattering at these walls, see Eq.\
(\ref{eq:tdsshear}). The values are from Ref.\ \protect\cite{WKN99}.}
\label{tab:visc}
\end{center}
\end{table}

Here $T_r^{u/d}$ stands for the reservoir temperature in nonequilibrium
computed from the assumption that there is no temperature slip at the
walls. Obviously, apart from the minus sign $dS$ is formally identical to the
first bracketed term in Eq.\ (\ref{eq:pscshear}). As in Section VII.D.1, the
issue is just whether one may identify the reservoir equilibrium temperature
$T^{u/d}$ with the reservoir temperature $T_r^{u/d}$ under nonequilibrium
conditions. Irrespective of this problem, the second term in Eq.\
(\ref{eq:pscshear}) yields a contribution to the phase space contraction at
the walls that has nothing to do with the thermodynamic entropy production of
Eq.\ (\ref{eq:epshear}).

We may inquire whether this term has a possible physical interpretation. For
the scattering rules Eq.\ (\ref{eq:tdsshear}) the average over in- and
outgoing velocities parallel to the walls $u_w:=(<v'_x>+<v_x>)/2$ is not
necessarily identical to $d$. By assuming that it is at least some linear
function of $d$ the second line in Eq.\ (\ref{eq:pscshear}) yields some
correction of order $d^2$. This contribution might be interpreted as a phase
space contraction due to a friction parallel to the walls and may be thought
of representing properties of a wall like roughness, or anisotropy of the wall
scatterers.

In this respect our shear flow collision rules are very close again to the
modifications that led to the non-ideal Gaussian and Nos\'e-Hoover thermostats
discussed in Section VI.A and B. Note that for the latter types of thermostats
the identity was destroyed due to what we called a {\em non-ideal coupling}
between subsystem and reservoir. That is, we generalized the way subsystem and
reservoir interact with each other by taking into account additional terms
going beyond standard Gaussian and Nos\'e-Hoover frictional forces. Similarly,
the above collision rules allow to incorporate further specificalities into
the microscopic subsystem-reservoir coupling. We conclude that such
generalizations may lead to phase space contraction rates that are arbitrarily
more complicated than the entropy production expected from thermodynamics.

That indeed the identity between phase space contraction and entropy
production does not hold is quantitatively assessed by Table
\ref{tab:epshear}. Here computer simulation results are presented for the fraction
$-dS/\kappa_b$ of the entropy production $dS$, Eq.\ (\ref{eq:epshear}),
divided by the average phase space contraction rate $\kappa_b$, Eq.\
(\ref{eq:pscshear}). Clearly, the absolute values of both quantities are not
the same, and the situation is getting even worse by approaching the
hydrodynamic limit.  

We remark that the entropy production in the bulk according to hydrodynamics
\cite{ChLe95,ChLe97} was also computed from the simulations. The result was
again compared to the entropy flux across the walls in terms of the Clausius
entropy yielding good agreement by approaching the hydrodynamic limit. In
order to sort out the possibility that the non-identity specifically depends
on the choice of the baker map in the collision rules Eqs.\
(\ref{eq:tdsshear}), the same simulations were carried out by replacing the
baker with the standard map mentioned in Section VII.B, which was tuned such
that its dynamics was approximately hyperbolic. However, instead of recovering
the identity here the mismatch between phase space contraction and entropy
production got even more profound \cite{WKN99}.

\begin{figure}[t]
\epsfxsize=12cm
\centerline{\rotate[r]{\epsfbox{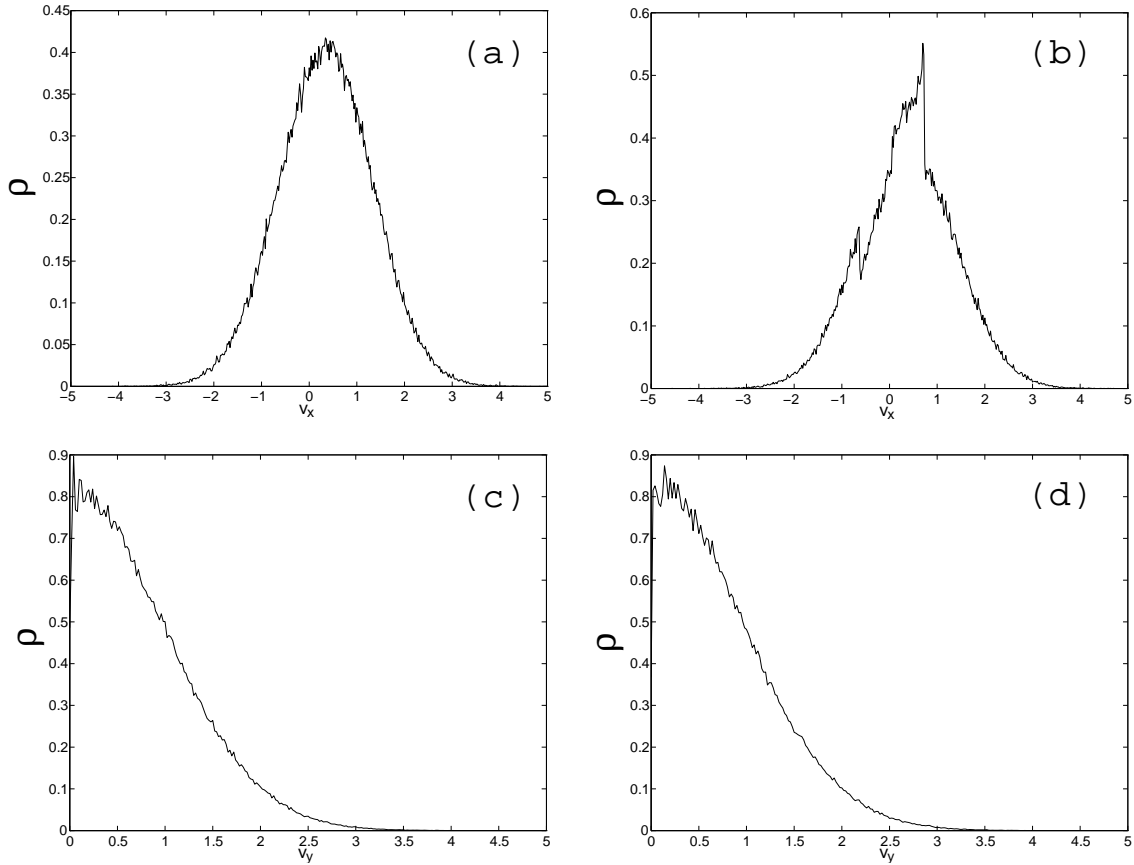}}}
\caption{Velocity distributions $\rho$ at the walls for a sheared fluid consisting
of $N$ hard disks. The system is driven by boundaries moving with velocities
$\pm d$ and thermostated by deterministic scattering at these walls, see Eq.\
(\ref{eq:tdsshear}). Shown are results for the velocity components $v_x$
parallel and $v_y$ perpendicular to the upper wall of the sheared system. (a)
and (c) depict the distributions of velocities before the collisions with the
wall, (b) and (d) the same after the collisions. The results are from Ref.\
\protect\cite{WKN99}.}
\label{fig:rho_shear}
\end{figure}

A more microscopic understanding concerning the origin of this inequality can
be obtained by looking at the velocity distributions of the in- and outgoing
particles at the walls. Fig.\ \ref{fig:rho_shear} shows that the two velocity
components coming in from the bulk are nice Gaussian distributions at the
wall. Indeed, the outgoing velocity distribution parallel to $y$ is
essentially indistiguishable from its ingoing counterpart. However, in sharp
contrast to that the outgoing velocity distribution parallel to $x$ exhibits a
very irregular structure with some discontinuities. The explanation is that,
according to the collision rules Eqs.\ (\ref{eq:tdsshear}), there will always
be more out- than ingoing particles with $v_x\ge d$ at the upper wall (and
with $v_x\le d$ at the lower wall). Thus, due to normalization the two
Gaussian halves in Fig.\ \ref{fig:rho_shear} (b) will never match to a full,
nice Gaussian even in the hydrodynamic limit. Correspondingly, the associated
in- and outgoing fluxes at the walls will never come close to a local
thermodynamic equilibrium.

This motivates to check for the importance of local thermodynamic equilibrium
at the walls concerning an indentity between phase space contraction and
entropy production. For this purpose a second set of time-reversible collision
rules was constructed \cite{WKN99} yielding local thermodynamic equilibrium at
the walls in the hydrodynamic limit. However, the analytical result for the
average phase space contraction rate already provided an equation that, again,
was not identical to the entropy production Eq.\ (\ref{eq:epshear}), not even
in the hydrodynamic limit. Computer simulations confirmed that,
quantitatively, there is no identity either. Hence, these collision rules
establish no identity even in case of local thermodynamic equilibrium at
the walls.
\begin{table}
\begin{center}
\begin{tabular}{l|llll}
&N=100&N=200&N=400&N=800 \\ \hline 
$d$=0.05&0.6761&0.5882&0.5023&0.4230\\ 
$d$=0.1&0.6457&0.5761&0.4934&0.4275\\
\end{tabular}
\vspace*{0.3cm}
\caption{Computer simulation results for the fraction $-dS/\kappa_b$ in a
sheared fluid consisting of $N$ hard disks. The system is driven by boundaries
moving with velocities $\pm d$ and thermostated by deterministic scattering at
these walls, see Eq.\ (\ref{eq:tdsshear}). The entropy production $dS$ was
numerically computed from Eq.\ (\ref{eq:epshear}), the average phase space
contraction rate $\kappa_b$ from Eq.\ (\ref{eq:pscshear}). The values are from
Ref.\ \protect\cite{WKN99}.}
\label{tab:epshear}
\end{center}
\end{table}

Eventually, a third set of collision rules was considered, again generating
local thermodynamic equilibrium at the walls, however, by breaking
time-reversal symmetry \cite{WKN99}. In this case both the analytical
expressions and the computer simulation results showed an identity between
phase space contraction and entropy production in the hydrodynamic
limit. Consequently, time-reversibility cannot be a necessary condition for an
identity between phase space contraction and entropy production. This was
already stated in Refs.\ \cite{ChLe95,ChLe97,NiDa98,DaNi99} but seems to be at
variance with conclusions drawn from the analysis of simple multibaker maps
\cite{VTB97,BTV98,VTB98,TVS00,Voll02}; see also Ref.\ \cite{MaNe03} for a
discussion of the relation between time-reversibility and entropy production
on the basis of Gibbs states.

Typically, the existence or non-existence of the identity will have direct
consequences for the validity of the Lyapunov sum rule as it was discussed in
Sections III.B.2, IV.C.1, VI.A and VI.B.  An exception to the rule was the
driven periodic Lorentz gas thermostated by deterministic scattering where the
identity was ambiguous, however, where in any case formally a Lyapunov sum
rule was recovered, see Section VII.D.1. In other words, the identity is a
sufficient condition for the validity of the Lyapunov sum rule but not
necessary.

For the heat flow case outlined above the identity holds, hence there is a
Lyapunov sum rule. On the other hand, a numerical computation of the Lyapunov
spectrum shows \cite{Wag00} that there is no conjugate pairing of Lyapunov
exponents, as expected for a system that is thermostated at the boundaries. A
computation of the Kaplan-Yorke dimension corroborates the existence of a
fractal attractor. It might be interesting to construct collision rules that
do not yield an identity for a heat flow, which should be possible along the
lines as performed for the shear flow.

As far as the latter is concerned, for sake of comparison we first state the
Lyapunov sum rule for the SLLOD equations. Here the identity holds and the
shear viscosity $\eta$ can be computed to \cite{ECM,Coh95}
\be
\eta(\gamma)=-\frac{T}{L^2\gamma^2}\sum_i\lambda_i(\gamma) \quad ,
\ee
where $T$ is the temperature of the thermal reservoir, respectively the
temperature pre-determined in the Gaussian thermostat, $\gamma:=dv_x(y)/dy$
stands for the shear rate and $L$ is again the system size.

For thermostating by deterministic scattering this relation may look rather
different, depending on whether the identity holds. Let us consider the model
for a boundary-driven shear flow Eqs.\ (\ref{eq:tdsshear}) explicitly
discussed above for which there is no identity.  By using the hydrodynamic
relation $dS=L^2\gamma^2\eta/T_r$ and Eqs.\ (\ref{eq:pscshear}),
(\ref{eq:epshear}) one can nevertheless relate the viscosity of this model to
the Lyapunov exponents arriving at
\be
\eta(\gamma)=-\frac{T}{L^2\gamma^2}
\left(\sum_i\lambda_i(\gamma)+O(d^2)\right) \quad . 
\ee
Here we have abbreviated the second line in Eq.\ (\ref{eq:pscshear}) by the
expression $O(d^2)$. Note that, in contrast to the non-ideal Gaussian and
Nos\'e-Hoover thermostats of Sections VI.A and B, in this case we could at
least establish an explicit relation between viscosity and Lyapunov
exponents. Still, the viscosity is no simple function of Lyapunov exponents
anymore thus providing a counterexample against a universality of the Lyapunov
sum rule in thermostated interacting many-particle systems.

\begin{figure}[t]
\epsfxsize=12cm
\centerline{\epsfbox{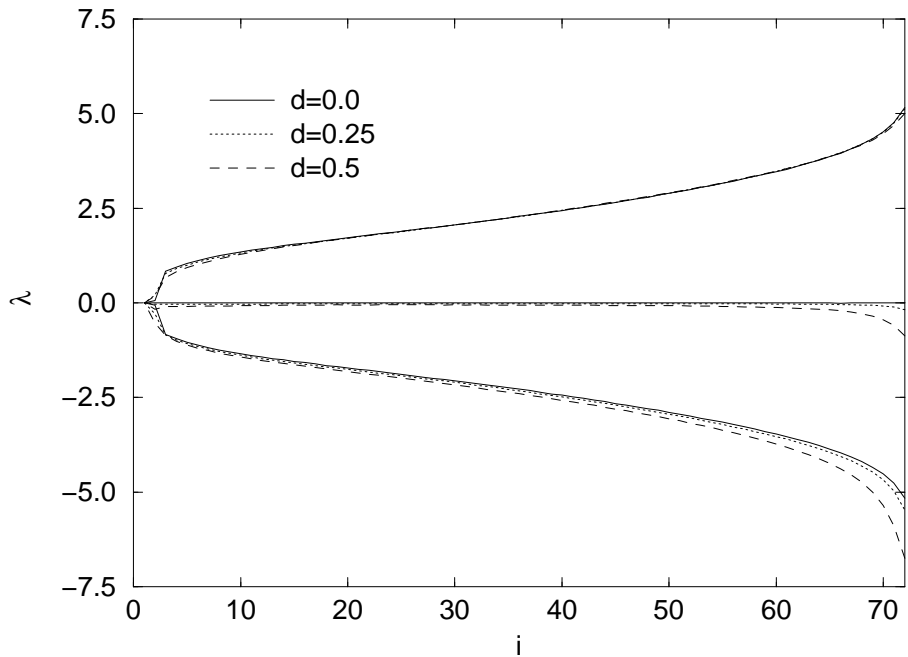}}
\caption{Spectra of Lyapunov exponents $\lambda$ for a sheared fluid
consisting of $36$ hard disks at a volume fraction of $\rho\simeq 0.47$. The
system is driven by boundaries moving with velocities $\pm d$ and thermostated
by deterministic scattering at these walls, see Eq.\
(\ref{eq:tdsshear}). Shown are results for three different driving forces as
indicated in the figure. The lines in the middle represent the sums of
conjugated pairs of Lyapunov exponents. The figure is from Ref.\
\protect\cite{Wag00}.}
\label{fig:lyap_shear}
\end{figure}

Fig.\ \ref{fig:lyap_shear} shows Lyapunov spectra for the shear flow generated
from Eq.\ (\ref{eq:tdsshear}) in case of a system with $36$ hard disks at a
volume fraction of $\rho\simeq 0.47$, where the particles are in a quadratic
box of length $L\simeq 7.746$. Depicted are results from simulations at three
different shear forces $d$ \cite{Wag00}. The overall shape of these spectra is
quite typical for sheared many-particle fluids
\cite{Morr89,Mo89a,PoHo89,ECM,SEM92,DePo97b,SEI98}. However, note the asymmetry
particularly between the largest positive and negative exponents. The sum of
pairs of respectively ordered exponents is also included in the figure showing
that, as in case of the heat flow, there is no conjugate pairing rule for the
Lyapunov exponents. 

Finally, we present results for the Kolmogorov-Sinai entropy of this system,
which, according to Pesin's theorem, is identical to the sum of positive
Lyapunov exponents. This quantity is plotted as a function of the shear rate
$\gamma$ for the same model parameters as in Fig.\ \ref{fig:lyap_shear}
above. The figure may be compared to the respective result for the driven
periodic Lorentz gas thermostated by deterministic scattering, see $\lambda_1$
in Fig.\ \ref{fig:lyaptds}. There is quite an analogy in that $h_{KS}$
exhibits a global minimum by increasing the driving force. Curiously, for the
non time-reversible collision rules of the third model discussed above there
is no such local minimum. Furthermore, the respective results for the heat
flow of both models also do not exhibit a local minimum, see Ref.\
\cite{Wag00} for further details.

Computations of the Kaplan-Yorke dimension for the first and the third shear
flow model yielded again evidence for the existence of fractal attractors. For
a total of $144$ phase space variables the associated loss of phase space
dimensionality went up to about $6$ at shear rates around $\gamma=0.5$.
Assessing the magnitude of the dimensionality loss became an active recent
topic particularly by applying Gaussian and Nos\'e-Hoover thermostats
restricted to boundary layers, after it was realized that the dimensionality
loss may significantly exceed the number of constrained variables
\cite{HPA+02,AK02,KLA02,PH03}.

\begin{figure}[t]
\epsfxsize=8cm
\centerline{\rotate[r]{\epsfbox{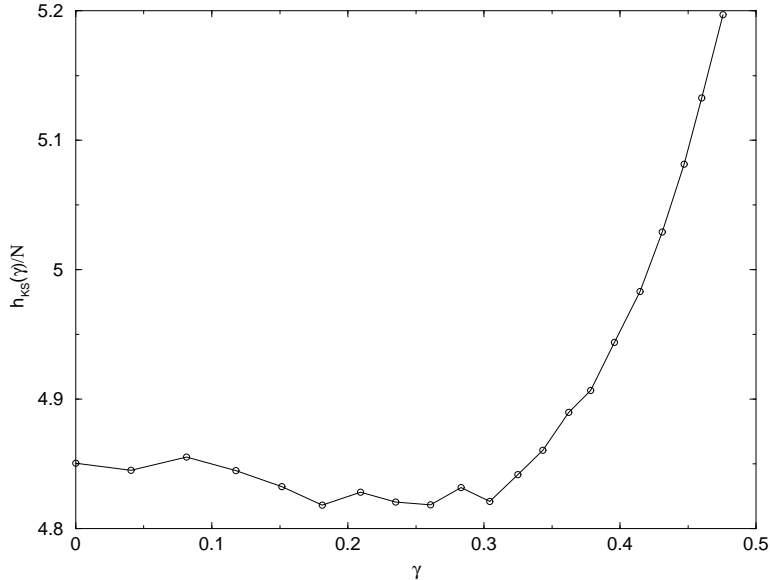}}}
  
\vspace*{0.3cm}
\caption{The Kolmogorov-Sinai entropy $h_{KS}$ per particle as a function of
the shear rate $\gamma$ for a fluid consisting of $N=36$ hard disks at a
volume fraction of $\rho\simeq 0.47$. The system is driven by boundaries
moving with velocities $\pm d$ and thermostated by deterministic scattering at
these walls, see Eq.\ (\ref{eq:tdsshear}).  The figure is from Ref.\
\protect\cite{Wag00}.}
\label{fig:hks_shear}
\end{figure}

\subsection{Summary}

\begin{enumerate}

\item  Modeling a stochastic thermal reservoir at the boundaries of a system is
well-known in form of {\em stochastic boundary conditions}. The precise
collision rules were briefly reviewed at the beginning of this chapter.

\item These stochastic collision rules can easily be made deterministic and
time-reversible by using a deterministic map. The map should exhibit certain
properties in order to enable a proper thermalization. The resulting
deterministic and time-reversible boundary thermostat has been called {\em
thermostating by deterministic scattering}.

\item Both types of boundary thermostats can be derived from first
principles in three steps: Firstly, the precise functional forms of the
equilibrium velocity distribution functions for arbitrarily many reservoir
degrees of freedom are required, as they have been calculated earlier
according to elementary equilibrium statistical mechanics. Secondly, one must
define some coupling between subsystem and thermal reservoir. One then has to
check whether the chosen coupling reproduces the correct velocity distribution
functions. For arbitrary couplings this will not hold, typically. In this case
one should enforce that the correct target velocity distributions are attained
by suitably modifying the collision rules. For a nonequilibrium situation one
needs to consider the limit of infinitely many reservoir degrees of freedom.
Following this construction step by step sheds light again onto the origin of
phase space contraction in thermostated systems, along similar lines as
already elucidated previously.

\item As an example for the application of a deterministic boundary
thermostat again the driven periodic Lorentz gas has been analyzed. Here the
identity between phase space contraction and entropy production is ambiguous
and depends on the definition of the temperature of the thermal reservoir. In
any case, there is a Lyapunov sum rule for the electrical conductivity.

The attractor for the driven periodic Lorentz gas deterministically
thermostated at the boundaries exhibits a fractal structure that is quite
analogous to the previous ones resulting from ideal and non-ideal Gaussian and
Nos\'e-Hoover thermostats. For stochastic boundaries this structure is not
present anymore suggesting that the corresponding attractor is possibly not
fractal. However, this conjecture remains to be verified quantitatively. In
contrast to previous thermostated driven Lorentz gases the bifurcation diagram
for the deterministic boundary thermostat is phase space covering for all
values of the electric field strength. The electrical conductivity shows a
nonlinear field dependence that is also different from the one obtained for
other thermostats, being more smooth than in previous cases. Still, there are
some irregularities on fine scales.

As is typical for boundary thermostated systems, there is no conjugate pairing
rule of Lyapunov exponents for this model. The Kolmogorov-Sinai entropy
related to the single positive Lyapunov exponents shows an interesting,
non-monotonous behavior as a function of the field strength which has not been
observed for bulk thermostats.

\item Finally, we have briefly summarized existing schemes that model shear
flows for interacting many-particle systems, such as sliding boundaries,
Less-Edwards `sliding brick' boundary conditions, applying the Gaussian
thermostat to shear in form of SLLOD equations of motion and refinements
leading to PUT thermostats. For ideal Gaussian thermostats applied to modeling
shear and heat flows there hold the usual relations between chaos and
transport such as the identity between phase space contraction and entropy
production and the Lyapunov sum rule. An alternative scheme was invented in
form of Maxwell daemon boundaries and was argued to share the same
properties. We also briefly outlined some conventional models of heat flows.

We then focused onto a hard disk fluid under shear and heat flow thermostated
by deterministic scattering. For a heat flow this system exhibited linear
response and yielded again an identity between phase space contraction and
entropy production. For modeling a shear flow three different versions of a
deterministic boundary thermostat were discussed. All of them generated a
linear response regime. Two of them did not yield an identity and consequently
no Lyapunov sum rule. One of them was not time reversible but reproduced the
identity. Two of the shear flow models were analyzed in further detail showing
that there was no conjugate pairing rule, as expected for thermostats acting
at the boundaries. However, both models exhibited a fractal attractor. The
latter properties are shared by the heat flow case. As in case of the
respectively thermostated driven periodic Lorentz gas, the Kolmogorov-Sinai
entropy displayed an interesting non-monotonous behavior as a function of the
shear rate.

\end{enumerate}

\section{$^*$Active Brownian particles and Nos\'e-Hoover thermostats}

In this chapter we return first to the Langevin equation reviewed in Section
II.B. Langevin's theory of Brownian motion presupposes that a Brownian particle is
only {\em passively} driven by collisions from the surrounding particles. This
input of energy is removed from the system by Stokes friction in the bulk
leading to a balance between molecular stochastic forces and friction as
formulated by the fluctation-dissipation theorem.

\begin{figure}[t]
\epsfxsize=8cm
\centerline{\rotate[r]{\epsfbox{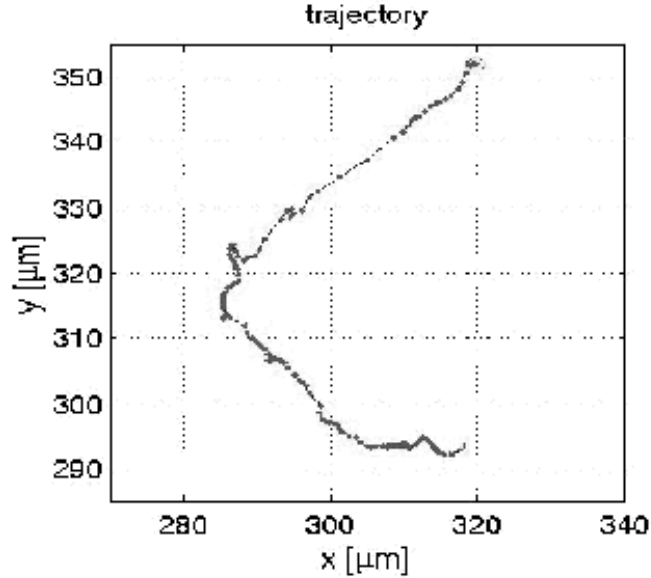}}}
  
\vspace*{0.3cm}
\caption{Experimental results for the migration of an isolated renal
epithelial MDCK-F (Madin-Darby canine kidney) cell on a substrate. The
positions $x$ and $y$ were extracted from microscopic phase contrast images
taken at time intervals of one minute and acquired for periods of two to four
hours \protect\cite{DDPKS03}.}
\label{fig:cell}
\end{figure}

Surprisingly, the Langevin equation can also be used to describe the motility
of biological cells crawling on substrates, at least if the cell dynamics is
sampled on large enough time scales. For experiments on moving cells such as
granulocytes\footnote{These cells are attracted to sites of inflammation to
destroy microorganisms and invaded cells.} and other types of crawling cells
see Refs.\ \cite{FrGr90,SLW91,SchGr93,HLCC94,DDPKS03}; see also Refs.\
\cite{GrBC94,RUGOS00,URGS01} for related work where the cells move in more
complex environments. In order to illustrate that cells may behave like
Brownian walkers Fig.\ \ref{fig:cell} depicts an experimentally measured
trajectory of an isolated epithelial\footnote{An epithelial cell is one of the
closely packed cells forming the epithelium, which is a thin layer of tissue
that covers organs, glands, and other structures within the body.} cell moving
on a substrate.

However, biological entities such as cells or bacteria may not really comply
with the physical assumption of being particles that move only because of
stochastic environmental forces. In order to sustain their motion such
organisms rather need some external supply of `fuel' that may be stored
internally, and one may think of some metabolic activity converting the fuel
into motion. This picture motivated to suitably amend conventional Langevin
equations arriving at {\em active} Brownian particles that are able to take up
energy from the environment, to store it onto internal degrees of freedom, and
to convert it into kinetic energy \cite{SET98,EST99,TSE99,EES+00}. Earlier
models of active Brownian particles featured a response of these entities to
environmental changes leading to pattern formation processes on a macroscopic
scale \cite{SSG94,SGMRM95}. Other recent work further generalizes active Brownian
particles in order to describe confined systems with rotational excitations
\cite{EEA02,EbRo02}.

In the following we will argue that limiting cases of Langevin equations
modeling active Brownian particles contain ingredients that trace back to the
conventional Nos\'e-Hoover thermostat discussed in Chapter IV. This may not
come too much as a surprise if one simply reinterprets the thermal reservoir
associated with the Nos\'e-Hoover dynamics as the internal energy of a
crawling cell. That is, instead of assuming the thermal reservoir to be
ubiquituous and stationary one now deems it to be locally co-moving with the
respective biological entity. Note that this merely amounts in suitably
adapting the physical interpretation of the equations of motion, without
requiring any reformulation of them.

Similarly to deterministic thermostats, in our view such equations for cell
motility may be considered as simple toy models that are proposed
``top-to-bottom'' on the basis of heuristic theoretical arguments rather than
starting ``bottom-up'' from a detailed analysis of biological
experiments. However, this approach may be useful for learning along which
lines ordinary Langevin equations need to be modified in order to model the
detailed dynamics of moving cells.

Here we review the formulation of active Brownian particles in terms of
generalized Langevin equations. We show that a particularly simple limiting
case of such equations exhibits formal analogies with the Nos\'e-Hoover
thermostat. On the basis of the existing literature we then briefly summarize
similarities and differences between active Brownian particles and
Nos\'e-Hoover thermostats from a dynamical systems point of view.

In the second section we outline some essential features of the velocity
distribution functions of active Brownian particles. We argue that the
appearance of crater-like structures in their profile may be understood in
connection with Nos\'e-Hoover thermostats. By employing the general functional
forms of the equilibrium velocity distribution functions discussed in Section
II.C we finally state some necessary conditions for the apearance of such
crater-like structures.

\subsection{Generalizing Langevin equations for modeling cell motility}

The ansatz for modeling active Brownian particles which we review here starts
from the generalized two-dimensional Langevin equation
\cite{SET98,EST99,TSE99,EES+00}
\bna
{\bf \dot{r}}&=&{\bf v}\nonumber\\
{\bf \dot{v}}&=&-\alpha{\bf v}+\mbox{{\boldmath ${\cal F}$}}(t) \quad , \label{eq:labp}
\ena
where $\mbox{{\boldmath ${\cal F}$}}$ is white noise, that is, a $\delta$-correlated
stochastic force of strength $S$ with zero mean value,
\be
<\mbox{{\boldmath ${\cal F}$}}(0) \mbox{{\boldmath ${\cal
F}$}}(t)>=S\delta(t) \quad . 
\ee
The noise strength $S$ can be straightforwardly associated with the friction
coefficient $\alpha$ according to the fluctuation-dissipation theorem Eq.\
(\ref{eq:fdt}) and with the diffusion coefficient $D$ by using the Einstein
relation Eq.\ (\ref{eq:einstvisc}),
\be
\frac{S}{2T}=\alpha=\frac{T}{D} \quad . \label{eq:sad}
\ee
Let us now assume that there is a generalized friction coefficient $\alpha$ in
form of
\be
\alpha=\alpha_0-d e \quad . \label{eq:alpabp}
\ee 
Here $\alpha_0$ holds for ordinary Stokes friction, the variable $e$ denotes
the internal energy of the moving entity, and the constant $d>0$ yields the
rate of conversion from internal into kinetic energy. $e$ is in turn obtained
from a suitable balance equation for which different choices have been
discussed. A general ansatz, denoted in the following as {\em case 1}, reads
\cite{SET98,EST99,TSE99}
\be
\dot{e}=\frac{1}{\mu}(q-e(c +d v^2)) \quad . \label{eq:ieabp}
\ee
In this equation $q\ge0$ describes a constant take-up of energy, which
generally may be space-dependent, and $c$ models some constant internal loss
of energy. Note that $d$ could be velocity-dependent, however, in Eq.\
(\ref{eq:ieabp}) we stick to the most simple assumption that the prefactor of
$e$ is only quadratic in $v$. $\mu$ determines the rate of change of the
internal energy $e$ and thus plays a role similar to $\tau^2$ in the
Nos\'e-Hoover thermostat, see Eq.\ (\ref{eq:anh}). In case of very fast
feedback $\mu\to 0$ Eq.\ (\ref{eq:ieabp}) boils down to the fixed point {\em
case 2},
\be
e=\frac{q}{c+dv^2} \quad . \label{eq:case2}
\ee
Combining this equation with Eq.\ (\ref{eq:alpabp}) yields the
velocity-dependent friction coefficient
\be
\alpha=\alpha_0\frac{v^2-v_0^2}{\frac{q}{\alpha_0}+v^2-v_0^2}
\ee
with $v_0^2:=q/ \alpha_0-c/d$. A special case of this type of friction is
obtained by assuming $v^2\ll v_0^2$, or alternatively $c/d>>v^2$, in the
denominator only leading to {\em case 3}
\be
\alpha=\alpha_1\frac{v^2-v_0^2}{v_0^2} \label{eq:abp3}
\ee
with $\alpha_1:=-\alpha_0+qd/c$. This equation is sometimes called the {\em
Rayleigh-type model}, because a similar ansatz for a velocity-dependent
friction coefficient has been introduced by Rayleigh in the context of the
theory of sound \cite{EES+00}. Eq.\ (\ref{eq:abp3}) looks strikingly similar
to Eq.\ (\ref{eq:anh}) defining the Nos\'e-Hoover thermostat. However, note
that the former represents the explicit functional form for $\alpha$, whereas
the latter is a differential equation with a time-derivative for $\alpha$ on
the left hand side.

Motivated by this formal similarity, one may ask to which extent models of
active Brownian particles can exhibit dynamical systems properties as outlined
in Chapter IV for the Nos\'e-Hoover thermostat. A fundamental difference
between active Brownian particles and Nos\'e-Hoover thermostats is that the
former have been introduced for modeling the energetic aspects of moving
biological entities on a microscopic level, without knowing in advance about
the velocity distributions generated by these models. In contrast to that,
Nos\'e-Hoover thermostats have been constructed for generating, under suitable
conditions, specifically canonical velocity distributions, see Section
IV.B.1. A further difference already resulting from the Stokes friction
coefficient is that the equations of motion for active Brownian particles are
irreversible. But even more, according to ordinary Langevin dynamics these
equations include noise, which is in contrast to the concept of Nos\'e-Hoover
and other deterministic thermostats.

In order to bring active Brownian particles and deterministic thermostats
closer together, thus making the three models introduced above more amenable
to methods of dynamical systems theory, one may replace the white noise term
by deterministic chaos, as we already briefly described in Section
II.B. Another option is to generate deterministic chaos by choosing a suitable
potential in configuration space such as the geometry of the periodic Lorentz
gas, see the beginning of Section III.A. Yet we do not know anything about the
velocity distributions for deterministic active Brownian particles and there
is still the lack of time-reversibility. Hence, there is no reason why for
these models one should expect an identity between phase space contraction and
entropy production to hold. Similarly, there is no reason why by default
equipartitioning of energy should be fulfilled. Whether fractal attractors
exist for this class of dissipative dynamical systems and whether Lyapunov sum
rules and conjugate pairing rules hold, as they typically do for ideal
Nos\'e-Hoover thermostats, are further interesting questions that have not
been studied so far. However, we remark that for so-called
canonical-dissipative systems containing active Brownian particles as a
special case generalized Hamiltonian equations have been constructed that
remind of the generalized Hamiltonian formalism developed for Gaussian and
Nos\'e-Hoover thermostats outlined in Section IV.C.2 \cite{Ebe00}.

\subsection{Crater-like velocity distributions}

Let us now summarize what is known about the velocity distribution functions
for the three different cases of active Brownian particles discussed above. In
this framework we shall also discuss connections with the Nos\'e-Hoover
thermostat.

In all cases, the most obvious solution for the velocity distribution is
obtained at $e=const$. This is the Stokes limit of constant friction for which
the velocity distribution is well-known to be purely canonical, see Section
II.C.

For {\em case 1}, Eqs.\ (\ref{eq:labp}), (\ref{eq:alpabp}) and
(\ref{eq:ieabp}), not much appears to be known regarding general combinations
of parameters in Eq.\ (\ref{eq:ieabp}).

\begin{figure}[t]
\epsfxsize=12cm
\centerline{\epsfbox{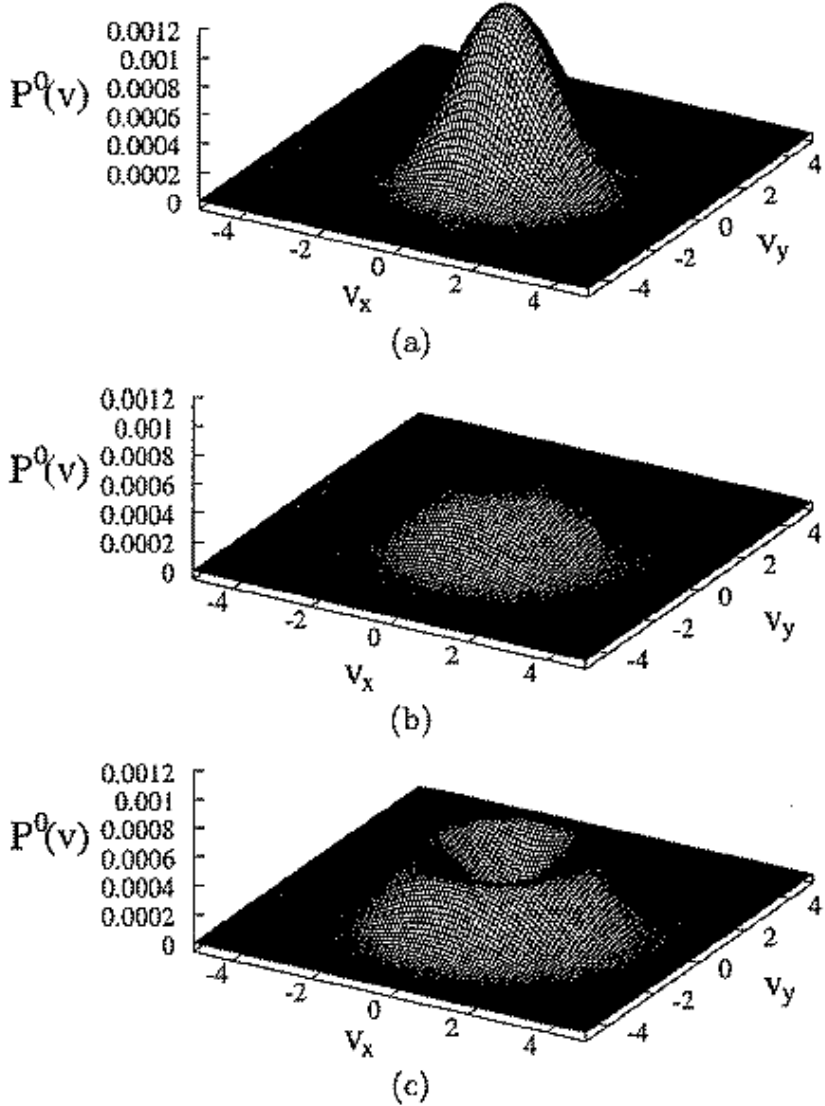}}
\caption{Transition from canonical to  {\em crater-like} velocity
distributions in the model of active Brownian particles called {\em case 2},
see Eq.\ (\ref{eq:case2}) in combination with Eqs.\ (\ref{eq:labp}),
(\ref{eq:alpabp}). The figure displays the analytical solution $\rho(v)\equiv
P^0(v)$ of Eq.\ (\ref{eq:vdfcase2}). The parameters are $c=1$, $q=10$, $S=4$,
$\alpha_0=2$ and (a) $d=0.07$, (b) $d=0.2$, (c) $d=0.7$. The figure is from
Ref.\ \protect\cite{EES+00}.}
\label{fig:crater}
\end{figure}

For the special {\em case 2} the situation is much better: Here the analytical
solution for the velocity distribution function can be obtained from solving
the corresponding Fokker-Planck equation and reads \cite{TSE99,EES+00}
\be
\rho(v)=C
\left(1+\frac{dv^2}{c}\right)^{q/S}\exp\left(-\frac{\alpha_0}{S}v^2\right)
\label{eq:vdfcase2} \quad ,
\ee
where $C$ is a normalization constant. Interestingly, as shown Fig.\
\ref{fig:crater} this velocity distribution displays a transition from 
canonical to more microcanonical-like under variation of $d$ or related
parameters. Such {\em crater-like} velocity distributions exhibiting a dip at
the place of the former maximum of the canonical distribution appear to be
quite typical for active Brownian particles
\cite{EES+00,SET01,ES01,EEA02,EbRo02}.

Fig.\ \ref{fig:crater} should be compared to our previous Fig.\
\ref{fig:rhonhtds} displaying velocity distributions for the
Nos\'e-Hoover thermostated driven periodic Lorentz gas. Apart from the
asymmetry due to the external electric field the scenario is completely the
same as in Fig.\ \ref{fig:crater}: By tuning the reservoir coupling parameter
$\tau$ the Nos\'e-Hoover velocity distribution clearly shows a transition from
canonical to microcanonical-like. The advantage of the Nos\'e-Hoover
thermostat is that due to its construction the origin of these different
functional forms is rather well-understood, as already outlined in Section
IV.B.2: In the limiting case of $\tau\to0$ the Nos\'e-Hoover thermostat
generates a microcanonical velocity distribution, whereas in the limit of
Stokes friction for $\tau\to\infty$ it yields a canonical one. Hence, there
must be a transition between canonical and microcanonical for intermediate
$\tau$ parameters. This transition is represented by a superposition of these
two different velocity distributions.

For Fig.\ \ref{fig:crater} and {\em case 2} a very similar explanation might
thus be employed: The canonical velocity distribution is probably reminiscent
of passive motion of the Brownian particles driven by the noise term in Eq.\
(\ref{eq:labp}). The appearance of microcanonical-like features must then be
due to the deterministic, velocity-dependent friction coefficient appearing in
the equations of motion. This interplay may result in a transition from a
canonical to a microcanonical(-like) distribution which is quite analogous to
the one displayed by the Nos\'e-Hoover system. In contrast to Nos\'e-Hoover,
however, for active Brownian particles this transition may rather reflect a
change from noise-driven, passive motion to more active dynamics as determined
by the deterministic, nonlinear part of the equations of motion.

On the other hand, we also have this formal analogy between the Nos\'e-Hoover
equations of motion and at least {\em case 3} of active Brownian particles as
noted in the previous section. If this link could be made more explicit one
might argue that, irrespective of stochastic contributions in the equations
governing active Brownian particles, there should as well be a Nos\'e-Hoover
like transition in the velocity distributions which is purely generated by a
tuning of the deterministic, velocity-dependent friction coefficient. This
leads us to the conclusion that in case of active Brownian particles there may
actually be two rather analogous transitions between canonical and
microcanonical distributions, one that is of a purely deterministic origin and
another one that is due to the interplay between deterministic and stochastic
forces. To which extent the one or the other is exhibited by the dynamics may
then depend on the specific choice of the control parameters.

We further remark at this point that an at first view analogous transition
from unimodal to bimodal distributions has been reported for the stationary
states of a nonlinear oscillator driven by L\'evy noise
\cite{CGKRT02,CKGMT03}. However, the detailed dynamical origin of this
transition appears to be very different from the two scenarios discussed above
thus demonstrating that there may actually be a larger number of different
dynamical mechanisms generating such bifurcations in the velocity
distributions.

In order to undermine these arguments and before discussing the velocity
distribution function of {\em case 3} in detail, let us analyze the transition
between microcanonical and canonical distributions for the Nos\'e-Hoover
thermostat more explicitly. Note that in the following we consider a purely
deterministic equlibrium dynamics without any external field. We recall that
the equation governing the velocity-dependent Nos\'e-Hoover friction
coefficient $\alpha$, Eq.\ (\ref{eq:anh}), read
\be
\dot{\alpha}=\frac{v^2-2T}{\tau^22T} \quad . \label{eq:anh2}
\ee
In the limit of $\tau\to0$ the Nos\'e-Hoover thermostat approaches the
Gaussian one for which the kinetic energy is strictly kept constant at any
time step. This suggests that the fluctuations of the friction coefficient
$\alpha$ triggered in Eq.\ (\ref{eq:anh2}) by the remaining equations of
motion should be getting smaller and smaller for larger times when the system
is evolving into a steady state. Expanding Eq.\ (\ref{eq:anh2}) linearly in
time yields
\be 
\alpha(t+\Delta t)=\frac{v^2-2T}{\tau^22T}\Delta t +\alpha(t) \quad
(\tau\to0\:,\:\Delta t\ll 1) \quad .
\label{eq:nhapp} 
\ee
As desired for thermalizing onto a steady steady this equation has a fixed
point that in the limit of $\tau\to0$ is associated with the Gaussian
thermostat constraint $v^2=2T$. However, if the first term on the right hand
side of Eq.\ (\ref{eq:nhapp}) were a non-zero constant the system cannot reach
a steady state, since $\alpha(t)\to\pm\infty$. Hence, in this case
$\alpha(t)=0$ is the only steady state solution. For $\tau\ll1$ small but
finite $\alpha(t)$ fluctuates around zero corresponding to the time-dependence
of $v$ as induced by the other equations of motion. According to our above
argument, for $\tau\ll1$ these fluctuations may be suitably approximated by
setting $\alpha(t)=0$ on the right hand side of Eq.\ (\ref{eq:nhapp}) thus
recovering the functional form of the Rayleigh friction Eq.\
(\ref{eq:abp3}). Note that in the other extreme of $\tau\to\infty$ the
friction coefficient $\alpha$ can be an arbitrary constant thus recovering
Stokes friction, which in this equilibrium situation corresponds to zero
velocity. We thus have a Hopf-like bifurcation scenario \cite{Ott} from $v=0$
for $\tau\to\infty$ to a limit cycle behavior with $v=const.$ for $\tau\to0$.

On the basis of this approximation, let us study what happens in the limit of
small but finite $\tau\ll1$ for the equilibrium velocity distribution function Eq.\
(\ref{eq:fans}) of Nos\'e-Hoover reading
\be
\rho(t,{\bf r},{\bf v},\alpha)= const.\
\exp\left[-\frac{v^2}{2T}-(\tau\alpha)^2\right] \quad .
\ee
The approximation Eq.\ (\ref{eq:nhapp}) enables us to eliminate $\alpha(t)$
from the above equation yielding
\bna
\rho(v)&=&C \exp\left(-\frac{v^2}{2T}-\frac{(v^2-2T)^2(\Delta t)^2}{4\tau^2T^2}\right)
\label{eq:nhdip1} \\
&=&C \:\delta_{\tau}\left(\frac{(v^2-2T)}{2T}\right) \exp\left(-\frac{v^2}{2T}\right)\quad
(\tau\ll1\:,\: \Delta t\ll 1) \label{eq:nhdip2}
\ena
For the second line we have used the definition of the $\delta$-function as a
series of exponentials in $\tau$ \cite{Reif}, where $C$ is a normalization
constant. We thus indeed arrive at the limiting case of a crater-like
Nos\'e-Hoover velocity distribution composed of a canonical and a
microcanonical contribution.

This solution may finally help to analyze the velocity distribution that was
obtained for the active Brownian particle {\em case 3} \cite{EES+00,EEA02},
\be
\rho(v)=C \exp\left(\frac{\alpha_1}{S}v^2-\frac{\alpha_2}{2S}v^4\right) 
\label{eq:vdfcase3}
\ee
with $\alpha_2:=\alpha_0 d/c$.  The functional form matches nicely to Eq.\
(\ref{eq:nhdip1}) and there should be a respective decomposition into a
canonical and into a microcanonical component in analogy to Eq.\
(\ref{eq:nhdip2}). However, there is the subtlety that Eq.\
(\ref{eq:vdfcase3}) represents the solution for a stochastic system as is
reminiscent by the strength $S$ of the stochastic force. We therefore have to
inquire whether this stochasticity is a crucial ingredient fundamentally
distinguishing this distribution from the formally completely analogous but
deterministic Nos\'e-Hoover solution Eq.\ (\ref{eq:nhdip1}).

Let us assume that in a suitable {\em case 3} system both the
fluctuation-dissipation theorem and the Einstein relation are valid.
According to Eq.\ (\ref{eq:sad}) $S$ is then a trival function of the
temperature $T$ and the diffusion coefficient $D$. The temperature $T$ may be
defined by furthermore assuming equipartitioning of energy, $T=<v_0^2>/2$,
where $<v_0^2>$ is the average kinetic energy of the active Brownian
particle. The diffusion coefficient $D$ might be thought of being generated
deterministically. We therefore argue that Eq.\ (\ref{eq:vdfcase3}) also holds
in a purely deterministic modeling of {\em case 3} thus establishing the
connection to the Nos\'e-Hoover dynamics analyzed before. Under these
conditions, {\em case 3} must exhibit the very same transition in the velocity
distributions as Nos\'e-Hoover.

For sake of completeness we remark that in the limit of small velocities and
by expanding the prefactor of the velocity distribution Eq.\
(\ref{eq:vdfcase2}) of {\em case 2}, Eq.\ (\ref{eq:vdfcase3}) of {\em case 3}
is recovered again. Along these lines one may understand a transition such as
the one depicted in Fig.\ \ref{fig:crater} also on the basis of Nos\'e-Hoover
dynamics.

To summarize, in all three cases of active Brownian particles discussed above
there exists Stokes limit of constant friction yielding canonical velocity
distributions irrespective of any stochastic contributions. We therefore
conjecture that, in analogy to Nos\'e-Hoover dynamics, for {\em purely
deterministic} active Brownian particles there exists a completely analogous
transition from canonical to microcanonical velocity distributions under
suitable parameter variation yielding intermediate crater-like velocity
distributions. One choice for such a parameter variation should be varying
$\mu$ in {\em case 1}, but other combinations of parameters, in analogy to
varying $\tau$ in Nos\'e-Hoover dynamics, should be possible as well. For
stochastic active Brownian particles defined in the framework of ordinary
Langevin dynamics, {\em in addition} there should be a transition from a
canonical velocity distribution function generated by the stochastic forcing
to a microcanonical-like counterpart that, again, is due to a suitably tuned
deterministic friction coefficient. This theoretical predicition still needs
to be verified by computer simulations.

Finally, we briefly comment on two important necessary conditions limiting the
range of existence of crater-like velocity distributions. For this purpose we
remind again of the derivation of velocity distributions via projection from a
microcanonical one outlined in Section II.C. Let us assume that the velocity
space of the moving particle has a dimension of $d\equiv d_s$, in order to be
consistent with our previous notations. Let us furthermore assume that the
$d_s$-dimensional velocity distribution of the moving particle is
microcanonical as, for example, resulting from the application of a Gaussian
thermostat, or in the limit of Nos\'e-Hoover dynamics for $\tau\to0$. The
central formula is then Eq.\ (\ref{eq:gaus}) telling us that {\em not} for any
dimensionality $d_s$ a one-dimensional projection $\rho(v_x)\equiv\rho(v_1)$
exhibits a crater-like minimum in its functional form.

More precisely, Fig.\ \ref{fig:d346} shows that for $d_s\ge3$ the particle's
one-component equilibrium velocity distribution does not display a global
minimum anymore. However, this eliminates any possibility for the generation
of crater-like velocity distributions. That is, creating a transition from a
projection of a microcanonical distrbution to a canonical one (by noise, or by
tuning $\tau$ in the Nos\'e-Hoover thermostat) for $d_s<3$ cannot generate any
non-monotonicities in intermediate distributions. In other words, for moving
particles with more than two degrees of freedom there cannot be any transition
between microcanonical and canonical one-component velocity distributions
anymore featuring crater-like intermediate functional forms.

An analogous argument holds if the dimensionality $d_r$ of an associated
thermal reservoir can be changed as discussed, e.g., for thermostating by
deterministic scattering in Section VII.C. For such a combination of subsystem
and reservoir one can straightforwardly check \cite{Kla03}, again by using
Eq.\ (\ref{eq:gaus}) with $d\equiv d_s$, respectively with $d\equiv d_s+d_r$,
that non-monotonous velocity distributions only appear if $d_s+d_r\ge3$. These
two bounds put therefore quite a window on the existence of crater-like
velocity distributions.

Our discussion thus shows how deterministic thermostats such as Gauss,
Nos\'e-Hoover and thermostating by deterministic scattering may help to
understand the origin of velocity distributions in, at first view, seemingly
quite different systems that may even be amenable to physical and biological
experiments.

\subsection{Summary}

\begin{enumerate}

\item Results from experimental measurements on the motility
of isolated cells moving on subtrates can be understood, to some extent, by
using conventional Langevin equations. Amended versions of such equations
modeling the storage of internal energy of biological cells lead to the
formulation of {\em active Browian particles}.

\item We briefly reviewed three fundamental versions of active Brownian
particles in which the friction coefficient of ordinary stochastic Langevin
equations has been suitably generalized. The most simple model shared a
striking formal similarity with the Nos\'e-Hoover dynamics discussed
before. However, {\em per se} active Brownian particles and deterministic
thermostats belong to the fundamentally different classes of stochastic,
respectively deterministic dynamical systems.

\item We then discussed the origin of crater-like velocity distributions both for
deterministic Nos\'e-Hoover dynamics and for stochastic active Brownian
particles in thermal equilibrium. In case of Nos\'e-Hoover this transition is
triggered by the tuning of the reservoir response time yielding a
superposition between a microcanonical and a canonical distribution, as
already mentioned in Section IV.B.2. Here we analyzed this scenario in more
detail starting from the Nos\'e-Hoover equations of motion. Based on formal
similarities of the underlying equations, we concluded that the same scenario
must be present for active Brownian particles.

However, in addition the latter class of systems appears to feature a second
mechanism leading to crater-like velocity distribution functions.  In this
case they result from a superposition between a stochastically generated
canonical distribution and a microcanonical(-like) one related to the action
of the deterministic friction coefficient.

Finally, we gave two necessary conditions concerning the involved number of
degrees of freedom which must be fulfilled in order to generate transitions
that are characterized by crater-like velocity distributions.

\end{enumerate}

\section{Concluding remarks}

We begin this final chapter by briefly summarizing the main contents of this
review. We then formulate our answer to the central question posed in this
work by which we were inquiring about a general theory of NSS in thermostated
dynamical systems starting from microscopic chaos. We furthermore list what we
believe are important open questions and conclude with some acknowledgements.

\subsection{A brief summary of this review}

There are currently two fundamental approaches aiming at a general theory of
NSS starting from microscopic chaos in the equations of motions of suitable
model systems: One of them employs purely Hamiltonian dynamics supplemented by
suitable (nonequilibrium) boundary conditions. The other one applies to
nonequilibrium situations where a coupling to some kind of thermal reservoir
is needed in order to sustain stationary solutions for statistical physical
quantities. Typically, the latter approach renders the dynamical systems under
consideration non-Hamiltonian due to a respective modeling of the thermal
reservoirs. We have briefly outlined these two theories of chaotic NSS in our
general introduction Chapter I. This review focused on the non-Hamiltonian
approach to nonequilibrium transport, that is, on the construction and
analysis of NSS in dissipative dynamical systems.

In Chapter II we motivated the use of thermal reservoirs both heuristically
and by employing the Langevin equation, a well-known example for a stochastic
modeling of thermostated dynamics. We sketched how to calculate the
equilibrium velocity distribution functions for a subsystem coupled to a
thermal reservoir and provided some general solutions. Finally, we introduced
the periodic Lorentz gas, a standard model in the field of chaos and
transport. In our review the model was connected to different thermal
reservoirs. By applying simple nonequilibrium conditions we inquired about
similarities and differences concerning chaos and transport properties in this
model.

We started with two paradigmatic modelings of thermal reservoirs, which are
known as the Gaussian and the Nos\'e-Hoover thermostat. These schemes go
beyond ordinary Langevin dynamics in that the resulting equations of motion
are still deterministic and time-reversible sharing fundamental features with
Newtonian dynamics. The purpose of the Gaussian method is to transform the
velocities of a subsystem onto microcanonical velocity distributions in
thermal equilibrium. The Nos\'e-Hoover method was constructed to yield
canonical velocity distributions. That is, for Gauss the total or kinetic
energy is kept constant at any time step, whereas for Nos\'e-Hoover it is
allowed to fluctuate canonically. If certain necessary conditions are
fulfilled, applications of these schemes in nonequilibrium situations lead to
well-defined NSS.

These two thermostats were introduced and analyzed in Chapters III and IV step
by step. In order to motivate Nos\'e-Hoover dynamics we derived the
(generalized) Liouville equation that is valid for dissipative dynamical
systems. As a demonstration, we applied both the Gaussian and the
Nos\'e-Hoover thermostat to the periodic Lorentz gas driven by an external
electric field. We then discussed the resulting chaos and transport properties
of these thermostated driven Lorentz gases. Additionally, we outlined
generalized Hamiltonian and Lagrangian formulations for Gaussian and
Nos\'e-Hoover equations of motion. This analysis was supplemented by reviewing
more recent work concerning the construction of thermal reservoirs that
generalize ordinary Nos\'e-Hoover dynamics.

Chapter V summarized our main conclusions concerning chaos and transport
properties of NSS according to Gaussian and Nos\'e-Hoover theory. These
preparations allowed us to start on the main question of our review, namely to
which extent the chaos and transport properties discussed so far are universal
for thermostated NSS altogether.

First answers to this fundamental question were provided in Chapter VI. Here
we presented straightforward modifications of conventional Gaussian and
Nos\'e-Hoover schemes leading to what we called non-ideal Gaussian and
Nos\'e-Hoover thermostats. Again, we applied these mechanisms to the driven
periodic Lorentz gas. It turned out that the non-ideal variants yield chaos
and transport properties that are in many respects very different from
conventional Gaussian and Nos\'e-Hoover dynamics.

All these thermal reservoirs acted directly upon the bulk of a subsystem to be
thermostated. Chapter VII was therefore devoted to thermostats defined at the
boundaries of a subsystem. A prominent example of this type of thermostats are
stochastic boundary conditions. We showed that they can be reformulated in
terms of deterministic and time-reversible collision rules making them
comparable to Gaussian and Nos\'e-Hoover dynamics. This scheme, denoted as
thermostating by deterministic scattering, was again applied to the driven
periodic Lorentz gas. As in case of non-ideal thermostats, the chaos and
transport properties obtained from this model were largely at variance with
respective results of Gaussian and Nos\'e-Hoover dynamics. Asking for possible
universalities of such properties in more complicated, interacting
many-particle systems lead us to investigate a hard disk fluid under heat and
shear. This system was also thermostated at the boundaries. Depending on the
precise definition of the coupling between subsystem and reservoir, relations
between chaos and transport known from ordinary Gaussian and Nos\'e-Hoover
dynamics were either recovered or not.

Our review concluded with Chapter VIII by outlining further generalizations of
ordinary Langevin dynamics. Here we discussed active Brownian particles
providing abstract models for the motility of biological entities. We argued
that there is a relationship between these models and conventional
Nos\'e-Hoover thermostats. This is particularly reflected in the emergence of
crater-like velocity distributions that are intermediate between canonical and
microcanonical velocity distributions.

We emphasize that this review includes results that have not been published
before. These are the construction and analysis of the non-ideal Gaussian
thermostat, see Section VI.A, as well as Chapter VIII discussing the analogy
between active Brownian particles and Nos\'e-Hoover thermostats.

\subsection{Do there exist universal chaos and transport properties of thermostated
dynamical systems?}

A central point of this review was to explore to which extent the description
of NSS generated by conventional Gaussian and Nos\'e-Hoover dynamics is
universal. In this endeavour Chapter V played a crucial role, where we
summarized what we consider to be the most important links between chaos and
transport resulting from Gaussian and Nos\'e-Hoover dynamics. On the basis of
the analysis provided in Chapters VI to VIII we now briefly go through this
list again and come to final conclusions.

\begin{enumerate}

\item {\bf non-Hamiltonian dynamics for NSS:} In our view, this type of
dynamics yields a rather natural description of NSS due to external forces. As
we argued on many occasions, the non-Hamiltonian character of thermostated
systems such as Langevin, Gaussian or Nos\'e-Hoover dynamics can be understood
with respect to projecting out spurious reservoir degrees of freedom. From
this point of view we do not see any reason to insist on a Hamiltonian
modeling of NSS.

\item {\bf phase space contraction and entropy production:} Conventional
Gaussian and Nos\'e-Hoover thermostats display a default identity between
these two quantities, which is at the heart of linking thermodynamics to
dynamical systems theory in dissipative dynamical systems. However, it is not
too difficult to construct counterexamples of deterministic and
time-reversible thermal reservoirs generating well-defined NSS but not
exhibiting this identity. We just mention non-ideal Gaussian and Nos\'e-Hoover
thermostats and sheared hard-disk fluids thermostated by deterministic
scattering. Care should therefore be taken exploring the second law of
thermodynamics on the basis of this identity.

\item {\bf transport coefficients and dynamical systems quantities:}
Conventional Gaussian and Nos\'e-Hoover dynamics features the Lyapunov sum
rule linking transport coefficients to Lyapunov exponents. However, the
existence of such a simple functional relationship is intimately connected
with the existence of the abovementioned identity. Hence, it is not surprising
that this relation between chaos quantities and transport coefficients is not
universal either. Non-ideal thermostats and thermostating by deterministic
scattering yield again explicit counterexamples. We furthermore emphasize that
there are broad classes of thermostated dynamical systems not exhibiting a
conjugate pairing of Lyapunov exponents.

\item {\bf fractal attractors characterizing NSS:} In all deterministically
thermostated dynamical systems analyzed in this review the resulting
attracting sets were characterized by fractal structures. This fractality is
thus a very promising candidate for a universal property of NSS associated
with deterministic and time-reversible thermal reservoirs. As it stands, it is
also the only surviving one of the list of possibly universal characteristics
discussed so far. However, note that the detailed topology of the attracting
sets still intimately depends on the type of thermal reservoir applied. This
is represented by the different types of bifurcation diagrams for the driven
periodic Lorentz gas connected to different thermal reservoirs.

\item {\bf nonlinear response in the thermostated driven periodic Lorentz
gas:} Despite a mathematical proof for the existence of linear response in the
Gaussian thermostated driven periodic Lorentz gas no clear signs of a regime
of linear response could be detected in computer simulations for the Gaussian
as well as for other thermostated Lorentz gases. Even worse, applying
different thermostats yielded significantly different results for the
field-dependent electrical conductivity of this model. We must therefore
conclude that there is no equivalence of nonequilibrium ensembles for the
thermostated driven periodic Lorentz gas in this regime of field strengths. In
case of a thermostated hard disk fluid the situation appears to be better in
that, at least, there is an equivalence of nonequilibrium ensembles as far as
thermodynamic properties are concerned. However, we emphasize that on the
level of chaos quantities there is no equivalence either.

We thus conclude that, from the point of view of nonequilibrium
thermodynamics, the periodic Lorentz gas seems to be a rather delicate
dynamical system. This appears to be due to the low dimensionality of the
dynamics featuring a single moving point particle only. In this case
statistical properties arise from the chaotic collisions of the particle with
the fixed scatterers and contain intricate dynamical correlations in time and
space. The different results for the field-dependent electrical conductivity
presented in this review demonstrate the limits of standard thermodynamic
descriptions applied to the thermostated model. However, in many other
respects both the periodic Lorentz gas with and without external fields still
exhibits ``nice'' thermodynamic behavior. And even the existence of
irregularities in the field-dependent electrical conductivity should not be
deemed ``unphysical'': Interacting many-particle systems typically display
thermodynamic behavior due to their intrinsic statistical properties, whereas
chaotically moving single particles may exhibit very intricate dynamical
phenomena reflecting specific nonlinearities in their equations of motion. To
us it appears that the periodic Lorentz gas, as well as multibaker maps, are
right in-between these two large classes of high and low-dimensional,
interacting and non-interacting many-particle systems. Such intermediate
dynamics may thus display both ordinary thermodynamic behavior as well as
specific chaotic dynamical properties. An adequate theoretical description
thus needs to develop a well-balanced combination of both statistical and
dynamical systems methods in order to adequately assess both the chaos and
transport properties of this highly interesting type of dynamical systems.

\end{enumerate}

\subsection{Some important open questions}

Here we only list a few basic problems that we consider worth to be studied in
further research. More detailed open questions were already mentioned in the
course of this review on various occasions.

\begin{enumerate}

\item {\bf defining a nonequilibrium entropy:} This is of course a central
question of nonequilibrium statistical mechanics altogether. From our point of
view it is not enough to construct a theory of NSS starting from the ordinary
Gibbs entropy. A proper definition of a nonequilibrium entropy should yield
results that are compatible with standard nonequilibrium thermodynamics. In
case of the driven Lorentz gas we think here of entropy production in terms of
Clausius' entropy combined with Joule's heat. To us it appears that, apart
from other problems, the Gibbs entropy may contain spurious contributions
compared to a thermodynamic entropy production, since it samples the complete
phase space which is at variance to an entropy production in terms of a heat
transfer. This problem may be eliminated by using a revised definition of the
Gibbs entropy based on a suitable coarse-graining. It might be interesting to
apply such recent methods of coarse-grained entropies to our examples of
thermostated dynamical systems not yielding an identity between phase space
contraction and entropy production. One may hope that starting from such an
amended Gibbs entropy ordinary thermodynamics can again be recovered even for
our more general examples.

\item {\bf fluctuation theorems:} We have said very little about this 
recent and very active research topic that concerns symmetry relations in the
fluctuations of nonequilibrium entropy production. Apart from the fractality
of attractors, fluctuation theorems currently feature as another important
candidate for universal results characterizing NSS in dissipative dynamical
systems. However, up to now fluctuation theorems have only been tested in
thermostated dynamical systems furnishing an identity between phase space
contraction and entropy production. It would thus be important to check for
their existence in thermostated systems not featuring this identity, such as
the ones described in our review. Interestingly, the existence of a certain
class of fluctation theorems is intimately linked to the validity of the
chaotic hypothesis mentioned in the introduction. Hence, testing these
fluctuation theorems one also studies, to some extent, the validity of the
chaotic hypothesis. We furthermore suggest to look for fluctuation theorems in
experiments on cell motility.

\item {\bf existence of fractal attractors in stochastically perturbed
dynamical systems:} As we have outlined on previous occasions, this important
question is still not settled. In order to solve this problem one may study,
for example, the driven periodic Lorentz gas or a hard disk fluid under shear
supplemented by stochastic boundaries. It might then be elucidating to compare
the respective results to the ones obtained from thermostated by deterministic
scattering.

\item {\bf analyzing NSS for generalized Nos\'e-Hoover thermostats and for active
Brownian particles:} Following the philosophy of Nos\'e-Hoover there emerged a
large collection of generalized deterministic and time-reversible thermostats
for which it is still not clear to which extent they are functioning under
nonequilibrium constraints. If they do, they may provide further access roads
towards an analysis of chaos and transport properties of NSS going beyond
ordinary Gaussian and Nos\'e-Hoover dynamics. The same reasoning applies to
active Brownian particles that may be considered as another type of
thermostated dynamical systems.

\item {\bf quantum-mechanical formulations of thermal reservoirs:}  In our
review we did not say anything about extensions of classical deterministic and
time-reversible thermostats towards the quantum regime. It seems that this
field is currently evolving, see Refs.\
\cite{GrTo89,Kus93,Kus95,MS01,KLA02,MS03} and further references
therein. Eventually, it might be interesting to study to which extent these
approaches furnish links between chaos and transport being analogous to their
classical counterparts.

\end{enumerate}

\subsection{Acknowledgements}

I am particularly indebted to Prof.\ G.Nicolis, who allowed me to work with
him on this subject during my 2 1/2 year-stay in Brussels. I wish to thank him
for many inspiring discussions and for his ongoing support. This project was
pursued together with Dr.\ K.Rateitschak, who obtained her Ph.D.\ in the course
of this cooperation and whom I thank for her hard and diligent work. Later on
Dr.\ C.Wagner joined this team as a postdoc, and I thank him for his
contributions that he partly finished independently. Financial support of my
research by postdoctoral fellowships of the DFG and of the EU during that time
is gratefully acknowledged.

Prof.\ C.Dellago kindly supported us with computer codes whenever needed, and
I am grateful to him for sharing his experience with us concerning intricate
details of nonequilibrium molecular dynamics computer simulations. Profs.\
P.Gaspard, J.R.Dorfman, S.Hess, H.van Beijeren, T.T\'el, W.G.Hoover, H.A.Posch
and D.J.Evans helped enormously with many valuable hints and discussions on
thermostats and related issues. Dr.\ D.Panja contributed with enlightening
discussions concerning the conjugate pairing rule. Thanks to all of them!

I wish to thank Prof.\ W.Ebeling for making me aware of the similarity between
Nos\'e-Hoover velocity distributions and the ones of active Brownian particles
as well as for ongoing inspiring discussions of this topic. I am furthermore
very much indebted to Dr.\ P.Dieterich for discussing with me his experiments
on cell motility as well as for studying together more intricate details of
Langevin dynamics as applied to moving cells.

Thanks go also to Profs.\ P.H\"anggi and G.Radons for their critical comments
on thermostated dynamical systems and for emphasizing the importance of
ordinary Langevin dynamics. I hope they will find that I have taken their
remarks suitably into account in this review.

Dr.\ A.Riegert and Mr.\ N.Korabel invested a lot of time for working
themselves through a first version of this review. I thank both of them for
their careful reading and for their many hints as well as Dr.\ K.Gelfert for
insightful discussions of mathematical subtleties. Interesting comments by
Prof.\ H.Kantz and Ms.\ N.Baba on thermal reservoirs are also gratefully
acknowledged.

Finally, I wish to express my sincere gratitude to the Max-Planck-Institute
for the Physics of Complex Systems in Dresden, in particular to Profs.\
P.Fulde, J.-M.Rost and F.J\"ulicher, for providing ongoing support of my
research, excellent working conditions and a fantastic working
atmosphere. This review was written while I was a {\em distinguished
postdoctoral fellow} at this institute. It is planned to be part of a
habilitation thesis at the Technical University of Dresden.


\newpage

\begin{thebibliography}{Mae03b}

\bibitem[Abr94]{AbMa94}
R.~Abraham and J.E. Marsden, {\em Foundations of mechanics}.
\newblock (Addison-Wesley, Reading, 1994).

\bibitem[Ald63]{AHW63}
B.J. Alder, W.G. Hoover and T.E. Wainwright, {\em Cooperative motion of hard
  disks leading to melting}.
\newblock Phys. Rev. Lett. {\bf 11}, 241--243 (1963).

\bibitem[All87]{AT87}
M.D. Allen and {D.J.} Tildesley, {\em Computer simulation of liquids}.
\newblock (Clarendon Press, Oxford, 1987).

\bibitem[All97]{ASY97}
K.T. Alligood, T.S. Sauer and J.A. Yorke, {\em Chaos - An introduction to
  dynamical systems}.
\newblock (Springer, New York, 1997).

\bibitem[Alo02]{ARV02}
D.~Alonso, A.~Ruiz and I.~de~Vega, {\em Polygonal billiards and transport:
  diffusion and heat conduction}.
\newblock Phys. Rev. E {\bf 66}, 066131--1/15 (2002).

\bibitem[And80]{And79}
H.C. Andersen, {\em Molecular dynamics simulations at constant pressure and/or
  temperature}.
\newblock J. Chem. Phys. {\bf 72}, 2384--3173 (1980).

\bibitem[And82]{And82}
L.~Andrey, {\em Ideal gas in empty space}.
\newblock Il Nuovo Cimento {\bf 69B}, 136--144 (1982).

\bibitem[And85]{And85}
L.~Andrey, {\em The rate of entropy change in non-Hamiltonian systems}.
\newblock Phys. Lett. {\bf 111A}, 45--46 (1985).

\bibitem[And86]{And86}
L.~Andrey, {\em Note concerning the paper ``The rate of entropy change in
  non-Hamiltonian systems''}.
\newblock Phys. Lett. {\bf 114A}, 183--184 (1986).

\bibitem[Aok02]{AK02}
K.~Aoki and D.~Kusnezov, {\em Lyapunov exponents, transport and the extensivity
  of dimensional loss}.
\newblock preprint arXiv.nlin.CD/0204015, 2002.

\bibitem[Arn68]{ArAv68}
V.I. Arnold and A.~Avez, {\em Ergodic problems of classical mechanics}.
\newblock (W.A. Benjamin, New York, 1968).

\bibitem[Asp98]{AGZ98}
T.~Aspelmeier, G.~Giese and A.~Zippelius, {\em Cooling dynamics of a dilute gas
  of inelastic rods: a many-particle simulation}.
\newblock Phys. Rev. E {\bf 57}, 857--865 (1998).

\bibitem[Ban92]{BBCDS92}
J.~Banks, J.~Brooks, G.~Cairns, G.~Davis and P.~Stacey, {\em On Devaney's
  definition of chaos}.
\newblock Am. Math. Monthly {\bf 99}, 332--334 (1992).

\bibitem[Bar93]{BarEC}
A.~Baranyai, {D.J.} Evans and {E.G.D.} Cohen, {\em Field-dependent conductivity
  and diffusion in a two-dimendional {L}orentz Gas}.
\newblock J. Stat. Phys. {\bf 70}, 1085--1098 (1993).

\bibitem[Bar03]{BLL03}
E.J. Barth, B.B. Laird and B.J. Leimkuhler, {\em Generating generalized
  distributions from dynamical simulation}.
\newblock J. Chem. Phys. {\bf 118}, 5759--5768 (2003).

\bibitem[Bec87]{BeRo87}
C.~Beck and G.~Roepstorff, {\em From dynamical systems to the {L}angevin
  equation.}
\newblock Physica A {\bf 145}, 1--14 (1987).

\bibitem[Bec93]{Beck}
Chr. Beck and F.~Schl\"{o}gl, {\em Thermodynamics of Chaotic Systems}, volume~4
  of Cambridge nonlinear science series.
\newblock (Cambridge University Press, Cambridge, 1993).

\bibitem[Bec95]{Bec95}
C.~Beck, {\em From the {P}erron-{F}robenius equation to the {F}okker-{P}lanck
  equation.}
\newblock J. Stat. Phys. {\bf 79}, 875--94 (1995).

\bibitem[Bec96]{Bec96}
C.~Beck, {\em Dynamical systems of {L}angevin type.}
\newblock Physica A {\bf 233}, 419--440 (1996).

\bibitem[Ben80]{BGGS80b}
G.~Benettin, L.~Galgani, A.~Giorgili and J.-M. Strelcyn, {\em Lyapunov
  characteristic exponents for smooth dynamical systems: A method for computing
  all of them. Part 2: numerical application}.
\newblock Meccanica {\bf 15}, 21--29 (1980).

\bibitem[Ben01]{BR01}
G.~Benettin and L.~Rondoni, {\em A new model for the transport of particles in
  a thermostatted system}.
\newblock Mathematical Physics Electronic Journal {\bf 7}, 1--22 (2001).

\bibitem[Bon97]{BGG97}
F.~Bonetto, G.~Gallavotti and P.L. Garrido, {\em Chaotic principle: an
  experimental test}.
\newblock Physica D {\bf 105}, 226--252 (1997).

\bibitem[Bon98a]{BCL98}
F.~Bonetto, N.I. Chernov and J.L. Lebowitz, {\em (Global and local)
  Fluctuations of phase space contraction in deterministic stationary
  nonequilibrium}.
\newblock Chaos {\bf 8}, 823--833 (1998).

\bibitem[Bon98b]{BCP98}
F.~Bonetto, {E.G.D.} Cohen and C.~Pugh, {\em On the validity of the conjugate
  pairing rule for Lyapunov exponents}.
\newblock J. Stat. Phys. {\bf 92}, 587--627 (1998).

\bibitem[Bon99]{BLL99}
S.D. Bond, B.J. Leimkuhler and B.B. Laird, {\em The Nos\'e-Poincar\'e method
  for constant temperature molecular dynamics}.
\newblock J. Comput. Phys. {\bf 151}, 114--134 (1999).

\bibitem[Bon00]{BDL00}
F.~Bonetto, D.~Daems and J.L. Lebowitz, {\em Properties of stationary
  nonequilibrium states in the thermostatted periodic {L}orentz gas. {I}. {T}he
  one particle system}.
\newblock J. Stat. Phys. {\bf 101}, 35--60 (2000).

\bibitem[Bon01]{BL01}
F.~Bonetto and J.L. Lebowitz, {\em Thermodynamic entropy production fluctuation
  in a two-dimensional shear flow model}.
\newblock Phys. Rev. E {\bf 64}, 056129/1--9 (2001).

\bibitem[Bon02]{BDL+02}
F.~Bonetto, D.~Daems, J.L. Lebowitz and V.~Ricci, {\em Properties of stationary
  nonequilibrium states in the thermostatted periodic {L}orentz gas: {T}he
  multiparticle system}.
\newblock Phys. Rev. E {\bf 65}, 051204/1--9 (2002).

\bibitem[Bra00a]{Bran00}
A.C. Branka, {\em Nos\'e-Hoover chain method for nonequilibrium molecular
  dynamics simulation}.
\newblock Phys. Rev. E {\bf 61}, 4769--4773 (2000).

\bibitem[Bra00b]{BrWo00}
A.C. Branka and K.W. Wojciechowski, {\em Generalization of Nos\'e and
  Nos\'e-Hoover isothermal dynamics}.
\newblock Phys. Rev. E {\bf 62}, 3281--3292 (2000).

\bibitem[Bre96]{BrTV96}
W.G. Breymann, T.~T\'el and J.~Vollmer, {\em Entropy production for open
  dynamical systems}.
\newblock Phys. Rev. Lett. {\bf 77}, 2945--2948 (1996).

\bibitem[Bre98]{BTV98}
W.~Breymann, T.~T\'el and J.~Vollmer, {\em Entropy balance, time reversibility,
  and mass transport in dynamical systems}.
\newblock Chaos {\bf 8}, 396--408 (1998).

\bibitem[Bri95]{Bric95}
J.~Bricmont, {\em Science of chaos or chaos in science?}
\newblock Phys. Mag. {\bf 17}, 159--212 (1995).

\bibitem[Bri96]{BSHP96}
N.V. Brillantov, F.~Spahn, J.-M. Hertzsch and Th. P\"oschel, {\em Model for
  collisions in granular gases}.
\newblock Phys. Rev. E {\bf 53}, 5382--5392 (1996).

\bibitem[Bri01]{BSF+01}
M.E. Briggs, J.V. Sengers, M.K. Francis, P.~Gaspard, R.W. Gammon, {J.R.}
  Dorfman and R.V. Calabrese, {\em Tracking a colloidal particle for the
  measurement of dynamic entropies.}
\newblock Physica A {\bf 296}, 42--59 (2001).

\bibitem[Bru76]{Bru76}
St.G. Brush, {\em The kind of motion we call heat}, volume 1/2.
\newblock (North Holland, Amsterdam, 1976).

\bibitem[Bul90a]{BuKu90}
A.~Bulgac and D.~Kusnezov, {\em Canonical ensemble averages from
  pseudomicrocanonical dynamics}.
\newblock Phys. Rev. A {\bf 42}, 5045--5048 (1990).

\bibitem[Bul90b]{BuKu90b}
A~Bulgac and D.~Kusnezov, {\em Deterministic and time-reversal invariant
  description of Brownian motion}.
\newblock Phys. Lett. A {\bf 151}, 122--128 (1990).

\bibitem[Bun80]{BuSi80a}
L.A. Bunimovich and Ya.G. Sinai, {\em Markov partitions for dispersed
  billards}.
\newblock Comm. Math. Phys. {\bf 78}, 247--280 (1980).

\bibitem[Bun81]{BuSi80b}
L.A. Bunimovich and Ya.G. Sinai, {\em Statistical properties of {L}orentz gas
  with periodic configuration of scatterers}.
\newblock Comm. Math. Phys. {\bf 78}, 479--497 (1981).

\bibitem[Bun91]{BuSiCh91}
L.A. Bunimovich, Ya.G. Sinai and N.I. Chernov, {\em Statistical properties of
  two-dimensional hyperbolic billiards}.
\newblock Russ. Math. Surveys {\bf 46}, 47--106 (1991).

\bibitem[Bun96]{BuSp}
L.A. Bunimovich and H.~Spohn, {\em Viscosity for a periodic two disk fluid: an
  existence proof}.
\newblock Commun. Math. Phys. {\bf 176}, 661--680 (1996).

\bibitem[Bun03]{BuKh03}
L.A. Bunimovich and M.A. Khlabystova, {\em One-dimensional Lorentz gas with
  rotating scatterers: exact solutions}.
\newblock J. Stat. Phys. {\bf 112}, 1207--1218 (2003).

\bibitem[Cec03]{CFVN02}
F.~Cecconi, M.~Falcioni, A.~Vulpiani and D.~del Castillo-Negrete, {\em The
  origin of diffusion: the case of non chaotic systems}.
\newblock Physica D {\bf 180}, 129--139 (2003).

\bibitem[Cen00]{CFOKV00}
M.~Cencini, M.~Falcioni, E.~Olbrich, H.~Kantz and A.~Vulpiani, {\em Chaos or
  noise: {D}ifficulties of a distinction.}
\newblock Phys. Rev. E {\bf 62}, 427--437 (2000).

\bibitem[Cha70]{ChCo}
S.~Chapman and T.G. Cowling, {\em The mathematical theory of non-uniform
  gases}.
\newblock (Cambridge University Press, Cambridge, 1970).

\bibitem[Che93a]{Ch1}
N.I. Chernov, G.L. Eyink, J.L. Lebowitz and Ya.G. Sinai, {\em Derivation of
  {O}hm's law in a deterministic mechanical model}.
\newblock Phys. Rev. Lett. {\bf 70}, 2209--2212 (1993).

\bibitem[Che93b]{Ch2}
N.I. Chernov, G.L. Eyink, J.L. Lebowitz and Ya.G. Sinai, {\em Steady-state
  electrical conduction in the periodic {L}orentz gas}.
\newblock Comm. Math. Phys. {\bf 154}, 569--601 (1993).

\bibitem[Che95]{ChLe95}
N.I. Chernov and J.L. Lebowitz, {\em Stationary shear flow in boundary driven
  {H}amiltonian systems}.
\newblock Phys. Rev. Lett. {\bf 75}, 2831--2834 (1995).

\bibitem[Che97]{ChLe97}
N.I. Chernov and J.L. Lebowitz, {\em Stationary nonequilibrium states in
  boundary-driven Hamiltonian systems: Shear flow}.
\newblock J. Stat. Phys. {\bf 86}, 953--990 (1997).

\bibitem[Che99]{Chern99}
N.I. Chernov, {\em Decay of correlations and dispersing billiards}.
\newblock J. Stat. Phys. {\bf 94}, 513--556 (1999).

\bibitem[Che02]{CGKRT02}
A.V. Chechkin, V.Yu. Gonchar, J.~Klafter, R.~Metzler and L.V. Tanatarov, {\em
  Stationary states of non-linear oscillators driven by L\'evy noise}.
\newblock Chem. Phys. {\bf 284}, 233--251 (2002).

\bibitem[Che03]{CKGMT03}
A.V. Chechkin, J.~Klafter, V.Yu. Gonchar, R.~Metzler and L.V. Tanatarov, {\em
  Bifurcation, bimodality, and finite variance in confied L\'evy flights}.
\newblock Phys. Rev. E {\bf 67}, 010102(R)/1--4 (2003).

\bibitem[Cho98]{Choq98}
Ph. Choquard, {\em Variational principles for thermostatted systems}.
\newblock Chaos {\bf 8}, 350--357 (1998).

\bibitem[Cic80]{CiTe80}
G.~Ciccotti and A.~Tenenbaum, {\em Canonical ensemble and nonequilibrium states
  by molecular dynamics}.
\newblock J. Stat. Phys. {\bf 23}, 767--772 (1980).

\bibitem[Cil98]{CiLa98}
S.~Ciliberto and C.~Laroche, {\em An experimental test of the Gallavotti-Cohen
  fluctuation theorem}.
\newblock J. Phys. IV France {\bf 8}, 215--219 (1998).

\bibitem[Cla02]{ClGa02}
I.~Claus and P.~Gaspard, {\em The fractality of the relaxation modes in
  reaction-diffusion systems}.
\newblock Physica D {\bf 168--169}, 266--291 (2002).

\bibitem[Coh92]{Coh92}
{E.G.D.} Cohen et~al., {\em Round table discussion (II): Irreversibility and
  {L}yapunov spectra}.
\newblock in: M.~Mareschal and B.L. Holian, Eds., Microscopic simulations of
  complex hydrodynamic phenomena, volume 292 of NATO ASI Series B: Physics,
  pages 327--343, Plenum Press, New York, 1992.

\bibitem[Coh95]{Coh95}
{E.G.D.} Cohen, {\em Transport coefficients and {L}yapunov exponents}.
\newblock Physica A {\bf 213}, 293--314 (1995).

\bibitem[Coh97]{Coh97}
{E.G.D.} Cohen, {\em Dynamical ensembles in statistical mechanics}.
\newblock Physica A {\bf 240}, 43--53 (1997).

\bibitem[Coh98]{CoRo98}
{E.G.D.} Cohen and L.~Rondoni, {\em Note on phase space contraction and entropy
  production in thermostatted Hamiltonian systems}.
\newblock Chaos {\bf 8}, 357--365 (1998).

\bibitem[Coh99]{CG99}
E.G.D. Cohen and G.~Gallavotti, {\em Note on two theorems in nonequilibrium
  statistical mechanics}.
\newblock J. Stat. Phys. {\bf 96}, 1343--1349 (1999).

\bibitem[Coh02]{CR02}
E.G.D. Cohen and L.~Rondoni, {\em Particles, maps and irreversible
  thermodynamics}.
\newblock Physica A {\bf 306}, 117--128 (2002).

\bibitem[Cvi92]{CvGS92}
P.~Cvitanovi{\'c}, P.~Gaspard and Th. Schreiber, {\em Investigation of the
  {L}orentz gas in terms of periodic orbits}.
\newblock Chaos {\bf 2}, 85--90 (1992).

\bibitem[Cvi95]{CEG91}
P.~Cvitanovi\'{c}, J.-P. Eckmann and P.~Gaspard, {\em Transport properties of
  the {L}orentz gas in terms of periodic orbits}.
\newblock Chaos, Solitons and Fractals {\bf 6}, 113--120 (1995).

\bibitem[Cvi03]{CAMTV01}
P.~Cvitanovi{\'c}, R.~Artuso, R.~Mainieri, G.~Tanner and G.~Vattay, {\em Chaos:
  Classical and Quantum}.
\newblock (Niels Bohr Institute, Copenhagen, 2003).
\newblock webbook under www.nbi.dk/ChaosBook/.

\bibitem[Dae99]{DaNi99}
D.~Daems and G.~Nicolis, {\em Entropy production and phase space volume
  contraction}.
\newblock Phys. Rev. E {\bf 59}, 4000--4006 (1999).

\bibitem[Del95a]{DeGP95}
C.~Dellago, L.~Glatz and H.A. Posch, {\em Lyapunov spectrum of the driven
  {L}orentz gas}.
\newblock Phys. Rev. E {\bf 52}, 4817--4826 (1995).

\bibitem[Del95b]{DGP95}
C.~Dellago, L.~Glatz and H.A. Posch, {\em Lyapunov spectrum of the driven
  {L}orentz gas}.
\newblock Phys. Rev. E {\bf 52}, 4817--4826 (1995).

\bibitem[Del96a]{DePo96}
C.~Dellago and H.A. Posch, {\em Lyapunov instability, local curvature, and the
  fluid-solid phase transition in two-dimensional particle systems}.
\newblock Physica A {\bf 230}, 364--387 (1996).

\bibitem[Del96b]{DePH96}
C.~Dellago, H.A. Posch and W.G. Hoover, {\em Lyapunov instability in a system
  of hard disks in equilibrium and nonequilibrium steady states}.
\newblock Phys. Rev. E {\bf 53}, 1485--1501 (1996).

\bibitem[Del97a]{DePo97b}
C.~Dellago and H.A. Posch, {\em Lyapunov instability of the boundary-driven
  Chernov-Lebowitz model for stationary shear flow}.
\newblock J. Stat. Phys. {\bf 88}, 825--842 (1997).

\bibitem[Del97b]{DePo97}
C.~Dellago and H.A. Posch, {\em Lyapunov spectrum and the conjugate pairing
  rule for a thermostatted random {L}orentz gas: Numerical simulations}.
\newblock Phys. Rev. Lett. {\bf 78}, 211--214 (1997).

\bibitem[Det95]{DMR95}
{C.P.} Dettmann, {G.P.} Morriss and L.~Rondoni, {\em Conjugate pairing in the
  three-dimensional periodic Lorentz gas}.
\newblock Phys. Rev. E {\bf 52}, R5746--R5748 (1995).

\bibitem[Det96a]{DeMo96a}
{C.P.} Dettmann and {G.P.} Morriss, {\em Crisis in the periodic Lorentz gas}.
\newblock Phys. Rev. E {\bf 54}, 4782--4790 (1996).

\bibitem[Det96b]{DeMo96}
{C.P.} Dettmann and {G.P.} Morriss, {\em Hamiltonian formulation of the
  Gaussian isokinetic thermostat}.
\newblock Phys. Rev. E {\bf 54}, 2495--2500 (1996).

\bibitem[Det96c]{DeMo96b}
C.P. Dettmann and G.P. Morriss, {\em Proof of Lyapunov exponent pairing for
  systems at constant kinetic energy}.
\newblock Phys. Rev. E {\bf 53}, R5545--R5548 (1996).

\bibitem[Det97a]{DeMo97a}
{C.P.} Dettmann and {G.P.} Morriss, {\em Hamiltonian reformulation and pairing
  of {L}yapunov exponents for Nose-Hoover dynamics}.
\newblock Phys. Rev. E {\bf 55}, 3693--3696 (1997).

\bibitem[Det97b]{DeMo97b}
{C.P.} Dettmann and {G.P.} Morriss, {\em Stability ordering of cycle
  expansions}.
\newblock Phys. Rev. Lett. {\bf 78}, 4201--4204 (1997).

\bibitem[Det97c]{DMR97}
{C.P.} Dettmann, {G.P.} Morriss and L.~Rondoni, {\em Irreversibility, diffusion
  and multifractal measures in thermostatted systems}.
\newblock Chaos, Solitons and Fractals {\bf 8}, 783--792 (1997).

\bibitem[Det99a]{Dett99}
C.P. Dettmann, {\em Hamiltonian for a restricted isoenergetic thermostat}.
\newblock Phys. Rev. E {\bf 60}, 7576--7577 (1999).

\bibitem[Det99b]{DC99}
{C.P.} Dettmann and {E.G.D.} Cohen, {\em Microscopic chaos from Brownian
  motion?}
\newblock Nature {\bf 401}, 875--875 (1999).

\bibitem[Det00a]{DettS00}
C.P. Dettmann, {\em The Lorentz gas: a paradigm for nonequilibrium steady
  states}.
\newblock in: Hard Ball Systems and the Lorentz Gas, pages 315--366. (2000).
\newblock see Ref.\ \cite{Sza00}.

\bibitem[Det00b]{DC00}
{C.P.} Dettmann and {E.G.D.} Cohen, {\em Microscopic chaos and diffusion}.
\newblock J. Stat. Phys. {\bf 101}, 775--817 (2000).

\bibitem[Dev89]{Dev89}
R.L. Devaney, {\em An introduction to chaotic dynamical systems}.
\newblock (Addison-Wesley, Reading, second edition, 1989).

\bibitem[dG62]{deGM84}
S.R. de~Groot and P.~Mazur, {\em Non-equilibrium thermodynamics}.
\newblock (North Holland, Amsterdam, 1962).
\newblock reprinted by Dover, New York, 1984.

\bibitem[Die03]{DDPKS03}
P.~Dieterich, V.~Dreval, R.~Preuss, R.~Klages and A.~Schwab, {\em
  Classification of the migratory process on different time-scales - evaluation
  of the Na+H+ exchanger in MDCK-F cell migration}.
\newblock in preparation, 2003.

\bibitem[Dob76a]{Dobb76a}
R.~Dobbertin, {\em On functional relations between reduced distribution
  functions and entropy production by non-Hamiltonian perturbations}.
\newblock Physica Scripta {\bf 14}, 85--88 (1976).

\bibitem[Dob76b]{Dobb76b}
R.~Dobbertin, {\em Vlasov equation and entropy}.
\newblock Physica Scripta {\bf 14}, 89--91 (1976).

\bibitem[Dol00]{DK00}
M.~Dolowschiak and Z.~Kovacs, {\em Breaking conjugate pairing in thermostated
  billiards by a magnetic field.}
\newblock Phys. Rev. E {\bf 62}, 7894--7897 (2000).

\bibitem[Dor95]{DoGa95}
{J.R.} Dorfman and P.~Gaspard, {\em Chaotic scattering theory of transport and
  reaction-rate coefficients}.
\newblock Phys. Rev. E {\bf 51}, 28--35 (1995).

\bibitem[Dor99]{Do99}
{J.R.} Dorfman, {\em An Introduction to Chaos in Nonequilibrium Statistical
  Mechanics}.
\newblock (Cambridge University Press, Cambridge, 1999).

\bibitem[Dor02]{DGG02}
J.R. Dorfman, P.~Gaspard and T.~Gilbert, {\em Entropy production of diffusion
  in spatially periodic deterministic systems}.
\newblock Phys. Rev. E {\bf 66}, 026110/1--9 (2002).

\bibitem[Dre88]{Dress88}
U.~Dressler, {\em Symmetry property of the Lyapunov spectra of a class of
  dissipative dynamical systems with viscous damping}.
\newblock Phys. Rev. A {\bf 38}, 2103--2109 (1988).

\bibitem[Dru00]{Drude00}
P.~Drude, {\em Zur Elektronentheorie der Metalle}.
\newblock Ann. Phys. {\bf 1}, 566--612 (1900).

\bibitem[Ebe99]{EST99}
W.~Ebeling, F.~Schweitzer and B.~Tilch, {\em Active Brownian particles with
  energy depots modeling animal mobility}.
\newblock BioSystems {\bf 49}, 17--29 (1999).

\bibitem[Ebe00]{Ebe00}
W.~Ebeling, {\em Canonical nonequililbrium statistics and applications to
  {F}ermi-{B}ose systems}.
\newblock Cond. Matt. Phys. {\bf 3}, 285--293 (2000).

\bibitem[Ebe01]{ES01}
W.~Ebeling and F.~Schweitzer, {\em Swarms of particle agents with harmonic
  interactions}.
\newblock Theory in Biosciences {\bf 120}, 207--224 (2001).

\bibitem[Ebe03]{EbRo02}
W.~Ebeling and G.~R\"opke, {\em Statistical mechanics of confined systems with
  rotational excitations}.
\newblock in print for the Physica D Special Issue Ref.\ \cite{chaotr03}, 2003.

\bibitem[Eck85]{ER}
J.-P. Eckmann and D.~Ruelle, {\em Ergodic theory of chaos and strange
  attractors}.
\newblock Rev. Mod. Phys. {\bf 57}, 617--656 (1985).

\bibitem[Eck99a]{EPRB99b}
J.-P. Eckmann, C.-A. Pillet and L.~Rey-Bellet, {\em Entropy production in
  non-linear, thermally driven Hamilatonian systems}.
\newblock J. Stat. Phys. {\bf 95}, 305--331 (1999).

\bibitem[Eck99b]{EPRB99}
J.-P. Eckmann, C.-A. Pillet and L.~Rey-Bellet, {\em Non-equilibrium statistical
  mechanics of anharmonic chains coupled to two heat baths at different
  temperatures}.
\newblock Commun. Math. Phys. {\bf 201}, 657--697 (1999).

\bibitem[Eck00]{EG00}
J.-P. Eckmann and O.~Gat, {\em Hydrodynamic Lyapunov modes in
  translation-invariant systems}.
\newblock J. Stat. Phys. {\bf 3/4}, 775--797 (2000).

\bibitem[Erd00]{EES+00}
U.~Erdmann, W.~Ebeling, L.~Schimansky-Geier and F.~Schweitzer, {\em Brownian
  particles far from equilibrium}.
\newblock Eur. Phys. J. B {\bf 15}, 105--113 (2000).

\bibitem[Erd02]{EEA02}
U.~Erdmann, W.~Ebeling and V.S. Anishchenko, {\em Excitation of rotational
  modes in two-dimensional systems of driven Brownian particles}.
\newblock Phys. Rev. E {\bf 65}, 061106--1/9 (2002).

\bibitem[Eva83a]{Ev83}
{D.J.} Evans, {\em Computer ``experiment'' for nonlinear thermodynamics of
  {C}ouette flow}.
\newblock J. Chem. Phys. {\bf 78}, 3297--3302 (1983).

\bibitem[Eva83b]{EH83}
{D.J.} Evans, W.G. Hoover, B.H. Failor, B.~Moran and A.J.C. Ladd, {\em
  Nonequilibrium molecular dynamcis via {G}auss's principle of least
  constraint}.
\newblock Phys. Rev. A {\bf 28}, 1016--1021 (1983).

\bibitem[Eva85a]{Ev85}
D.J. Evans, {\em Response theory as a free-energy extremum}.
\newblock Phys. Rev. A {\bf 32}, 2923--2925 (1985).

\bibitem[Eva85b]{EvHo85}
D.J. Evans and B.L. Holian, {\em The Nos\'e-Hoover thermostat}.
\newblock J. Chem. Phys. {\bf 83}, 4069--4074 (1985).

\bibitem[Eva86]{EvMo86}
D.J. Evans and {G.P.} Morriss, {\em Shear Thickening and Turbulence in Simple
  Fluids}.
\newblock Phys. Rev. Lett. {\bf 56}, 2172--2175 (1986).

\bibitem[Eva90a]{ECM}
{D.J.} Evans, {E.G.D.} Cohen and {G.P.} Morriss, {\em Viscosity of a simple
  fluid from its maximal {L}yapunov exponents}.
\newblock Phys. Rev. A {\bf 42}, 5990--5997 (1990).

\bibitem[Eva90b]{EvMo90}
{D.J.} Evans and {G.P.} Morriss, {\em Statistical Mechanics of Nonequilibrium
  Liquids}.
\newblock Theoretical Chemistry. (Academic Press, London, 1990).

\bibitem[Eva93a]{ECM93}
{D.J.} Evans, {E.G.D.} Cohen and {G.P.} Morriss, {\em Probability of second law
  violations in shearing steady flows}.
\newblock Phys. Rev. Lett. {\bf 71}, 2401--2404 (1993).

\bibitem[Eva93b]{EvSa93}
D.J. Evans and S.S. Sarman, {\em Equivalence of thermostatted nonlinear
  response}.
\newblock Phys. Rev. E {\bf 48}, 65--70 (1993).

\bibitem[Eva98]{ESHH98}
D.J. Evans, D.J. Searles, W.G. Hoover, B.L. Holian, H.A. Posch and G.P.
  Morriss, {\em Comment on ``Modified nonequilibrium molecular dynamics for
  fluid flows with energy conservation''}.
\newblock J. Chem. Phys. {\bf 108}, 4351--4352 (1998).

\bibitem[Eva00]{ECS+00}
D.J. Evans, E.G.D. Cohen, D.J. Searles and F.~Bonetto, {\em Note on the
  {K}aplan-{Y}orke dimension and linear transport coefficients}.
\newblock J. Stat. Phys. {\bf 101}, 17--34 (2000).

\bibitem[Eva02a]{ER02}
D.J. Evans and L.~Rondoni, {\em Comments on the entropy of nonequilibrium
  steady states}.
\newblock J. Stat. Phys. {\bf 109}, 895--920 (2002).

\bibitem[Eva02b]{EvSe02}
D.J. Evans and D.J. Searles, {\em The Fluctuation Theorem}.
\newblock Advances in Physics {\bf 51}, 1529--1585 (2002).

\bibitem[Eyi92]{EyLe92}
G.L. Eyink and J.L. Lebowitz, {\em Generalized Gaussian dynamics, phase-space
  reduction and irreversibility: a comment}.
\newblock in: M.~Mareschal and B.L. Holian, Eds., Microscopic simulations of
  complex hydrodynamic phenomena, volume 292 of NATO ASI Series B: Physics,
  pages 323--326, Plenum Press, New York, 1992.

\bibitem[Fle92]{FGK92}
R.~Fleischmann, T.~Geisel and R.~Ketzmerick, {\em Magnetoresistance due to
  chaos and nonlinear resonance in lateral surface superlattices}.
\newblock Phys. Rev. Lett. {\bf 68}, 1367--1370 (1992).

\bibitem[For87]{FK87}
G.W. Ford and M.~Kac, {\em On the quantum Langevin equation}.
\newblock J. Stat. Phys. {\bf 46}, 803--810 (1987).

\bibitem[For03]{FHPH03}
C.~Forster, R.~Hirschl, H.A. Posch and W.G. Hoover, {\em Perturbed phase-space
  dynamics of hard-disk fluids}.
\newblock in print for the Physica D Special Issue Ref.\ \cite{chaotr03}, 2003.

\bibitem[Fra90]{FrGr90}
K.~Franke and H.~Gruler, {\em Galvanotaxis of human granulocytes: electric
  field jump studies}.
\newblock Europ. Biophys. J. {\bf 18}, 335--346 (1990).

\bibitem[Gal77]{Galt77}
F.~Galton, {\em Typical laws of heredity}.
\newblock Nature {\bf 15}, 492--495, 512--514, 532--533 (1877).

\bibitem[Gal95a]{GaCo95a}
G.~Gallavotti and {E.G.D.} Cohen, {\em Dynamical ensembles in nonequilibrium
  statistical mechanics}.
\newblock Phys. Rev. Lett. {\bf 74}, 2694--2697 (1995).

\bibitem[Gal95b]{GaCo95b}
G.~Gallavotti and {E.G.D.} Cohen, {\em Dynamical ensembles in stationary
  states}.
\newblock J. Stat. Phys. {\bf 80}, 931--970 (1995).

\bibitem[Gal96a]{Gall96b}
G.~Gallavotti, {\em Equivalence of dynamical ensembles and Navier-Stokes
  equations}.
\newblock Phys. Lett. A {\bf 223}, 91--95 (1996).

\bibitem[Gal96b]{Gall96}
G.~Gallavotti, {\em Extension of Onsager's reciprocity to large fields and the
  chaotic hypothesis}.
\newblock Phys. Rev. Lett. {\bf 77}, 4334--4337 (1996).

\bibitem[Gal97]{Gall97}
G.~Gallavotti, {\em Dynamical ensembles equivalence in fluid mechanics}.
\newblock Physica D {\bf 105}, 163--184 (1997).

\bibitem[Gal98]{Gall98}
G.~Gallavotti, {\em Chaotic dynamics, fluctuations, nonequilibrium ensembles}.
\newblock Chaos {\bf 8}, 384--393 (1998).

\bibitem[Gal99]{Gall99}
G.~Gallavotti, {\em Statistical mechanics - a short treatise}.
\newblock (Springer, Berlin, 1999).
\newblock Chapter 9.

\bibitem[Gal03]{Gall03b}
G.~Gallavotti, {\em Nonequilibrium thermodynamics?}
\newblock preprint cond-mat/0301172, 2003.

\bibitem[Gar02]{GaOl02}
P.~Garbaczewski and R.~Olkiewicz, Eds.
\newblock {\em Dynamics of dissipation}, volume 597 of Lecture Notes in
  Physics, Springer, Berlin, 2002.

\bibitem[Gas71]{G78}
D.~Gass, {\em Enskog theory for a rigid disk fluid}.
\newblock J. Chem. Phys. {\bf 54}, 1898--1902 (1971).

\bibitem[Gas90]{GN}
P.~Gaspard and G.~Nicolis, {\em Transport properties, {L}yapunov exponents, and
  entropy per unit time}.
\newblock Phys. Rev. Lett. {\bf 65}, 1693--1696 (1990).

\bibitem[Gas92a]{PG1}
P.~Gaspard, {\em Diffusion, effusion, and chaotic scattering}.
\newblock J. Stat. Phys. {\bf 68}, 673--747 (1992).

\bibitem[Gas92b]{GB1}
P.~Gaspard and F.~Baras, {\em Dynamical Chaos underlying diffusion in the
  {L}orentz Gas}.
\newblock in: M.~Mareschal and B.L. Holian, Eds., Microscopic simulations of
  complex hydrodynamic phenomena, volume 292 of NATO ASI Series B: Physics,
  pages 301--322, Plenum Press, New York, 1992.

\bibitem[Gas93]{Gas93}
P.~Gaspard, {\em What is the role of chaotic scattering in irreversible
  processes?}
\newblock Chaos {\bf 3}, 427--442 (1993).

\bibitem[Gas95a]{GaBa95}
P.~Gaspard and F.~Baras, {\em Chaotic scattering and diffusion in the Lorentz
  gas}.
\newblock Phys. Rev. E {\bf 51}, 5332--5352 (1995).

\bibitem[Gas95b]{GaDo95}
P.~Gaspard and {J.R.} Dorfman, {\em Chaotic scattering theory, thermodynamic
  formalism, and transport coefficients}.
\newblock Phys. Rev. E {\bf 52}, 3525--3552 (1995).

\bibitem[Gas96]{Gasp96}
P.~Gaspard, {\em Hydrodynamic modes as singular eigenstates of the
  {L}iouvillian dynamics: Deterministic diffusion}.
\newblock Phys. Rev. E {\bf 53}, 4379--4401 (1996).

\bibitem[Gas97]{Gasp97a}
P.~Gaspard, {\em Entropy production in open volume-preserving systems}.
\newblock J. Stat. Phys. {\bf 88}, 1215--1240 (1997).

\bibitem[Gas98a]{Gasp}
P.~Gaspard, {\em Chaos, Scattering, and Statistical Mechanics}.
\newblock (Cambridge University Press, Cambridge, 1998).

\bibitem[Gas98b]{GBFS+98}
P.~Gaspard, M.E. Briggs, M.K. Francis, J.V. Sengers, R.W. Gammons, {J.R.}
  Dorfman and R.V. Calabrese, {\em Experimental evidence for microscopic
  chaos}.
\newblock Nature {\bf 394}, 865--868 (1998).

\bibitem[Gas98c]{GaKl}
P.~Gaspard and R.~Klages, {\em Chaotic and fractal properties of deterministic
  diffusion-reaction processes}.
\newblock Chaos {\bf 8}, 409--423 (1998).

\bibitem[Gas01]{GCGD01}
P.~Gaspard, I.~Claus, T.~Gilbert and {J.R.} Dorfman, {\em Fractality of the
  hydrodynamic modes of diffusion}.
\newblock Phys. Rev. Lett. {\bf 86}, 1506--1509 (2001).

\bibitem[Gas02]{Gasp02}
P.~Gaspard, {\em Dynamical theory of relaxation in clasical and quantum
  systems}.
\newblock in: Dynamics of dissipation, pages 111--163. (2002).
\newblock see Ref.\ \cite{GaOl02}.

\bibitem[Gas03]{GND03}
P.~Gaspard, G.~Nicolis and J.R. Dorfman, {\em Diffusive Lorentz gases and
  multibaker maps are compatible with irreversible thermodynamics}.
\newblock Physica A {\bf 323}, 294--322 (2003).

\bibitem[Gei90]{Geis90}
T.~Geisel, J.~Wagenhuber, P.~Niebauer and G.~Obermair, {\em Chaotic dynamics of
  ballistic electrons in lateral superlattices and magnetic fields}.
\newblock Phys. Rev. Lett. {\bf 64}, 1581--1584 (1990).

\bibitem[Ger73]{Gerl73}
G.~Gerlich, {\em Die verallgemeinerte Liouville-Gleichung}.
\newblock Physica {\bf 69}, 458--466 (1973).

\bibitem[Ger99]{stat99}
A.~Gervois, D.~Iagolnitzer, M.~Moreau and Y.~Pomeau, Eds.
\newblock {\em Statistical Physics - invited papers from STATPHYS 20}, volume
  263 of Physica A, North Holland, Amsterdam, 1999.

\bibitem[Gil99a]{GiDo99}
T.~Gilbert and {J.R.} Dorfman, {\em Entropy production: from open
  volume-preserving to dissipative systems}.
\newblock J. Stat. Phys. {\bf 96}, 225--269 (1999).

\bibitem[Gil99b]{GFD99}
T.~Gilbert, C.D. Ferguson and {J.R.} Dorfman, {\em Field driven thermostated
  systems: A nonlinear multibaker map}.
\newblock Phys. Rev. E {\bf 59}, 364--371 (1999).

\bibitem[Gil00a]{GD00}
T.~Gilbert and J.R. Dorfman, {\em Entropy production in a persistent random
  walk}.
\newblock Physica A {\bf 282}, 427--449 (2000).

\bibitem[Gil00b]{GDG00}
T.~Gilbert, J.R. Dorfman and P.~Gaspard, {\em Entropy production, fractals, and
  relaxation to equilibrium}.
\newblock Phys. Rev. Lett. {\bf 85}, 1606--1609 (2000).

\bibitem[Gil01]{GDG01}
T.~Gilbert, {J.R.} Dorfman and P.~Gaspard, {\em Fractal dimensions of the
  hydrodynamic modes of diffusion}.
\newblock Nonlinearity {\bf 14}, 339--358 (2001).

\bibitem[Gil03]{GilDo03}
T.~Gilbert and J.R. Dorfman, {\em On the parametric dependences of a class of
  non-linear singular maps}.
\newblock to appear in Discrete and Continuous Dynamical Systems, Series B,
  2003.

\bibitem[Gle88]{Gle88}
J.~Gleick, {\em Chaos - Making a New Science}.
\newblock (Penguin, New York, 1988).

\bibitem[Gol85]{GKI85}
S.~Goldstein, C.~Kipnis and N.~Ianiro, {\em Stationary states for a mechanical
  system with stochastic boundary conditions}.
\newblock J. Stat. Phys. {\bf 41}, 915--939 (1985).

\bibitem[Gra99]{GS99}
P.~Grassberger and T.~Schreiber, {\em Microscopic chaos from Brownian motion?}
\newblock Nature {\bf 401}, 875--876 (1999).

\bibitem[Gra02]{GNY02}
P.~Grassberger, W.~Nadler and L.~Yang, {\em Heat conduction and entropy
  production in a one-dimensional hard-particle gas}.
\newblock Phys. Rev. Lett. {\bf 89}, 180601/1--4 (2002).

\bibitem[Gre84]{GOPY84}
C.~Grebogi, E.~Ott, S.~Pelikan and J.A. Yorke, {\em Strange attractors that are
  not chaotic}.
\newblock Physica D {\bf 13}, 261--268 (1984).

\bibitem[Gri89]{GrTo89}
M.~Grilli and E.~Tosatti, {\em Exact canonical averages from microcanonical
  dynamics for quantum systems}.
\newblock Phys. Rev. Lett. {\bf 62}, 2889--2892 (1989).

\bibitem[Gro02]{GrKl02}
J.~Groeneveld and R.~Klages, {\em Negative and nonlinear response in an exactly
  solved dynamical model of particle transport}.
\newblock J. Stat. Phys. {\bf 109}, 821--861 (2002).

\bibitem[Gru94]{GrBC94}
H.~Gruler and A.~de~Boisfleury-Chevance, {\em Directed cell movement and
  cluster formation: physical principles}.
\newblock J. Phys. I: France {\bf 4}, 1085--1105 (1994).

\bibitem[Guc90]{GH}
J.~Guckenheimer and P.~Holmes, {\em Nonlinear Oscillations, Dynamical Systems,
  and Bifurcations of Vector Fields}, volume~42 of Applied mathematical
  sciences.
\newblock (Springer, Berlin, 3rd edition, 1990).

\bibitem[Gup94]{GKC94}
D.~Gupalo, A.S. Kaganovich and E.G.D. Cohen, {\em Symmetry of Lyapunov
  spectrum}.
\newblock J. Stat. Phys. {\bf 74}, 1145--1159 (1994).

\bibitem[Ham90]{Ham90}
I.P. Hamilton, {\em Modified Nos\'e-Hoover equation for a one-dimensional
  oscillator: enforcement of the virial theorem}.
\newblock Phys. Rev. A {\bf 42}, 7467--7470 (1990).

\bibitem[Har94]{HLCC94}
R.S. Hartmann, K.~Lau, W.~Chou and T.D. Coates, {\em The fundamental motor of
  the human neutrophil is not random: evidence for local non-Markov movement in
  neutrophils}.
\newblock Biophys. J. {\bf 67}, 2535--2545 (1994).

\bibitem[Har01]{HaGa01}
T.~Harayama and P.~Gaspard, {\em Diffusion of particles bouncing on a
  one-dimensional periodically corrugated floor}.
\newblock Phys. Rev. E {\bf 64}, 036215/1--16 (2001).

\bibitem[Har02]{HaKlGa02}
T.~Harayama, R.~Klages and P.~Gaspard, {\em Deterministic diffusion in
  flower-shaped billiards}.
\newblock Phys. Rev. E {\bf 66}, 026211/1--7 (2002).

\bibitem[Has09]{Bo09}
F.~Hasen\"ohrl, Eds., {\em Wissenschaftliche Abhandlungen von L. Boltzmann},
  volume~2.
\newblock (J.A. Barth Verlag, Leipzig, 1909).

\bibitem[Hes90]{HeLo90}
S.~Hess and W.~Loose, {\em Flow properties and shear-induced structural changes
  in fluids: a case study on the interplay between theory, simulation,
  experiment and application}.
\newblock Ber. Bunsenges. Phys. Chem. {\bf 94}, 216--222 (1990).

\bibitem[Hes96a]{Hess96}
S.~Hess, {\em Constraints in molecular dynamics, nonequilibrium processes in
  fluids via computer simulations}.
\newblock in: K.H. Hoffmann and M.~Schreiber, Eds., Computational physics,
  pages 268--293, Springer, Berlin, 1996.

\bibitem[Hes96b]{HKL96}
S.~Hess, M.~Kr\"oger, W.~Loose, C.~Pereira Borgmeyer, R.~Schramek, H.~Voigt and
  T.~Weider, {\em Simple and complex fluids under shear}.
\newblock in: K.~Binder and G.~Ciccotti, Eds., Monte Carlo and Molecular
  Dynamics of Condensed Matter Systems, volume~49 of IPS Conf. Proc., pages
  825--841, Bologna, 1996.

\bibitem[Hes97]{HABK97}
S.~Hess, C.~Aust, L.~Bennett, M.~Kr{\"o}er, C.~Pereira-Borgmeyer and T.~Weider,
  {\em Rheology: from simple and to complex fluids}.
\newblock Physica A {\bf 240}, 126--144 (1997).

\bibitem[Hes02]{HeMo02}
S.~Hess and G.P. Morriss, {\em Regular and chaotic rotation of a polymer
  molecule subjected to a shear flow}.
\newblock preprint, 2002.

\bibitem[Hes03]{He03}
S.~Hess, {\em Construction and test of thermostats and twirlers for molecular
  rotations}.
\newblock Z. Naturforsch. {\bf 58a}, 377--391 (2003).

\bibitem[Hol86]{HoHo86}
B.L. Holian and W.G. Hoover, {\em Numerical test of the Liouville equation}.
\newblock Phys. Rev. {\bf A}, 4229--4237 (1986).

\bibitem[Hol87]{HHP87}
B.L. Holian, W.G. Hoover and H.A. Posch, {\em Resolution of {L}oschmidt's
  paradox: the origin of irreversible behavior in reversible atomistic
  dynamics}.
\newblock Phys. Rev. Lett. {\bf 59}, 10--13 (1987).

\bibitem[Hol95]{HVR95}
B.L. Holian, A.F. Voter and R.~Ravelo, {\em Thermostatted molecular dynamics:
  How to avoid the Toda demon hidden in Nos\'e-Hoover dynamics}.
\newblock Phys. Rev. E {\bf 52}, 2338--2347 (1995).

\bibitem[Hoo75]{HoAs75}
W.G. Hoover and W.T. Ashurst, {\em Nonequilibrium Molecular Dynamics}.
\newblock in: H.~Eyring and D.~Henderson, Eds., Theoretical Chemistry,
  volume~1, Academic Press, New York, 1975.

\bibitem[Hoo82]{HLM82}
W.G. Hoover, A.J.C. Ladd and B.~Moran, {\em High-strain-rate plastic flow
  studied via nonequilibrium molecular dynamics}.
\newblock Phys. Rev. Lett. {\bf 48}, 1818--1821 (1982).

\bibitem[Hoo84]{HMH84}
W.G. Hoover, B.~Moran and J.M. Haile, {\em Homogeneous periodic heat flow via
  nonequilibrium molecular dynamics}.
\newblock J. Stat. Phys. {\bf 37}, 109--121 (1984).

\bibitem[Hoo85]{Hoov85}
W.G. Hoover, {\em Canonical dynamics: equilibrium phase-space distributions}.
\newblock Phys. Rev. A {\bf 31}, 1695--1697 (1985).

\bibitem[Hoo86]{HoKr86}
W.G. Hoover and K.W. Kratky, {\em Heat conductivity of three periodic hard
  disks via nonequilibrium molecular dynamics}.
\newblock J. Stat. Phys. {\bf 42}, 1103--1114 (1986).

\bibitem[Hoo87]{HoPo87}
W.G. Hoover and H.A. Posch, {\em Direct measurement of equilibrium and
  nonequilibrium {L}yapunov spectra}.
\newblock Phys. Lett. A {\bf 123}, 227--230 (1987).

\bibitem[Hoo88a]{HMHE88}
W.G. Hoover, B.~Moran, C.G. Hoover and D.J. Evans, {\em Irreversibility in the
  Galton Board via conservative and quantum Hamiltonian and Gaussian Dynamics}.
\newblock Phys. Lett. A {\bf 133}, 114--120 (1988).

\bibitem[Hoo88b]{HTP88}
W.G. Hoover, C.G. Tull and H.A. Posch, {\em Negative {L}yapunov exponents for
  dissipative systems}.
\newblock Phys. Lett. A {\bf 131}, 211--215 (1988).

\bibitem[Hoo89a]{Ho89}
W.G. Hoover, {\em Generalization of Nos\'e's isothermal molecular dynamics:
  Non-Hamiltonian dynamics for the canonical ensemble}.
\newblock Phys. Rev. {\bf 40}, 2814--2815 (1989).

\bibitem[Hoo89b]{HoMo89}
W.G. Hoover and B.~Moran, {\em Phase-space singularities in atomistic planar
  diffusive flow}.
\newblock Phys. Rev. A {\bf 40}, 5319--5326 (1989).

\bibitem[Hoo91]{Hoo91}
W.G. Hoover, {\em Computational Statistical Mechanics}, volume~11 of Studies in
  Modern Thermodynamics.
\newblock (Elsevier, Amsterdam, 1991).

\bibitem[Hoo92]{HoMo92}
W.G. Hoover and B.~Moran, {\em Viscous attractor for the Galton board}.
\newblock Chaos {\bf 2}, 599--602 (1992).

\bibitem[Hoo94]{HoPo94}
W.G. Hoover and H.A. Posch, {\em Second-law irreversibility and phase-space
  dimensionality loss from time-reversible nonequilibrium steady-state Lyapunov
  spectra}.
\newblock Phys. Rev. E {\bf 49}, 1913--1920 (1994).

\bibitem[Hoo96a]{HoHo96}
W.G. Hoover and B.L. Holian, {\em Kinetic moments method for the canonical
  ensemble distribution}.
\newblock Phys. Lett. {\bf 211}, 253--257 (1996).

\bibitem[Hoo96b]{HKP96}
W.G. Hoover, O.~Kum and H.A. Posch, {\em Time-reversible dissipative ergodic
  maps}.
\newblock Phys. Rev. E {\bf 53}, 2123--2129 (1996).

\bibitem[Hoo97]{HoKu97}
W.G. Hoover and O.~Kum, {\em Ergodicity, mixing, and time reversibility for
  atomistic nonequilibrium steady states}.
\newblock Phys. Rev. E {\bf 56}, 5517--5523 (1997).

\bibitem[Hoo98a]{HEPHM98}
W.G. Hoover, D.J. Evans, H.A. Posch, B.L. Holian and G.P. Morriss, {\em Comment
  on "{T}oward a statistical thermodynamics of steady states"}.
\newblock Phys. Rev. Lett. {\bf 80}, 4103 (1998).

\bibitem[Hoo98b]{HP98b}
W.G. Hoover and H.A. Posch, {\em Chaos and irreversibility in simple model
  systems.}
\newblock Chaos {\bf 8}, 366--373 (1998).

\bibitem[Hoo98c]{HP98}
W.G. Hoover and H.A. Posch, {\em Multifractals from stochastic many-body
  molecular dynamics}.
\newblock Phys. Lett. A {\bf 246}, 247--251 (1998).

\bibitem[Hoo99]{HoB99}
W.G. Hoover, {\em Time Reversibility, Computer Simulation, and Chaos}.
\newblock (World Scientific, Singapore, 1999).

\bibitem[Hoo01]{HHI01}
W.G. Hoover, C.G. Hoover and D.I. Isbister, {\em Chaos, ergodic convergence,
  and fractal instability for a thermostated canonical harmonic oscillator}.
\newblock Phys. Rev. E {\bf 63}, 026209--1/5 (2001).

\bibitem[Hoo02a]{Hoo02}
W.G. Hoover, 2002.
\newblock private communication.

\bibitem[Hoo02b]{HPA+02}
W.G. Hoover, H.A. Posch, K.~Aoki and D.~Kusnezov, {\em Remarks on
  non-{H}amiltonian statistical mechanics: {L}yapunov exponents and phase-space
  dimensionality loss.}
\newblock Europhys. Lett. {\bf 60}, 337--341 (2002).

\bibitem[Hoo02c]{HPF+02}
W.G. Hoover, H.A. Posch, C.~Forster, C.~Dellago and M.~Zhou, {\em Lyapunov
  modes of two-dimensional many-body systems; soft disks, hard disks, and
  rotors.}
\newblock J. Stat. Phys. {\bf 109}, 765--776 (2002).

\bibitem[Hoo03]{HAHG03}
W.G. Hoover, K.~Aoki, C.G. Hoover and S.V.~De Groot, {\em Time-reversible
  deterministic thermostats}.
\newblock in print for the Physica D Special Issue Ref.\ \cite{chaotr03}, 2003.

\bibitem[Hop37]{Ho37}
E.~Hopf, {\em Ergodentheorie}.
\newblock (Springer, Berlin, 1937).

\bibitem[Hua87]{Huang}
K.~Huang, {\em Statistical mechanics}.
\newblock (Wiley, New York, 1987).

\bibitem[Iba96]{IbLu96}
H.~Ibach and H.~L\"uth, {\em Solid state physics}.
\newblock (Springer, Berlin, 1996).

\bibitem[Jar00]{Jar00}
C.~Jarzynski, {\em Hamiltonian derivation of a detailed fluctuation theorem}.
\newblock J. Stat. Phys. {\bf 98}, 77--102 (2000).

\bibitem[Jel88a]{Jell88}
J.~Jellinek, {\em Dynamics for nonconservative systems: ergodicity beyond the
  microcanonical ensemble}.
\newblock J. Phys. Chem. {\bf 92}, 3163--3173 (1988).

\bibitem[Jel88b]{JeBe88}
J.~Jellinek and R.~S. Berry, {\em Generalization of Nos\'e's isothermal
  molecular dynamics}.
\newblock Phys. Rev. A {\bf 38}, 3069--3072 (1988).

\bibitem[Jel89]{JeBe89}
J.~Jellinek and R.~S. Berry, {\em Generalization of Nos\'e's isothermal
  molecular dynamics: necessary and sufficient conditions of dynamical
  simulations of statistical ensembles}.
\newblock Phys. Rev. A {\bf 40}, 2816--2818 (1989).

\bibitem[Jep03]{JES03}
O.~Jepps, D.J. Evans and D.J. Searles, {\em The fluctuation theorem and
  Lyapunov weights}.
\newblock in print for the Physica D Special Issue Ref.\ \cite{chaotr03}, 2003.

\bibitem[Jus01]{JKRH01}
W.~Just, H.~Kantz, C.~R\"odenbeck and M.~Helm, {\em Stochastic modelling:
  replacing fast degrees of freedom by noise}.
\newblock J. Phys. A: Math. Gen. {\bf 34}, 3199--3213 (2001).

\bibitem[Jus03]{JGBRK03}
W.~Just, K.~Gelfert, N.~Baba, A.~Riegert and H.~Kantz, {\em Elimination of fast
  chaotic degrees of freedom: on the accuracy of the {B}orn approximation}.
\newblock J. Stat. Phys. {\bf 112}, 277--292 (2003).

\bibitem[Kan03]{KJBGR03}
H.~Kantz, W.~Just, N.~Baba, K.~Gelfert and A.~Riegert, {\em Replacing fast
  chaotic degrees of freedom by noise: a formally exact result}.
\newblock in print for the Physica D Special Issue Ref.\ \cite{chaotr03}, 2003.

\bibitem[Kar00]{Kark00}
J.~Karkheck, Eds.
\newblock {\em Dynamics: Models and Kinetic Methods for Non-equilibrium Many
  Body Systems}, volume 371 of NATO Science Series E: Applied Sciences, Kluwer,
  Dordrecht, 2000.

\bibitem[Kla]{Kla03}
R.~Klages.
\newblock unpublished.

\bibitem[Kla95]{RKD}
R.~Klages and {J.R.} Dorfman, {\em Simple maps with fractal diffusion
  coefficients}.
\newblock Phys. Rev. Lett. {\bf 74}, 387--390 (1995).

\bibitem[Kla96]{RKdiss}
R.~Klages, {\em Deterministic diffusion in one-dimensional chaotic dynamical
  systems}.
\newblock (Wissenschaft \& Technik-Verlag, Berlin, 1996).

\bibitem[Kla99a]{KlDo99}
R.~Klages and {J.R.} Dorfman, {\em Simple deterministic dynamical systems with
  fractal diffusion coefficients}.
\newblock Phys. Rev. E {\bf 59}, 5361--5383 (1999).

\bibitem[Kla99b]{KR99}
R.~Klages and K.~Rateitschak.
\newblock unpublished, 1999.

\bibitem[Kla00a]{KlDe00}
R.~Klages and C.~Dellago, {\em Density-dependent diffusion in the periodic
  {L}orentz gas.}
\newblock J. Stat. Phys. {\bf 101}, 145--159 (2000).

\bibitem[Kla00b]{KRN00}
R.~Klages, K.~Rateitschak and G.~Nicolis, {\em Thermostating by deterministic
  scattering: construction of nonequilibrium steady states.}
\newblock Phys. Rev. Lett. {\bf 84}, 4268--4271 (2000).

\bibitem[Kla02a]{RKla01a}
R.~Klages, {\em Suppression and enhancement of deterministic diffusion in
  disordered dynamical systems}.
\newblock Phys. Rev. E {\bf 65}, 055203(R)/1--4 (2002).

\bibitem[Kla02b]{RKla01b}
R.~Klages, {\em Transitions from deterministic to stochastic diffusion}.
\newblock Europhys. Lett. {\bf 57}, 796--802 (2002).

\bibitem[Kla02c]{KlKo02}
R.~Klages and N.~Korabel, {\em Understanding deterministic diffusion by
  correlated random walks}.
\newblock J. Phys. A: Math. Gen. {\bf 35}, 4823--4836 (2002).

\bibitem[Kla03]{chaotr03}
R.~Klages, H.~van Beijeren, J.R. Dorfman and P.~Gaspard, Eds.
\newblock {\em Microscopic chaos and transport in many-particle systems},
  Physica D, North Holland, Amsterdam, 2003.
\newblock see
  http://www.mpipks-dresden.mpg.de/\~{}chao\-tran/proceedings/chaotran\_proc.html, to be published as a Special Issue in Physica D.

\bibitem[Kor02]{KoKl02}
N.~Korabel and R.~Klages, {\em Fractal structures of normal and anomalous
  diffusion in nonlinear nonhyperbolic dynamical systems}.
\newblock Phys. Rev. Lett. {\bf 89}, 214102/1--4 (2002).

\bibitem[Kub92]{KTH92}
R.~Kubo, M.~Toda and N.~Hashitsume, {\em Statistical Physics}, volume~2 of
  Solid State Sciences.
\newblock (Springer, Berlin, 2 edition, 1992).

\bibitem[Kur98]{Kur98}
J.~Kurchan, {\em Fluctuation theorem for stochastic dynamics}.
\newblock J. Phys. A: Math. Gen. {\bf 31}, 3719--3729 (1998).

\bibitem[Kus90]{KBB90}
D.~Kusnezov, A.~Bulgac and W.~Bauer, {\em Canonical ensembles from chaos}.
\newblock Ann. Phys. {\bf 204}, 155--185 (1990).

\bibitem[Kus93]{Kus93}
D.~Kusnezov, {\em Quantum ergodic wavefunctions from a thermal non-linear
  Schr\"odinger equation}.
\newblock Phys. Lett. A {\bf 184}, 50--56 (1993).

\bibitem[Kus95]{Kus95}
D.~Kusnezov, {\em Dimensional loss in nonequilibrium quantum systems}.
\newblock Phys. Rev. Lett. {\bf 74}, 246--249 (1995).

\bibitem[Kus02]{KLA02}
D.~Kusnezov, E.~Lutz and K.~Aoki, {\em Non-equilibrium statistical mechanics of
  classical and quantum systems}.
\newblock in: Dynamics of dissipation, pages 83--108. (2002).
\newblock see Ref.\ \cite{GaOl02}.

\bibitem[Lad85]{LaHo83}
A.J.C. Ladd and W.G. Hoover, {\em Lorentz gas shear viscosity via
  nonequilibrium molecular dynamics and Boltzmann's equation}.
\newblock J. Stat. Phys. {\bf 38}, 973--988 (1985).

\bibitem[Lai03]{LaLe03}
B.B. Laird and B.J. Leimkuhler, {\em Generalized dynamical thermostating
  technique}.
\newblock Phys. Rev. E {\bf 68}, 016704/1--6 (2003).

\bibitem[Lan08]{Lang08}
P.~Langevin, {\em Sur la th\'eorie du mouvement brownien}.
\newblock C.R. Acad. Sci. (Paris) {\bf 146}, 530--533 (1908).
\newblock see also the English translation in Am. J. Phys. {\bf 65}, 1079
  (1997).

\bibitem[Lar03]{LLM02}
H.~Larralde, F.~Leyvraz and C.~Mejia-Monasterio, {\em Transport properties of a
  modified Lorentz gas}.
\newblock J. Stat. Phys. {\bf 113}, 197--231 (2003).

\bibitem[Lat97]{LvBD97}
A.~Latz, H.~van {B}eijeren and J.R. Dorfman, {\em Lyapunov spectrum and the
  conjugate pairing rule for a thermostatted random {L}orentz gas: Kinetic
  theory}.
\newblock Phys. Rev. Lett. {\bf 78}, 207--210 (1997).

\bibitem[Leb78]{LeSp78}
J.L. Lebowitz and H.~Spohn, {\em Transport properties of the {L}orentz gas:
  {F}ourier's law}.
\newblock J. Stat. Phys. {\bf 19}, 633--654 (1978).

\bibitem[Leb99]{LS99}
J.L. Lebowitz and H.~Spohn, {\em A {G}allavotti-{C}ohen-type symmetry in the
  large deviation functional for stochastic dynamics}.
\newblock J. Stat. Phys. {\bf 95}, 333--365 (1999).

\bibitem[Lep00]{LRB00}
S.~Lepri, L.~Rondoni and G.~Benettin, {\em The {G}allavotti-{C}ohen fluctuation
  theorem for a nonchaotic model.}
\newblock J. Stat. Phys. {\bf 99}, 857--872 (2000).

\bibitem[Lep03]{LLP03}
S.~Lepri, S.~Livi and A.~Politi, {\em Thermal conduction in classical
  low-dimensional lattices}.
\newblock Phys. Rep. {\bf 377}, 1--80 (2003).

\bibitem[Lev89]{Lev}
R.W. Leven, B.-P. Koch and B.~Pompe, {\em Chaos in dissipativen {S}ystemen}.
\newblock (Vieweg, Braunschweig, 1989).

\bibitem[L'H93]{LHHa93}
I.~L'Heureux and I.~Hamilton, {\em Canonically modified Nos\'e-Hoover equation
  with explicit inclusion of the virial}.
\newblock Phys. Rev. E {\bf 47}, 1411--1414 (1993).

\bibitem[Lie92]{LBC92}
S.Y. Liem, D.~Brown and J.H.R. Clarke, {\em Investigation of the
  homogeneous-shear nonequilibrium-molecular-dynamics method}.
\newblock Phys. Rev. A {\bf 45}, 3706--3713 (1992).

\bibitem[Lie99a]{Lieb99}
E.~Lieb, {\em Some problems in statistical mechanics that I would like to see
  solved}.
\newblock Physica A {\bf 263}, 491--499 (1999).

\bibitem[Lie99b]{LiYn99}
E.H. Lieb and J.~Yngvason, {\em The physics and mathematics of the second law
  of thermodynamics}.
\newblock Phys. Rep. {\bf 310}, 1--96 (1999).

\bibitem[Lio38]{Liou38}
J.~Liouville, {\em Sur la th\'eorie de la variation des constantes
  arbitraires}.
\newblock J. Math. Pures Appl. {\bf 3}, 342--349 (1838).

\bibitem[Liu00]{LiTu00}
Y.~Liu and M.E. Tuckerman, {\em Generalized Gaussian moment thermostatting: a
  new continuous dynamical approach to the canonical ensemble}.
\newblock J. Chem. Phys. {\bf 112}, 1685--1700 (2000).

\bibitem[Llo94]{LRM94}
J.~Lloyd, L.~Rondoni and {G.P.} Morriss, {\em Breakdown of ergodic behavior in
  the {L}orentz gas}.
\newblock Phys. Rev. E {\bf 50}, 3416--3421 (1994).

\bibitem[Llo95]{LNRM95}
J.~Lloyd, M.~Niemeyer, L.~Rondoni and {G.P.} Morriss, {\em The nonequilibrium
  {L}orentz gas}.
\newblock Chaos {\bf 5}, 536--551 (1995).

\bibitem[Loo90]{LoHe90}
W.~Loose and S.~Hess, {\em Shear-induced ordering revisited}.
\newblock in: M.~Mareschal, Eds., Microscopic Simulations of Complex Flows,
  volume 236 of NATO Asi Series B, Plenum Press, New York, 1990.

\bibitem[Lor05]{Lo05}
H.A. Lorentz, {\em The motion of electrons in metallic bodies}.
\newblock Proc. Roy. Acad. Amst. {\bf 7}, 438--453 (1905).

\bibitem[Lor91]{LKP91}
A.~Lorke, J.P. Kotthaus and K.~Ploog, {\em Magnetotransport in two-dimensional
  superlattices}.
\newblock Phys. Rev. B {\bf 44}, 3447--3450 (1991).

\bibitem[Lue93]{LuBr93}
A.~Lue and H.~Brenner, {\em Phase flow and statistical structure of
  {G}alton-board systems}.
\newblock Phys. Rev. E {\bf 47}, 3128--3144 (1993).

\bibitem[Mac83]{MaZw83}
J.~Machta and R.~Zwanzig, {\em Diffusion in a periodic {L}orentz gas}.
\newblock Phys. Rev. Lett. {\bf 50}, 1959--1962 (1983).

\bibitem[Mae99]{Mae99}
C.~Maes, {\em The fluctuation theorem as a {G}ibbs property.}
\newblock J. Stat. Phys. {\bf 95}, 367--392 (1999).

\bibitem[Mae03a]{MaNe03}
C.~Maes and K.~Netocny, {\em Time-reversal and entropy}.
\newblock J. Stat. Phys. {\bf 110}, 269--310 (2003).

\bibitem[Mae03b]{MNV03}
C.~Maes, K.~Netocny and M.~Verschuere, {\em Heat conduction networks}.
\newblock J. Stat. Phys. {\bf 111}, 1219--1244 (2003).

\bibitem[Mar92a]{MaHo92}
M.~Mareschal and B.L. Holian, Eds.
\newblock {\em Microscopic simulations of complex hydrodynamic phenomena},
  volume 292 of NATO ASI Series B: Physics, Plenum Press, New York, 1992.

\bibitem[Mar92b]{MKT92}
G.J. Martyna, M.L. Klein and M.~Tuckerman, {\em Nos\'e-Hoover chains: the
  canonical ensemble via continuos dynamics}.
\newblock J. Chem. Phys. {\bf 97}, 2635--2643 (1992).

\bibitem[Mar97]{Mare97}
M.~Mareschal, Eds.
\newblock {\em The microscopic approach to complexity in non-equilibrium
  molecular simulations}, volume 240 of Physica A, North Holland, Amsterdam,
  1997.

\bibitem[Mar03]{MaNa03}
M.~Mareschal and S.~McNamara, {\em Lyapunov hydrodynamics in the dilute limit}.
\newblock in print for the Physica D Special Issue Ref.\ \cite{chaotr03}, 2003.

\bibitem[Mat97]{MaMa97}
H.~Matsuoka and R.F. Martin, {\em Long-time tails of the velocity
  autocorrelation functions for the triangular periodic Lorentz gas}.
\newblock J. Stat. Phys {\bf 88}, 81--103 (1997).

\bibitem[Mat01]{MTV01}
L.~Matyas, T.~Tel and J.~Vollmer, {\em Multibaker map for shear flow and
  viscous heating}.
\newblock Phys. Rev. E {\bf 64}, 056106/1--11 (2001).

\bibitem[Max79]{Ma1879}
J.C. Maxwell, {\em On Boltzmann's Theorem on the average distribution in a
  system of material points}.
\newblock Cam. Phil. Trans. {\bf 12}, 547 (1879).

\bibitem[McN01]{NaMa01}
S.~McNamara and M.~Mareschal, {\em Origin of the hydrodynamic Lyapunov modes}.
\newblock Phys. Rev. E {\bf 64}, 051103/1--14 (2001).

\bibitem[Mei92]{Meis92}
J.D. Meiss, {\em Symplectic maps, variational principles, and transport}.
\newblock Rev. Mod. Phys. {\bf 64}, 795--848 (1992).

\bibitem[Men01]{MS01}
D.~Mentrup and J.~Schnack, {\em Nos\'e-Hoover dynamics for coherent states}.
\newblock Physica A {\bf 297}, 337--347 (2001).

\bibitem[Men03]{MS03}
D.~Mentrup and J.~Schnack, {\em Nos\'e-Hoover sampling of quantum entangled
  distribution functions}.
\newblock Physica A {\bf 326}, 370--383 (2003).

\bibitem[Met00]{MeKl00}
R.~Metzler and J.~Klafter, {\em The random walk's guide to anomalous diffusion:
  a fractional dynamics approach}.
\newblock Phys. Rep. {\bf 339}, 1--77 (2000).

\bibitem[Mil98a]{MPH98}
L.~Milanovic, H.A. Posch and W.G. Hoover, {\em Lyapunov instability of
  two-dimensional fluids: Hard dumbbells}.
\newblock Chaos {\bf 8}, 455--461 (1998).

\bibitem[Mil98b]{MPT98}
L.J. Milanovi\'c, H.A. Posch and W.~Thirring, {\em Statistical mechanics and
  computer simulation of systems with attractive positive power-law
  potentials}.
\newblock Phys. Rev. E {\bf 57}, 2763--2775 (1998).

\bibitem[Mil02]{MP02}
L.~Milanovic and H.A. Posch, {\em Localized and delocalized modes in the
  tangent-space dynamics of planar hard dumbbell fluids}.
\newblock J. Molec. Liqu. {\bf 96-97}, 221--244 (2002).

\bibitem[MM01]{MLL01}
C.~Mejia-Monasterio, H.~Larralde and F.~Leyvraz, {\em Coupled normal heat and
  matter transport in a simple model system}.
\newblock Phys. Rev. Lett. {\bf 86}, 5417--5420 (2001).

\bibitem[Mor87a]{MH87}
B.~Moran and W.G. Hoover, {\em Diffusion in a periodic {L}orentz gas}.
\newblock J. Stat. Phys. {\bf 48}, 709--726 (1987).

\bibitem[Mor87b]{Morr87}
{G.P.} Morriss, {\em The information dimension of the nonequilibrium
  distribution function}.
\newblock Phys. Lett. A {\bf 122}, 236--240 (1987).

\bibitem[Mor89a]{Morr89}
{G.P.} Morriss, {\em Dimensional contraction in nonequilibrium systems}.
\newblock Phys. Lett. A {\bf 134}, 307--313 (1989).

\bibitem[Mor89b]{Mo89a}
{G.P.} Morriss, {\em Phase-space singularities in a planar Couette flow}.
\newblock Phys. Rev. A {\bf 39}, 4811--4816 (1989).

\bibitem[Mor89c]{MECB89}
G.P. Morriss, D.J. Evans, E.G.D. Cohen and H.~van Beijeren, {\em Linear
  response of phase-space trajectories to shearing}.
\newblock Phys. Rev. Lett. {\bf 14}, 1579--1582 (1989).

\bibitem[Mor94]{MoRo94}
{G.P.} Morriss and L.~Rondoni, {\em Periodic orbit expansions for the {L}orentz
  gas}.
\newblock J. Stat. Phys. {\bf 75}, 553--584 (1994).

\bibitem[Mor96a]{MDI96}
{G.P.} Morriss, {C.P.} Dettmann and {D.J.} Isbister, {\em Field dependence of
  Lyapunov exponents for nonequilibrium systems}.
\newblock Phys. Rev. E {\bf 54}, 4748--4754 (1996).

\bibitem[Mor96b]{MoRo96}
{G.P.} Morriss and L.~Rondoni, {\em Equivalence of ''nonequilibrium'' ensembles
  for simple maps}.
\newblock Physica A {\bf 233}, 767--784 (1996).

\bibitem[Mor98]{MoDe98}
{G.P.} Morriss and {C.P.} Dettmann, {\em Thermostats: Analysis and
  application}.
\newblock Chaos {\bf 8}, 321--336 (1998).

\bibitem[Mor99]{MR99}
G.P. Morriss and L.~Rondoni, {\em Definition of temperature in equilibrium and
  nonequilibrium systems}.
\newblock Phys. Rev. E {\bf 59}, R5--R8 (1999).

\bibitem[Mor02]{Mor02}
G.P. Morriss, {\em Conjugate pairing of {L}yapunov exponents for isokinetic
  shear flow algorithms.}
\newblock Phys. Rev. E {\bf 65}, 017201/1--3 (2002).

\bibitem[MT99]{TMM99}
C.J.~Mundy M.E.~Tuckerman and G.J. Martyna, {\em On the classical statistical
  mechanics of non-Hamiltonian systems}.
\newblock Europhys. Lett. {\bf 45}, 149--155 (1999).

\bibitem[Mun00]{Mund00}
C.J. Mundy, S.~Balasubramanian, K.~Bagchi, M.E. Tuckerman, G.J. Martyna and
  M.L. Klein, {\em Nonequilibrium molecular dynamics}, volume~14 of Reviews in
  Computational Chemistry, chapter~5.
\newblock (Wiley-VCH, New York, 2000).

\bibitem[Nic96]{NiDa96}
G.~Nicolis and D.~Daems, {\em Nonequilibrium thermodynamics of dynamical
  systems}.
\newblock J. Phys. Chem. {\bf 100}, 19187--19191 (1996).

\bibitem[Nic98]{NiDa98}
G.~Nicolis and D.~Daems, {\em Probabilistic and thermodynamic aspects of
  dynamical systems}.
\newblock Chaos {\bf 8}, 311--320 (1998).

\bibitem[Nos84a]{Nose84a}
S.~Nos\'e, {\em A molecular dynamics method for simulations in the canonical
  ensemble}.
\newblock Mol. Phys. {\bf 52}, 255--268 (1984).

\bibitem[Nos84b]{Nose84b}
S.~Nos\'e, {\em A unified formulation of the constant temperature molecular
  dynamics methods}.
\newblock J. Chem. Phys. {\bf 81}, 511--519 (1984).

\bibitem[Nos91]{Nose91}
S.~Nos\'e, {\em Molecular dynamics simulations at constant temperature and
  pressure}.
\newblock in: M.~Meyer and V.~Pontikis, Eds., Computer Simulation in material
  science, pages 21--41. (Kluwer Academic Publishers, Netherlands, 1991).

\bibitem[Nos93]{Nose93}
S.~Nos\'e, {\em Dynamical behavior of a thermostated isotropic harmonic
  oscillator}.
\newblock Phys. Rev. E {\bf 47}, 164--177 (1993).

\bibitem[Ott93]{Ott}
E.~Ott, {\em Chaos in Dynamical Systems}.
\newblock (Cambridge University Press, Cambridge, 1993).

\bibitem[Pan02a]{Pan02}
D.~Panja, {\em An elementary proof of {L}yapunov exponent pairing for
  hard-sphere systems at constant kinetic energy.}
\newblock J. Stat. Phys. {\bf 109}, 705--727 (2002).

\bibitem[Pan02b]{PvZ02b}
D.~Panja and R.~van Zon, {\em Lyapunov exponent pairing for a thermostatted
  hard-sphere gas under shear in the thermodynamic limit.}
\newblock Phys. Rev. E {\bf 65}, 060102(R)/1--4 (2002).

\bibitem[Pan02c]{PvZ02}
D.~Panja and R.~van Zon, {\em Pairing of {L}yapunov exponents for a hard-sphere
  gas under shear in the thermodynamic limit.}
\newblock Phys. Rev. E {\bf 66}, 021101/1--12 (2002).

\bibitem[Pat88]{Path88}
R.K. Pathria, {\em Statistical mechanics}, volume~45 of International series in
  natural philosophy.
\newblock (Pergamon press, Oxford, 1988).

\bibitem[Pen79]{Penr79}
O.~Penrose, {\em Foundations of statistical mechanics}.
\newblock Rep. Prog. Phys. {\bf 42}, 1937--2006 (1979).

\bibitem[Pet94]{PIM94}
J.~Petravic, D.J. Isbister and G.P. Morriss, {\em Correlation dimension of the
  sheared hard-disk Lorentz gas}.
\newblock J. Stat. Phys. {\bf 76}, 1045--1063 (1994).

\bibitem[Pos86]{PHV86}
H.A. Posch, W.G. Hoover and F.J. Vesely, {\em Canonical dynamics of the
  {N}os\'e oscillator: stability, order, and chaos}.
\newblock Phys. Rev. A {\bf 33}, 4253--4265 (1986).

\bibitem[Pos87]{PoHo87}
H.A. Posch and W.G. Hoover, {\em Direct measurement of equilibrium and
  nonequilibrium {L}yapunov spectra}.
\newblock Phys. Lett. A {\bf 123}, 227--230 (1987).

\bibitem[Pos88]{PoHo88}
H.A. Posch and W.G. Hoover, {\em Lyapunov instability of dense
  {L}ennard-{J}ones fluids}.
\newblock Phys. Rev. A {\bf 38}, 473--482 (1988).

\bibitem[Pos89]{PoHo89}
H.A. Posch and W.G. Hoover, {\em Equilibrium and nonequilibrium {L}yapunov
  spectra for dense fluids and solids}.
\newblock Phys. Rev. A {\bf 39}, 2175--2188 (1989).

\bibitem[Pos97]{PH97}
H.A. Posch and W.G. Hoover, {\em Time-reversible dissipative attractors in
  three and four phase-space dimensions.}
\newblock Phys. Rev. E {\bf 55}, 6803--6810 (1997).

\bibitem[Pos98]{PH98}
H.A. Posch and W.G. Hoover, {\em Heat conduction in one-dimensional chains and
  nonequilibrium {L}yapunov spectrum.}
\newblock Phys. Rev. E {\bf 58}, 4344--4350 (1998).

\bibitem[Pos00a]{PoHiS00}
H.A. Posch and R.~Hirschl, {\em Simulation of billiards and of hard body
  fluids}.
\newblock in: Hard Ball Systems and the Lorentz Gas, pages 279--314. 2000).
\newblock see Ref.\ \cite{Sza00}.

\bibitem[Pos00b]{PHH00}
H.A. Posch, R.~Hirschl and W.G. Hoover, {\em Multifractal phase-space
  distributions for stationary nonequilibrium systems}.
\newblock in: J.~Karkheck, Eds., Dynamics: Models and Kinetic Methods for
  Non-equilibrium Many Body Systems, volume 371 of NATO Science Series E:
  Applied Sciences, pages 169--189, Kluwer, Dordrecht, 2000.

\bibitem[Pos03]{PH03}
H.A. Posch and W.G. Hoover, {\em Large-system phase-space dimensionality loss
  in stationary heat flows}.
\newblock in print for the Physica D Special Issue Ref.\ \cite{chaotr03}, 2003.

\bibitem[Pre92]{PFTV92}
W.H. Press, B.P. Flanery, S.A. Teukolsky and W.T. Vetterling, {\em Numerical
  recipes in FORTRAN}.
\newblock (Cambridge University Press, 2nd edition edition, 1992).

\bibitem[Rad96]{Rado96}
G.~Radons, {\em Suppression of chaotic diffusion by quenched disorder}.
\newblock Phys. Rev. Lett. {\bf 77}, 4748--4751 (1996).

\bibitem[Ram86]{Rams86}
J.D. Ramshaw, {\em Remarks on entropy and irreversibility in non-Hamiltonian
  systems}.
\newblock Phys. Lett. A {\bf 116}, 110--114 (1986).

\bibitem[Ram02]{Rams02}
J.D. Ramshaw, {\em Remarks on non-Hamiltonian statistical mechanics}.
\newblock Europhys. Lett. {\bf 59}, 319--323 (2002).

\bibitem[Rat00a]{RKH00}
K.~Rateitschak, R.~Klages and WG. Hoover, {\em The {N}ose-{H}oover thermostated
  {L}orentz gas.}
\newblock J. Stat. Phys. {\bf 101}, 61--77 (2000).
\newblock see also chao-dyn/9912018 for further details.

\bibitem[Rat00b]{RKN00}
K.~Rateitschak, R.~Klages and G.~Nicolis, {\em Thermostating by deterministic
  scattering: the periodic {L}orentz gas.}
\newblock J. Stat. Phys. {\bf 99}, 1339--1364 (2000).

\bibitem[Rat02]{RaKl02}
K.~Rateitschak and R.~Klages, {\em Lyapunov instability for a periodic Lorentz
  gas thermostated by deterministic scattering}.
\newblock Phys. Rev. E {\bf 65}, 036209/1--11 (2002).

\bibitem[Rei65]{Reif}
F.~Reif, {\em Fundamentals of statistical and thermal physics}.
\newblock (McGraw-Hill, Auckland, 1965).
\newblock see also the amended German translation: Statistische Physik und
  Theorie der W\"{a}rme, 3rd edition, de Gruyter, Berlin, 1987.

\bibitem[Rei98]{Reim98}
P.~Reimann, {\em Comment on "{T}oward a statistical thermodynamics of steady
  states"}.
\newblock Phys. Rev. Lett. {\bf 80}, 4104 (1998).

\bibitem[Rie00]{RUGOS00}
J.P. Rieu, A.~Upadhyaya, J.A. Glazier, N.B. Ouchi and Y.~Sawada, {\em Diffusion
  and deformations of single Hydra cells in cellular aggregates}.
\newblock Biophys. J. {\bf 79}, 1903--1914 (2000).

\bibitem[Ris96]{RC96}
D.~Risso and P.~Cordero, {\em Two-dimensional gas of disks: thermal
  conductivity}.
\newblock J. Stat. Phys {\bf 82}, 1453--1466 (1996).

\bibitem[Rob92]{Robe}
J.A.G. Roberts and G.R.W. Quispel, {\em Chaos and time-reversal symmetry: order
  and chaos in reversible dynamical systems}.
\newblock Phys. Rep. {\bf 216}, 63--177 (1992).

\bibitem[Ron00a]{RC00}
L.~Rondoni and E.G.D. Cohen, {\em Gibbs entropy and irreversible
  thermodynamics.}
\newblock Nonlinearity {\bf 13}, 1905--1924 (2000).

\bibitem[Ron00b]{RTV00}
L.~Rondoni, T.~T\'el and J.~Vollmer, {\em Fluctuation theorems for entropy
  production in open systems}.
\newblock Phys. Rev. E {\bf 61}, R4679--R4682 (2000).

\bibitem[Ron02a]{Rond02}
L.~Rondoni, {\em Deterministic thermostats and fluctuation relations}.
\newblock in: Dynamics of dissipation, pages 35--61. (2002).
\newblock see Ref.\ \cite{GaOl02}.

\bibitem[Ron02b]{RC02}
L.~Rondoni and E.G.D. Cohen, {\em On some derivations of irreversible
  thermodynamics from dynamical systems theory.}
\newblock Physica D {\bf 168-169}, 341--55 (2002).

\bibitem[Rue71]{RT71}
D.~Ruelle and F.~Takens, {\em On the nature of turbulence}.
\newblock Commun. Math. Phys. {\bf 20}, 167--192 (1971).

\bibitem[Rue96]{Rue96}
D.~Ruelle, {\em Positivity of entropy production in nonequilibrium statistical
  mechanics}.
\newblock J. Stat. Phys. {\bf 85}, 1--23 (1996).

\bibitem[Rue97a]{Rue97b}
D.~Ruelle, {\em Entropy production in nonequilibrium statistical mechanics}.
\newblock Commun. Math. Phys. {\bf 189}, 365--371 (1997).

\bibitem[Rue97b]{Rue97}
D.~Ruelle, {\em Positivity of entropy production in the presence of a random
  thermostat}.
\newblock J. Stat. Phys. {\bf 86}, 935--951 (1997).

\bibitem[Rue99a]{Rue99b}
D.~Ruelle, {\em Gaps and new ideas in our understanding of nonequilibrium}.
\newblock Physica A {\bf 263}, 540--544 (1999).

\bibitem[Rue99b]{Rue99}
D.~Ruelle, {\em Smooth dynamics and new theoretical ideas in nonequilibrium
  statistical mechanics}.
\newblock J. Stat. Phys. {\bf 95}, 393--468 (1999).

\bibitem[Rue03]{Rue03}
D.~Ruelle, {\em Extending the definition of entropy to nonequilibrium steady
  states}.
\newblock Proc. Natl. Acad. Sci. {\bf 100}, 3054--3058 (2003).

\bibitem[Rug97]{Rugh97}
H.H. Rugh, {\em Dynamical approach to temperature}.
\newblock Phys. Rev. Lett. {\bf 78}, 772--775 (1997).

\bibitem[San92]{SGB92}
A.~Santos, V.~Garz\'o and J.J. Brey, {\em Comparison between the
  homogeneous-shear and the sliding-boundary methods to produce shear flow}.
\newblock Phys. Rev. A {\bf 46}, 8018--8020 (1992).

\bibitem[Sar92]{SEM92}
S.S. Sarman, D.J. Evans and G.P. Morriss, {\em Conjugate-pairing rule and
  thermal-transport coefficients}.
\newblock Phys. Rev. E {\bf 45}, 2233--2242 (1992).

\bibitem[Sar98]{SEC98}
S.S. Sarman, D.J. Evans and P.T. Cummings, {\em Recent developments in
  non-Newtonian molecular dynamics}.
\newblock Phys. Rep. {\bf 305}, 1--92 (1998).

\bibitem[Sch78]{SchSt78}
T.~Schneider and E.~Stoll, {\em Molecular-dynamics study of a three-dimensional
  one-component model for distortive phase transitions}.
\newblock Phys. Rev. B {\bf 17}, 1302--1322 (1978).

\bibitem[Sch89]{Schu}
H.G. Schuster, {\em Deterministic Chaos}.
\newblock (VCH Verlagsgesellschaft mbH, Weinheim, 2nd edition, 1989).

\bibitem[Sch93]{SchGr93}
M.~Schienbein and H.~Gruler, {\em Langevin equation, Fokker-Planck equation and
  cell migration}.
\newblock Bull. Math. Biol. {\bf 55}, 585--608 (1993).

\bibitem[Sch94]{SSG94}
F.~Schweitzer and L.~Schimansky-Geier, {\em Clustering of ``active'' walkers in
  a two-component system}.
\newblock Physica A {\bf 206}, 359--379 (1994).

\bibitem[Sch98]{SET98}
F.~Schweitzer, W.~Ebeling and B.~Tilch, {\em Complex motion of {B}rownian
  particles with energy depots}.
\newblock Phys. Rev. Lett. {\bf 80}, 5044--5047 (1998).

\bibitem[Sch01]{SET01}
F.~Schweitzer, W.~Ebeling and B.~Tilch, {\em Statistical mechanics of
  canonical-dissipative systems and applications to swarm dynamics}.
\newblock Phys. Rev. E {\bf 64}, 021110/1--12 (2001).

\bibitem[Sch03]{SE03}
J.~Schumacher and B.~Eckhardt, {\em Fluctuations of energy injection rate in a
  shear flow}.
\newblock in print for the Physica D Special Issue Ref.\ \cite{chaotr03}, 2003.

\bibitem[Sea98]{SEI98}
{D.J.} Searles, {D.J.} Evans and {D.J.} Isbister, {\em The conjugate-pairing
  rule for non-Hamiltonian systems}.
\newblock Chaos {\bf 8}, 337--349 (1998).

\bibitem[Sei33]{Sei33}
W.~Seidel, {\em Note on a metrically transitive system}.
\newblock Proc. Nat. Acad. Sci. {\bf 19}, 453--456 (1933).

\bibitem[Ser01]{SeFe01}
A.~Sergi and M.~Ferrario, {\em Non-Hamiltonian equations of motion with a
  conserved energy}.
\newblock Phys. Rev. E {\bf 64}, 056125/1--9 (2001).

\bibitem[Ser03]{Serg03}
A.~Sergi, {\em Non-Hamiltonian equilibrium statistical mechanics}.
\newblock Phys. Rev. E {\bf 67}, 021101/1--7 (2003).

\bibitem[SG95]{SGMRM95}
L.~Schimansky-Geier, M.~Mieth, H.~Rost and H.~Malchow, {\em Structure formation
  by active Brownian particles}.
\newblock Phys. Lett. A {\bf 207}, 140--146 (1995).

\bibitem[Sin70]{Sin70}
Ya.G. Sinai, {\em Dynamical systems with elastic reflections. Ergodic
  properties of dispersing billiards}.
\newblock Russ. Math. Surv. {\bf 25}, 137--189 (1970).

\bibitem[Sma80]{Sma80}
S.~Smale, {\em The mathematics of time}.
\newblock (Springer, Berlin, 1980).

\bibitem[Sta89]{Sta89}
J.~Stachel, Eds., {\em The collected papers of Albert Einstein}, volume~2.
\newblock (Princeton University Press, Princeton, 1989).

\bibitem[Ste79]{Stee79}
W.-H. Steeb, {\em Generalized Liouville equation, entropy, and dynamic systems
  containing limit cycles}.
\newblock Physica {\bf 95A}, 181--190 (1979).

\bibitem[Ste80]{Stee80}
W.-H. Steeb, {\em A comment on the generalized Liouville equation}.
\newblock Found. Phys. {\bf 10}, 485--493 (1980).

\bibitem[Sto91]{SLW91}
C.L. Stokes and S.K.~Williams D.A.~Lauffenburger, {\em Migration of individual
  microvessel endothelial cells: stochastic model and parameter measurement}.
\newblock J. Cell Science {\bf 99}, 419--430 (1991).

\bibitem[Stu99]{SW99}
A.M. Stuart and J.O. Warren, {\em Analysis and experiments for a computational
  model of a heat bath}.
\newblock J. Stat. Phys. {\bf 97}, 687--723 (1999).

\bibitem[Sza00]{Sza00}
D.~Szasz, Eds., {\em Hard-Ball Systems and the Lorentz gas}, volume 101 of
  Encyclopedia of mathematical sciences.
\newblock (Springer, Berlin, 2000).

\bibitem[Tan02a]{TDM02}
T.~Taniguchi, C.P. Dettmann and G.P. Morriss, {\em {L}yapunov spectra of
  periodic orbits for a many-particle system}.
\newblock J. Stat. Phys. {\bf 109}, 747--764 (2002).

\bibitem[Tan02b]{TM02b}
T.~Taniguchi and G.M. Morriss, {\em Master equation approach to the conjugate
  pairing rule of Lyapunov spectra for many-particle thermostated systems}.
\newblock Phys. Rev. E {\bf 66}, 066203/1--11 (2002).

\bibitem[Tan02c]{TM02}
T.~Taniguchi and G.P. Morriss, {\em Stepwise structure of {L}yapunov spectra
  for many-particle systems using a random matrix dynamics}.
\newblock Phys. Rev. E {\bf 65}, 056202/1--15 (2002).

\bibitem[Tas95]{TG2}
S.~Tasaki and P.~Gaspard, {\em Fick's law and fractality of nonequilibrium
  stationary states in a reversible multibaker map}.
\newblock J. Stat. Phys. {\bf 81}, 935--987 (1995).

\bibitem[Tas99]{TG99}
S.~Tasaki and P.~Gaspard, {\em Thermodynamic behavior of an area-preserving
  multi-baker map}.
\newblock Theoret. Chem. Acc. {\bf 102}, 385--396 (1999).

\bibitem[Tas00]{TG00}
S.~Tasaki and P.~Gaspard, {\em Entropy production and transports in a
  conservative multibaker map with energy.}
\newblock J. Stat. Phys. {\bf 101}, 125--144 (2000).

\bibitem[Tel96]{TeVB96}
T.~Tel, J.~Vollmer and W.~Breymann, {\em Transient chaos: The origin of
  transport in driven systems}.
\newblock Europhys. Lett. {\bf 35}, 659--664 (1996).

\bibitem[Tel98]{TGN98}
T.~Tel, P.~Gaspard and G.~Nicolis, Eds.
\newblock {\em Chaos and Irreversibility}, volume~8 of Chaos, American
  Institute of Physics, College Park, 1998.

\bibitem[Tel00]{TVS00}
T.~Tel and J.~Vollmer, {\em Entropy balance, multibaker maps, and the dynamics
  of the Lorentz gas}.
\newblock in: Hard Ball Systems and the Lorentz Gas, pages 367--420. 2000).
\newblock see Ref.\ \cite{Sza00}.

\bibitem[Tel01]{TVM01}
T.~Tel, J.~Vollmer and L.~Matyas, {\em Shear flow, viscous heating, and entropy
  balance from dynamical systems}.
\newblock Europhys. Lett. {\bf 53}, 458--464 (2001).

\bibitem[Tel02]{MTV02}
T.~Tel, J.~Vollmer and L.~Matyas, {\em Comments on the paper `Particles, maps
  and irreversible thermodynamics' by E.G.D. Cohen and L.Rondoni, Physica A 306
  (2002) 117}.
\newblock Physica A {\bf 323}, 323--326 (2002).

\bibitem[Ten82]{TCG82}
A.~Tenenbaum, G.~Ciccotti and R.~Gallico, {\em Stationary nonequilibrium states
  by molecular dynamics. {F}ourier's law}.
\newblock Phys. Rev. A {\bf 25}, 2778--2787 (1982).

\bibitem[Til99]{TSE99}
B.~Tilch, F.~Schweitzer and W.~Ebeling, {\em Directed motion of {B}rownian
  particles with internal energy depot}.
\newblock Physica A {\bf 273}, 294--314 (1999).

\bibitem[Tuc97]{TMK97}
M.E. Tuckerman, C.J. Mundy and M.L. Klein, {\em {T}oward a statistical
  thermodynamics of steady states}.
\newblock Phys. Rev. Lett. {\bf 78}, 2042--2045 (1997).

\bibitem[Tuc98a]{TMBK98}
M.E. Tuckerman, C.J. Mundy, S.~Balasubramanian and M.L. Klein, {\em Response to
  ``Comment on `Modified nonequilibrium molecular dynamics for fluid flows with
  energy conservation' ''}.
\newblock J. Chem. Phys. {\bf 108}, 4353--4354 (1998).

\bibitem[Tuc98b]{TMK98}
M.E. Tuckerman, C.J. Mundy and M.L. Klein, {\em Reply to comments on "{T}oward
  a statistical thermodynamics of steady states"}.
\newblock Phys. Rev. Lett. {\bf 80}, 4105--4106 (1998).

\bibitem[Tuc00]{TuMa00}
M.E. Tuckerman and G.J. Martyna, {\em Understanding modern molecular dynamics:
  techniques and applications}.
\newblock J. Phys. Chem. B {\bf 104}, 159--178 (2000).

\bibitem[Uff01]{Uff01}
J.~Uffink, {\em Bluff your way in the Second Law of Thermodynamics}.
\newblock Stud. Hist. Philos. M. P. {\bf 32B}, 305--394 (2001).

\bibitem[Upa01]{URGS01}
A.~Upadhyaya, J.P. Rieu, J.A. Glazier and Y.~Sawada, {\em Anomalous diffusion
  and non-Gaussian velocity distributions of Hydra cells in cellular
  aggregates}.
\newblock Physica A {\bf 293}, 549--558 (2001).

\bibitem[Van92]{Vanc}
W.N. Vance, {\em Unstable periodic orbits and transport properties of
  nonequilibrium steady states}.
\newblock Phys. Rev. Lett. {\bf 69}, 1356--1359 (1992).

\bibitem[vB95]{BD95}
H.~van Beijeren and J.R. Dorfman, {\em {L}yapunov exponents and {KS} entropy
  for the {L}orentz gas at low densities}.
\newblock Phys. Rev. Lett. {\bf 74}, 4412--4415 (1995).

\bibitem[vB96]{vBDCP96}
H.~van Beijeren, J.R. Dorfman, {E.G.D.} Cohen, H.A. Posch and C.~Dellago, {\em
  Lyapunov exponents from kinetic theory for a dilute, field-driven Lorentz
  gas}.
\newblock Phys. Rev. Lett. {\bf 77}, 1974--1977 (1996).

\bibitem[vB97]{BDPD97}
H.~van Beijeren, {J.R.} Dorfman, H.A. Posch and C.~Dellago, {\em
  Kolmogorov-Sinai entropy for dilute gases in equilibrium}.
\newblock Phys. Rev. E {\bf 56}, 5272--5277 (1997).

\bibitem[vB98]{BLD98}
H.~van Beijeren, A.~Latz and {J.R.} Dorfman, {\em Chaotic properties of dilute
  two- and three-dimensional random Lorentz gases: Equilibrium systems}.
\newblock Phys. Rev. E {\bf 57}, 4077--4094 (1998).

\bibitem[vB00a]{vBD00}
H.~van Beijeren and J.R. Dorfman, {\em On thermostats and entropy production}.
\newblock Physica A {\bf 279}, 21--29 (2000).

\bibitem[vB00b]{vBLD00}
H.~van Beijeren, A.~Latz and J.R. Dorfman, {\em Chaotic properties of dilute
  two- and three-dimensional random {L}orentz gases. {II}. {O}pen systems.}
\newblock Phys. Rev. E {\bf 63}, 016312/1--14 (2000).

\bibitem[vK71]{vK71}
N.~van Kampen, {\em The case against linear response theory}.
\newblock Physica Norvegica {\bf 5}, 279--284 (1971).

\bibitem[vK92]{vK}
N.~van Kampen, {\em Stochastic processes in physics and chemistry}.
\newblock (North Holland, Amsterdam, 1992).

\bibitem[Vol97]{VTB97}
J.~Vollmer, T.~T\'el and W.~Breymann, {\em Equivalence of irreversible entropy
  production in driven systems: an elementary chaotic map approach}.
\newblock Phys. Rev. Lett. {\bf 79}, 2759--2762 (1997).

\bibitem[Vol98]{VTB98}
J.~Vollmer, T.~T\'el and W.~Breymann, {\em Entropy balance in the presence of
  drift and diffusion currents: an elementary chaotic map approach}.
\newblock Phys. Rev. E {\bf 58}, 1672--1684 (1998).

\bibitem[Vol00]{VTM00}
J.~Vollmer, T.~T\'el and L.~Matyas, {\em Modeling thermostating, entropy
  currents, and cross effects by dynamical systems}.
\newblock J. Stat. Phys. {\bf 101}, 79--105 (2000).

\bibitem[Vol02]{Voll02}
J.~Vollmer, {\em Chaos, spatial extension, transport, and non-equilibrium
  thermodynamics}.
\newblock Phys. Rep. {\bf 372}, 131--267 (2002).

\bibitem[Vol03]{VTB03}
J.~Vollmer, T.~Tel and W.~Breymann, {\em Dynamical-system models of transport:
  chaos characteristics, the macroscopic limit, and irreversibility}.
\newblock in print for the Physica D Special Issue Ref.\ \cite{chaotr03}, 2003.

\bibitem[vZ98]{vZvBD98}
R.~van Zon, R.~van Beijeren and C.~Dellago, {\em Largest {L}yapunov {E}xponent
  for many particle systems at low densities}.
\newblock Phys. Rev. Lett. {\bf 80}, 2035--2038 (1998).

\bibitem[vZ99]{vZ99}
R.~van Zon, {\em Kinetic approach to the {G}aussian thermostat in a dilute
  sheared gas in the thermodynamic limit}.
\newblock Phys. Rev. E {\bf 60}, 4158--4163 (1999).

\bibitem[Wag99]{WKN99}
C.~Wagner, R.~Klages and G.~Nicolis, {\em Thermostating by deterministic
  scattering: {H}eat and shear flow.}
\newblock Phys. Rev. E {\bf 60}, 1401--1411 (1999).

\bibitem[Wag00]{Wag00}
C.~Wagner, {\em Lyapunov instability for a hard-disk fluid in equilibrium and
  nonequilibrium thermostated by deterministic scattering}.
\newblock J. Stat. Phys. {\bf 98}, 723--742 (2000).

\bibitem[Wan66]{Wan66}
G.H. Wannier, {\em Statistical Physics}.
\newblock (Dover, New York, 1966).

\bibitem[Wan02]{WSM+02}
G.M. Wang, E.M. Sevick, E.~Mittag, D.J. Searles and D.J. Evans, {\em
  Experimental demonstration of violations of the second law of thermodynamics
  for small systems and short time scales}.
\newblock Phys. Rev. Lett. {\bf 89}, 050601/1--4 (2002).

\bibitem[Wax54]{Wax54}
N.~Wax, {\em Selected Papers on Noise and Stochastic Processes}.
\newblock (Dover, New York, 1954).

\bibitem[Wei91]{Weis91}
D.~Weiss, M.L. Roukes, A.~Menschig, P.~Grambow, K.~von Klitzing and G.~Weimann,
  {\em Electron pinball and commensurate orbits in a periodic array of
  scatterers}.
\newblock Phys. Rev. Lett. {\bf 66}, 2790--2793 (1991).

\bibitem[Wei97]{WLR97}
D.~Weiss, G.~Lutjering and K.~Richter, {\em Chaotic electron motion in
  macroscopic and mesoscopic antidot lattices}.
\newblock Chaos, Solitons and Fractals {\bf 8}, 1337--1357 (1997).

\bibitem[Win92]{Wink92}
R.G. Winkler, {\em Extended-phase-space isothermal molecular dynamics:
  canonical harmonic oscillator}.
\newblock Phys. Rev. A {\bf 45}, 2250--2255 (1992).

\bibitem[Win95]{WKR95}
R.G. Winkler, V.~Kraus and P.~Reineker, {\em Time reversible and phase-space
  conserving molecular dynamics at constant temperature}.
\newblock J. Chem. Phys. {\bf 102}, 9018--9025 (1995).

\bibitem[Woj98]{WL98}
M.P. Wojtkowski and C.~Liverani, {\em Conformally symplectic dynamics and
  symmetry of the Lyapunov spectrum}.
\newblock Commun. Math. Phys. {\bf 194}, 47--60 (1998).

\bibitem[Woj02]{WD02}
D.K. Wojcik and J.R. Dorfman, {\em Quantum multibaker maps: extreme quantum
  regime}.
\newblock Phys. Rev. E {\bf 66}, 36110/1--16 (2002).

\bibitem[Woj03]{WD03}
D.K. Wojcik and J.R. Dorfman, {\em Diffusive-ballistic crossover in 1D quantum
  walks}.
\newblock Phys. Rev. Lett. {\bf 90}, 230602/1--4 (2003).

\bibitem[Wol85]{WSSV85}
A.~Wolf, J.B. Swift, H.L. Swinney and J.A. Vastano, {\em Determinig the
  Ljapunov exponents from a time series}.
\newblock Physica D {\bf 16}, 285--317 (1985).

\bibitem[You02]{Young02}
L.-S. Young, {\em What are SRB measures, and which dynamical systems have
  them?}
\newblock J. Stat. Phys. {\bf 108}, 733--754 (2002).

\bibitem[Zha92]{ZhZh92}
K.~Zhang and K.~Zhang, {\em Mechanical models of Maxwell's demon with
  noninvariant phase volume}.
\newblock Phys. Rev. A {\bf 46}, 4598--4605 (1992).

\bibitem[Zwa73]{Zwan73}
R.~Zwanzig, {\em Nonlinear generalized Langevin equations}.
\newblock J. Stat. Phys. {\bf 9}, 215--220 (1973).

\bibitem[Zwa01]{Zwan01}
R.~Zwanzig, {\em Nonequilibrium statistical mechanics}.
\newblock (Oxford University Press, Oxford, 2001).

\end{thebibliography}
\end{document}